\documentclass[aoas]{imsart}
\RequirePackage[OT1]{fontenc}
\RequirePackage{amsthm,amsmath}
\RequirePackage{natbib}
\RequirePackage[colorlinks,citecolor=blue,urlcolor=blue]{hyperref}

\startlocaldefs
\numberwithin{equation}{section}
\theoremstyle{plain}

\endlocaldefs

\usepackage[mathscr]{eucal}

\usepackage{amsfonts,amsmath,amssymb,bm,latexsym,graphicx,mathtools}

\usepackage[mathscr]{eucal}

\usepackage{centernot}
\usepackage{lipsum}
\usepackage{dlfltxbcodetips,bbm,mathtools,subfigure,amsfonts,amsmath,amssymb,fancybox,graphicx,bm,latexsym}
\usepackage[usenames,dvipsnames]{color}

\usepackage{tikz}

\usetikzlibrary{bayesnet}
\usetikzlibrary{fit,positioning}

\newcommand{\logit}{\mbox{logit}}

\usepackage[tikz]{bclogo}

\newcommand{\bt}{\begin{bclogo}[couleur={rgb:orange,0;yellow,0;white,1},arrondi=0.1,logo=\bcplume,ombre=true]}
\newcommand{\et}{\end{bclogo}\s}
\newcommand{\btt}{\begin{box}}
\newcommand{\ett}{\end{box}}

\newcommand{\btheorem}{\begin{bclogo}[couleur={rgb:orange,0;yellow,0;white,1},arrondi=0.1,logo=\bcplume,ombre=true]{Theorem}}
\newcommand{\ettheorem}{\end{bclogo}}

\newcommand{\bst}{\begin{bclogo}[couleur={rgb:orange,1;yellow,1;white,0.5},arrondi=0.1,logo=\bcpanchant]}
\newcommand{\est}{\end{bclogo}}

\newcommand{\define}{\mathop{=}\limits^{\mbox{\tiny def}}}

\newcommand{\benum}{\begin{enumerate}}
\newcommand{\eenum}{\end{enumerate}}

\newcommand{\bq}{\begin{quote}\em}
\newcommand{\eq}{\end{quote}}
\newcommand{\bbq}{\begin{quote}\bf\em}
\newcommand{\ebq}{\end{quote}}

\newcommand{\ind}{\msim\limits^{\mbox{\tiny ind}}}
\newcommand{\iid}{\msim\limits^{\mbox{\tiny iid}}}

\newcommand{\mR}{\mathbb{R}}

\newcommand{\mbE}{\mathbb{E}}

\newcommand{\mbP}{\mathbb{P}}

\newcommand{\hide}[1]{}
\newcommand{\ba}{\begin{array}{llllllllll}}
\newcommand{\ea}{\end{array}}
\newcommand{\bea}{\begin{equation}\begin{array}{llllllllll}}
\newcommand{\eea}{\end{array}\end{equation}}
\newcommand{\be}{\begin{equation}\begin{array}{lllllllllllllllll}}
\newcommand{\beno}{\begin{equation}\begin{array}{lllllllllllll}\nonumber}
\newcommand{\ee}{\end{array}\end{equation}}
\newcommand{\bel}{\begin{equation}\begin{array}{lllllllllllll}\nonumber}
\newcommand{\eel}{\Box\end{array}\end{equation}}
\newcommand{\bi}{\begin{itemize}}
\newcommand{\ei}{\end{itemize}}
\newcommand{\ben}{\begin{enumerate}}
\newcommand{\een}{\end{enumerate}}
\newcommand{\alert}{\textcolor{black}}

\newcommand{\dsum}{\displaystyle\sum\limits}
\newcommand{\dint}{\displaystyle\int\limits}
\newcommand{\dprod}{\displaystyle\prod\limits}
\newcommand{\dd}{\mathop{\mbox{d}}\nolimits}

\newcommand{\s}{\vspace{0.25cm}}

\newcommand{\bv}{\bm{v}}

\newcommand{\mT}{\mathscr{T}}

\newcommand{\mY}{\mathscr{Y}}

\newcommand{\mB}{\mathbb{B}}

\newcommand{\bY}{\bm{Y}}

\newcommand{\by}{\bm{y}}

\newcommand{\bbeta}{\bm{\beta}}

\newcommand{\bmu}{\mbox{\boldmath$\mu$}}

\newcommand{\bTheta}{\mbox{\boldmath$\Theta$}}
\newcommand{\bta}{\mbox{\boldmath$\eta$}}
\newcommand{\btheta}{\boldsymbol{\theta}}

\newcommand{\msim}{\mathop{\rm \sim}}

\newcounter{comment}
\newenvironment{comment}[1][]{\refstepcounter{comment}\vspace{0.25cm}\par\medskip\noindent%
\textbf{Comment~\thecomment #1}:\vspace{0.25cm}\\ \rmfamily}{\medskip}
\newcommand{\bc}{\begin{comment}\em}
\newcommand{\ec}{\end{comment}}

\newcounter{ex}

\newcounter{counterexample}

\newcounter{definition}

\newcounter{theorem}

\newcounter{proposition}

\newcounter{result}

\newcounter{ttproof}
\newcounter{pproof}
\newcounter{corollary}

\newcounter{ccproof}
\newcounter{llproof}
\newcounter{lemma}

\newcounter{example}

\newcounter{com}

\newcounter{lproof}
\newcounter{assumption}

\newcommand{\yy}{\by}

\renewcommand{\mB}{\mathscr{B}}

\newcommand{\mw}{\bm{a}}

\newcommand{\la}{\mbox{\boldmath $\lambda$}}

\def\th{\mbox{\boldmath $\theta$}}

\def\Pr{\mathbb{P}}

\def\logit{\mbox{\rm logit}}

\makeatletter

\usepackage[outerbars,color]{changebar}
\ifx\pdfoutput\undefined
\else\ifnum\pdfoutput>0
  \usepackage{pdfcolmk}
\fi\fi
\cbcolor{black}

\usepackage{tikz} 
\usetikzlibrary{positioning,arrows.meta,quotes}
\usetikzlibrary{bayesnet}
\usetikzlibrary{fit,positioning}

\begin{document}

\begin{frontmatter}

\title{A latent process model for monitoring progress towards hard-to-measure targets,
with applications to mental health and online educational assessments}

\runtitle{A latent process model for monitoring progress}

\thankstext{T1}{Equal contributions.}

\begin{aug}
\author{\fnms{Minjeong} \snm{Jeon}\ead[label=e1]{mjjeon@g.ucla.edu}}
\and
\author{\fnms{Michael} \snm{Schweinberger}\ead[label=e2]{mus47@psu.edu}}

\runauthor{M.\ Jeon et al.}

\affiliation{University of California, Los Angeles and Penn State University}

\address{Minjeong Jeon\\
Department of Education\\
University of California, Los Angeles\\
Moore Hall 3141\\
405 Hilgard Avenue\\
Los Angeles, CA 90095-1521\\
E-mail:\ mjjeon@g.ucla.edu\\
\mbox{}\\
Michael Schweinberger\\
Department of Statistics\\
Penn State University\\
326 Thomas Building\\
University Park, PA 16802\\
E-mail:\ mus47@psu.edu}

\end{aug}

\begin{abstract}
The recent shift to remote learning and work has aggravated long-standing problems, such as the problem of monitoring the mental health of individuals and the progress of students towards learning targets. We introduce a novel latent process model with a view to monitoring the progress of individuals towards a hard-to-measure target of interest, measured by a set of variables. The latent process model is based on the idea of embedding both individuals and variables measuring progress towards the target of interest in a shared metric space, interpreted as an interaction map that captures interactions between individuals and variables. The fact that individuals are embedded in the same metric space as the target helps assess the progress of individuals towards the target. \alert{We demonstrate, with the help of simulations and applications, that the latent process model enables a novel look at mental health and online educational assessments in disadvantaged subpopulations.}
\end{abstract}



\end{frontmatter}

\section{Introduction}
\label{sec:intro}

The recent shift to remote learning and work has aggravated long-standing problems,
such as the problem of monitoring the mental health of individuals \citep[e.g.,][]{daly:20, holmes:20} and the progress of students towards learning targets \citep[e.g.,][]{engzell:21, kuhfeld:20, bansak:20}.

We introduce a novel approach to monitoring the progress of individuals towards a hard-to-measure target of interest.
Examples are measuring the progress of individuals with mental health problems or the progress of students towards learning targets.
Both examples have in common that there is a target of interest (e.g., improving mental health or the understanding of mathematical concepts) and measuring progress towards the target is more challenging than measuring changes in physical quantities (e.g., temperature) or medical conditions (e.g., cholesterol levels),
but a set of variables is available for measuring progress towards the target. 
If, 
e.g., 
the goal is to monitor the progress of students towards learning targets,
measurements can be collected by paper-and-pencil or computer-assisted educational assessments,
whereas progress in terms of mental health can be monitored by collecting data on mental well-being by using surveys along with physical measurements related to stress by using wearable devices.

We propose a novel latent process model with a view to monitoring the progress of individuals towards a target of interest,
measured by a set of variables. 
The latent process model is based on the idea of embedding both individuals and variables measuring progress towards the target of interest in a shared metric space and can be considered as a longitudinal extension of the \citet{jeon:21} model.
\alert{The fact that individuals are embedded in the same metric space as the target helps capture
\bi
\item interactions between individuals and variables arising from unobserved variables,
such as cultural background, 
upbringing, 
and mentoring of students,
which may affect responses;\vspace{.1cm}
\item whether individuals make progress towards the target;\vspace{.1cm}
\item how much progress individuals make;\vspace{.1cm}
\item whether individuals make more progress during some periods than others;\vspace{.1cm}
\item how much more progress individuals can make in the future.
\ei
We first demonstrate that the latent process model enables a novel look at mental health and online educational assessments in disadvantaged subpopulations.
}

\subsection{Motivating example}
\label{example}

\begin{figure}[htbp]
    \centering
    \includegraphics[width=.45 \textwidth]{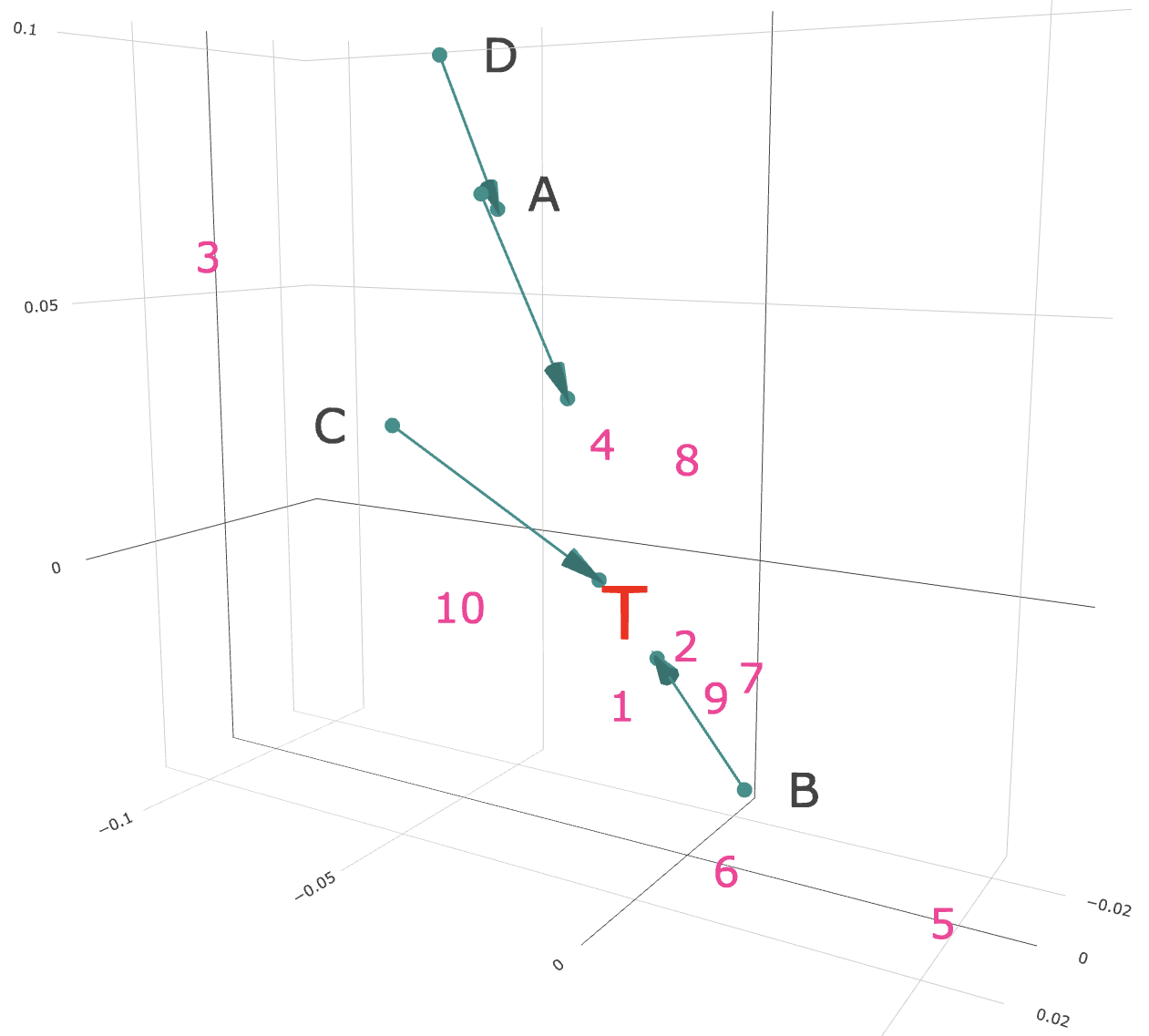}
\caption{Mental health:
An interaction map $(\mathbb{M},\, d)$ with $\mathbb{M} \coloneqq \mR^3$ and $d(\bm{a},\, \bm{b}) \coloneqq |\!|\bm{a}-\bm{b}|\!|_2$ ($\bm{a} \in \mathbb{M}$, $\bm{b} \in \mathbb{M}$) shows the progress of selected mothers $A$, $B$, $C$, $D$ in low-income communities towards the target of interest $\mT$ (improving mental health),
measured by items $1, \dots, 10$ (questions about depression).
The interaction map is estimated by a Bayesian approach to the proposed latent process model.
The interaction map reveals interactions between individuals (mothers) and items (questions about depression):
e.g.,
item $5$ deviates from the bulk of the items,
and mother $B$ is closest to item $5$.
It turns out that mother $B$ agreed with item $5$ at the first assessment (``feeling hopeful"),
whereas mothers $A$, $C$, $D$ did not.
In addition to revealing interactions,
the interaction map suggests that mothers $B$ and $C$ have made strides towards improving mental health,
while mothers $A$ and $D$ may need to make more progress.
}
\label{fg:dep_progress}
\end{figure}

\hide{

> X1[p,]
    cesd1 cesd5 cesd6 cesd7 cesd8 cesd9 cesd12 cesd13 cesd19 cesd20
A     1     0     1     0     0     0      1      1      0      0
B     0     1     1     0      1     1      1      0      1      1
C     0     0     1     0     0     1      1      1      0      1
D     0     0     1     0     0     0      0      1      0      0
> X2[p,]
    cesd1 cesd5 cesd6 cesd7 cesd8 cesd9 cesd12 cesd13 cesd19 cesd20
A     1     1     1     0     0     1      1      1      0      1
B     1     1     1     1     1     1      1      1      1      1
C     1     1     1     1     1     1      1      1      1      1
D     0     0     1     0     1     0      0      1      0      1
> score
   t1  t2
A 0.4 0.7
B 0.7 1.0
C 0.5 1.0
D 0.2 0.4

}

To demonstrate the proposed latent process model,
we assess the progress of $257$ mothers with infants in low-income communities towards improving mental health.
The data are taken from \cite{santos:18} and are described in more detail in Section \ref{sec:depression}.

\begin{figure}[htbp]
    \centering
     \includegraphics[width=.4\textwidth]{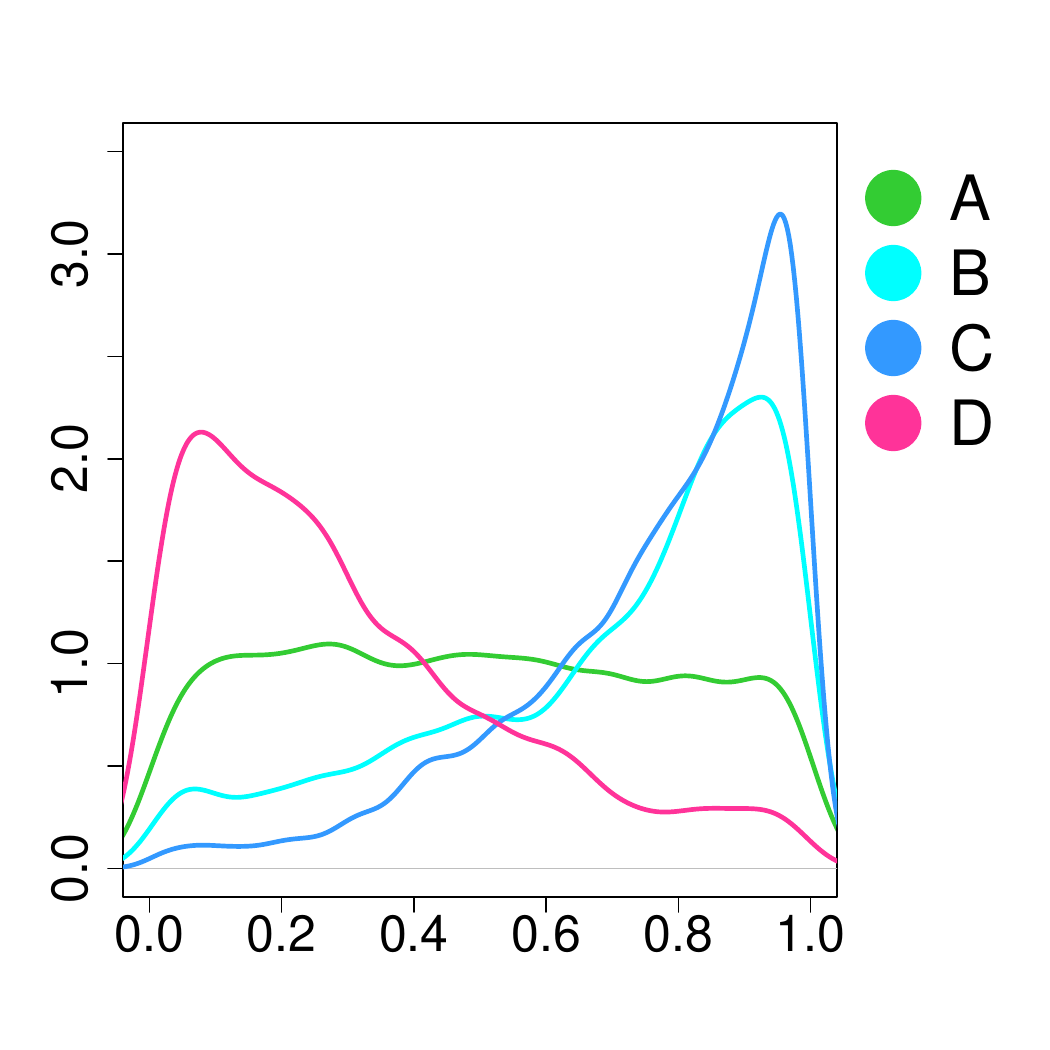}
     \\
\caption{Mental health:
Marginal posteriors of the rates of progress $\lambda_A$, $\lambda_B$, $\lambda_C$, $\lambda_D \in [0,\, 1]$ of mothers $A$, $B$, $C$, $D$ towards target $\mT$ (improving mental health).
The progress of mother $A$ is unclear,
but the marginal posteriors of the rates of progress $\lambda_B$ and $\lambda_C$ of mothers $B$ and $C$ have modes close to $1$,
confirming that mothers $B$ and $C$ have made strides towards improving mental health.
By contrast,
the marginal posterior of the rate of progress $\lambda_D$ of mother $D$ has a mode close to $0$,
underscoring that mother $D$ may need additional assistance.
}
\label{fg:dep_kernel}
\end{figure}

Figure \ref{fg:dep_progress} presents an interaction map, 
based on a Bayesian approach to the proposed latent process model.
The interaction map embeds individuals (mothers) and items (questions about depression) into Euclidean space $(\mathbb{M},\, d)$.
The interaction map offers at least three insights:
\bi
\item Out of the $10$ items used to assess the mental health of the $257$ mothers,
some of the items (e.g., items $3$ and $5$) deviate from the bulk of the items.\vspace{.1cm}
\item There are interactions between individuals (mothers) and items (questions about depression):
e.g.,
mother $B$ is closest to item $5$.
It turns out that mother $B$ agreed with item $5$ at the first assessment (``feeling hopeful"),
whereas mothers $A$, $C$, and $D$ did not.
\item Mothers $B$ and $C$ have made strides towards improving mental health,
whereas mothers $A$ and $D$ may need to make more progress in the future.
\ei
Figure \ref{fg:dep_kernel} clarifies that the progress of mother $A$ is unclear,
but confirms the conclusions regarding mothers $B$, $C$, and $D$.
We describe the results in more detail in Section \ref{sec:depression}.

\subsection{Existing approaches}
\label{sec:literature}

A classic approach to longitudinal educational assessments is based on \citeauthor{andersen:85}'s (\citeyear{andersen:85}) model.
\citeauthor{andersen:85}'s (\citeyear{andersen:85}) model assumes that binary responses $Y_{i,j,t} \in \{0, 1\}$ are independent $\mbox{Bernoulli}(\mu_{i,j,t})$ random variables with $\logit(\mu_{i,j,t}) \coloneqq \alpha_{i,t} + \beta_j$, 
where $\alpha_{i,t} \in \mR$ can be interpreted as the ability of student $i$ at time $t$ and $\beta_j \in \mR$ can be interpreted as the easiness of item $j$.
Based on \citeauthor{andersen:85}'s (\citeyear{andersen:85}) model,
the progress of student $i$ between time $t=1$ and $t=2$ can be quantified by $\alpha_{i,2} - \alpha_{i,1}$.
\citet{embretson:91} reparameterized \citeauthor{andersen:85}'s (\citeyear{andersen:85}) model by modeling changes in abilities via a simplex structure. 
\citeauthor{andersen:85}'s (\citeyear{andersen:85}) model has been extended with a view to capturing temporal dependence 
\citep[e.g.,][]{cai:10},
and addressing multiple learning topics at each time point \citep[e.g.,][]{wang:20, huang:15}.  
Other approaches model a linear change in abilities as a function of time \cite[e.g.,][]{pastor:06, wilson:12} and incorporate first-order autoregressive structure \citep{jeon:2016,segawa:05}.

\hide{
In summary,
\citeauthor{andersen:85}'s (\citeyear{andersen:85}) model and its extensions help educators estimate the progress of students $i$ by estimating differences in abilities $\alpha_{i,2} - \alpha_{i,1}$ (model estimation).
That said,
these models do not help educators determine whether students make progress at all (model selection),
which would enable educators identify students who need more support than others.
In addition,
these models assume that all students $i$ with the same ability $\alpha_{i,t}$ have the same response probability for all items $j$ with the same easiness $\beta_j$.
Such assumptions may well be violated in applications,
because some students $i$ with the same ability $\alpha_{i,t}$ may respond to items $j$ with the same easiness $\beta_j$ differently,
owing to unobserved variables such as cultural background,
upbringing, 
and mentoring.
By contrast,
the proposed latent process captures interactions among individuals (students) and variables (items) and provides a visual interaction map to reveal such interactions,
in addition to helping assess whether individuals make progress;
how much progress individuals make;
and how much more progress individuals can make in the future.
}

\alert{We provide a comparison of the proposed latent process model with the \citet{Rasch60} model and the \citet{jeon:21} model of cross-sectional educational assessment data and \citeauthor{andersen:85}'s (\citeyear{andersen:85}) model of longitudinal educational assessment data in Section \ref{sec:data.model}.
In addition, 
we demonstrate by simulations in Section \ref{sec:interaction} that the latent process model can capture interactions between individuals and variables,
whereas \citeauthor{andersen:85}'s (\citeyear{andersen:85}) model does not capture interactions.
}

\subsection{Outline}

We introduce the proposed latent process model in Section \ref{sec:probabilistic.framework} and outline a Bayesian approach to statistical inference in Section \ref{sec:inference}.
Simulation results and applications can be found in Sections \ref{sec:simulations}, 
\ref{sec:depression},
and \ref{sec:math}, 
respectively.

\section{Latent process model}
\label{sec:probabilistic.framework}

We consider responses $Y_{i,j,t} \in \mathcal{Y}_{i,j,t}$ of individuals $i \in \{1, \dots, n\}$ ($n \geq 1$) to variables $j \in \{1, \dots, p\}$ ($p \geq 1$) at times $t \in \{1, \dots, T\}$ ($T \geq 2$).
To accommodate data from multiple sources (e.g., self-reported mental health assessments collected by surveys and physical measurements related to stress collected by wearable devices),
we allow responses $Y_{i,j,t}$ to be binary ($\mathcal{Y}_{i,j,t} = \{0, 1\}$),
count-valued ($\mathcal{Y}_{i,j,t} = \{0, 1, \dots\}$),
or real-valued ($\mathcal{Y}_{i,j,t} = \mR$).

To monitor the progress of individuals towards a target of interest $\mT$,
we assume that the individuals and the variables measuring progress towards target $\mT$ have positions in a shared metric space $(\mathbb{M},\, d)$, 
consisting of a set $\mathbb{M}$ and a distance function $d: \mathbb{M}^2 \mapsto [0, +\infty)$.
We assume that the set $\mathbb{M}$ is convex and allow the metric space $(\mathbb{M},\, d)$ to be Euclidean or non-Euclidean.
In the domain of statistical network analysis \citep{HuKrSc12,SmAsCa19},
two broad classes of latent space models can be distinguished,
based on the geometry of the underlying metric space:
Euclidean latent space models \citep{Hoff:2002} and latent models with intrinsic hierarchical structure,
based on ultrametric space \citep{ScSn03} or hyperbolic space \citep{hyperbolic}.
The proposed probabilistic framework can accommodate these and other metric spaces.
A discussion of the non-trivial issue of choosing the geometry of the metric space $(\mathbb{M},\, d)$ can be found in Section \ref{select.geo}.
Given a metric space $(\mathbb{M},\, d)$,
we assume that individuals $i$ have positions $\bm{a}_{i,t} \in \mathbb{M}$ at time $t$ and move towards the target of interest $\mT \in \mathbb{M}$,
measured by variables $j$ with positions $\bm{b}_{j} \in \mathbb{M}$.
The position of the target $\mT$ is assumed to be time-invariant. 
It is possible to extend the proposed latent process model to time-varying targets,
provided that the data at hand warrant the resulting increase in model complexity,
but we do not consider such extensions here.

We assume that the responses $Y_{i,j,t} \in \mY_{i,j,t}$ are independent conditional on the positions $\bm{a}_{i,t}$ of individuals $i$ at time $t$ and the positions $\bm{b}_j$ of variables $j$ measuring progress towards target $\mT$,
and are distributed as
\beno
Y_{i,j,t} \mid \btheta,\, \bm{a}_{i,t},\, \bm{b}_j 
&\ind& \mbP_{\btheta,\bm{a}_{i,t},\bm{b}_j},
\ee
where $\mbP_{\btheta,\bm{a}_{i,t},\bm{b}_j}$\, is a probability distribution with support $\mY_{i,j,t}$ and $\btheta \in \bTheta$ is a vector of parameters.

We divide the description of the probabilistic framework into 
\bi
\item the data model:
the model that generates the responses $Y_{i,j,t}$ conditional on the positions $\bm{a}_{i,t}$ of individuals $i$ at time $t$ and the positions $\bm{b}_j$ of variables $j$ measuring progress towards the target of interest $\mT$ (Section \ref{sec:data.model});\vspace{.1cm}
\item the process model:
the process that determines whether and how much progress individuals $i$ make towards the target of interest $\mT$ (Section \ref{sec:latent.process.model}).
\ei
The non-trivial issue of selecting the geometry of the metric space is discussed in Section \ref{select.geo}.
We mention other possible approaches to assessing progress in Section \ref{sec:other.approaches}.
Priors are reviewed in Section \ref{sec:prior} and identifiability issues are discussed in Section \ref{sec:id}.

\subsection{Data model}
\label{sec:data.model}

The data model describes how the responses $Y_{i,j,t}$ are generated conditional on the positions $\bm{a}_{i,t}$ of individuals $i$ at time $t$ and the positions $\bm{b}_j$ of variables $j$ measuring progress towards the target of interest $\mT$. 

To leverage data from multiple sources (e.g., binary, count-, and real-valued responses),
we assume that the responses $Y_{i,j,t}$ are generated by generalized linear models \citep{Su19,efron22}.
Let
\beno
\mu_{i,j,t}(\btheta,\, \bm{a}_{i,t},\, \bm{b}_j)
&\coloneqq& \mbE_{\btheta,\,\bm{a}_{i,t},\, \bm{b}_j}\; Y_{i,j,t}
\ee
be the mean response of individual $i$ to variable $j$ at time $t$ and $\eta_{i,j,t}$ be a link function,
which links the mean response $\mu_{i,j,t}$ to a linear predictor:
\beno 
\eta_{i,j,t}(\mu_{i,j,t}(\btheta,\, \bm{a}_{i,t},\, \bm{b}_j))
&\coloneqq&
\begin{cases} 
\alpha_i + \beta_j - \gamma\; d(\bm{a}_{i,t}, \bm{b}_j)
& \mbox{if } t = 1\s
\\
\alpha_i + \beta_j - \gamma\; d(\bm{a}_{i,t}, \mT) 
& \mbox{if } t = 2, \dots, T,
\end{cases}
\ee
where $\mT \coloneqq (1/p) \sum_{j=1}^p \bm{b}_j$ is the target of interest,
measured by variables $j$ with positions $\bm{b}_j$,
and $\btheta \in \bTheta$ is the vector of weights $\alpha_i \in \mR$ ($i = 1, \dots, n$),\,
$\beta_j \in \mR$ ($j = 1, \dots, p$),\,
and $\gamma \in [0, +\infty)$.
Note that the position $\bm{b}_j$ of variable $j$ affects the linear predictor $\eta_{i,j,t}$ and hence the mean response $\mu_{i,j,t}$ of individual $i$ to variable $j$ at each time point $t \in \{1, \dots, T\}$,
because the distance $d(\bm{a}_{i,t}, \mT)$ depends on $\mT$ and $\mT \coloneqq (1/p) \sum_{j=1}^p \bm{b}_j$ depends on the position $\bm{b}_j$ of variable $j$ at each time point $t \in \{2, \dots, T\}$.

The terms $\alpha_i$,
$\beta_j$,
$\gamma$,
and $\gamma\, d(\bm{a}_{i,t}, \mT)$ can be interpreted as follows:
\bi
\item {\bf Weight $\alpha_i$.} 
The weight $\alpha_i$ quantifies the potential (ability) of individual $i$.
The potential (ability) $\alpha_i$ takes on values in $\mR$ and can therefore be as large as desired.
The potential (ability) $\alpha_i$ will be estimated from the observed responses $Y_{i,j,t}$ along with the weights $\beta_j$ and $\gamma$ and the distances $d(\bm{a}_{i,t}, \mT)$,
using the Bayesian approach described in Section \ref{sec:inference}.\vspace{.1cm}
\item {\bf Weight $\beta_j$.}
The weight $\beta_j$ indicates how agreeable variable $j$ is:
The higher $\beta_j$ is,
the higher is the linear predictor $\eta_{i,j,t}$ and hence the mean response $\mu_{i,j,t}$,
holding everything else fixed.\vspace{.1cm}
\item {\bf Weight $\gamma$.}
If $\gamma > 0$,
the distance term $\gamma\, d(\bm{a}_{i,t}, \mT)$ quantifies how far individual $i$ is below her potential (ability) $\alpha_i$.
For example,
if $d(\bm{a}_{i,t}, \mT) = 0$,
then individual $i$ has reached her potential (ability) $\alpha_i$ and cannot improve with respect to target $\mT$.
By contrast,
if $d(\bm{a}_{i,t}, \mT) > 0$,
then individual $i$ is below her potential (ability) $\alpha_i$ and can improve.\vspace{.1cm}
\item {\bf Making progress by reducing distance $d(\bm{a}_{i,t}, \mT)$.} 
The fact that the weights $\alpha_i$ and $\beta_j$ do not depend on time $t$ implies that the distance term $\gamma \, d(\bm{a}_{i,t}, \mT)$ captures the progress of individual $i$ towards target $\mT$.
Thus,
the more individual $i$ reduces the distance $d(\bm{a}_{i,t}, \mT)$ to target $\mT$,
the more progress $i$ makes,
provided that $\gamma > 0$.
A mathematical description of the rate of progress as a function of the distance of individual $i$ to target $\mT$ is provided by Equations \eqref{d.reduction} and \eqref{rate.progress} in Section \ref{special.cases}.
\ei

It is worth noting that the definition $\mT \coloneqq (1/p) \sum_{j=1}^p \bm{b}_j$ of target $\mT$ as a function of the positions $\bm{b}_j$ of variables $j$ is motivated by the fact that the variables measure progress towards target $\mT$.
Therefore,
it makes sense to specify $\mT$ as a function of the variable positions $\bm{b}_j$.
The specification of target $\mT$ as the mean $(1/p) \sum_{j=1}^p \bm{b}_j$ of the variable positions $\bm{b}_j$ is simple and convenient,
although other specifications are possible.

\subsubsection{Comparison with the \citet{Rasch60} and \citet{jeon:21} models}

The proposed model can be viewed as an extension of the \citet{Rasch60} model and the \citet{jeon:21} model to longitudinal data.
The \citet{Rasch60} and \citet{jeon:21} models consider binary responses $Y_{i,j} \in \{0, 1\}$ observed at $T = 1$ time point and assume that $Y_{i,j} \mid \mu_{i,j}\, \ind\, \mbox{Bernoulli}(\mu_{i,j})$,
where $\logit(\mu_{i,j}) \coloneqq \alpha_i + \beta_j$ \citep{Rasch60} and $\logit(\mu_{i,j}) \coloneqq \alpha_i + \beta_j - \gamma\, d(\bm{a}_i, \bm{b}_j)$ \citep{jeon:21}.
The proposed model can be viewed as an extension of these models to binary and non-binary responses $Y_{i,j,t}$ observed at $T \geq 2$ time points $t \in \{1, \dots, T\}$,
and reduces to
\bi
\item the \citet{Rasch60} model when binary responses $Y_{i,j} \in \{0, 1\}$ are observed at $T = 1$ time point,
the link function is the logit link,
and $\gamma = 0$;\vspace{.1cm}
\item the \citet{jeon:21} model when binary responses $Y_{i,j} \in \{0, 1\}$ are observed at $T = 1$ time point,
the link function is the logit link,
and $\gamma \in [0, +\infty)$.
\ei
As a result,
the proposed model inherits the advantages of the \citet{jeon:21} model:
e.g.,
in educational assessments,
\bi
\item $\alpha_i$ can be interpreted as the potential (ability) of student $i$;\vspace{.1cm}
\item $\beta_j$ can be interpreted as the easiness of item $j$;\vspace{.1cm}
\item the metric space $(\mathbb{M},\, d)$ can be interpreted as an interaction map that captures interactions between students $i$ and items $j$;\vspace{.1cm}
\item for any given student $i$ with a given potential (ability) $\alpha_i$ and any given item $j$ with a given easiness $\beta_j$,
the lower the distance $d(\bm{a}_{i,t}, \mT)$ between individual $i$ and target $\mT$ is,
the higher is the probability of a correct response to item $j$.
\ei

In addition to inheriting the advantages of the \citet{jeon:21} model,
the proposed model helps assess
\bi
\item whether student $i$ makes progress towards learning target $\mT$,
based on changes of the distance $d(\bm{a}_{i,t}, \mT)$ as a function of time $t$;\vspace{.1cm}
\item how much progress student $i$ makes towards learning target $\mT$,
based on changes of the distance $d(\bm{a}_{i,t}, \mT)$,
and how much uncertainty is associated with such assessments;\vspace{.1cm}
\item whether student $i$ makes more progress during some periods than others;\vspace{.1cm}
\item how much more progress student $i$ can make in the future.
\ei

\subsubsection{Comparison with \citeauthor{andersen:85}'s (\citeyear{andersen:85}) model}
\label{sec:interaction.map}

To compare the proposed model with \citeauthor{andersen:85}'s (\citeyear{andersen:85}) model,
consider binary responses $Y_{i,j,t} \mid \mu_{i,j,t}\, \ind\, \mbox{Bernoulli}(\mu_{i,j,t})$.
\citeauthor{andersen:85}'s (\citeyear{andersen:85}) model assumes that $\logit(\mu_{i,j,t}) \coloneqq \alpha_{i,t} + \beta_j$ is additive in the ability $\alpha_{i,t}$ of student $i$ at time $t$ and the easiness $\beta_j$ of item $j$ and therefore does not capture interactions between students $i$ and items $j$ arising from unobserved variables,
such as cultural background, 
upbringing, 
and mentoring of students.
By contrast,
the proposed model does 
capture interactions between students $i$ and items $j$:
e.g.,
a large distance $d(\bm{a}_{i,1}, \bm{b}_j)$ between student $i$ and item $j$ at time $1$ indicates that the mean response $\mu_{i,j,1}$ of student $i$ to item $j$ is lower than would be expected based on the potential (ability) $\alpha_i$ of student $i$ and the easiness $\beta_j$ of item $j$.
The simulation results in Section \ref{sec:interaction} demonstrate that the latent process model can capture interactions between individuals and variables,
whereas \citeauthor{andersen:85}'s (\citeyear{andersen:85}) model does not capture interactions.

\subsubsection{Assessing progress over time}

To assess whether an individual $i$ makes more progress during some periods than others,
one can inspect changes in the linear predictor $\eta_{i,j,t}$,
under \citeauthor{andersen:85}'s (\citeyear{andersen:85}) model and the proposed model.
To demonstrate,
consider binary responses $Y_{i,j,t} \in \{0, 1\}$ at time $t \in \{2, 3, 4\}$ and assume that $Y_{i,j,t} \mid \mu_{i,j,t}\, \ind\, \mbox{Bernoulli}(\mu_{i,j,t})$ with log odds $\eta_{i,j,t}$.
We write henceforth $\mu_{i,j,t}$ and $\eta_{i,j,t}$ instead of $\mu_{i,j,t}(\btheta,\, \bm{a}_{i,t},\, \bm{b}_j)$ and $\eta_{i,j,t}(\mu_{i,j,t}(\btheta,\, \bm{a}_{i,t},\, \bm{b}_j))$,
respectively.

Under \citeauthor{andersen:85}'s (\citeyear{andersen:85}) model with log odds
\beno
\eta_{i,j,t}
&=& \log\dfrac{\mu_{i,j,t}}{1-\mu_{i,j,t}} 
&=& 
\begin{cases}
\alpha_{i,2} + \beta_j & \mbox{ if $t=2$}
\\
\alpha_{i,3} + \beta_j & \mbox{ if $t=3$}
\\
\alpha_{i,4} + \beta_j & \mbox{ if $t=4$,}
\end{cases}
\ee
one can assess whether individual $i$ makes more progress between time $2$ and $3$ than between time $3$ and $4$ by inspecting changes in log odds:
\beno
\eta_{i,j,3} - \eta_{i,j,2} &=& \alpha_{i,3} - \alpha_{i,2}\s
\\
\eta_{i,j,4} - \eta_{i,j,3} &=& \alpha_{i,4} - \alpha_{i,3}.
\ee
If, 
e.g.,
\beno
\alpha_{i,3} - \alpha_{i,2}
&>& \alpha_{i,4} - \alpha_{i,3},
\ee
then individual $i$ makes more progress between time $2$ and $3$ than between time $3$ and $4$.

Under the proposed model with log odds
\beno
\eta_{i,j,t}
&=&
\log \dfrac{\mu_{i,j,t}}{1-\mu_{i,j,t}}
&=& \begin{cases}
\alpha_i + \beta_j - \gamma\, d(\bm{a}_{i,2}, \mT)
& \mbox{if $t=2$}
\\
\alpha_i + \beta_j - \gamma\, d(\bm{a}_{i,3}, \mT)
& \mbox{if $t=3$}
\\
\alpha_i + \beta_j - \gamma\, d(\bm{a}_{i,4}, \mT)
& \mbox{if $t=4$,}
\end{cases}
\ee
one can assess whether individual $i$ makes more progress between time $2$ and $3$ than between time $3$ and $4$ along the same lines,
by inspecting changes in log odds:
\beno
\eta_{i,j,3} - \eta_{i,j,2}
&=& \gamma\, d(\bm{a}_{i,3}, \mT) - \gamma\, d(\bm{a}_{i,2}, \mT)\s
\\
\eta_{i,j,4} - \eta_{i,j,3}
&=& \gamma\, d(\bm{a}_{i,4}, \mT) - \gamma\, d(\bm{a}_{i,3}, \mT).
\ee
If,
e.g.,
\beno
\gamma\, d(\bm{a}_{i,3}, \mT) - \gamma\, d(\bm{a}_{i,2}, \mT)
&>& \gamma\, d(\bm{a}_{i,4}, \mT) - \gamma\, d(\bm{a}_{i,3}, \mT),
\ee
which is equivalent to
\beno
d(\bm{a}_{i,3}, \mT) - d(\bm{a}_{i,2}, \mT)
&>& d(\bm{a}_{i,4}, \mT) - d(\bm{a}_{i,3}, \mT)
\ee
when $\gamma > 0$,
then individual $i$ makes more progress between time 2 and 3 than between time 3 and 4,
because the log odds increases more during the first period than the second one.

In short,
one can assess whether individual $i$ makes more progress during some periods than others by inspecting changes in log odds,
under \citeauthor{andersen:85}'s (\citeyear{andersen:85}) model and the proposed model:
\bi
\item Under \citeauthor{andersen:85}'s (\citeyear{andersen:85}) model,
assessing progress over time by inspecting changes in log odds amounts to inspecting changes of abilities $\alpha_{i,t}$.\vspace{.1cm}
\item Under the proposed model,
assessing progress over time by inspecting changes in log odds amounts to inspecting changes of distances towards target $\mT$.
\ei
Both approaches to assessing progress are legitimate.
That being said,
the proposed model captures interactions between students and items whereas \citeauthor{andersen:85}'s (\citeyear{andersen:85}) model does not capture interactions,
as explained in Section \ref{sec:interaction.map} and demonstrated by simulations in Section \ref{sec:interaction}.

\subsection{Process model}
\label{sec:latent.process.model}

The process model determines whether and how much progress individuals $i$ make towards the target of interest $\mT$.
The process model assumes that a metric space $(\mathbb{M},\, d)$ has been chosen.
We discuss the non-trivial issue of selecting the geometry of the metric space $(\mathbb{M},\, d)$ in Section \ref{select.geo} and review special cases of metric spaces in Section \ref{special.cases} (normed vector spaces), 
Section \ref{example1} (Euclidean space), 
and \ref{example2} (hyperbolic space).

The latent process model assumes that the variables $j$ measuring progress towards target $\mT$ are located at positions
\[
\begin{array}{ccc}
\bm{b}_j &\iid& G,
\end{array}
\]
where $G$ is a distribution with support $\mathbb{M}$.
The target $\mT$ is defined as $\mT \coloneqq (1/p) \sum_{j=1}^p \bm{b}_j$.

The positions $\bm{a}_{i,t}$ of individuals $i$ at time $t$ are generated as follows.
First,
the position $\bm{a}_{i,1}$ of individual $i$ at time $t = 1$ is generated by sampling
\beno
\bm{a}_{i,1} &\iid& H,
\ee
where $H$ is a distribution with support $\mathbb{M}$.
The position $\bm{a}_{i,t}$ of individual $i$ at time $t \in \{2, \dots, T\}$ is a convex combination of $i$'s position $\bm{a}_{i,t-1}$ at time $t-1$ and the target's position $\mT \coloneqq (1/p) \sum_{j=1}^p \bm{b}_j$:
\be
\label{aposition}
\bm{a}_{i,t}
&\coloneqq& (1-\lambda_{i,t})\; \bm{a}_{i,t-1} + \lambda_{i,t}\; \mT,
\ee
where $\bm{a}_{i,t} \in \mathbb{M}$ provided that $\bm{a}_{i,t-1} \in \mathbb{M}$ and $\mT \in \mathbb{M}$,
because the set $\mathbb{M}$ is convex. 
The quantity $\lambda_{i,t} \in [0, 1]$ can be interpreted as the rate of progress of individual $i$ towards target $\mT$ between time $t-1$ and $t$.
In other words,
if individual $i$ makes progress,
$i$ moves towards target $\mT$ on the shortest path between $\bm{a}_{i,t-1}$ and $\mT$.
A random term can be added to the right-hand side of \eqref{aposition} to allow individuals $i$ to deviate from the shortest path between $\bm{a}_{i,t-1}$ and $\mT$.
That said,
we prefer to keep the model simple and do not consider deviations of individuals $i$ from the shortest path between $\bm{a}_{i,t-1}$ and $\mT$,
because the moves of individuals $i$ in $\mathbb{M}$ are unobserved. Covariates can be incorporated into the rates of progress $\lambda_{i,t}$ of individuals $i$ between time $t-1$ and $t$ by using a suitable link function.

\setcounter{figure}{1}

\begin{figure}[t]
\centering
\begin{tikzpicture}
[scale=1, 
  transform shape, node distance=2cm,
  roundnode/.style={circle, draw=black, thick, minimum size=10mm},
  squarednode/.style={rectangle, draw=black, thick, minimum size=6mm},
  arrow/.style = {semithick,-Stealth},
  dotnode/.style={fill,inner sep=0pt,minimum size=2pt,circle} 
]

\draw   (0,0) -- (10, 0) ;
\draw   (0,0) -- (0,7) ;

\draw [fill=green!10,draw=none] (3,5) circle (0.5cm) node [scale=1.2]  {$\mT$} ; 

\draw [fill] (2.9,5.7) circle (.6mm)  ;
\draw [fill] (2.6,4.5) circle (.6mm)  ;
\draw [fill] (3.5,4.9) circle (.6mm)  ;
\draw [fill] (3.7,5.8) circle (.6mm)  ;
\draw [fill] (2.1,5) circle (.6mm)  ;

\draw [fill] (2,1.3) circle (1mm)  ;
\draw (1.3,1.3) node [scale=1] {$\bm{a}_{A,1}$} ;

\draw[->,dashed] (2,1.3) -- (3,4.8) ; 

\draw [fill] (2.55,3.3) circle (1mm)  ;
\draw (1.8,3.3) node [scale=1] {$\bm{a}_{A,2}$} ;

\draw[->, thick,dashed] (2,1.3) -- (2.55,3.2) ; 

\draw [fill] (7,1) circle (1mm)  ;
\draw (7.3,.6) node [scale=1] {$\bm{a}_{B,1}$} ;

\draw [fill] (6,1.9) circle (1mm)  ;
\draw (6.6,2.1) node [scale=1] {$\bm{a}_{B,2}$} ;

\draw[->, dashed] (7,1) -- (3,4.8); 
\draw[->, thick,dashed] (7,1) -- (6, 1.9); 

\draw (2,1.3) to [out=10,in =-20]  (2.55,3.3); 
\draw (3.5,2.3) node [scale=1]  {$\lambda_{A,2}$} ;

\draw (7,1) to [out=-120,in =-120]  (6, 1.9); 
\draw (5.8,.8) node [scale=1] {$\lambda_{B,2}$} ;

\end{tikzpicture}
\caption{Interaction map:
Two individuals $A$ and $B$ with positions $\bm{a}_{A,1}$ and $\bm{a}_{B,1}$ at time $1$ and positions $\bm{a}_{A,2}$ and $\bm{a}_{B,2}$ at time $2$ make progress towards a target of interest $\mT \coloneqq (1/p) \sum_{j=1}^p \bm{b}_j$,
measured by variables $j$ with positions $\bm{b}_j$ (unlabeled points).
The interaction map represents interactions between individuals and variables along with the progress of individuals towards target $\mT$.
The rates of progress $\lambda_{A,2}$ and $\lambda_{B,2}$ determine how much the distances of $A$ and $B$ to target $\mT$ are reduced between time $1$ and $2$,
respectively. 
\label{fg:target}
}
\end{figure}
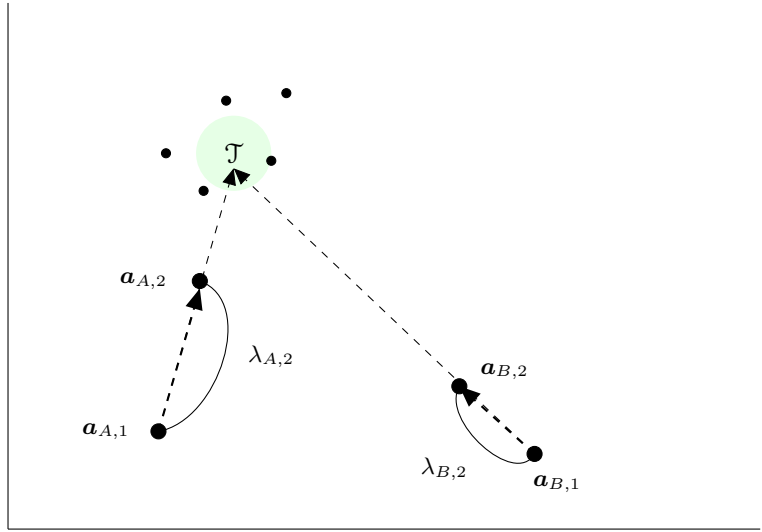

\subsubsection{Special cases}
\label{special.cases}

In general,
the process model makes two assumptions.
First,
the set $\mathbb{M}$ is convex,
so that the positions $\bm{a}_{i,t} \coloneqq (1-\lambda_{i,t})\; \bm{a}_{i,t-1} + \lambda_{i,t}\; \mT$ of individuals $i$ at time $t$ are contained in the set $\mathbb{M}$,
provided that $\bm{a}_{i,t-1} \in \mathbb{M}$ and $\mT \in \mathbb{M}$.
Second,
the set $\mathbb{M}$ is equipped with a distance function $d$,
so that the distances between individuals $i$ and variables $j$ measuring progress towards target $\mT$ can be quantified.

Despite the fact that the process model---at least in its most general form---does not require more than two assumptions,
the interpretation and application of the process model is facilitated by additional assumptions.
For example,
the interpretation of the process model is facilitated if the set $\mathbb{M}$ is endowed with a norm $|\!|.|\!|$ (Euclidean or non-Euclidean) and $d(\bm{a}_{i,t}, \mT) \coloneqq |\!|\bm{a}_{i,t} - \mT|\!|$,
because the distance $d(\bm{a}_{i,t},\, \mT)$ can then be expressed as a function of the distance $d(\bm{a}_{i,t-1},\, \mT)$:
\be
\label{d.reduction}
d(\bm{a}_{i,t},\, \mT)
&=& |\!|((1 - \lambda_{i,t})\, \bm{a}_{i,t-1} + \lambda_{i,t}\, \mT) - \mT|\!|
\hide{
&=& |\!|(1 - \lambda_{i,t})\, \bm{a}_{i,t-1} - (1 - \lambda_{i,t})\, \mT|\!|\s
\\
&&
}
&=& (1 - \lambda_{i,t})\; d(\bm{a}_{i,t-1},\, \mT).
\ee
As a consequence,
the rate of progress $\lambda_{i,t}$ of individual $i$ between time $t-1$ and $t$ can be expressed as a function of the distances $d(\bm{a}_{i,t-1}, \mT)$ and $d(\bm{a}_{i,t}, \mT)$:
\be
\label{rate.progress}
\lambda_{i,t}
&=& 1 - \dfrac{d(\bm{a}_{i,t}, \mT)}{d(\bm{a}_{i,t-1}, \mT)},
\ee
provided that $d(\bm{a}_{i,t-1}, \mT) > 0$.
In other words,
the rate of progress $\lambda_{i,t}$ of individual $i$ between time $t-1$ and $t$ reveals how much the distance between individual $i$'s position and the target's position $\mT$ is reduced between time $t-1$ and $t$.

\subsubsection{Special case 1: Euclidean space}
\label{example1}

In the special case $\mathbb{M} \coloneqq \mR^q$ ($q \geq 1$) and $d(\bm{a}_{i,t}, \mT) \coloneqq |\!|\bm{a}_{i,t} - \mT|\!|_2$,
it is convenient to choose $G$ and $H$ to be multivariate Gaussians.
A demonstration of the process model in the special case $\mathbb{M} \coloneqq \mR^2$ and $d(\bm{a}_{i,t}, \mT) \coloneqq |\!|\bm{a}_{i,t} - \mT|\!|_2$ is provided by Figure \ref{fg:target}.

\subsubsection{Special case 2: hyperbolic space}
\label{example2}

An alternative to Euclidean space is a space with an intrinsic hierarchical structure,
such as hyperbolic space.
For example,
consider the two-dimensional Poincar\'e disk with radius $\rho \in (0, +\infty)$,
that is,
$\mathbb{M} \coloneqq \{\bm{x} \in \mathbb{R}^2:\, |\!|\bm{x}|\!|_2 < \rho\}$.
The distance between the position $\bm{a}_{i,t} \in \mathbb{M}$ of individual $i$ at time $t$ and target $\mT \in \mathbb{M}$ on the Poincar\'e disk with radius $\rho$ is defined by
\beno
d(\bm{a}_{i,t}, \mT)
&\coloneqq& \mathop{\mbox{arcosh}}\left(1 + \dfrac{2\; \rho^2\; |\!|\bm{a}_{i,t} - \mT|\!|_2^2}{\left(\rho^2-|\!|\bm{a}_{i,t}|\!|_2^2\right)\, \left(\rho^2-|\!|\mT|\!|_2^2\right)}\right).
\ee
It is then convenient to choose $G$ and $H$ to be the Uniform distribution on $\{\bm{x} \in \mathbb{R}^2:\, |\!|\bm{x}|\!|_2 < \rho\}$.

\subsection{Selecting the geometry of the metric space}
\label{select.geo}

An open issue is how to select the geometry of the metric space $(\mathbb{M},\, d)$.
In the statistical analysis of network data,
recent work by \citet{LuChMc23} introduced a promising approach to selecting the geometry of latent space models of network data,
although the approach of \citet{LuChMc23} falls outside of the Bayesian framework considered here.
That said,
we expect that the pioneering work of \citet{LuChMc23} paves the way for Bayesian approaches to selecting the geometry of the underlying space,
for models of network data and models of educational data.
In the special case $\mathbb{M} \coloneqq \mR^q$,
an additional issue may arise,
in that the dimension $q$ of $\mR^q$ may be unknown.
Since we are interested in dimension reduction (embedding both individuals and variables in a low-dimensional space) and helping professionals (e.g., educators, medical professionals) assess the progress of individuals by using an easy-to-interpret interaction map,
it is tempting to choose a low-dimensional Euclidean space,
e.g.,
$\mR^2$ or $\mR^3$.
In the simulations and applications in Sections \ref{sec:simulations},
\ref{sec:depression}, 
and \ref{sec:math},
we compare metric spaces $(\mathbb{M},\, d)$ with $\mathbb{M} \coloneqq \mR^q$ and $q \in \{1, 2, 3, 4\}$ by using the Watanabe–Akaike information criterion \citep{waic}.

\subsection{\alert{Comparison with alternative approaches to measuring progress}}
\label{sec:other.approaches}

\alert{There are alternative approaches to measuring progress:
e.g.,
one can measure progress based on expected scores.

To demonstrate,
recall that the proposed statistical framework builds on generalized linear models.
In other words,
the responses $Y_{i,j,t}$ have statistical exponential-family distributions with canonical parameters $\eta_{i,j,t}$ \citep{Br86,Su19,efron22}.
It can be shown that the statistical exponential-family distributions of the responses $Y_{i,j,t}$ imply that the vectors of responses $\bm{Y}_i \coloneqq (Y_{i,j,t})_{1 \leq j \leq p,\, 1 \leq t \leq T}$ have statistical exponential-family distributions with canonical parameter vector $\bta_i$ with coordinates $\eta_{i,j,t}$ and mean-value parameter vector $\bmu_i(\bta_i) \coloneqq \mbE_{\bta_i}\, \bY_i$ with coordinates $\mu_{i,j,t}(\bta_i) \coloneqq \mbE_{\bta_i}\, Y_{i,j,t}$:
Mathematical details can be found in Supplement \ref{sec:expected.scores}. 
In statistical exponential families,
the map $\bta_i \mapsto \bmu_i(\bta_i)$ is one-to-one:
see,
e.g., 
Theorem 3.6 on page 74 of \citet{Br86}.
The one-to-one relationship of $\bta_i$ and $\bmu_i(\bta_i) \coloneqq \mbE_{\bta_i}\, \bY_i$ implies that one can specify models and assess progress based on one of two parameterizations:
\bi
\item[1.] One can specify models and assess progress by specifying the canonical parameter vector $\bta_i$.\vspace{.1cm}
\item[2.] One can specify models and assess progress by specifying the mean-value parameter vector $\bmu_i(\bta_i) \coloneqq \mbE_{\bta_i}\, \bY_i$,
that is,
the expected score vector.
\ei
Due to the one-to-one relationship of $\bta_i$ and $\bmu_i(\bta_i)$,
one can go back and forth between $\bta_i$ and $\bmu_i(\bta_i)$,
regardless of whether one assesses progress based on natural parameter vector $\bta_i$ or mean-value parameter vector $\bmu_i(\bta_i)$,
i.e.,
expected scores \citep[although the map $\bta_i \mapsto \bmu_i(\bta_i)$ may be nonlinear and may not be available in closed form: see, e.g.,][]{ScSt19}.

That said,
when covariates are available,
it makes sense to specify the model by specifying the canonical parameter vector $\bta_i$.
To demonstrate,
consider binary responses\break 
$Y_{i,j,t} \mid \mu_{i,j,t}\, \ind\, \mbox{Bernoulli}(\mu_{i,j,t})$ along with a time-dependent covariate $x_{i,t} \in \mR$:
e.g., 
the time-dependent covariate $x_{i,t} \in \mR$ may be the income of the parents of student $i$ at time $t$,
which may affect the access of student $i$ to educational resources and mentoring and may therefore affect the mean responses $\mu_{i,j,t}$ of student $i$ at time $t$.
To specify models of binary responses $Y_{i,j,t} \mid \mu_{i,j,t}\, \ind\, \mbox{Bernoulli}(\mu_{i,j,t})$ with time-dependent covariates $x_{i,t} \in \mR$,
it is common practice to leverage the canonical parameterization $\bta_i$ of generalized linear models,
that is,
the logit link:
\beno
\eta_{i,j,t}
&=& \log\dfrac{\mu_{i,j,t}}{1-\mu_{i,j,t}}
&=&
\begin{cases}
\alpha_i + \beta_j - \gamma\; d(\bm{a}_{i,t}, \bm{b}_j) + \delta\, x_{i,t}
& \mbox{if } t = 1\s
\\
\alpha_i + \beta_j - \gamma\; d(\bm{a}_{i,t}, \mT) + \delta\, x_{i,t}
& \mbox{if } t = 2, \dots, T,
\end{cases}
\ee
where $\delta \in \mR$ is the weight of covariate $x_{i,t} \in \mR$.
By contrast,
it is less convenient to specify models of binary responses $Y_{i,j,t} \mid \mu_{i,j,t}\, \ind\, \mbox{Bernoulli}(\mu_{i,j,t})$ with time-dependent covariates $x_{i,t} \in \mR$ by using the mean-value parameterization $\bmu_i(\bta_i)$:
e.g.,
a linear model of the form 
\beno
\mu_{i,j,t}(\bta_i) 
&\coloneqq& \mbE_{\bta_i}\, Y_{i,j,t} 
&=& \alpha_i + \beta_j - \gamma\; d(\bm{a}_{i,t}, \bm{b}_j) + \delta\, x_{i,t}
\ee
would not be appropriate,
not least because the mean response $\mu_{i,j,t}$ could be less than $0$ or greater than $1$---despite the fact that the response $Y_{i,j,t}$ is either $0$ or $1$.

In addition to having advantages for specifying models and estimating models \citep[see advantages 1--5 in Section 1 of][pp.~628--629]{ScKrBu17},
the canonical parameterization facilitates the assessment of progress compared with the mean-value parametrization (i.e., expected scores) when there are time-dependent covariates $x_{i,t} \in \mR$ that may affect the mean responses $\mu_{i,j,t} \coloneqq \mbE_{\bta_i}\, Y_{i,j,t}$.
To demonstrate,
consider assessing the progress of individual $i$ based on the change in the expected score of individual $i$ between time $1$ and $2$:
\beno
\mbox{progress of individual $i$}
&\coloneqq& \dsum_{j=1}^p \mu_{i,j,2}(\bm\eta_i) - \dsum_{j=1}^p \mu_{i,j,1}(\bm\eta_i)\s
\\
&=& \dsum_{j=1}^p \mbE_{\bm\eta_i}\, Y_{i,j,2} - \dsum_{j=1}^p \mbE_{\bm\eta_i}\, Y_{i,j,1}\s
\\
&=& \mbE_{\bm\eta_i} \dsum_{j=1}^p Y_{i,j,2} - \mbE_{\bm\eta_i} \dsum_{j=1}^p Y_{i,j,1}.
\ee
A change in the expected score of individual $i$ between time $1$ and $2$ could then be due to
\bi 
\item[(a)] a change in the time-dependent covariate $x_{i,t} \in \mR$ between time $1$ and $2$,
e.g.,
a change in the income of the parents of student $i$ between time $1$ and $2$;\vspace{.1cm}
\item[(b)] the progress $\lambda_{i,2}$ of individual $i$ between time $1$ and $2$;\vspace{.1cm}
\item[(c)] a combination of (a) and (b).
\ei
If,
e.g.,
the time-dependent covariate $x_{i,t} \in \mR$ of student $i$ changed between time $1$ and $2$ and the expected score of student $i$ changed between time $1$ and $2$,
it is not straightforward to disentangle whether---and to which extent---the change in the expected score is due to the change in the time-dependent covariate $x_{i,t} \in \mR$ or the change in the ability of student $i$ or both.
As a result,
it is challenging to assess progress based on expected scores when there are time-dependent covariates that affect the mean responses $\mu_{i,j,t}$ of student $i$.
The challenge of disentangling the source of change in expected scores increases when there are two or more time-dependent covariates.

While both approaches to specifying models and assessing progress (based on the canonical and mean-value parameterization, i.e., expected scores) are legitimate,
we choose the canonical route.
A more technical discussion can be found in Supplement \ref{sec:expected.scores}.
}

\subsection{Priors}
\label{sec:prior}

We assume that $\alpha_i \mid \mu_\alpha,\, \sigma_\alpha^2\; \iid\; N(\mu_\alpha,\, \sigma_\alpha^2)$,\;
$\beta_j \mid \mu_\beta,\, \sigma_\beta^2\; \iid\; N(\mu_\beta,\, \sigma_\beta^2)$,\;
and $\gamma \mid \sigma_\gamma^2$ $\sim$ Half-Normal$(0,\, \sigma^2_\gamma)$.
The hyperparameters $\mu_\alpha \in \mR$ and $\mu_\beta \in \mR$ are set to $0$ to address identifiability issues (see Section \ref{sec:id}),
while hyperparameter $\sigma_\alpha^2 \in (0, +\infty)$ is assigned hyperprior $\sigma_\alpha^2 \mid a_{\sigma_\alpha},\, b_{\sigma_\alpha} \sim \mbox{Inverse-Gamma}(a_{\sigma_\alpha},\,  b_{\sigma_\alpha})$.
The hyperparameters $\sigma_\beta^2 \in (0, +\infty)$, 
$\sigma_\gamma^2 \in (0, +\infty)$,
$a_{\sigma_\alpha} \in (0, +\infty)$ and $b_{\sigma_\alpha} \in (0, +\infty)$ need to be specified by users.
A flexible prior of $\lambda_{i,t}$ can be specified by
\beno
\logit(\lambda_{i,t})
\mid \pi_{i,t},\, \mu_0,\, \sigma_0^2,\, \mu_1,\, \sigma_1^2
&\ind& (1-\pi_{i,t})\; N(\mu_0,\, \sigma_0^2) + \pi_{i,t}\; N(\mu_1,\, \sigma_1^2),
\ee
where $\pi_{i,t} \in (0,\, 1)$.
The hyperparameters $(\mu_0,\, \sigma_0^2) \in \mR \times (0,\, +\infty)$ and $(\mu_1,\, \sigma_1^2) \in \mR \times (0,\, +\infty)$ are chosen so that the distribution of $\lambda_{i,t}$ is a mixture of two distributions,
one of them placing 95\% of its mass between .01 and .21 and the other placing 95\% of its mass between .04 and .96.
The resulting two-component mixture prior of $\lambda_{i,t}$ is motivated by the observation that often some individuals make negligible progress while others make non-negligible progress.
The two-component mixture prior of $\lambda_{i,t}$ is shown in Figure \ref{fg:priors}. 
The mixing proportions $\pi_{i,t}$ have priors $\pi_{i,t} \mid a_\pi,\, b_\pi\, \iid\, \mbox{Beta}(a_\pi,\, b_\pi)$,
with hyperparameters $a_\pi \in (0, +\infty)$ and $b_\pi \in (0, +\infty)$.

\begin{figure}[htbp]
    \centering
\begin{tabular}{cc}
(a) & (b)  \\
     \includegraphics[width=0.35 \textwidth]{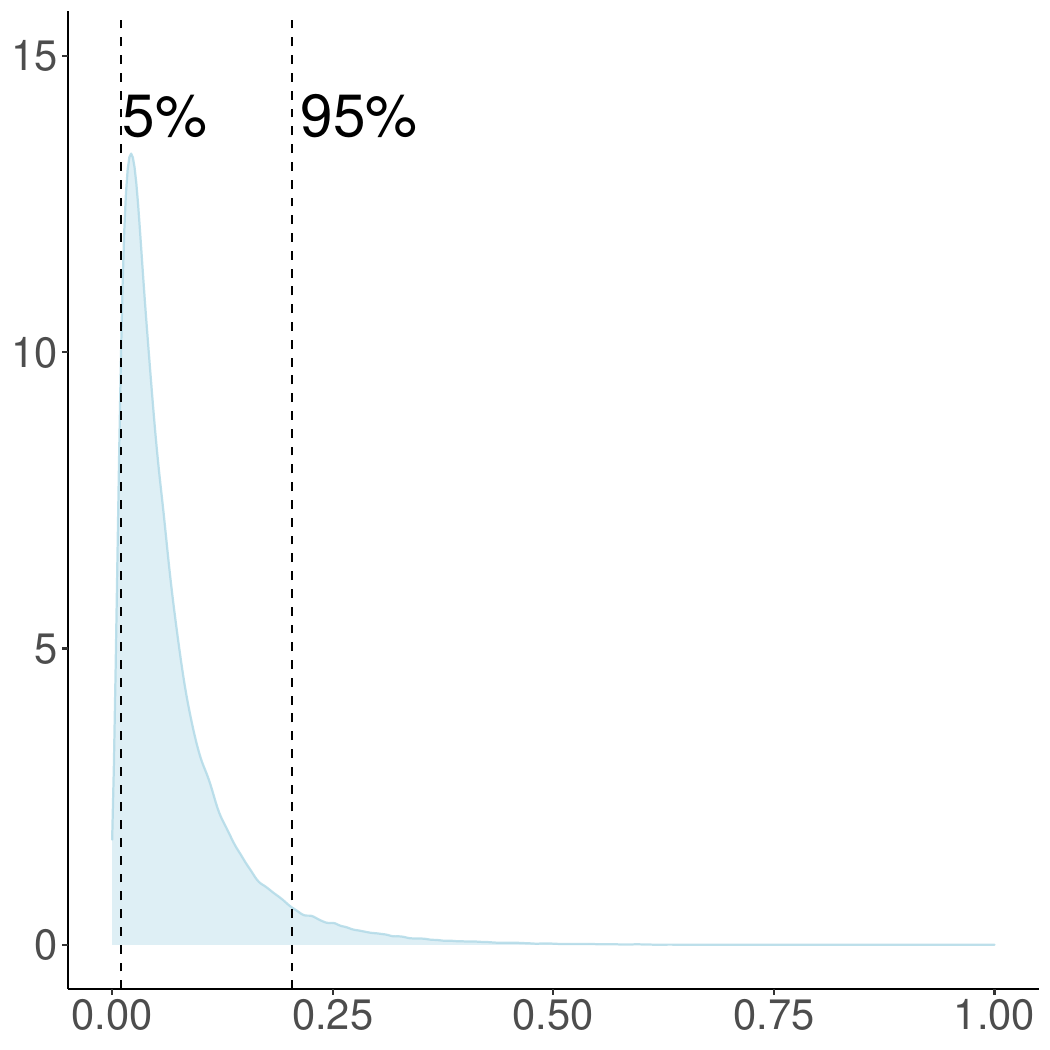}
     &  \includegraphics[width=0.35 \textwidth]{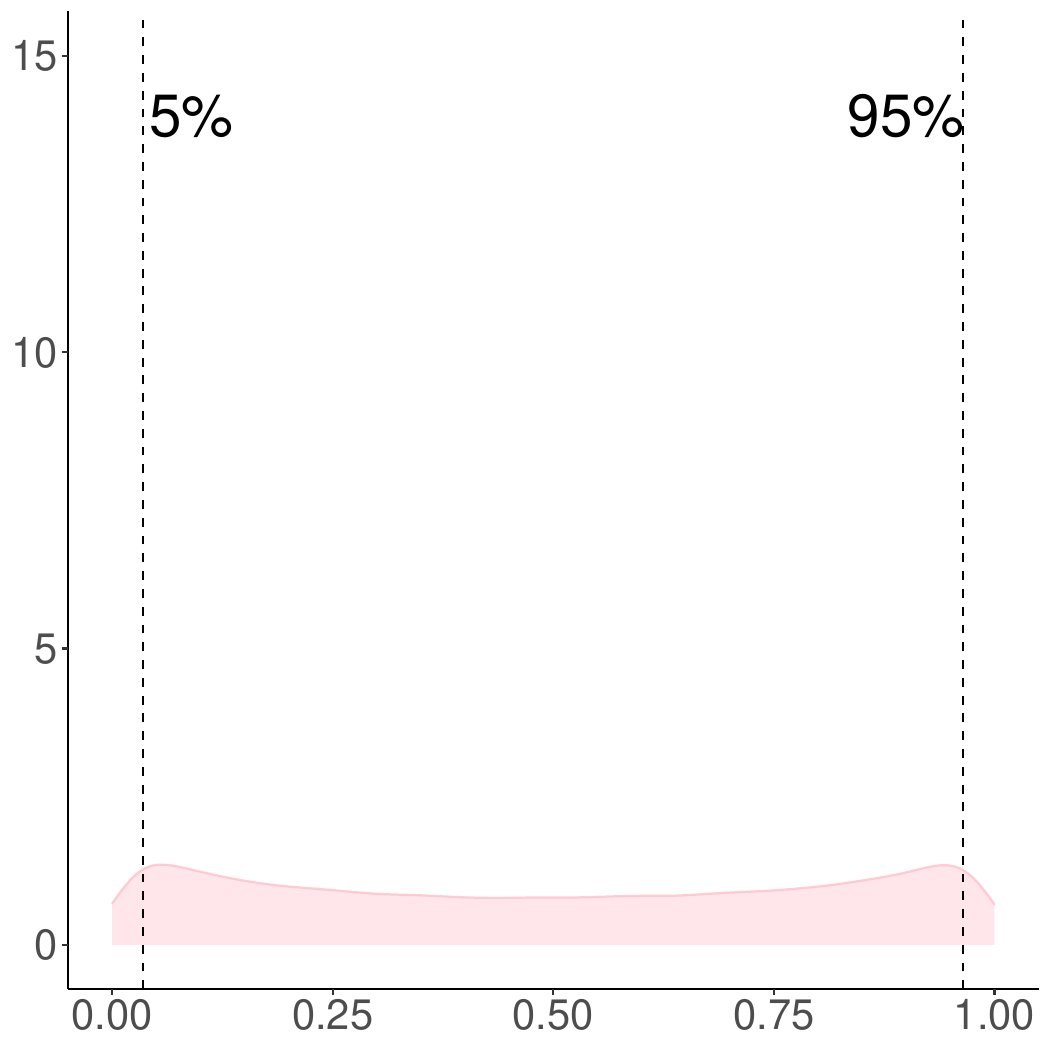}  \\
\end{tabular}
\caption{The two component distributions of the two-component mixture prior for the rates of progress $\lambda_{i,t}$,
with 5\% and 95\% percentiles indicated by dashed vertical lines.
}
\label{fg:priors}
\end{figure}

\subsection{\alert{Identifiability issues}}
\label{sec:id}

\alert{We discuss identifiability issues along with possible solutions,
following a Bayesian approach.
A Bayesian approach addresses identifiability issues via the prior,
because the posterior is proportional to the likelihood times the prior.
We follow a standard approach to addressing identifiability issues in Bayesian statistics:
see,
e.g.,
the monograph on Bayesian statistics by \citet{GeHi07},
the Bayesian approaches to item response models by \citet{Al92} and \citet{Cu10},
and the Bayesian approaches to latent space models for network data by \citet{Hoff:2002},
\citet{HaRaTa07},
\citet{KrHaRaHo07},
\citet{sewell2015latent},
and others.

A well-known issue of the classic \citet{Rasch60} model and its extensions---including the proposed latent process model---is that the weights $\alpha_i$ and $\beta_j$ cannot be all estimated,
unless additional restrictions are imposed.
We follow convention in Bayesian statistics by constraining the means of the priors of $\alpha_i \mid \mu_\alpha,\, \sigma_\alpha^2\; \iid\; N(\mu_\alpha,\, \sigma_\alpha^2)$ and $\beta_j \mid \mu_\beta,\, \sigma_\beta^2\; \iid\; N(\mu_\beta,\, \sigma_\beta^2)$ by setting $\mu_\alpha \coloneqq 0$ and $\mu_\beta \coloneqq 0$:
see,
e.g.,
page 316 of \citet{GeHi07}.
Such constraints are widely used for addressing the identifiability issues of the \citet{Rasch60} model and other item response models in a Bayesian setting:
see,
e.g.,
\citet{Al92} and \citet{Cu10}.
The chosen constraints help ensure that $\alpha_i$ and $\beta_j$ can be estimated and that $\alpha_i$ and $\beta_j$ can be separated from the distance term $\gamma\, d(\bm{a}_{i,t}, \mT)$.

In the special case $\mathbb{M} \coloneqq \mR^2$ and $d(\bm{a}_{i,t}, \mT) \coloneqq |\!|\bm{a}_{i,t} - \mT|\!|_2$,
the latent process model has the same identifiability issues as other Euclidean latent space models,
e.g.,
the Euclidean latent space models of \citet{Hoff:2002}:
The distances $d(\bm{a}_{i,t}, \mT)$ are invariant to reflection,
translation,
and rotation of the positions $\bm{a}_{i,t}$ of individuals $i$ at time $t$ and target $\mT$.
We follow \citet{Hoff:2002} and address such identifiability issues by basing statistical inference on equivalence classes of positions using Procrustes matching:
see pages 1092--1093 of \citet{Hoff:2002}.

An additional identifiability issue arises from the fact that,
for all $c \in (0, 1) \cup (1, +\infty)$,
\beno
\eta_{i,j,t}(\mu_{i,j,t}(\btheta,\,  \bm{a}_{i,t},\, \bm{b}_j))
&\coloneqq& \alpha_i + \beta_j - \gamma\, |\!|\bm{a}_{i,t} - \mT|\!|_2
&=& \alpha_i + \beta_j - \dfrac{\gamma}{c}\, |\!|c\, \bm{a}_{i,t} - c\, \mT|\!|_2,
\ee
where $\mT \coloneqq (1/p) \sum_{j=1}^p \bm{b}_j$.
The same identifiability issue arises in the latent space model for network data by \citet{HaRaTa07}.
We address it by following \citet{HaRaTa07} and constraining the positions $\bm{b}_j$ of variables $j$,
without constraining the positions $\bm{a}_{i,t}$ of individuals $i$ at time $t$:
\beno
\sqrt{\dfrac{1}{p} \dsum_{j=1}^p |\!|\bm{b}_j|\!|_2^2}
&=& 1.
\ee
A related constraint is discussed on page 304 of \citet{HaRaTa07}.
}

\section{Bayesian inference}
\label{sec:inference}

Define $\bm{y} \coloneqq (y_{i,j,t})_{1 \leq i \leq n,\, 1 \leq j \leq p,\, 1 \leq t \leq T}$,\,
$\bm\alpha \coloneqq (\alpha_i)_{1 \leq i \leq n}$,\,
$\bm\beta \coloneqq (\beta_j)_{1 \leq j \leq p}$,\,
$\bm\lambda \coloneqq (\lambda_{i,t})_{1 \leq i \leq n,\, 2 \leq t \leq T}$,
$\bm\pi \coloneqq (\pi_{i,t})_{1 \leq i \leq n,\, 2 \leq t \leq T}$,\,
$\bm{A} \coloneqq (\bm{a}_{i,t})_{1 \leq i \leq n,\, 1 \leq t \leq T}$,\,
and $\bm{B} \coloneqq (\bm{b}_j)_{1 \leq j \leq p}$.
Then the posterior is proportional to
\beno
f(\bm\alpha,\, \bm\beta,\, \gamma,\, \bm{A},\, \bm{B},\, \bm\lambda,\, \bm\pi \mid \bm{y})
&\propto& \left[\dprod_{i=1}^n\, \dprod_{j=1}^{p}\, \dprod_{t=1}^T f(y_{i,j,t} \mid \alpha_i,\, \beta_j,\, \gamma,\, \bm{a}_{i,t},\, \bm{b}_1, \dots, \bm{b}_p)\right]\s
\\
&\times& \left[\, \dprod_{i=1}^n\, f(\alpha_i \mid \sigma_{\alpha}^2)\right] \times\, f(\sigma_{\alpha}^2)\, \times \left[\, \dprod_{j=1}^p\, f(\beta_j)\right] \times\, f(\gamma)\s
\\
&\times& \left[\, \dprod_{i=1}^n\, f(\bm{a}_{i,1})\, \dprod_{t=2}^T\, f(\bm{a}_{i,t} \mid \bm{a}_{i,t-1},\, \bm{b}_1, \dots, \bm{b}_p,\, \lambda_{i,t})\right]\s
\\
&\times& \left[\, \dprod_{j=1}^p\, f(\bm{b}_j)\right] \times \left[\, \dprod_{i=1}^n\, \dprod_{t=2}^T\, f(\lambda_{i,t} \mid \pi_{i,t})\; f(\pi_{i,t})\right].
\ee

Here,
in an abuse of notation, 
$f$ denotes a probability mass or density function with suitable support.
It is worth noting that the conditional densities $f(\bm{a}_{i,t} \mid \bm{a}_{i,t-1},\, \bm{b}_1, \dots, \bm{b}_p,\, \lambda_{i,t})$ are point masses,
because the positions $\bm{a}_{i,t} \coloneqq (1-\lambda_{i,t})\; \bm{a}_{i,t-1} + \lambda_{i,t}\; \mT$ of individuals $i$ at time $t$ are non-random functions of $\bm{a}_{i,t-1}$, $\mT \coloneqq (1/p) \sum_{j=1}^p \bm{b}_j$, and $\lambda_{i,t}$ for fixed $\bm{a}_{i,t-1}$, $\bm{b}_1, \dots, \bm{b}_p$, and $\lambda_{i,t}$.

We approximate the posterior by Markov chain Monte Carlo:
e.g.,
if all responses $Y_{i,j,t} \in \{0, 1\}$ are binary,
we use the P\'{o}lya-Gamma sampler of \citet{polson:13} for updating unknown quantities in the data model and Gibbs samplers and Metropolis-Hasting steps for updating unknown quantities in the process model.
Details are provided in Supplement \ref{appendix:mcmc}.

\section{Simulations}
\label{sec:simulations}

\alert{We conduct two simulation studies: 
\bi
\item In Section \ref{sec:misspecification}, 
we investigate whether the proposed statistical framework can capture interesting features of real-world data that have not been generated by the model.
In other words, 
we investigate the behavior of the proposed statistical framework under model misspecification.\vspace{.1cm}
\item In Section \ref{sec:interaction}, 
we demonstrate how the proposed model can capture interactions between individuals and variables,
whereas \citeauthor{andersen:85}'s (\citeyear{andersen:85}) model does not.
\ei
}

Throughout Sections \ref{sec:simulations},  
\ref{sec:depression}, 
and \ref{sec:math},
we focus on binary responses $Y_{i,j,t} \in \{0, 1\}$ at $T = 2$ time points $t \in \{1, 2\}$ and write $\lambda_i$ instead of $\lambda_{i,2}$.
All results are based on the proposed latent process model with a logit link function and the special case $\mathbb{M} \coloneqq \mR^q$ and $d(\bm{a}_{i,t}, \mT) \coloneqq |\!|\bm{a}_{i,t} - \mT|\!|_2$.
The prior and Markov chain Monte Carlo algorithm are described in Supplement \ref{appendix:estimation}.

\hide{
For Study 1, we compare results based on $\mathbb{M} \coloneqq \mR$,\, $\mR^2$, $\mR^3$, and $\mR^4$ using the Watanabe–Akaike information criterion \citep{waic} along with interaction maps based on $\mR$,\, $\mR^2$, and $\mR^3$. 
\textcolor{blue}{For Study 2, we focus on $\mathbb{M} \coloneqq \mR^2$ for the sake of simplicity. }
}

\subsection{Simulation Study I: Model Misspecification}\label{sec:misspecification}

\subsubsection{Scenario $n=300$, $p=10$, $T=2$} 
\label{sec:sim_scenario1}

\begin{table}[htb]
    \centering
    \begin{tabular}{l|cccr}
    \hline 
$\mathbb{M}$  & 10\% Percentile & Median & 90\% Percentile & Minimizer of WAIC\\
      \hline 
$\mathbb{M} \coloneqq \mR$ & 17937& 18170& 19228  & 2.0\% \\ 
$\mathbb{M} \coloneqq \mR^2$ & 17639& 17815& 18016 & 73.2\%   \\
$\mathbb{M} \coloneqq \mR^3$ & 17613& 18129& 18391  & 21.6\% \\
$\mathbb{M} \coloneqq \mR^4$ &18075& 18304& 18512  &  3.2\% \\
\hline
\end{tabular}
\caption{Simulation results ($n = 300$):
Watanabe–Akaike information criterion (WAIC) based on 250 simulated data sets.
The last column indicates how often the latent process model with $\mathbb{M} \coloneqq \mR$,\, $\mR^2$, $\mR^3$, $\mR^4$ minimized the WAIC. 
}
\label{tab:sim2_fit}
\end{table}

We first consider a scenario in which the progress of individuals towards a target of interest $\mT \in \mathbb{M} \coloneqq \mR^q$ is assessed based on binary responses $Y_{i,j,t} \in \{0, 1\}$ of $n = 300$ individuals to $p=10$ variables at $T = 2$ time points $t \in \{1, 2\}$.
We assume that there are three groups of individuals,
called G1, G2, and G3.
Each group is comprised of 100 individuals. 
At time 1,
the success probabilities of all individuals are .2.
At time 2, 
individuals have success probabilities .25 (G1), .5 (G2), and .75 (G3).  
We generated 250 data sets and estimated the latent process model with $\mathbb{M} \coloneqq \mR^q$ and $q \in \{1, 2, 3, 4\}$. 
\textcolor{blue}{The resulting interaction maps of those three groups of individuals are presented in Supplement \ref{appendix:add}.}

The Watanabe–Akaike information criterion \citep{waic} in Table \ref{tab:sim2_fit} suggests that the latent process model with $\mathbb{M} \coloneqq \mR^2$ is more adequate than $\mathbb{M} \coloneqq \mR$,\,
$\mR^3$,
and $\mR^4$.

\begin{figure}[htbp]
\begin{tabular}{cccc}
$\mathbb{M} \coloneqq \mR$ & $\mathbb{M} \coloneqq \mR^2$ & $\mathbb{M} \coloneqq \mR^3$\\
     \includegraphics[width=.31\textwidth]{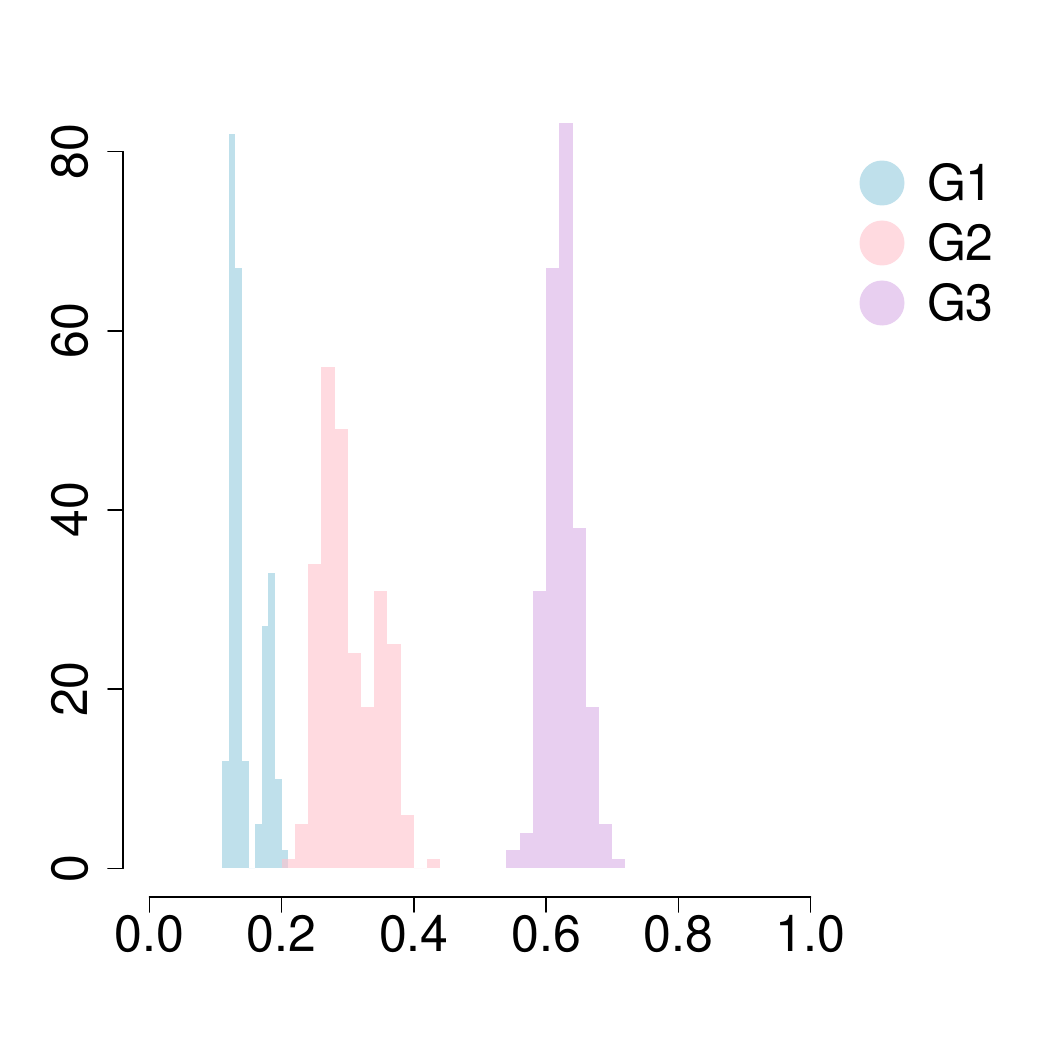} 
     &
 \includegraphics[width=.31\textwidth]{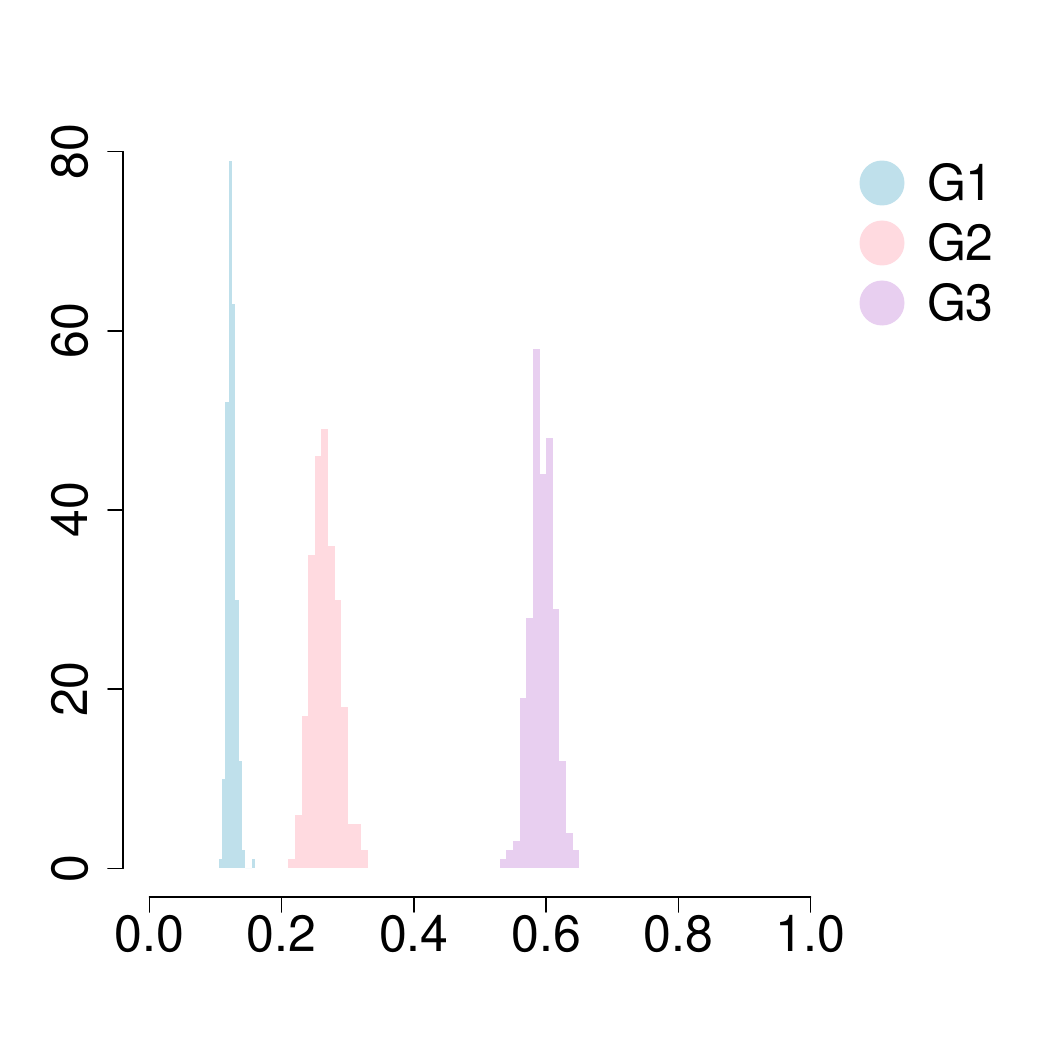} 
 &
      \includegraphics[width=.31\textwidth]{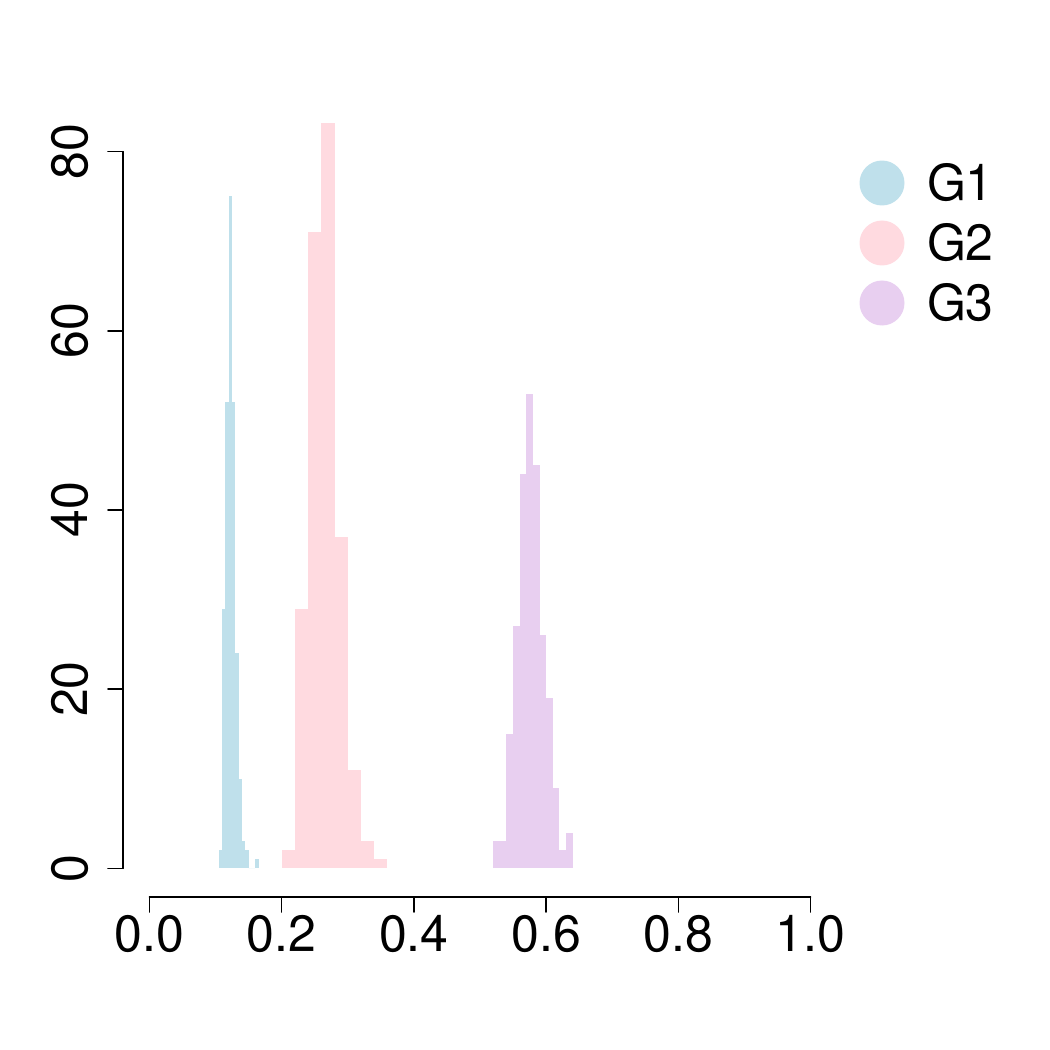} 
     \\
 \end{tabular}
\caption{\alert{Simulation results ($n = 300$): 
Histogram of the posterior medians of rates of progress $\lambda_i$ of individuals $i$.      
For each individual $i$,
the mean of 250 posterior medians of $\lambda_i$ based on 250 simulated data sets is plotted.}}
\label{fg:sim2_progress}
\end{figure}

A natural question to ask is whether the latent process model can separate groups G1, G2, and G3 based on data.
Figure \ref{fg:sim2_progress} indicates that the latent process model with $\mathbb{M} \coloneqq \mR^q$ and $q \in \{1, 2, 3\}$ can distinguish groups G1, G2, and G3,
despite the fact that the data were not generated by the model. 
That said,
the latent process model with $\mathbb{M} \coloneqq \mR^2$ or $\mR^3$ can better distinguish groups G1, G2, and G3 than the latent process model with $\mathbb{M} \coloneqq \mR$,
reinforcing the findings in Table \ref{tab:sim2_fit}.

\begin{figure}[htbp]
    \centering
\begin{tabular}{cccc}
$\mathbb{M} \coloneqq \mR$ & $\mathbb{M} \coloneqq \mR^2$ & $\mathbb{M} \coloneqq \mR^3$\\ 
     \includegraphics[width=.31\textwidth]{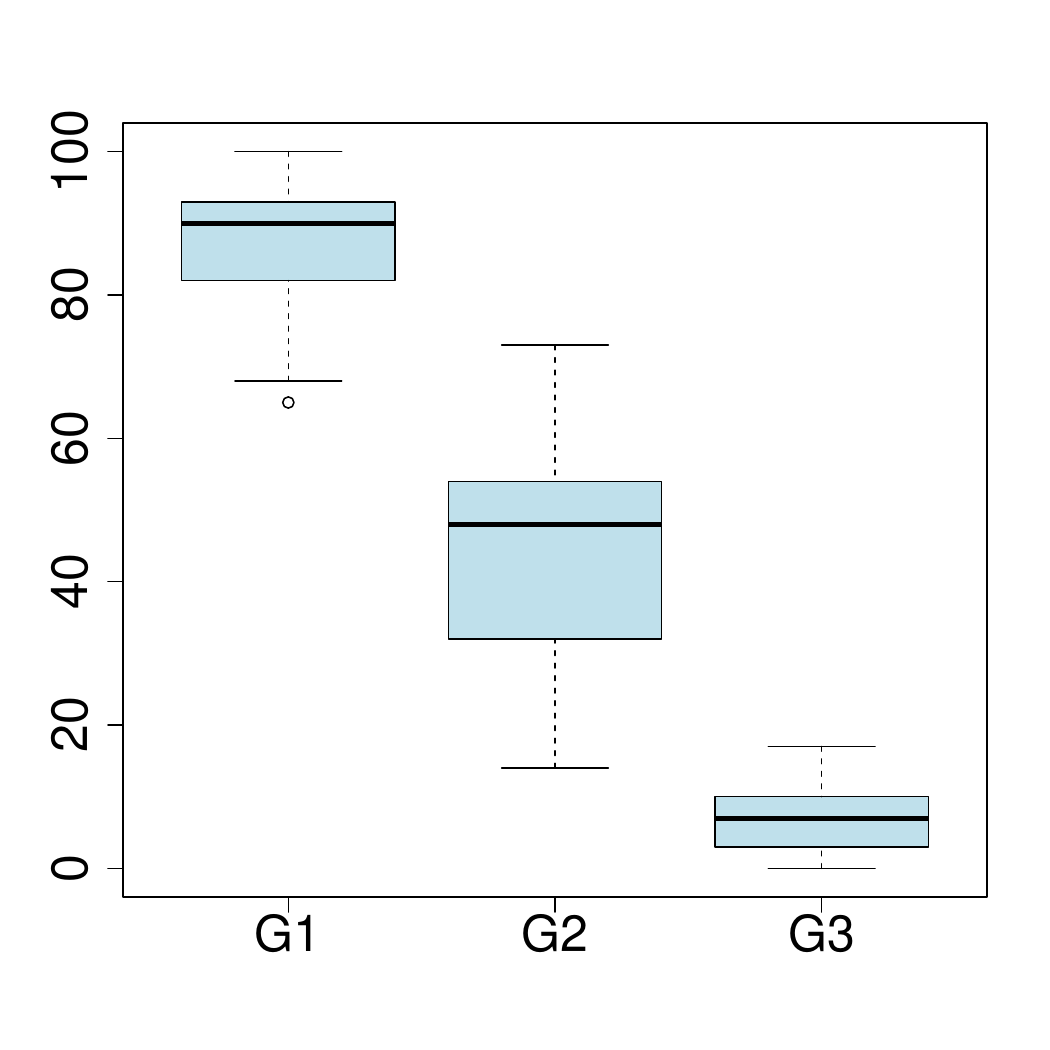} 
     &
 \includegraphics[width=.31\textwidth]{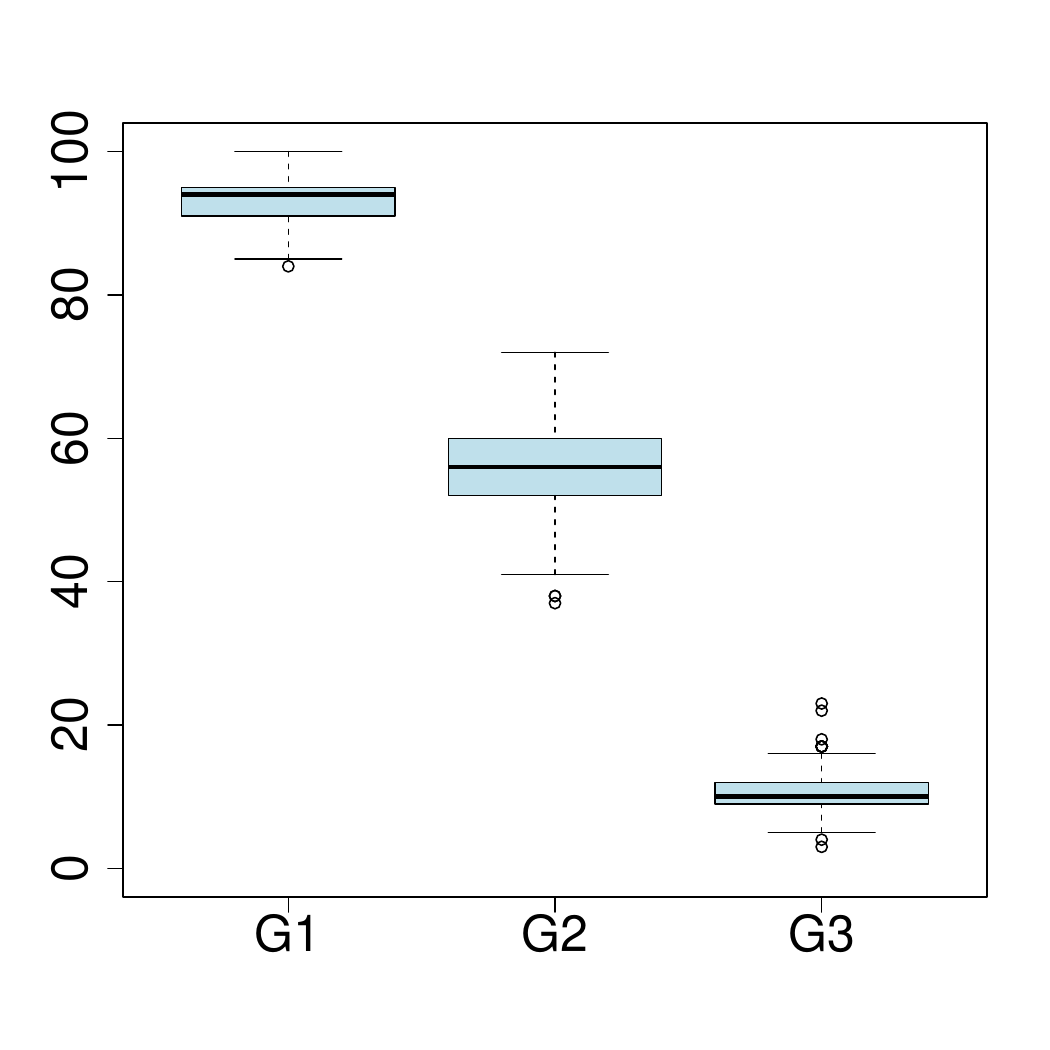} 
&
    \includegraphics[width=.31\textwidth]{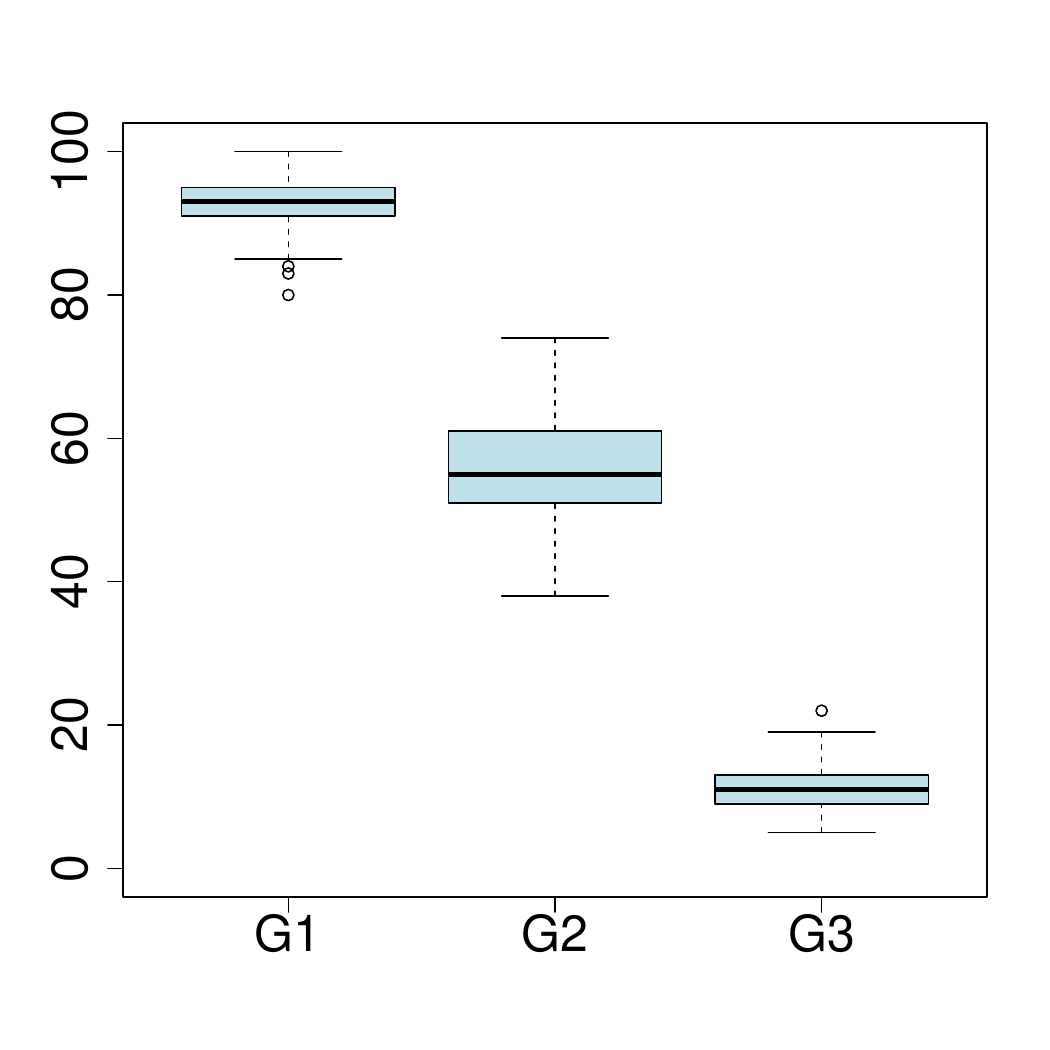}    
     \\
 \end{tabular}
\caption{Simulation results ($n = 300$):
Boxplots of the number of individuals in groups G1, G2, G3 deemed to have made negligible progress.}
\label{fg:sim2_count}
\end{figure}

A related question is whether the latent process model can distinguish individuals with non-neglible progress from those with negligible progress,
based on the two-component mixture prior for the rates of progress $\lambda_i$ of individuals $i$ described in Section \ref{sec:prior}.
Figure \ref{fg:sim2_count} reveals that---among the 100 individuals in groups G1, G2, and G3---the percentage of individuals deemed to have made negligible progress is more than 80\% in the low-progress group G1;
is between 40\% and 80\% in the moderate-progress group G2;
and is less than 20\% in the high-progress group G3.
The latent process model with $\mathbb{M} \coloneqq \mR^2$ or $\mR^3$ seems to lead to more accurate assessments,
compared with the latent process model with $\mathbb{M} \coloneqq \mR$.

\subsubsection{Scenario $n=600$, $p=10$, $T=2$}
\label{sec:sim_scenario2}

\begin{table}[htb]
    \centering
    \begin{tabular}{l|cccr}
    \hline 
$\mathbb{M}$  & 10\% Percentile & Median & 90\% Percentile & Minimizer of WAIC\\

      \hline 
$\mathbb{M} \coloneqq \mR$ & 35703 & 36567& 38665   & 10.0\% \\ 
$\mathbb{M} \coloneqq \mR^2$ & 35256 &35567 &35920  & 79.6\%  \\
$\mathbb{M} \coloneqq \mR^3$ & 35581 &36017 &36639  & 10.4\%   \\
$\mathbb{M} \coloneqq \mR^4$ & 36243 &36976 &38083   & 0.0\%   \\
\hline
    \end{tabular}
\caption{Simulation results ($n = 600$):
Watanabe–Akaike information criterion (WAIC) based on 250 simulated data sets.
The last column indicates how often the latent process model with $\mathbb{M} \coloneqq \mR$,\, $\mR^2$, $\mR^3$, $\mR^4$ minimized the WAIC. 
}
    \label{tab:sim3_fit}
\end{table}

We double the sample size from $n = 300$ to $n = 600$ to explore the behavior of the latent process model as the sample size increases.
We assume that there are four groups of individuals,
called G1, G2, G3 and G4.
Each group is comprised of 150 individuals. 
At time 1 the success probabilities of all individuals are .2, 
while at time 2 individuals have success probabilities .3 (G1), .5 (G2), .7 (G3), and .9 (G4). 
We generated 250 data sets and estimated the latent process model with $\mathbb{M} \coloneqq \mR^q$ and $q \in \{1, 2, 3, 4\}$. 
\textcolor{blue}{The resulting interaction maps of those three groups of individuals are presented in Supplement \ref{appendix:add}.}

\begin{figure}[t]
    \centering
\begin{tabular}{cccc}
$\mathbb{M} \coloneqq \mR$ & $\mathbb{M} \coloneqq \mR^2$ & $\mathbb{M} \coloneqq \mR^3$\\ 
     \includegraphics[width=.31\textwidth]{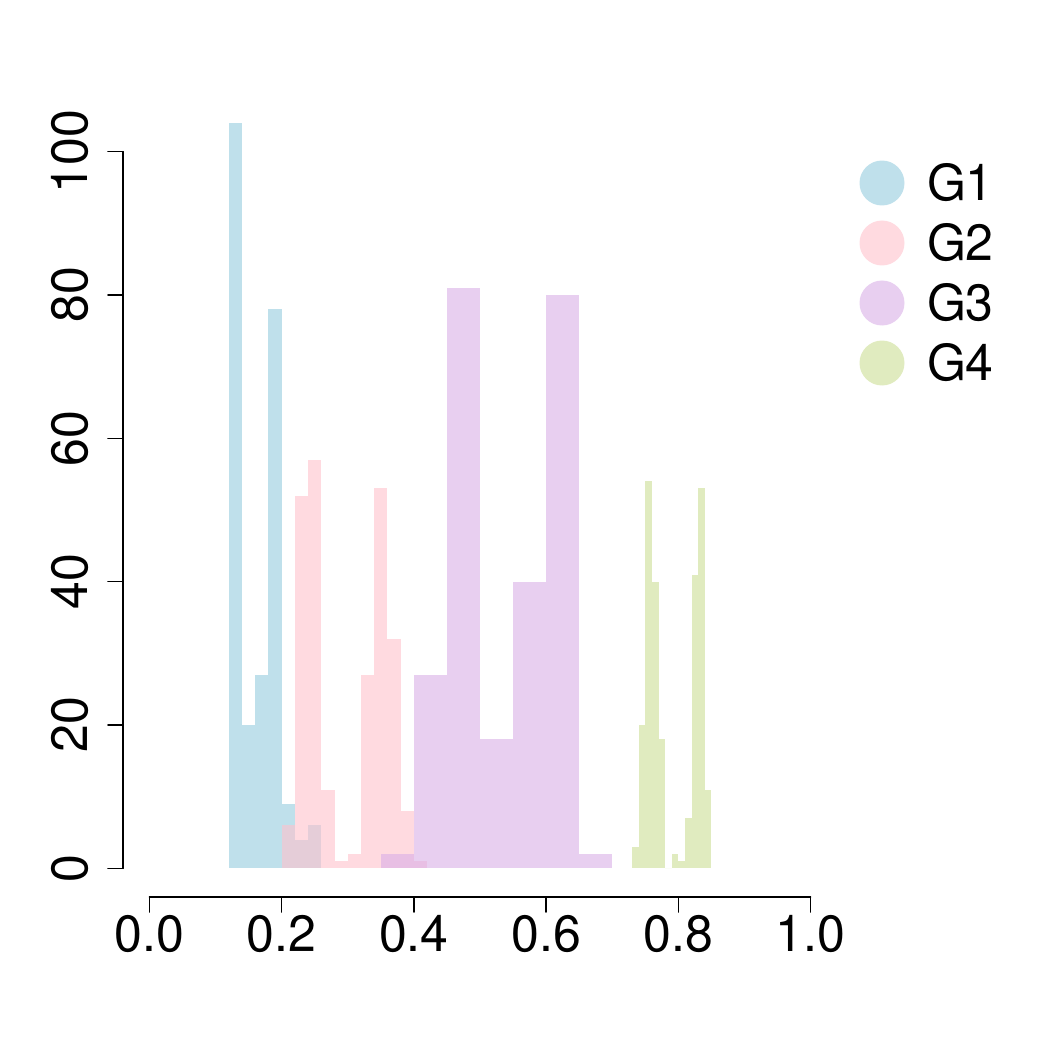} 
     &
 \includegraphics[width=.31\textwidth]{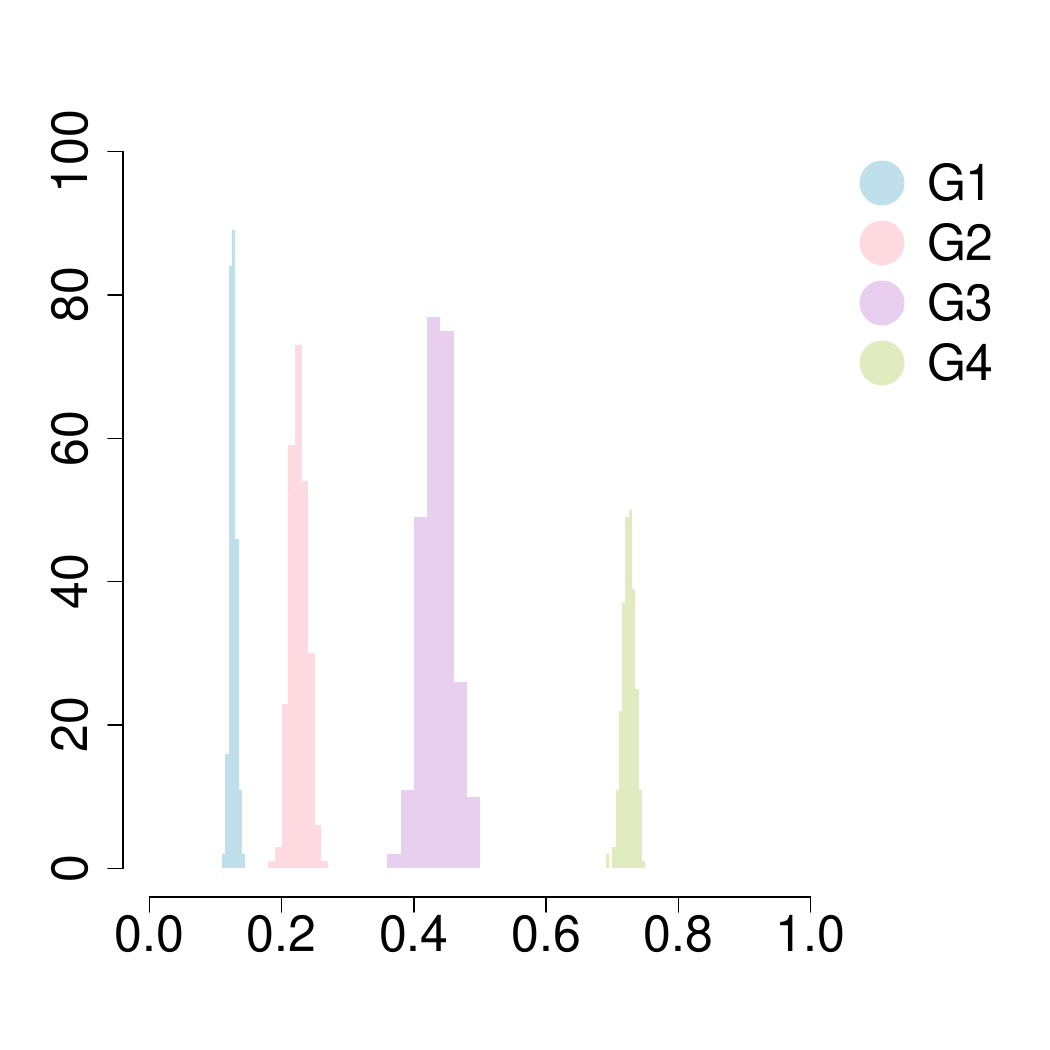} 
 &
 \includegraphics[width=.31\textwidth]{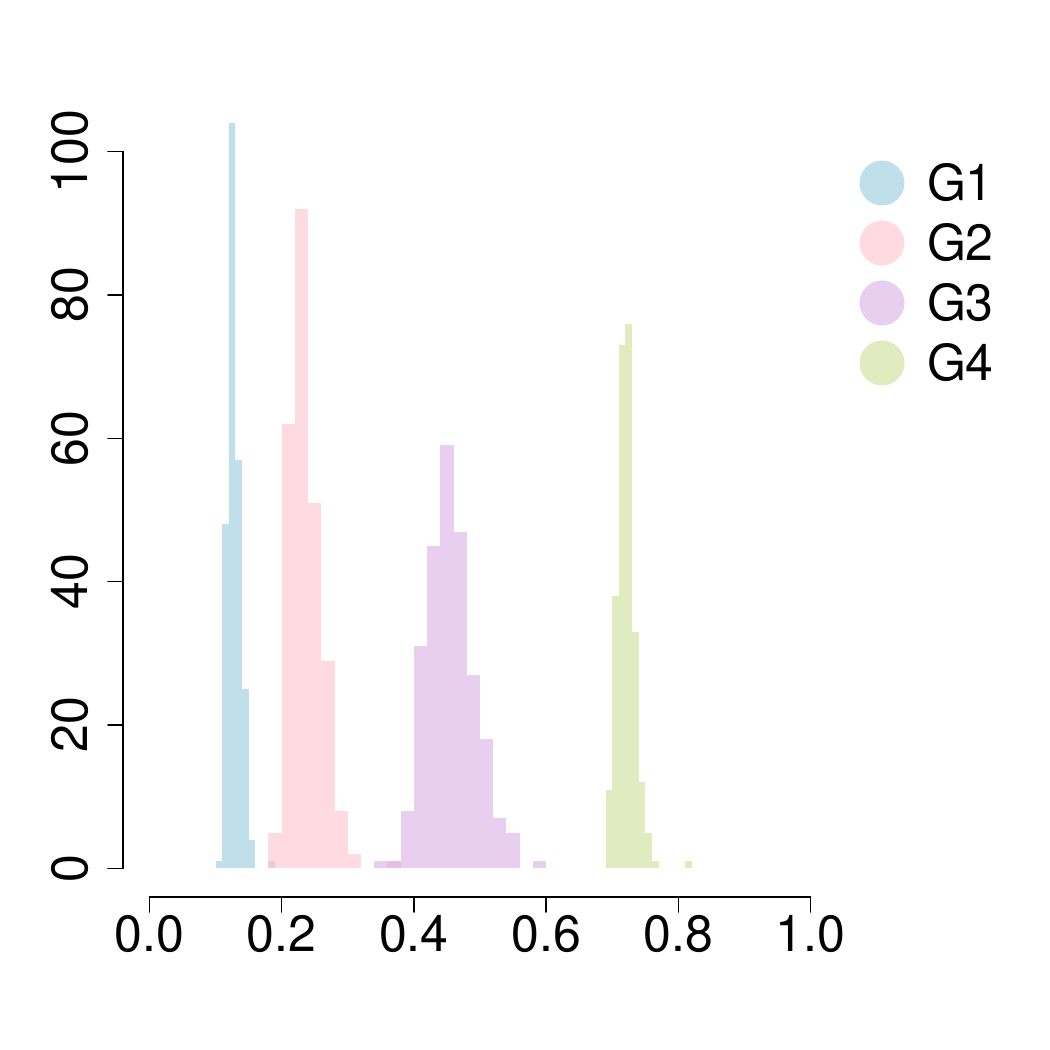}      
     \\
 \end{tabular}
\caption{\alert{Simulation results ($n = 600$):
Histogram of the posterior medians of rates of progress $\lambda_i$ of individuals $i$.      
For each individual $i$,
the mean of 250 posterior medians of $\lambda_i$ based on 250 simulated data sets is plotted.}}
\label{fg:sim3_progress}
\end{figure}

The results dovetail with the results in Section \ref{sec:sim_scenario1},
but suggest that the one-dimensional space $\mathbb{M} \coloneqq \mR$ is even less appealing when $n$ is large:
First, 
the Watanabe–Akaike information criterion \citep{waic} in Table \ref{tab:sim3_fit} favors $\mathbb{M} \coloneqq \mR^2$ over $\mR$,\, $\mR^3$, and $\mR^4$.
Second,
the latent process model can distinguish the groups G1, G2, G3, and G4 when $\mathbb{M} \coloneqq \mR^q$ and $q \geq 2$ according to Figure \ref{fg:sim3_progress},
but the groups G1, G2, G3, and G4 are less well-separated when $\mathbb{M} \coloneqq \mR$.

\subsection{Simulation Study II: Interactions}
\label{sec:interaction}

\begin{figure}[htb]
    \centering
    \begin{tabular}{cc}
     G1 &  G2  \\
 \includegraphics[width=0.35 \textwidth]{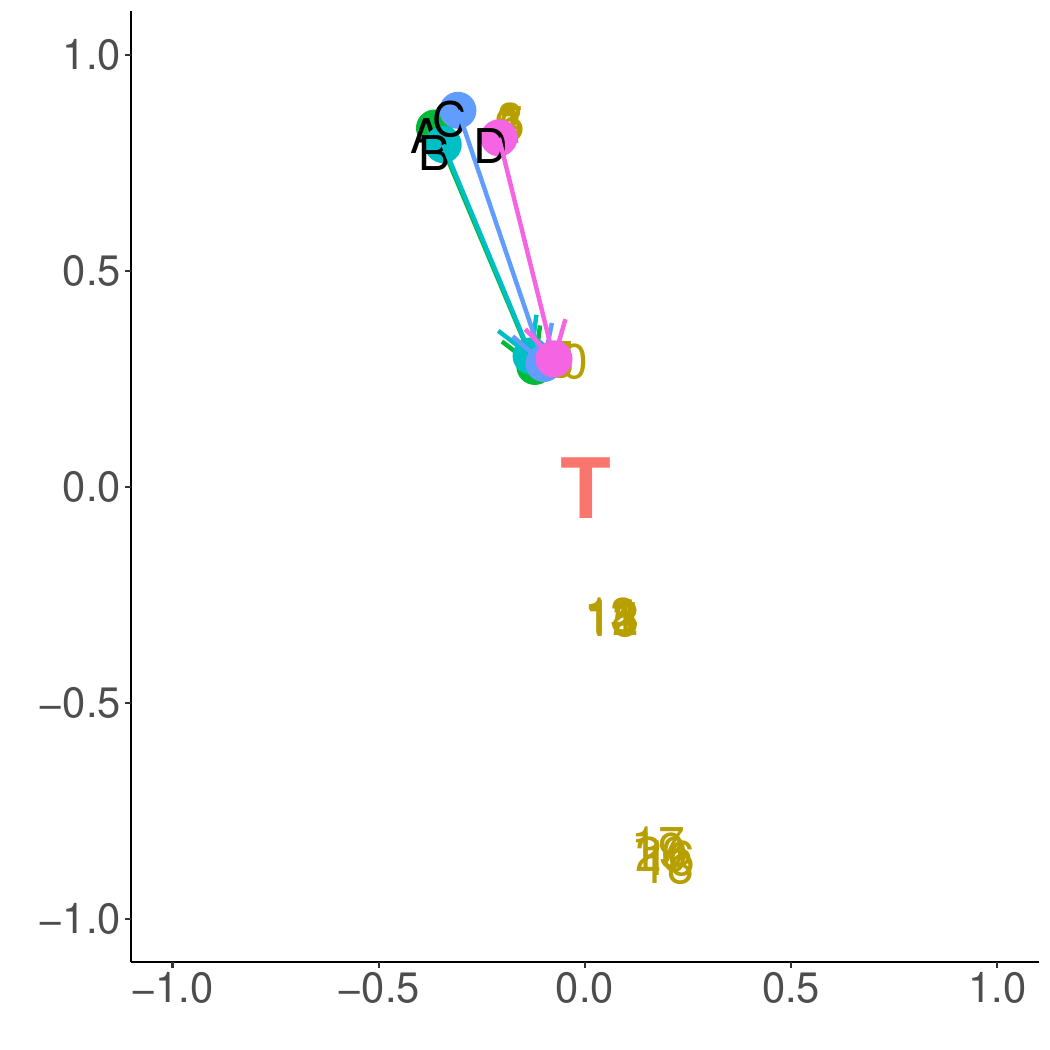}
  &   \includegraphics[width=0.35 \textwidth]{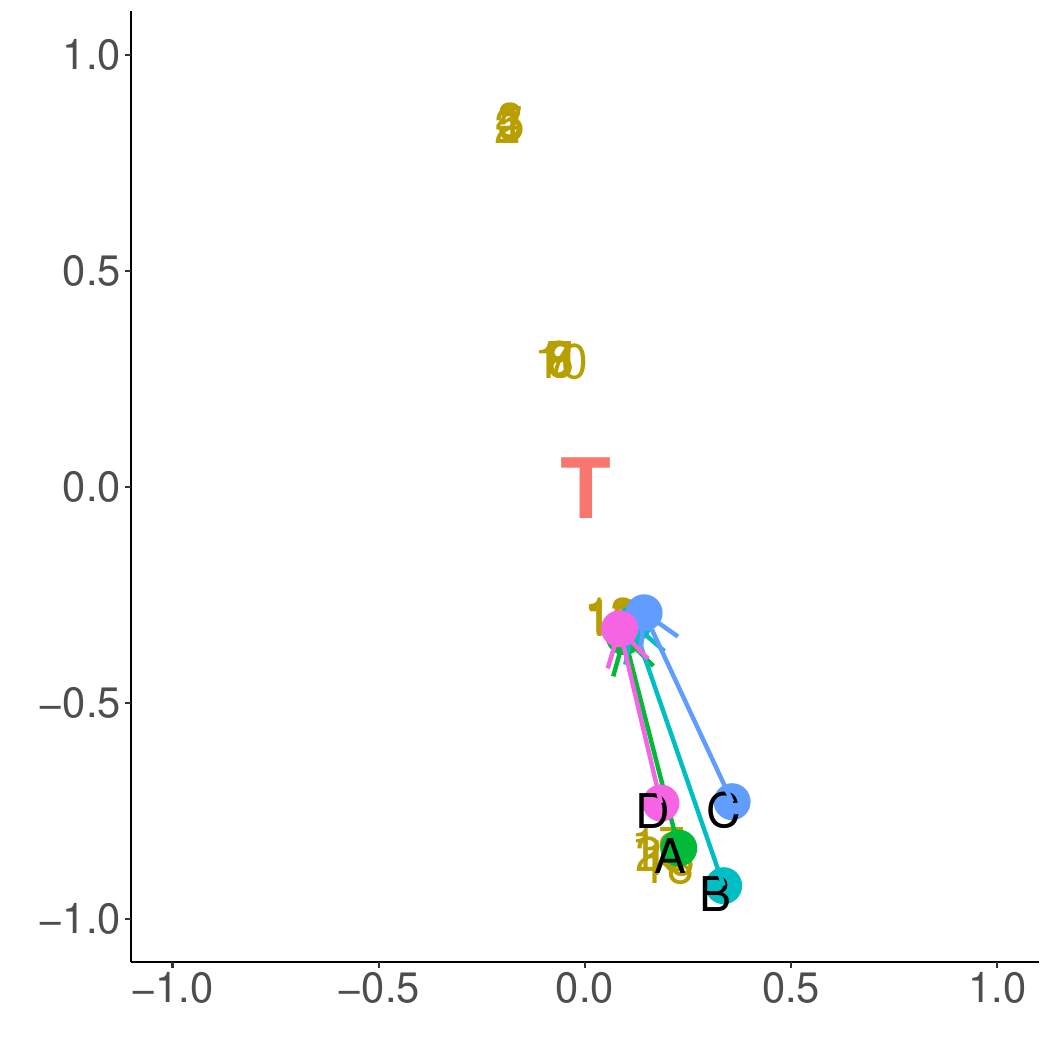} \\
       \end{tabular}
\caption{Latent process model: The estimated interaction maps $(\mathbb{M}, d)$ with $\mathbb{M} \coloneqq \mR^2$ show the progress of four selected members of groups G1 (left) and G2 (right) towards target $\mT$.
Members of group G1 provided correct responses to variables 1--5 at time $1$ and variables 1--10 at time $2$,
while members of G2 provided correct responses to variables 16--20 at time $1$ and variables 11--20 at time $2$.
The estimated interaction maps reveal these group-variable interactions.
}
    \label{fig:progress1}
\end{figure}

To demonstrate that the latent process model can capture interactions between individuals and variables while \citeauthor{andersen:85}'s (\citeyear{andersen:85}) model does not,
we consider $n = 200$ individuals divided into two groups G1 and G2 of the same size,
with the same progress but distinct responses to $p = 20$ variables.
Members of group G1 provide correct responses to variables 1--5 at time $1$ and variables 1--10 at time $2$,
while members of G2 provide correct responses to variables 16--20 at time $1$ and variables 11--20 at time $2$.
Both groups G1 and G2 make the same progress,
by increasing the number of correct responses from 5 to 10 between time $1$ and $2$,
but exhibit distinct response patterns.

\begin{figure}[htb]
    \centering 
    \begin{tabular}{cc} 
    G1 &  G2  \\
 \includegraphics[width=0.35 \textwidth]{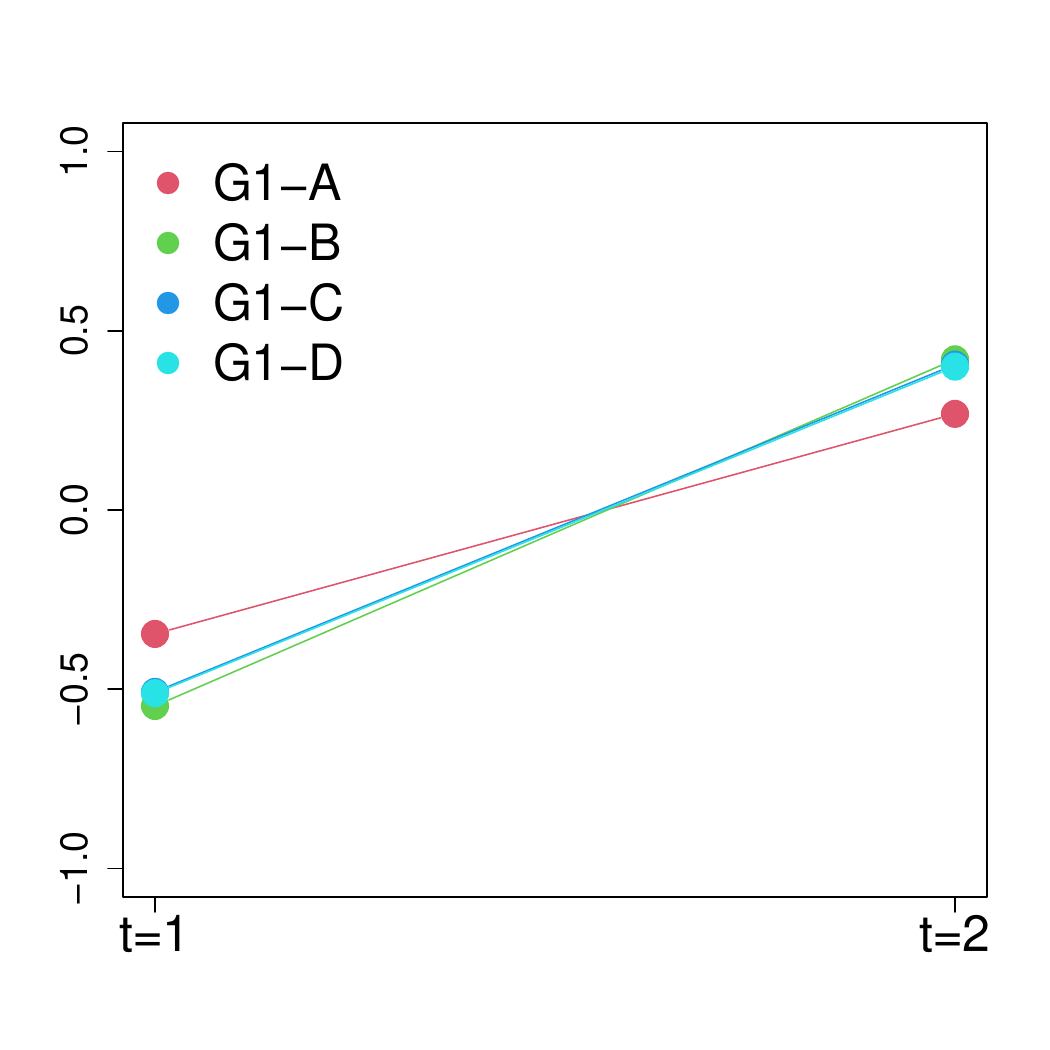}
  &   \includegraphics[width=0.35 \textwidth]{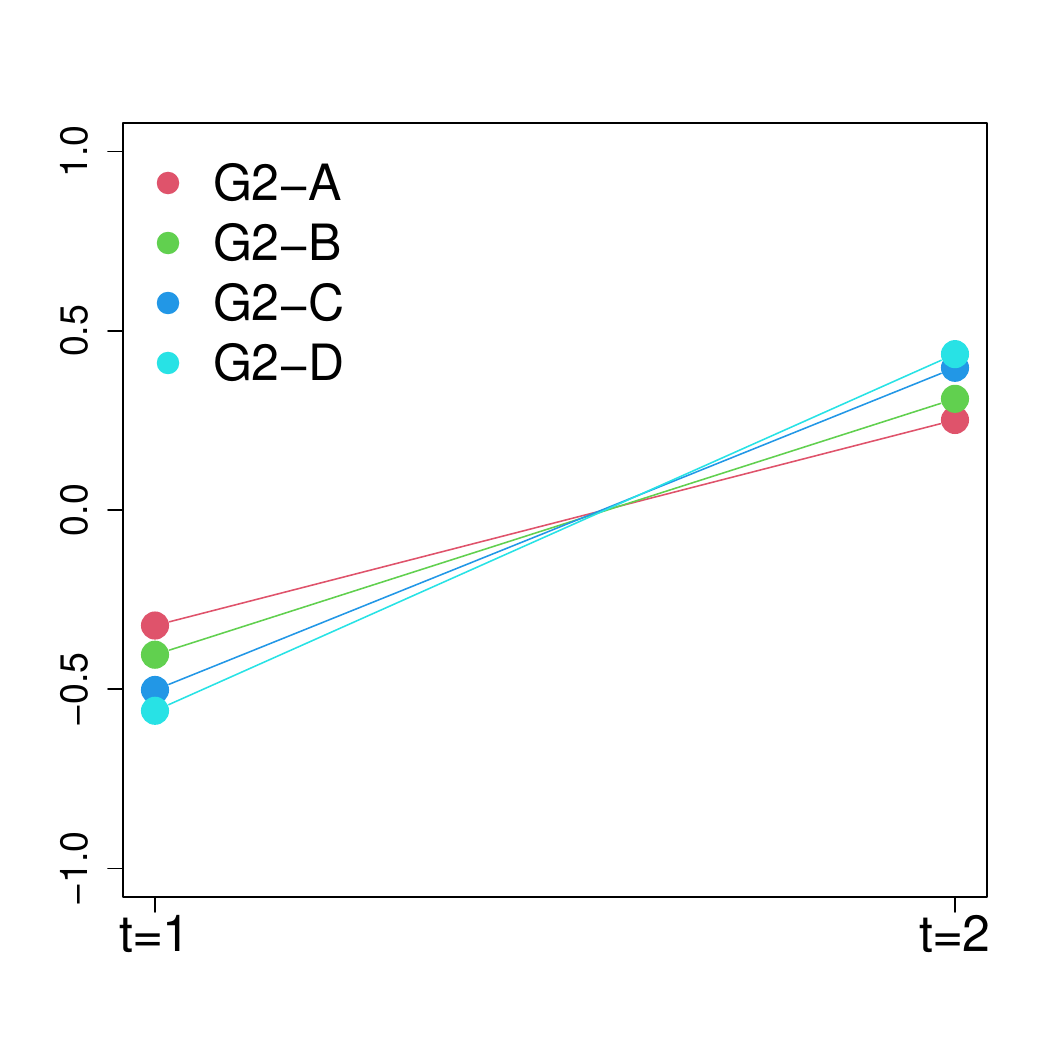} \\

       \end{tabular}
\caption{\citeauthor{andersen:85}'s (\citeyear{andersen:85}) model:
The estimated progress of selected members of groups G1 (left) and G2 (right),
which are identical to the members of groups G1 and G2 in Figure \ref{fig:progress1}.
The vertical axis represents posterior means of $\alpha_{i,t}$.
Members of group G1 provided correct responses to variables 1--5 at time $1$ and variables 1--10 at time $2$,
while members of G2 provided correct responses to variables 16--20 at time $1$ and variables 11--20 at time $2$.
The estimated progress maps reveal that all group members make similar progress,
but do not reveal the group-variable interactions.
}
    \label{fig:progress2}
\end{figure}

Figure \ref{fig:progress1} shows estimated interaction maps $\mathbb{M} \coloneqq \mR^2$ of four selected members of groups G1 (left) and G2 (right).
These interaction maps reveal that the members of group G1 are located in a neighborhood of variables 1--5 at time $1$,
while the members of group G2 are located in a neighborhood of variables 16--20 at time $1$.
The placements of groups G1 and G2 aligns with the response patterns of groups G1 and G2.

We compare the results in Figure \ref{fig:progress1} to the results based on \citeauthor{andersen:85}'s (\citeyear{andersen:85}) model.
The posterior means of abilities $\alpha_{i,t}$ of \citeauthor{andersen:85}'s (\citeyear{andersen:85}) model with $\logit(\mu_{i,j,t}) = \alpha_{i,t} + \beta_j$ are shown in Figure \ref{fig:progress2}.
Figure \ref{fig:progress2} suggests that the selected members of groups G1 and G2 make similar progress,
but obscures the fact that the response patterns of groups G1 and G2 are distinct.
In other words,
\citeauthor{andersen:85}'s (\citeyear{andersen:85}) model does not capture interactions between individuals and variables.

\section{Application: mental health}
\label{sec:depression}

We describe the motivating example introduced in Section \ref{example} in more detail. 
The motivating example is concerned with assessing the progress of vulnerable population members in terms of mental health.
We focus on a vulnerable population of special interest:
mothers with infants in low-income communities.

\subsection{Data}
\label{sec:mental.health.data}

\cite{santos:18} conducted between 2003 and 2010 large-scale randomized clinical trials in low-income communities in the U.S.\ states of North Carolina and New York \citep{beeber:14}. 
The data set consists of 306 low-income mothers with infants aged 6 weeks to 36 months,
enrolled in the Early Head Start Program in North Carolina or New York. 
The mental health of these mothers was assessed four times (baseline, 14 weeks, 22 weeks, and 26 weeks).
We focus on the $n = 257$ mothers who participated in the first two assessments. 
The Center for Epidemiological Studies Depression (CES-D) scale was used to measure depression of the mothers.
We focus on $p=10$ items:
1 (``bothered"), 
2 (``trouble in concentration"), 
3 (``feeling depressed"), 
4 (``effort"), 
5 (``not feeling hopeful"), 
6 (``failure"), 
7 (``not happy"), 
8 (``talking less"), 
9 (``people dislike"), 
and 10 (``get going"),
described in Supplement \ref{depression_items}. 
To measure progress toward the target of interest (improving mental health),
we recoded the $p=10$ items by assigning ``depression" $\mapsto$ 0 and ``no depression" $\mapsto$ 1.  
The proportions of positive responses (``no depression") range from .436 to .747 at the pre-assessment (median .630), 
and from .621 to .859 at the post-assessment (median .777). 



\subsection{Results}
\label{sec:dep_spike}

\begin{table}[tbh]
    \centering
    \begin{tabular}{l|cccr}
$\mathbb{M}$  & Minimum & Median & Maximum & Minimizer of WAIC\\
      \hline  
$\mathbb{M} \coloneqq \mR$ & 14650      &14713    & 14897 & 0\% \\ 
$\mathbb{M} \coloneqq \mR^2$ & 14432    & 14606  &  14679 & 0\%   \\
$\mathbb{M} \coloneqq \mR^3$ & 14351   & 14450    & 14509 & 100\% \\
$\mathbb{M} \coloneqq \mR^4$ & 14499   & 14623    & 14741 & 0\% \\
\hline
    \end{tabular}
\caption{Mental health: 
Watanabe–Akaike information criterion (WAIC) based on five Markov chain Monte Carlo runs.
The last column indicates how often the latent process model with $\mathbb{M} \coloneqq \mR$,\, $\mR^2$, $\mR^3$, $\mR^4$ minimized the WAIC. 
}
    \label{tab:waic_dep}
\end{table}
We assess the progress of $n = 257$ low-income mothers towards the target of interest $\mT$ (improving mental health),
measured by $p = 10$ items.
The priors and Markov chain Monte Carlo algorithm can be found in Supplement \ref{appendix:estimation}.
To detect signs of non-convergence,
we used trace plots along with the multivariate Gelman-Rubin potential scale reduction factor of \citet{vats:18a}.
These convergence diagnostics are reported in Supplement \ref{appendix:convergence} and do not reveal signs of non-convergence.

Table \ref{tab:waic_dep} shows the Watanabe–Akaike information criterion \citep{waic} based on the latent process model with $\mathbb{M} \coloneqq \mR^q$ and $q \in \{1, 2, 3, 4\}$,
suggesting that $\mathbb{M} \coloneqq \mR^3$ is most appropriate.

\begin{figure}[htbp]
    \centering
\begin{tabular}{ccc}
$\mathbb{M} \coloneqq \mR$ & $\mathbb{M} \coloneqq \mR^2$ & $\mathbb{M} \coloneqq \mR^3$ \\ 
     \includegraphics[width=.31\textwidth]{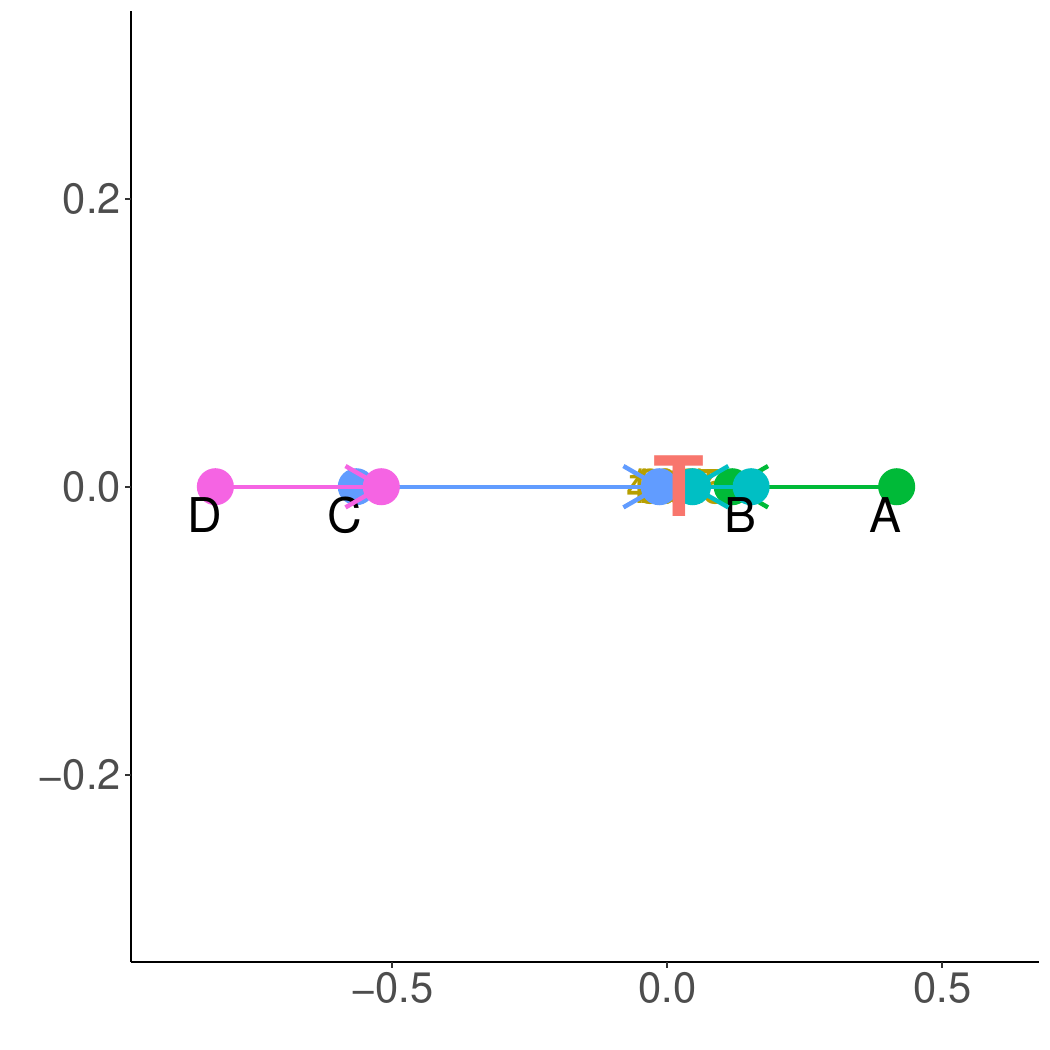} 
     &  \includegraphics[width=.31 \textwidth]{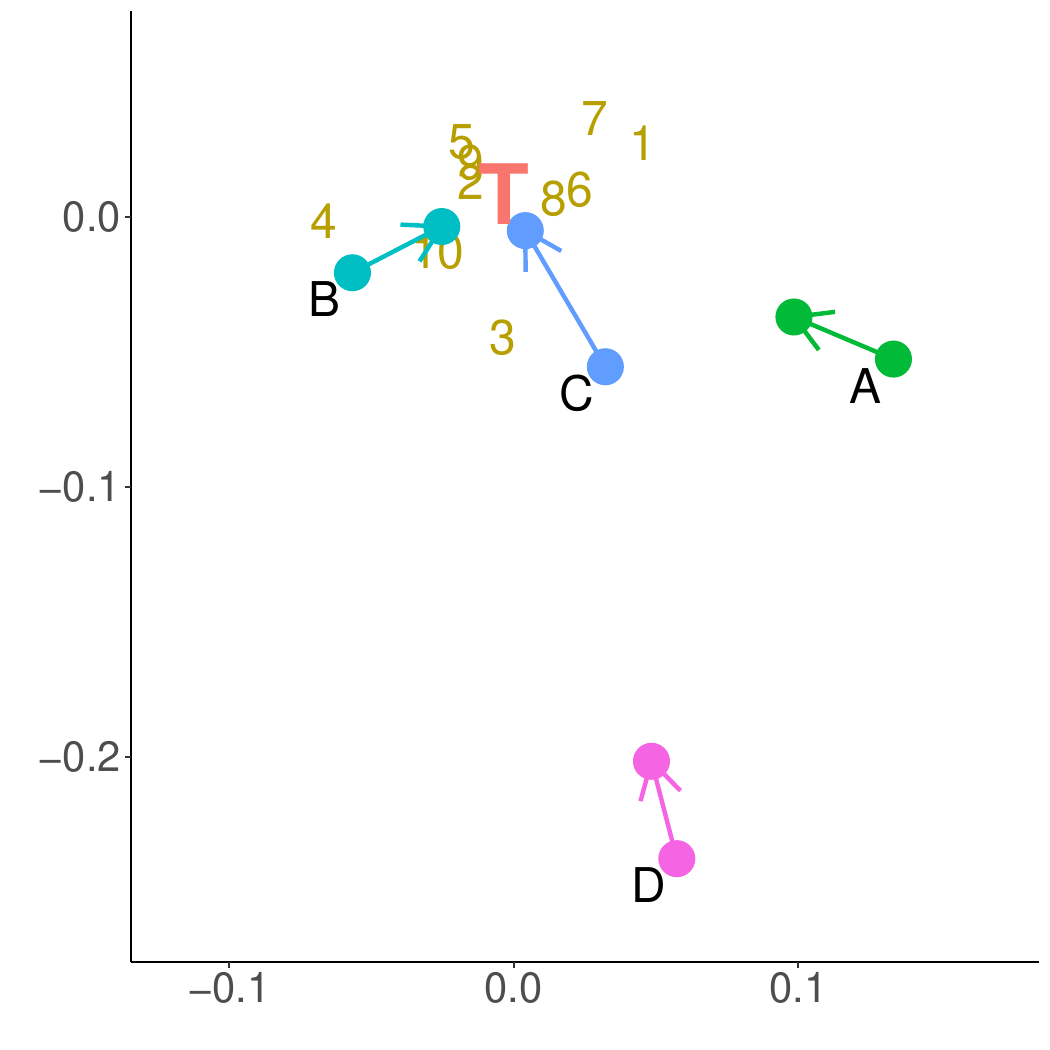} 
     &  \includegraphics[width=.31 \textwidth]{figure/dep_one_4_poly_spike_3d_1_sharp_v2.png}  
     \\
\end{tabular}
\caption{Mental health: 
Interaction maps $(\mathbb{M},\, d)$ with $\mathbb{M} \coloneqq \mR^q$ and $q \in \{1, 2, 3\}$ show the progress of low-income mothers $A$, $B$, $C$, $D$ towards target $\mT$ (improving mental health),
measured by items $1, \dots, 10$ (questions about depression).
The positions of mothers and items are estimated by posterior medians.
While the one-dimensional interaction map ($q=1$) suggests that all items are close to target $\mT$, 
the three-dimensional interaction map ($q = 3$) reveals interactions between mothers and items:
e.g.,
item $5$ deviates from the bulk of the items,
and mother $B$ is closest to item $5$.
It turns out that mother $B$ agreed with item $5$ at the first assessment (``feeling hopeful"),
whereas mothers $A$, $C$, $D$ did not.
In addition,
the interaction map suggests that mothers $B$ and $C$ have made strides towards improving mental health,
whereas mothers $A$ and $D$ may need to make more progress.
}
\label{fg:dep_progress2}
\end{figure}

\begin{figure}[htbp]
    \centering
\begin{tabular}{cccc}
$\mathbb{M} \coloneqq \mR$ & $\mathbb{M} \coloneqq \mR^2$ & $\mathbb{M} \coloneqq \mR^3$\\ 
        \includegraphics[width=.31 \textwidth]{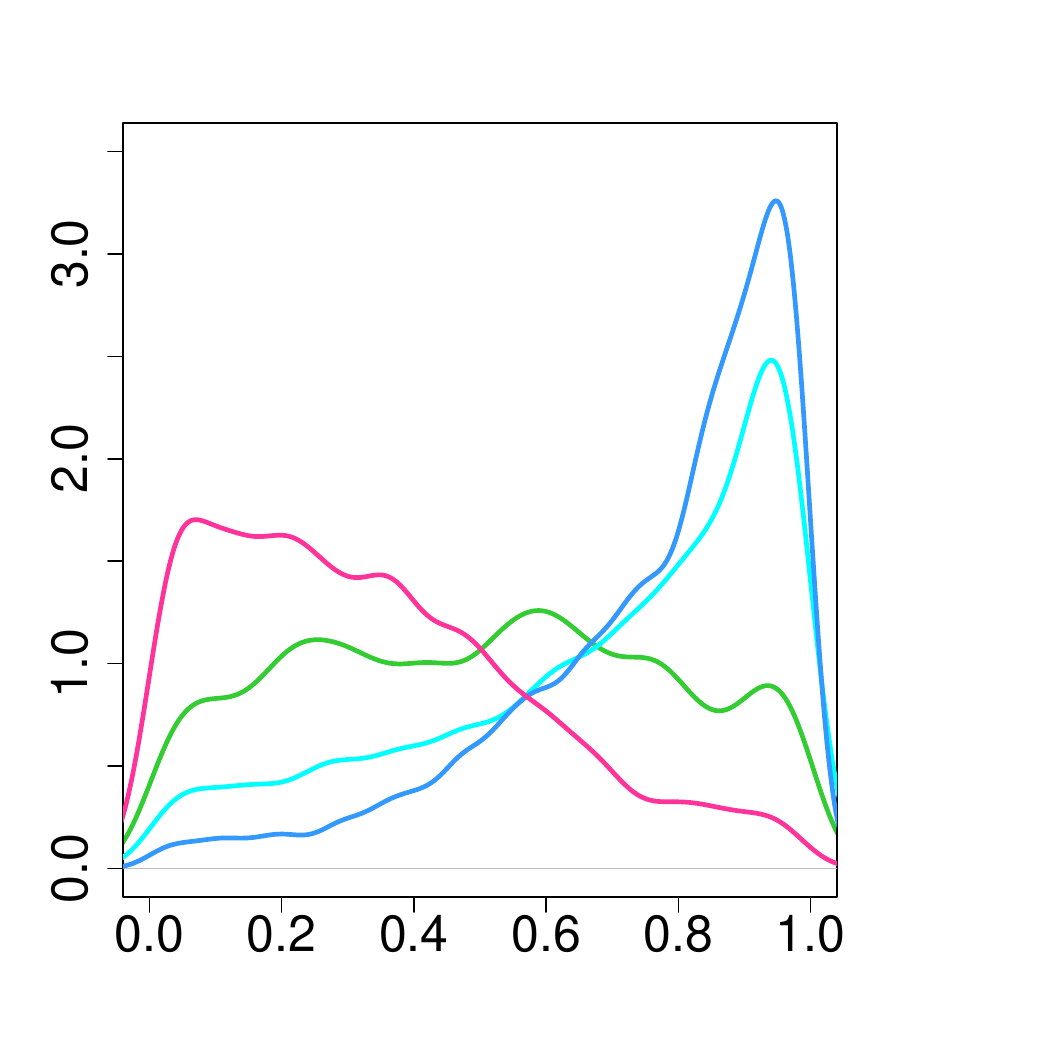}
     &  \includegraphics[width=.31 \textwidth]{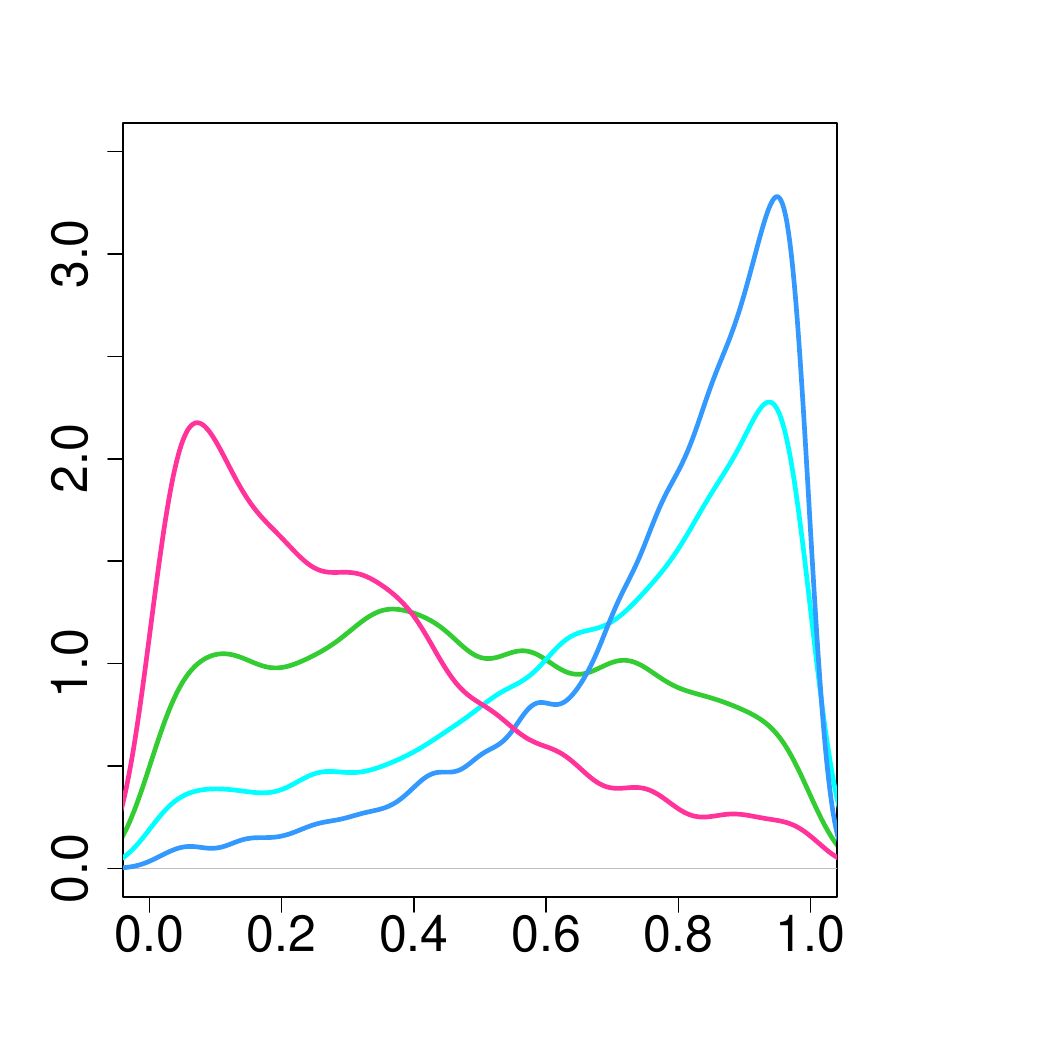}
     &  \includegraphics[width=.31 \textwidth]{figure/dep_kernel_poly_spike_3d_1_sharp.pdf}
     \\
\end{tabular}
\caption{Mental health:
Marginal posteriors of the rates of progress $\lambda_A$, $\lambda_B$, $\lambda_C$, $\lambda_D$ of mothers $A$, $B$, $C$, $D$ towards target $\mT$ (improving mental health).
The progress of mother $A$ is unclear,
but the marginal posteriors of the rates of progress $\lambda_B$ and $\lambda_C$ of mothers $B$ and $C$ have modes close to $1$,
confirming that mothers $B$ and $C$ have made strides towards improving mental health.
By contrast,
the mode of the marginal posterior of the rate of progress $\lambda_D$ of mother $D$ is close to $0$,
suggesting that mother $D$ may need additional assistance.
}
\label{fg:dep_kernel2}
\end{figure}
We present interaction maps in Figure \ref{fg:dep_progress2}.
While the one-dimensional interaction map suggests that all items are close to target $\mT$, 
the three-dimensional interaction map (which, according to the Watanabe–Akaike information criterion, is more appropriate than the one-dimensional interaction map) reveals that there are interactions between individuals (mothers) and items (questions about depression):
e.g.,
item $5$ deviates from the bulk of the items,
and mother $B$ is closest to item $5$.
It turns out that mother $B$ agreed with item $5$ at the first assessment (``feeling hopeful"),
whereas mothers $A$, $C$, and $D$ did not.
In addition,
the interaction map suggests that mothers $B$ and $C$ have made strides towards improving mental health,
whereas mothers $A$ and $D$ may need to make more progress in the future.

The posterior summaries in Table \ref{tab:dep_lambda} confirms that,
more likely than not,
mothers $B$ and $C$ have made progress,
while mother $D$ has not.
To gain more insight into the uncertainty about the progress of mothers $A$, $B$, $C$, and $D$, 
we present the marginal posteriors of $\lambda_A$,
$\lambda_B$,
$\lambda_C$,
and $\lambda_D$ in Figure \ref{fg:dep_kernel2}.
The marginal posteriors of $\lambda_A$,
$\lambda_B$,
$\lambda_C$,
and $\lambda_D$ caution that there is non-negligible uncertainty about the progress of some of the mothers.
A case in point is mother $A$:
The marginal posterior of mother $A$'s rate of progress $\lambda_A$ resembles the Uniform$(0, 1)$ distribution.
As a consequence, 
it is unclear whether mother $A$ has made progress,
and how much.
By contrast,
the marginal posteriors of the rates of progress $\lambda_B$ and $\lambda_C$ of mothers $B$ and $C$ have modes close to $1$,
suggesting that mothers $B$ and $C$ have made strides towards improving mental health.
The marginal posterior of the rate of progress $\lambda_D$ of mother $D$ has a mode close to $0$,
which underscores that mother $D$ may need additional assistance.

\begin{figure}[htbp]
    \centering
\begin{tabular}{cccc}
$\mathbb{M} \coloneqq \mR$ & $\mathbb{M} \coloneqq \mR^2$ & $\mathbb{M} \coloneqq \mR^3$\\ 
     \includegraphics[width=.31\textwidth]{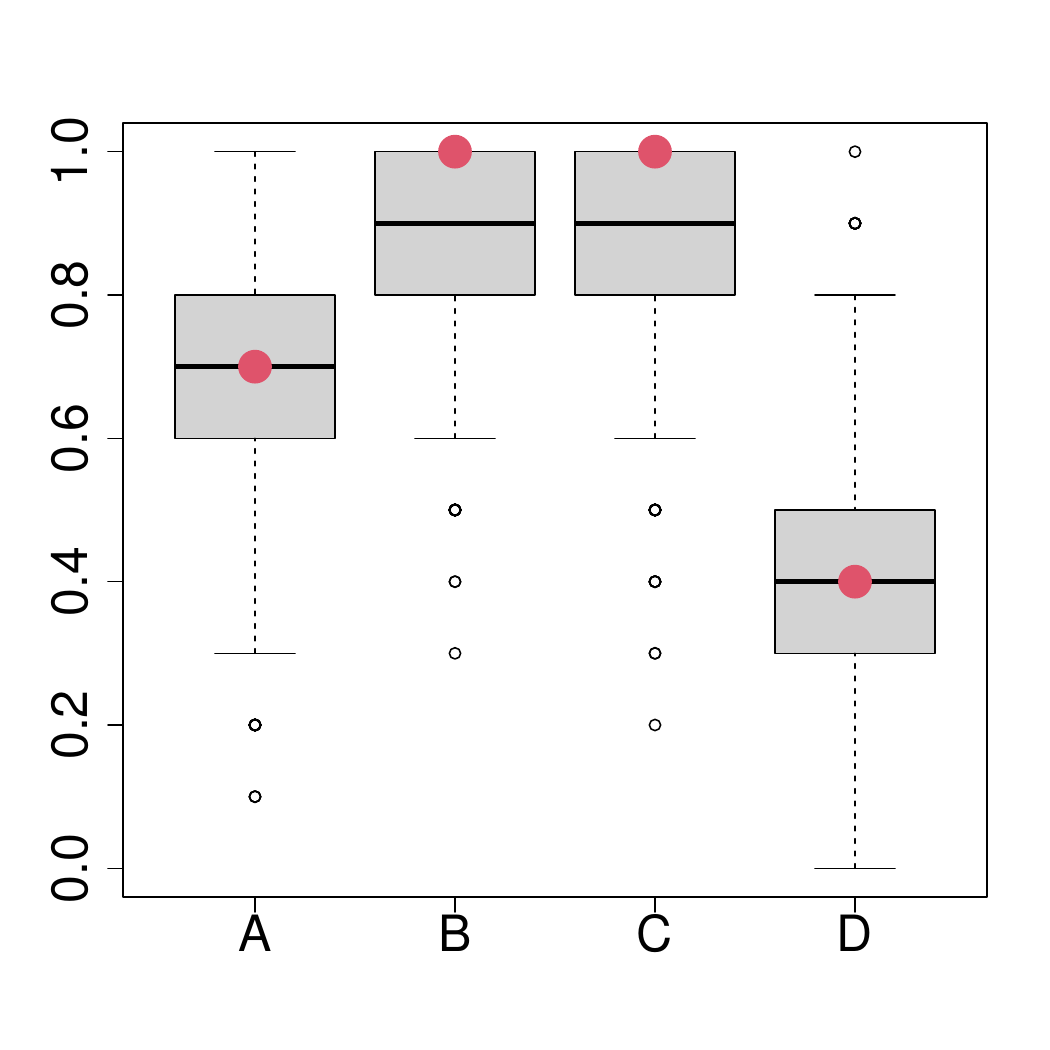} 
     &  \includegraphics[width=.31 \textwidth]{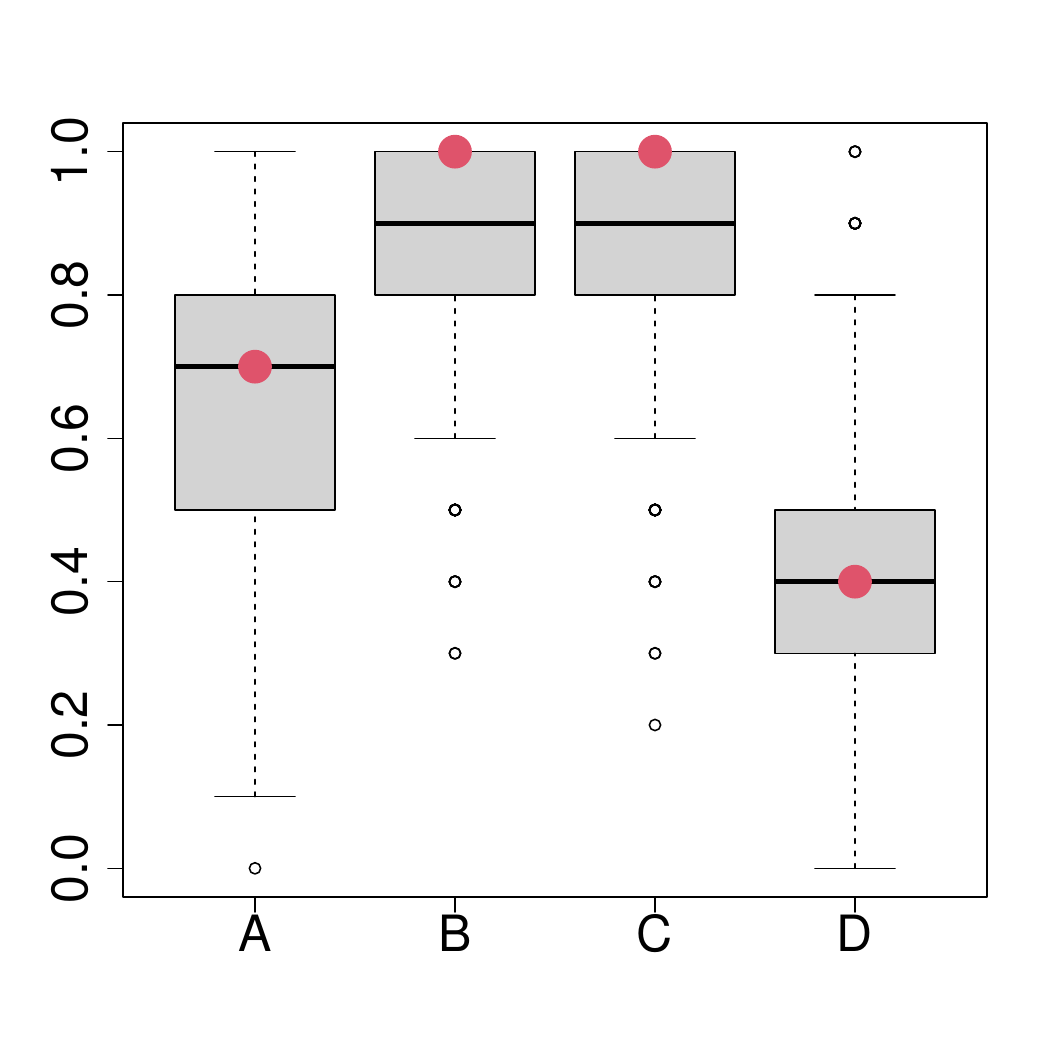} 
     &  \includegraphics[width=.31 \textwidth]{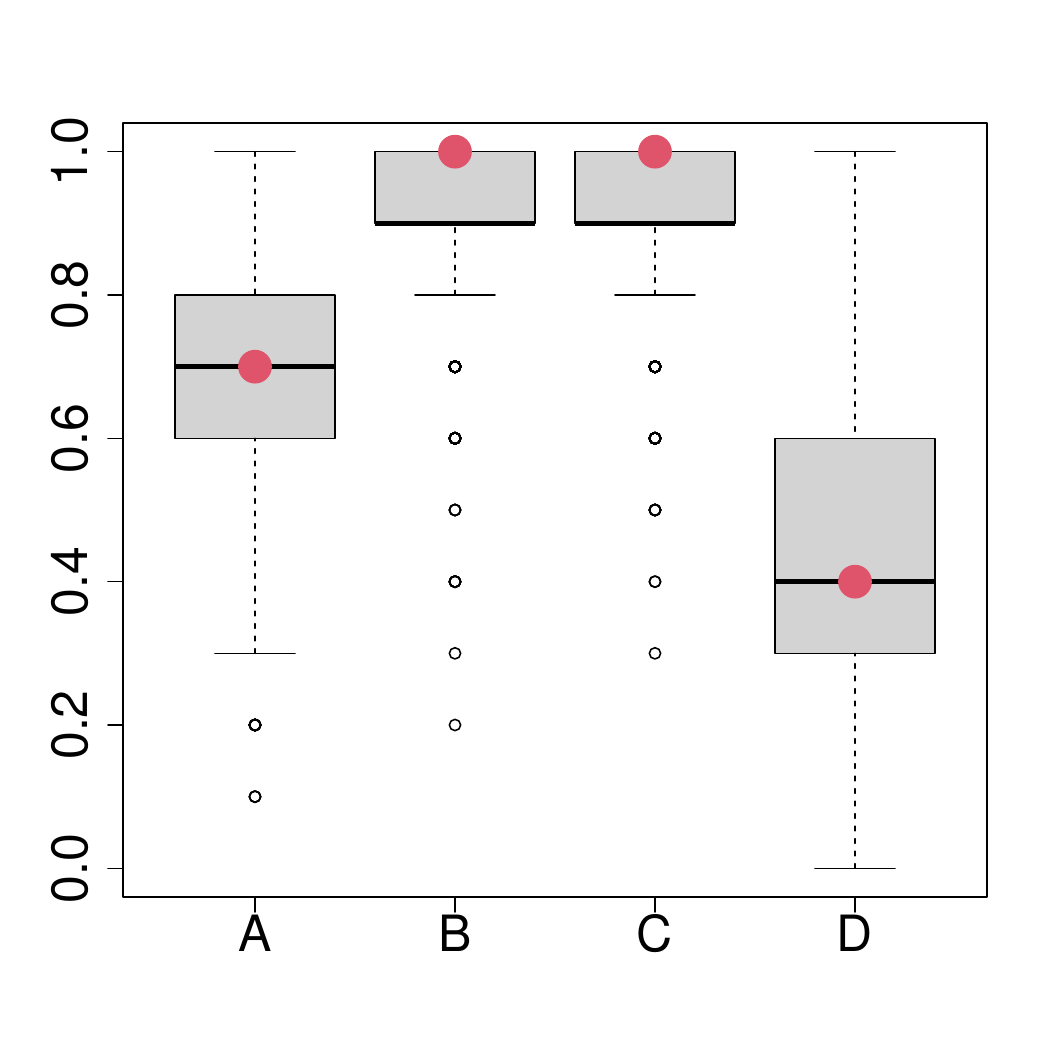}  
     \\
\end{tabular}
\caption{Mental health: 
Posterior predictions of the proportions of positive responses by mothers $A$, $B$, $C$, $D$ at the second assessment.
The observed proportions are indicated by red circles.
}
\label{fg:dep_ppc}
\end{figure}

To assess the goodness-of-fit of the model,
we generated 1,000 posterior predictions of the proportions of positive responses at the second assessment.
Figure \ref{fg:dep_ppc} shows the proportions of predicted and observed positive responses. 
By and large, 
the model predictions agree with the observations.

\begin{table}[tbh]
\begin{tabular}{c cccc}
 & 10\% Percentile & Median & 90\% Percentile & Probability: Progress? \\ 
  \hline
 $\mathbb{M} \coloneqq \mR$:  &&&& \\   
  $\lambda_A$ & .060 & .290 & .790 & .550 \\ 
  $\lambda_B$ & .085 & .578 & .950 & .707 \\ 
  $\lambda_C$ & .213 & .775 & .965 & .847 \\ 
  $\lambda_D$ & .041 & .179 & .517 & .415 \\  
    \hline
 $\mathbb{M} \coloneqq \mR^2$:  &&&& \\  
  $\lambda_A$ & .055 & .255 & .757 & .518 \\ 
  $\lambda_B$ & .092 & .590 & .946 & .718 \\ 
  $\lambda_C$ & .278 & .790 & .965 & .866 \\ 
  $\lambda_D$ & .039 & .147 & .463 & .377 \\ 
      \hline
 $\mathbb{M} \coloneqq \mR^3$:  &&&& \\ 
  $\lambda_A$ & .058 & .258 & .806 & .527 \\ 
  $\lambda_B$ & .090 & .599 & .944 & .718 \\ 
  $\lambda_C$ & .280 & .774 & .969 & .861 \\ 
  $\lambda_D$ & .038 & .148 & .464 & .385 \\ 
      \hline
\end{tabular}
\caption{Mental health: 
Marginal posteriors of the rates of progress $\lambda_A$, $\lambda_B$, $\lambda_C$, $\lambda_D$ of mothers $A$, $B$, $C$, $D$.
The last column (``Probability: Progress?") shows the posterior probability of non-negligible progress.
}
\label{tab:dep_lambda}
\end{table}

\section{Application: online educational assessments}
\label{sec:math}

We present an application to online educational assessments, 
using the My Math Academy data \citep{bang:22}.

\begin{figure}[hptb]
\centering
\includegraphics[scale = .325]{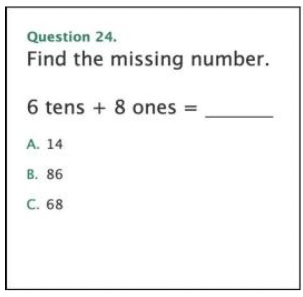}
\includegraphics[scale = .325]{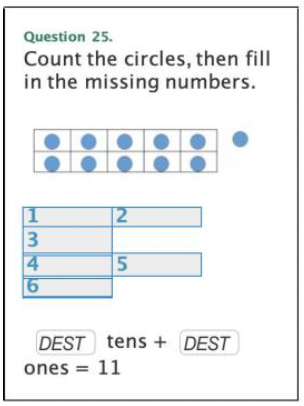}
\includegraphics[scale = .325]{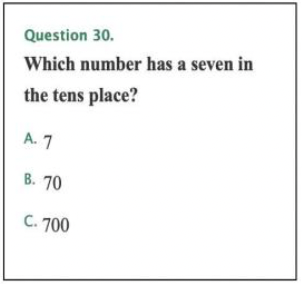}
\\
\caption{\label{fig:math_academy0}
My Math Academy: 
Selected questions 24, 25, 30 corresponding to items 6, 7, 8.
Each question has a single correct response.
Items 6, 7, 8 are indicators of whether students choose the correct response to questions 24, 25, 30.
}
\end{figure}

\subsection{Data}

The My Math Academy data set consists of 479 kindergarten children, first- and second-grade students who participated in the My Math Academy effectiveness study in 2019 \citep{bang:22}. 
The students worked on pre- and post-assessments before and after working on the online learning platform My Math Academy. 
The pre- and post-assessments were separated by three months,
and both included a common set of 31 problems.
We focus on $p = 8$ items measuring numerical understanding. 
Three selected items are presented in Figure \ref{fig:math_academy0}.
We exclude 58 of the 479 children who had no responses at either the pre- or post-assessment or who reported no background information. 
55\% of the remaining $n = 421$ students are kindergarten children,
48\% are female, 
87\% are low-income students eligible to receive free lunches, 
87\% are Hispanic, 
and 6\% are African-American. 
The proportions of correct responses range from .114 to .556 at the pre-assessment (median .397),
and from .216 to .755 at the post-assessment (median .544).

\subsection{Results}
\label{sec:math_spike}

\begin{table}[tbh]
    \centering
    \begin{tabular}{l|cccr}
$\mathbb{M}$  & Minimum & Median & Maximum & Minimizer of WAIC\\
      \hline  
$\mathbb{M} \coloneqq \mR$ & 20209 & 20310  & 20597  & 0\% \\ 
$\mathbb{M} \coloneqq \mR^2$ &  19838  &  20035  & 20077 & 60\%   \\
$\mathbb{M} \coloneqq \mR^3$ &  19841   & 20032  & 20071 & 40\%  \\
$\mathbb{M} \coloneqq \mR^4$ &   20166  & 20436  &  20678 & 0\%  \\
\hline
    \end{tabular}
\caption{My Math Academy: 
Watanabe–Akaike information criterion (WAIC) based on five Markov chain Monte Carlo runs.
The last column indicates how often the latent process model with $\mathbb{M} \coloneqq \mR$,\, $\mR^2$, $\mR^3$, $\mR^4$ minimized the WAIC. 
}
    \label{tab:waic_math}
\end{table}

We assess the progress of $n = 421$ students towards the target of interest (numerical understanding),
measured by $p = 8$ items.
The prior and Markov chain Monte Carlo algorithm are described in Supplement \ref{appendix:estimation}.
To detect signs of non-convergence,
we used trace plots along with the multivariate Gelman-Rubin potential scale reduction factor of \citet{vats:18a}.
These convergence diagnostics can be found in Supplement \ref{appendix:convergence} and do not reveal signs of non-convergence.

Table \ref{tab:waic_math} shows the Watanabe–Akaike information criterion \citep{waic} based on the latent process model with $\mathbb{M} \coloneqq \mR^q$ and $q \in \{1, 2, 3, 4\}$.
The Watanabe–Akaike information criterion suggests that $\mathbb{M} \coloneqq \mR^2$ or $\mR^3$ is most appropriate.

\begin{figure}[htbp]
    \centering
\begin{tabular}{ccc}
$\mathbb{M} \coloneqq \mR$ & $\mathbb{M} \coloneqq \mR^2$ & $\mathbb{M} \coloneqq \mR^3$\\ 
     \includegraphics[width=.31\textwidth]{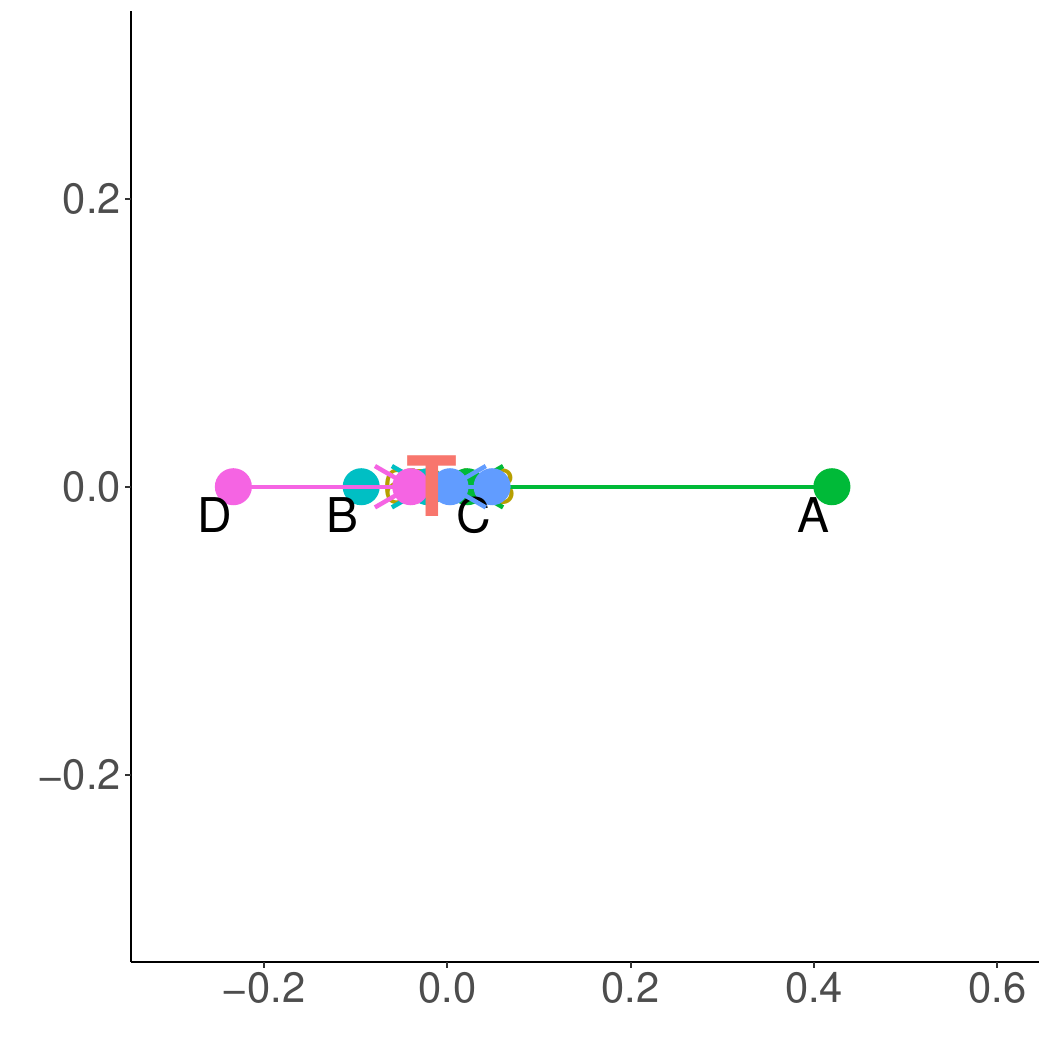} 
     &  \includegraphics[width=.31 \textwidth]{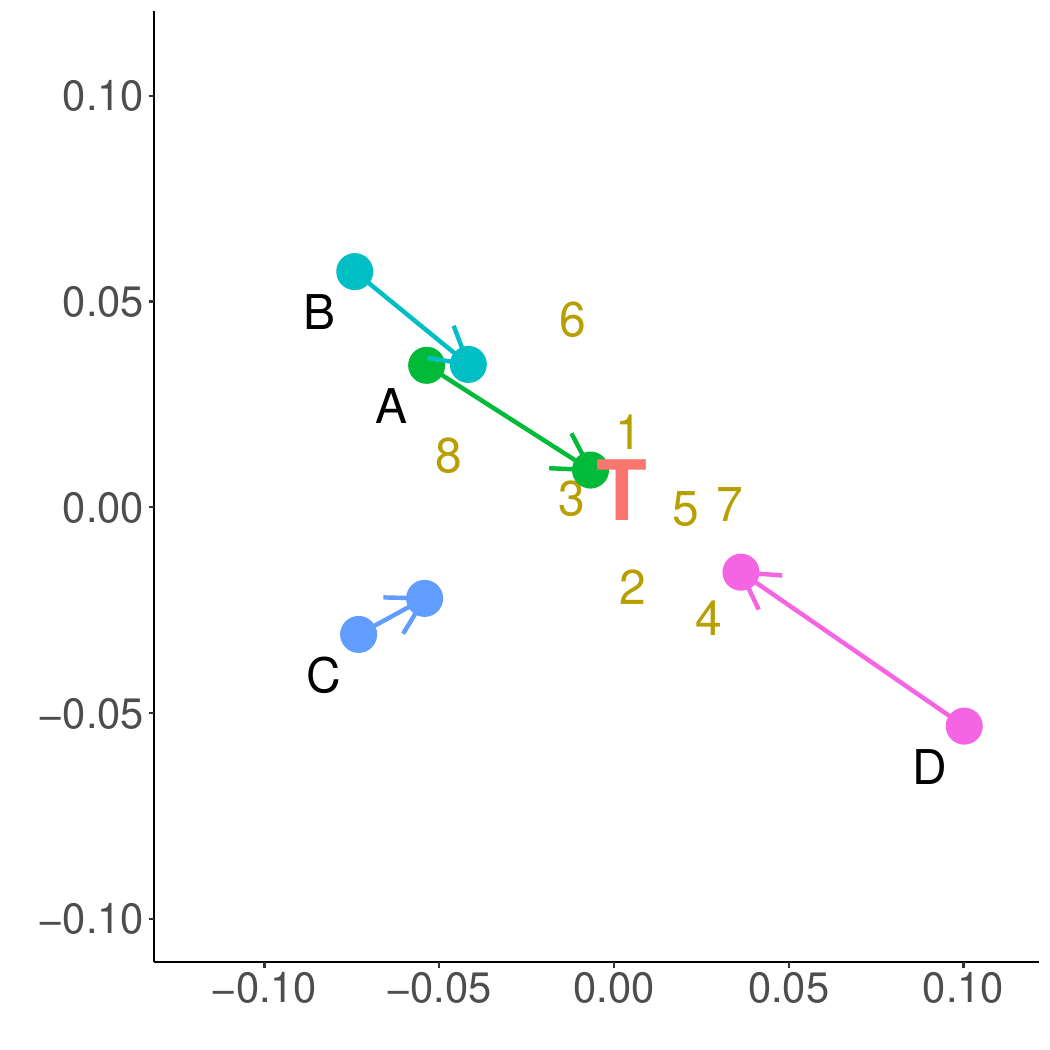} 
     &  \includegraphics[width=.31 \textwidth]{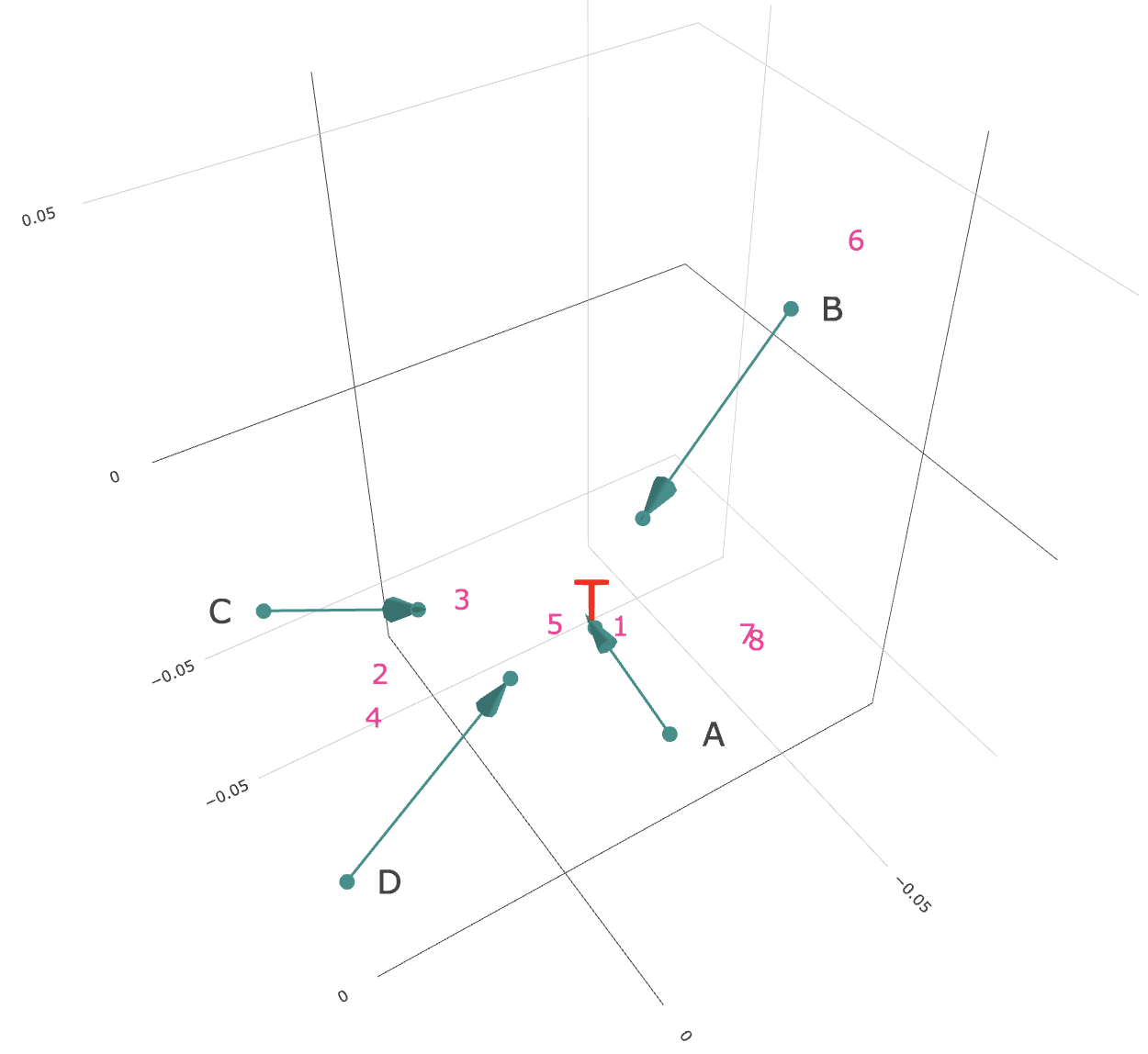}  
     \\
\end{tabular}
\caption{My Math Academy: 
Interaction maps $(\mathbb{M},\, d)$ with $\mathbb{M} \coloneqq \mR^q$ and $q \in \{1, 2, 3\}$ show the progress of Hispanic students $A$, $B$, $C$, $D$ in low-income families towards learning target $\mT$ (numerical understanding),
measured by items $1, \dots, 8$. 
The positions of students and items are estimated by posterior medians.
While the one-dimensional interaction map ($q = 1$) suggests that all items are close to learning target $\mT$,
the two- and three-dimensional interaction maps ($q \in \{2, 3\}$) indicate that some of the items deviate from the bulk of the items:
e.g.,
item 6 is the most abstract item among the counting-related items 5, 6, 7 and the three-dimensional interaction map reveal that item 6 deviates from learning target $\mT$ more than any other item.
Student $B$ is closest to item 6 and provided a correct response to item 6 at the pre-assessment,
whereas students $A$, $C$, $D$ did not.
}
\label{fg:math_progress}
\end{figure}

Figure \ref{fg:math_progress} shows interaction maps,
focusing on selected Hispanic students $A$, $B$, $C$, and $D$ in low-income families.
While the one-dimensional interaction map suggests that all items are close to learning target $\mT$,
the two- and three-dimensional interaction maps (which, according to the Watanabe–Akaike information criterion, are more appropriate than the one-dimensional interaction map) indicate that some of the items deviate from the bulk of the items:
e.g.,
item 6 is the most abstract item among the counting-related items 5, 6, and 7 and \alert{the three-dimensional interaction map reveals that item 6 deviates from learning target $\mT$ more than any other item.
}
Student $B$ is closest to item 6 and provided a correct response to item 6 at the pre-assessment,
whereas students $A$, $C$, and $D$ did not.

\begin{figure}[htbp]
    \centering
\begin{tabular}{cccc}
$\mathbb{M} \coloneqq \mR$ & $\mathbb{M} \coloneqq \mR^2$ & $\mathbb{M} \coloneqq \mR^3$\\ 
     \includegraphics[width=.31\textwidth]{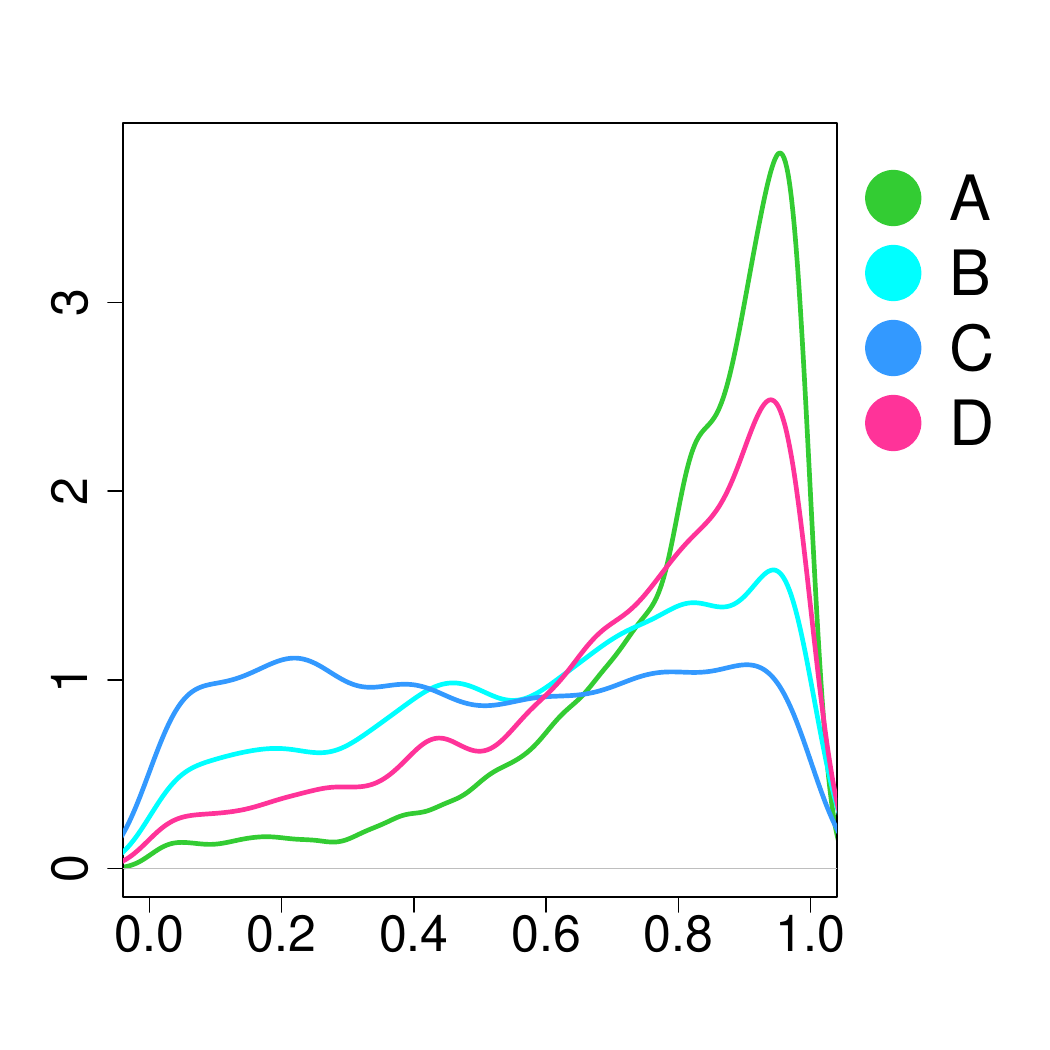} 
     &  \includegraphics[width=.31 \textwidth]{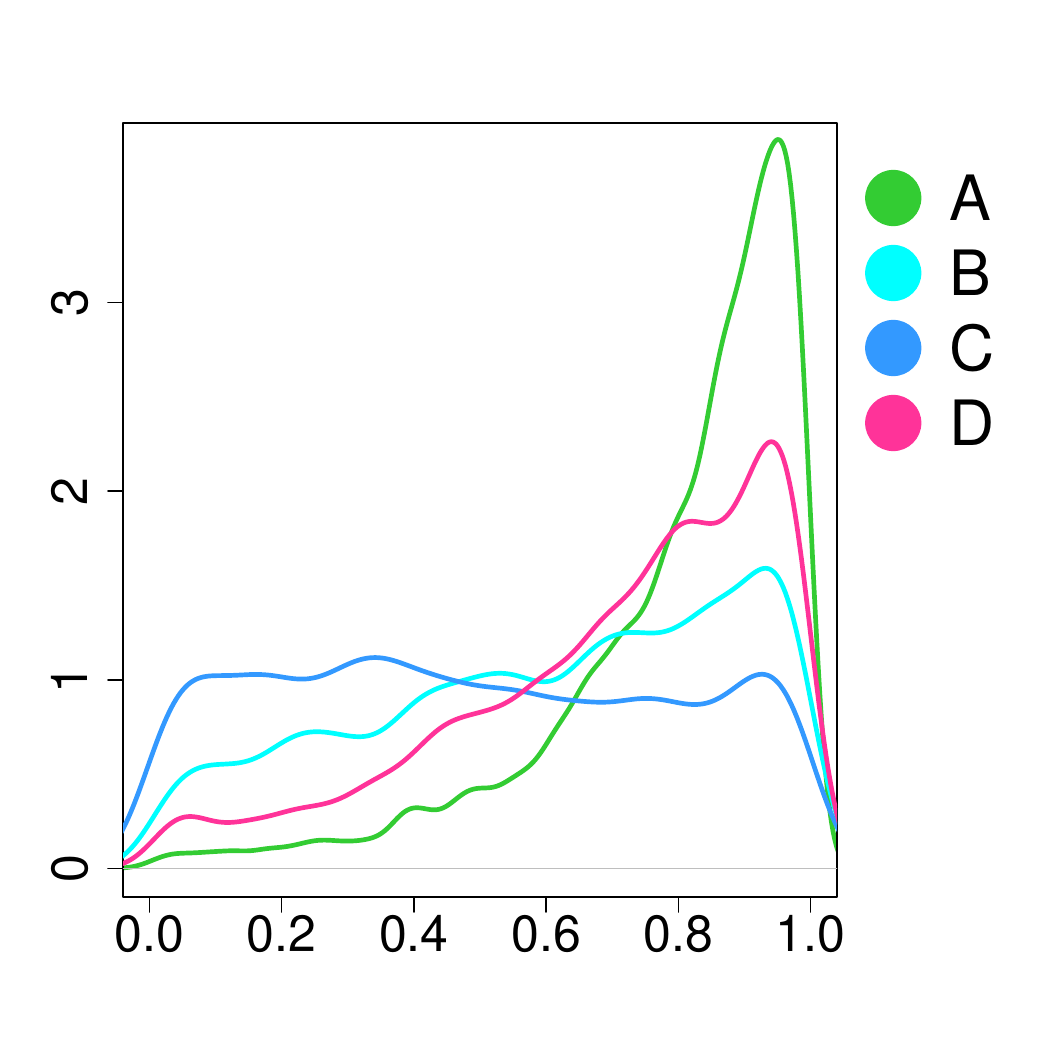} 
     &  \includegraphics[width=.31 \textwidth]{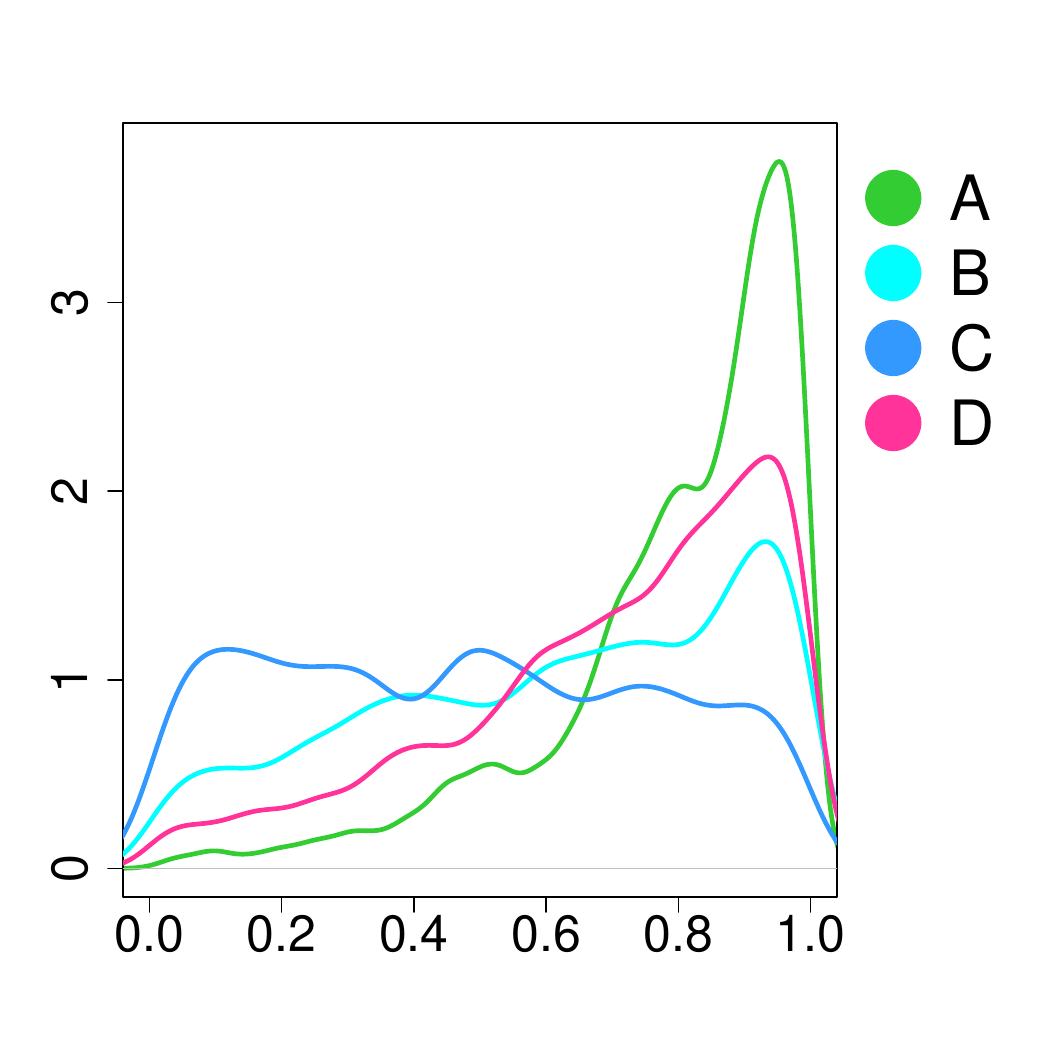}  
     \\
\end{tabular}
\caption{My Math Academy:
Marginal posteriors of the rates of progress $\lambda_A$, $\lambda_B$, $\lambda_C$, $\lambda_D$ of Hispanic students $A$, $B$, $C$,$D$ towards learning target $\mT$ (numerical understanding).
}
\label{fg:math_kernel}
\end{figure}

Posterior summaries of the rates of progress $\lambda_A$, $\lambda_B$, $\lambda_C$, and $\lambda_D$ of students $A$, $B$, $C$, and $D$ are shown in Table \ref{tab:math_lambda}.
These posterior summaries confirm that,
with high posterior probability,
students $A$ and $D$ have made non-negligible progress,
but there is more uncertainty about students $B$ and $C$. 
To gain more insight into the uncertainty about the progress of students $A$, $B$, $C$, and $D$,
we present the marginal posteriors of the rates of progress $\lambda_A$, $\lambda_B$, $\lambda_C$, and $\lambda_D$ of students $A$, $B$, $C$, and $D$ in Figure \ref{fg:math_kernel}.
The marginal posteriors reveal non-negligible uncertainty:
e.g.,
the marginal posteriors of the rates of progress $\lambda_A$ and $\lambda_D$ of students $A$ and $D$ have modes close to $1$,
but both of them have long tails.
Assessing the progress of students $B$ and $C$ is harder still.

\begin{figure}[htbp]
    \centering
\begin{tabular}{cccc}
$\mathbb{M} \coloneqq \mR$ & $\mathbb{M} \coloneqq \mR^2$ & $\mathbb{M} \coloneqq \mR^3$\\ 
     \includegraphics[width=.31\textwidth]{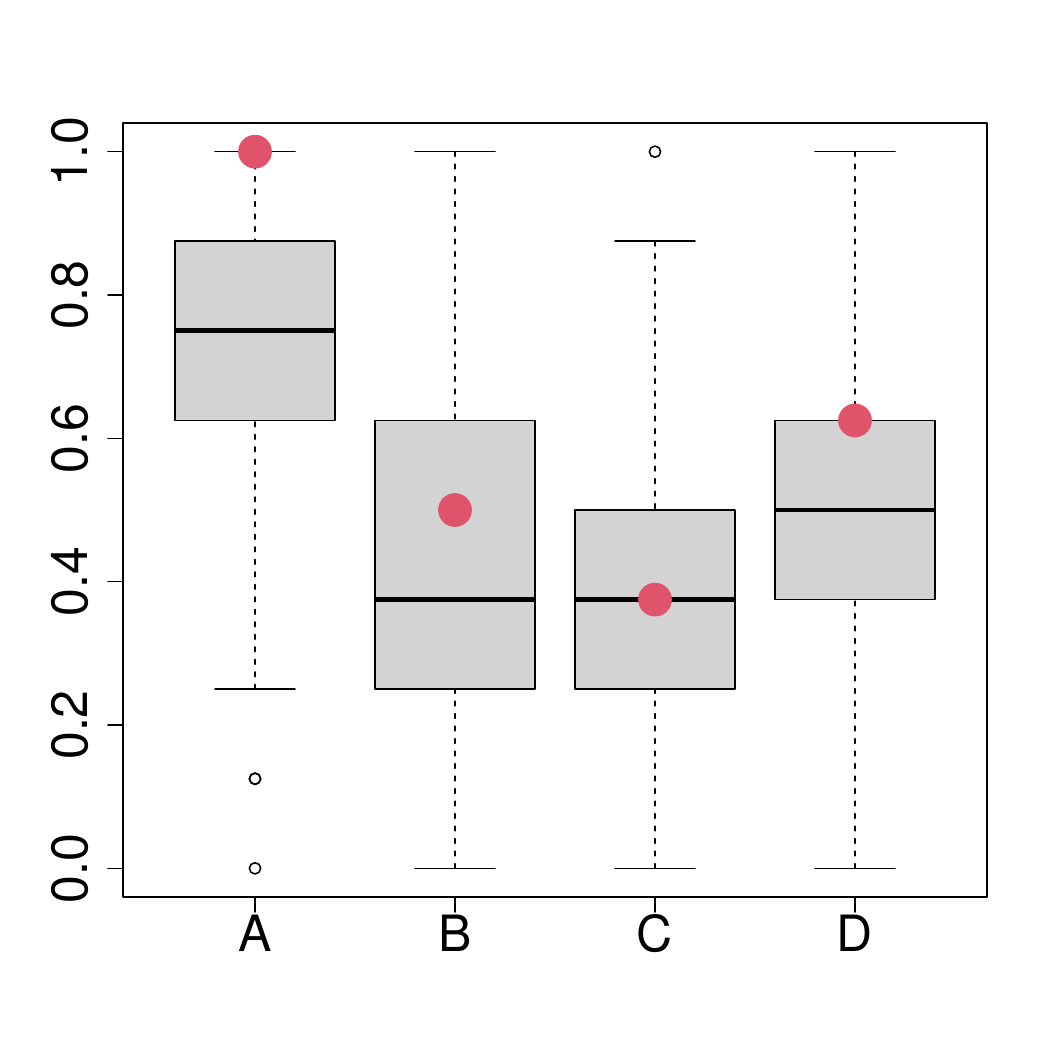} 
     &  \includegraphics[width=.31 \textwidth]{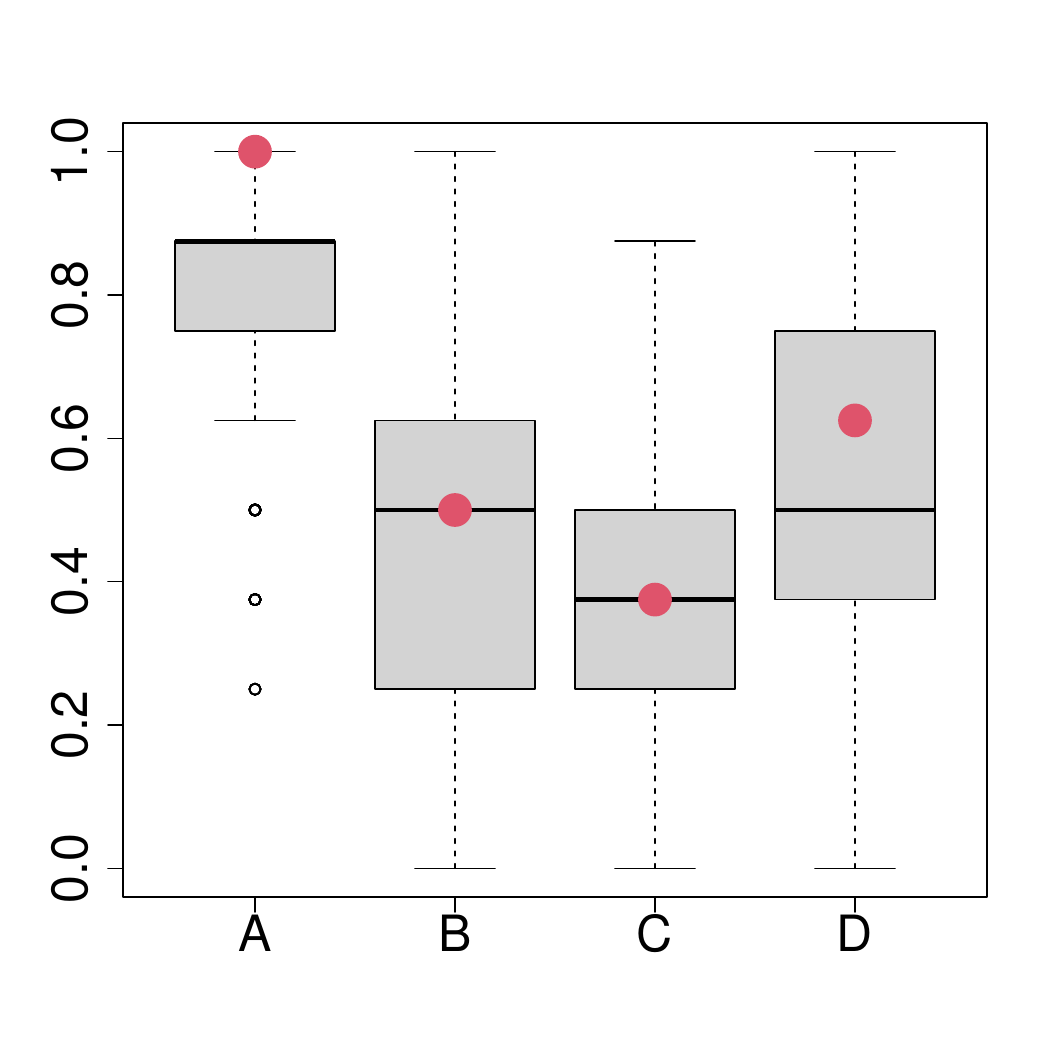} 
     &  \includegraphics[width=.31 \textwidth]{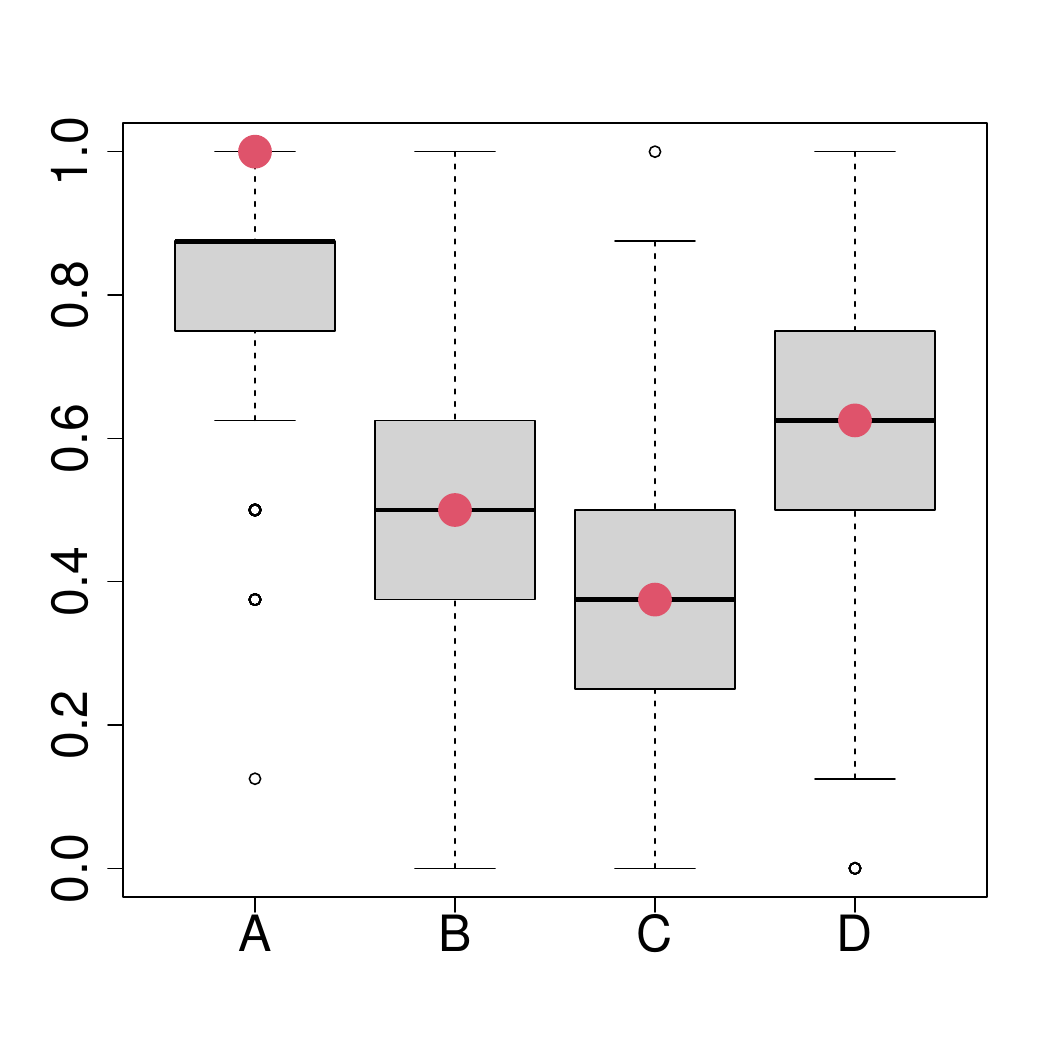}  
     \\
\end{tabular}
\caption{My Math Academy: 
Posterior predictions of the proportions of correct responses by Hispanic students $A$, $B$, $C$, $D$ at the post-assessment.
The observed proportions are indicated by red circles.
}
\label{fg:math_ppc}
\end{figure}

To assess the goodness-of-fit of the model,
we generated 1,000 posterior predictions of the proportions of correct responses at the post-assessment.
Figure \ref{fg:math_ppc} suggests that the posterior predictions by and large match the observed proportions of correct responses.

\begin{table}[ht]
\centering
\begin{tabular}{c cccc}
 & 10\% Percentile & Median & 90\% Percentile & Probability: Progress? \\ 
  \hline
$\mathbb{M} \coloneqq \mR$:  &&&& \\   
$\lambda_A$ & .196 & .817 & .970 & .863 \\ 
$\lambda_B$ & .072 & .419 & .918 & .635 \\ 
$\lambda_C$ & .055 & .246 & .826 & .523 \\ 
$\lambda_D$ & .107 & .648 & .953 & .743 \\ 
    \hline
$\mathbb{M} \coloneqq \mR^2$:  &&&& \\  
$\lambda_A$ & .352 & .837 & .970 & .900 \\ 
$\lambda_B$ & .080 & .424 & .913 & .642 \\ 
$\lambda_C$ & .051 & .250 & .814 & .515 \\ 
$\lambda_D$ & .116 & .653 & .948 & .770 \\ 
      \hline
$\mathbb{M} \coloneqq \mR^3$:  &&&& \\  
$\lambda_A$ & .322 & .814 & .970 & .885 \\ 
$\lambda_B$ & .076 & .406 & .920 & .634 \\ 
$\lambda_C$ & .051 & .227 & .771 & .498 \\ 
$\lambda_D$ & .133 & .639 & .947 & .765 \\ 
      \hline
\end{tabular}
\caption{My Math Academy: 
Summaries of the marginal posteriors of the rates of progress $\lambda_A$, $\lambda_B$, $\lambda_C$, $\lambda_D$ of students $A$, $B$, $C$, $D$.
The last column (``Probability: Progress?") shows the posterior probability of non-negligible progress.
}
\label{tab:math_lambda}
\end{table}

\clearpage 

\section{Discussion}
\label{sec:discussion}

We have introduced a latent process model for monitoring progress towards a hard-to-measure target of interest,
with a number of possible extensions.

For example,
it may be of interest to monitor the progress of individuals towards two or more targets, 
which may or may not be related. If the targets are known to be unrelated (e.g., improving the command of English language and the understanding of geometry),
the targets could be analyzed by separate latent process models, 
with separate latent spaces.
By contrast,
if the targets are related (e.g., improving the understanding of random variables and of stochastic processes, i.e., collections of random variables),
it may be of interest to monitor the progress of individuals towards both targets.
Extensions of the latent process model for tackling multiple targets constitute an interesting direction for future research.
A second example is regress,
the opposite of progress.
Capturing regress is of interest in mental health applications,
because the mental health of vulnerable individuals may deteriorate rather than improve.

\alert{In addition to model extensions,
it would be of interest to investigate the identifiability issues of the latent process model and other latent variable models in more depth.
}

\s



\clearpage 

\begin{appendix}

\begin{center}
\Large\bf\textsc\bf
Supplement\s\s
\end{center}

\thispagestyle{empty}

\noindent
{Supplement \ref{sec:expected.scores}: Expected scores}\dotfill\pageref{sec:expected.scores}\s
\\
{Supplement \ref{depression_items}: Mental health data}\dotfill\pageref{depression_items}\s
\\
{Supplement \ref{appendix:mcmc}: Markov chain Monte Carlo algorithm}\dotfill\pageref{appendix:mcmc}\s
\\
Supplement \ref{appendix:estimation}: Details on prior and algorithm specification\dotfill\pageref{appendix:estimation}\s
\\
{Supplement \ref{appendix:convergence}: Convergence diagnostics}\dotfill\pageref{appendix:convergence}\s
\\
Supplement \ref{appendix:add}: Additional results\dotfill\pageref{appendix:add}\s

\hide{

\section{Two-component mixture prior of the rates of progress}
\label{mixture_prior}

\begin{figure}[htbp]
    \centering
\begin{tabular}{cc}
(a) & (b)  \\ 
     \includegraphics[width=0.35 \textwidth]{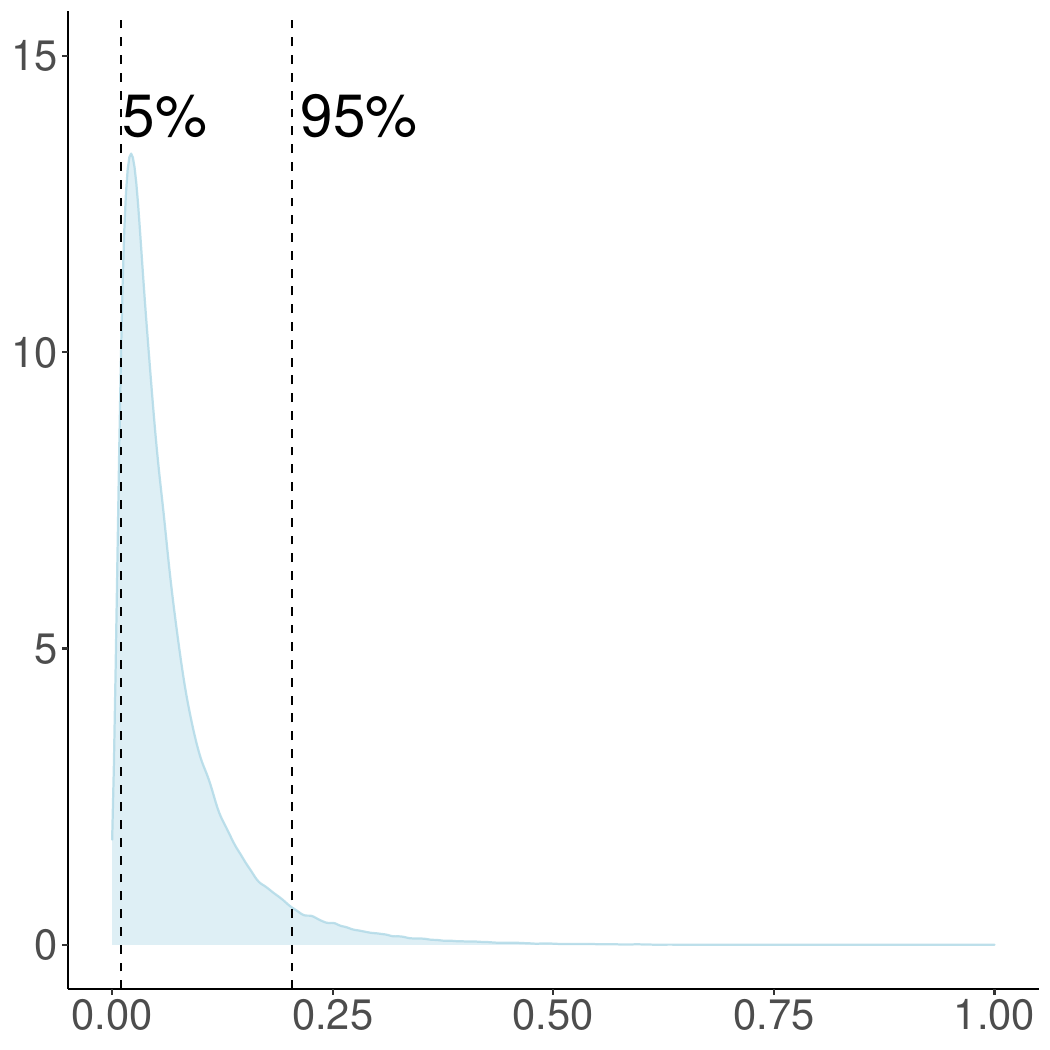} 
     &  \includegraphics[width=0.35 \textwidth]{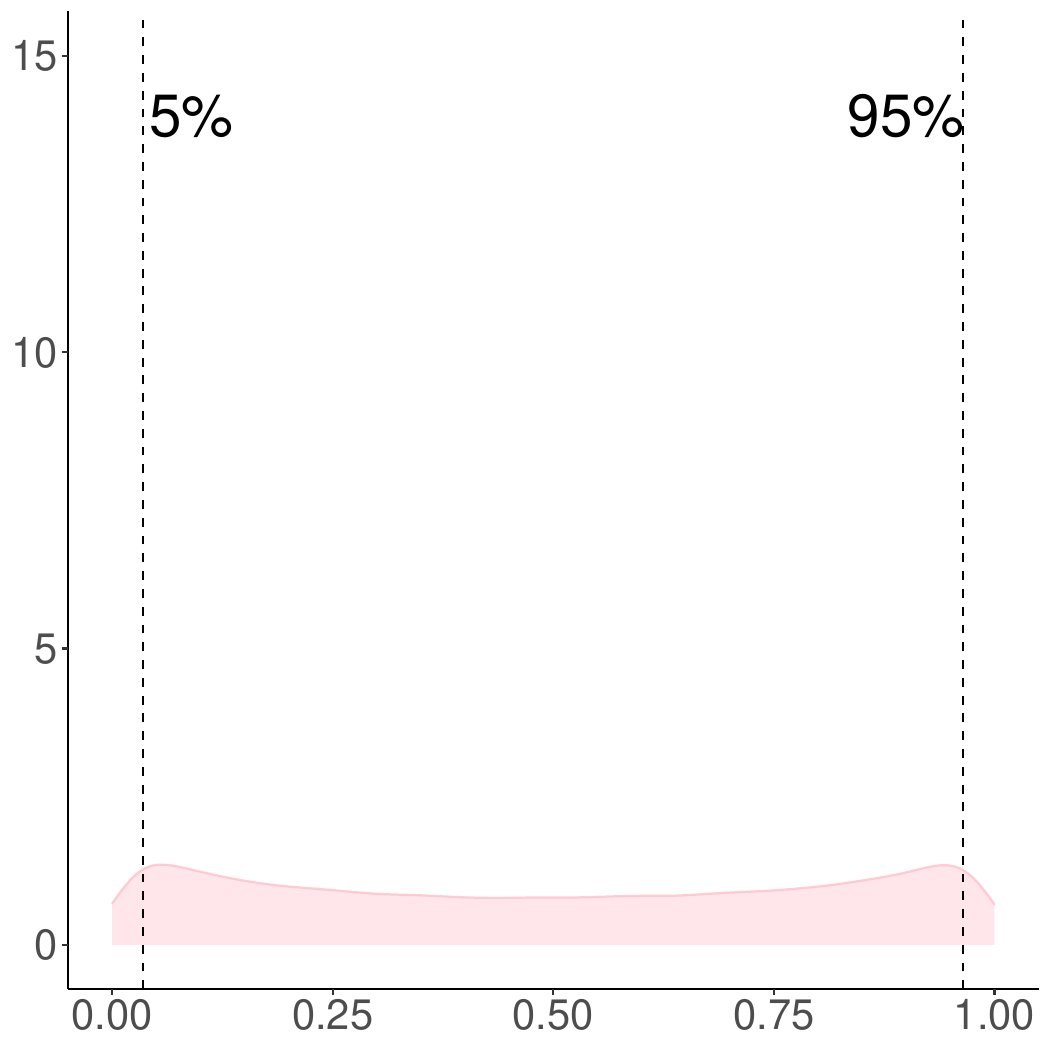}  \\
\end{tabular}
\caption{The two component distributions of the two-component mixture prior of the rates of progress $\lambda_{i,t}$ described in Section \ref{sec:prior}.
}\label{fg:priors}
\end{figure}

\clearpage 

}

\section{Expected scores}
\label{sec:expected.scores}

\alert{There are other possible approaches to measuring progress:
e.g.,
one can base the assessment of progress on expected scores.

To demonstrate,
recall that the proposed statistical framework builds on generalized linear models.
In other words,
the responses $Y_{i,j,t}$ have statistical exponential-family distributions with canonical parameters $\eta_{i,j,t}$ \citep{Br86,Su19,efron22}.
In the language of statistical exponential families,
the proposed statistical framework measures progress based on the canonical parameterization of statistical exponential families.
An alternative would be to measure progress based on the mean-value parameterization of statistical exponential families,
that is,
based on expected scores.

To compare these alternative approaches to assessing progress,
let $\bm{Y}_i \coloneqq (Y_{i,j,t})_{1 \leq j \leq p,\, 1 \leq t \leq T}$\, be the vector of responses $Y_{i,j,t}$ of individual $i$ and let the distributions $\mbP_{\btheta,\bm{a}_{i,t},\bm{b}_j}$ of responses $Y_{i,j,t}$ be one-parameter exponential-family distributions (e.g., $Y_{i,j,t} \mid \mu_{i,j,t}$ $\ind$ $\mbox{Bernoulli}(\mu_{i,j,t})$ with mean $\mu_{i,j,t} \in (0, 1)$,\,
$Y_{i,j,t} \mid \mu_{i,j,t}\; \ind\; \mbox{Poisson}(\mu_{i,j,t})$ with mean $\mu_{i,j,t} \in (0, +\infty)$,\,
or $Y_{i,j,t} \mid \mu_{i,j,t},\, \sigma_{i,j,t}^2\; \ind\; N(\mu_{i,j,t},\, \sigma_{i,j,t}^2)$ with mean $\mu_{i,j,t} \in \mR$ and known variance $\sigma_{i,j,t}^2 \in (0, +\infty)$).
Then the probability density function of an exponential-family probability measure $\mbP_{\btheta,\bm{a}_{i,t},\bm{b}_j}$ dominated by a $\sigma$-finite measure $\nu$ can be represented as
\beno
f(y_{i,j,t} \mid \eta_{i,j,t})
&=& a_{i,j,t}(y_{i,j,t})\,  \exp(\eta_{i,j,t}\, s_{i,j,k}(y_{i,j,t}) - \psi_{i,j,t}(\eta_{i,j,t})),
\ee
where $a_{i,j,t}(y_{i,j,t}) \in [0, +\infty)$ is a function of $y_{i,j,t}$,\,
$\eta_{i,j,t} \equiv \eta_{i,j,t}(\mu_{i,j,t}(\btheta,\, \bm{a}_{i,t},\, \bm{b}_j)) \in \mR$ is a canonical parameter,
$s_{i,j,t}(y_{i,j,t}) \in \mR$ is a sufficient statistic,
and
\beno
\psi_{i,j,t}(\eta_{i,j,t})
&\coloneqq& \dint_{\mY_{i,j,t}} \exp(\eta_{i,j,t}\, s_{i,j,k}(y_{i,j,t})) \dd \nu(y_{i,j,t})
\ee
ensures that $f(y_{i,j,t} \mid \eta_{i,j,t})$ integrates to $1$;
note that the general formulation above covers both the discrete setting (in which case $\nu$ may be counting measure) and the continuous setting (in which case $\nu$ may be Lebesgue measure).
As a result,
the probability density function of the response vector $\bm{Y}_i$ of individual $i$ can be represented as
\beno
f(\bm{y}_i \mid \bm\eta_i)
&=& \dprod_{j=1}^p\, \dprod_{t=1}^T f(y_{i,j,t} \mid \eta_{i,j,t})
&=& a_i(\bm{y}_i)\, \exp\left(\langle\bm\eta_i, s_i(\bm{y}_i)\rangle - \psi_i(\bta_i)\right),
\ee
where $a_i(\bm{y}_i) \coloneqq \prod_{j=1}^p \prod_{t=1}^T a_{i,j,t}(y_{i,j,t})$,\,
$\psi_i(\bta_i) \coloneqq \sum_{j=1}^p \sum_{t=1}^T \psi_{i,j,t}(\eta_{i,j,t})$,\,
and $\langle\bm\eta_i, s_i(\bm{y}_i)\rangle$ denotes the inner product of the vector of canonical parameters $\bm\eta_i \coloneqq (\eta_{i,j,t})_{1 \leq j \leq p,\, 1 \leq t \leq T}$ and the vector of sufficient statistics $s_i(\bm{y}_i) \coloneqq (s_{i,j,t}(y_{i,j,t}))_{1 \leq j \leq p,\, 1 \leq t \leq T}$.
In other words,
if the responses $Y_{i,j,t}$ have exponential-family distributions,
so does the response vector $\bm{Y}_i$ of individual $i$.
The parameter vector
\beno
\bm\mu_i(\bm\eta_i)
&\coloneqq& \mbE_{\bm\eta_i}\, s_i(\bm{Y}_i)
\ee
with coordinates $\mu_{i,j,t}(\bm\eta_i) \coloneqq \mbE_{\bm\eta_i}\, s_{i,j,t}(Y_{i,j,t})$ is known as the mean-value parameter vector of the exponential family.
Since the map $\bm\eta_i \mapsto \bm\mu_i(\bm\eta_i)$ is a homeomorphism and is therefore one-to-one \citep[][Theorem 3.6, p.\ 74]{Br86},
one can specify models and assess progress based on one of two parameterizations:
\bi
\item[1.] One can specify models and assess progress by specifying the canonical parameter vector $\bta_i$.\vspace{.1cm}
\item[2.] One can specify models and assess progress by specifying the mean-value parameter vector $\bmu_i(\bta_i) \coloneqq \mbE_{\bta_i}\, \bY_i$,
that is,
the expected score vector.
\ei
As a consequence,
one can measure progress based on either the canonical parameterization or the mean-value parameterization:
e.g.,
if the responses $Y_{i,j,t}$ of individual $i$ to variables $j$ have been recorded at $T = 2$ time points,
one could measure progress based on differences in mean-value parameters $\mu_{i,j,2}(\bm\eta_i)$ and $\mu_{i,j,1}(\bm\eta_i)$,
summed over all variables $j$:
\beno
\mbox{progress of individual $i$}
&\coloneqq& \dsum_{j=1}^p \mu_{i,j,2}(\bm\eta_i) - \dsum_{j=1}^p \mu_{i,j,1}(\bm\eta_i)\s
\\
&=& \dsum_{j=1}^p \mbE_{\bm\eta_i}\, Y_{i,j,2} - \dsum_{j=1}^p \mbE_{\bm\eta_i}\, Y_{i,j,1}\s
\\
&=& \mbE_{\bm\eta_i} \dsum_{j=1}^p Y_{i,j,2} - \mbE_{\bm\eta_i} \dsum_{j=1}^p Y_{i,j,1}.
\ee
That being said,
it is common practice to specify generalized linear models by specifying the canonical parameters $\eta_{i,j,t}$ rather than the mean-value parameters $\mu_{i,j,t}(\bm\eta_i)$,
because the canonical parameterization helps incorporate covariates,
as explained in Section \ref{sec:other.approaches}.
While both approaches to specifying generalized linear models are legitimate,
we choose the canonical route.
}

\s

\section{Mental health data}
\label{depression_items}

\s\s


The following ten symptoms were used for data analysis. 
\begin{enumerate}
    \item ``I was bothered by things that usually don’t bother me.''
    \item ``I had trouble keeping my mind on
what I was doing.''
    \item ``I felt depressed.''
    \item ``I felt that everything I did was an
effort.''
    \item ``I did not feel hopeful about the future.''
    \item ``I thought my life had been a failure.''
    \item ``I was not happy.''
    \item ``I talked less than usual.''
    \item ``I felt that people dislike me.''
    \item ``I could not get ``going''. ''
\end{enumerate}

The duration of each symptom over the last seven days was asked based on four response categories (1: rarely or none of the time (less than 1 day); 2: some or a
little of the time (1-2 days); 3: occasionally or a moderate amount of time (3-4 days); 4: most or all of the time (5-7 days). 
We dichotomized the responses such that response 1 became 1 and responses 2--4 became 0.

\section{Markov chain Monte Carlo algorithm}
\label{appendix:mcmc}

We approximate the posterior by combining the following Markov chain Monte Carlo steps by cycling or mixing \citep*{Tl94}:

\begin{enumerate}
  
    \item Sample $\beta_{j}$ from its full conditional distribution: 
    \[ 
\beta_{j} \sim \mbox{N}\left(m_\beta, {v_\beta^2} \right),
   \]
\begin{align*}
         m_\beta & =   \left(\frac{1}{\sigma^2_\beta} + \sum_i \sum_t  \, \omega_{i,j,t} \right)^{-1} \, \left( \sum_i \sum_t\, K_{i,j,t} - \sum_i \sum_t\, \omega_{i,j,t} \, \alpha_{i,t} + \sum_i \sum_t\, \omega_{i,j,t} \, \gamma \, d_{i,j,t}  \right) \\
         {v_\beta^2} &=  \left(\frac{1}{\sigma^2_\beta} + \sum_i \sum_t\, \omega_{i,j,t} \right)^{-1}, 
  \end{align*} 
  where $K_{i,j,t} = Y_{i,j,t} - {1}/{2}$, and ${K_{i,j,t}}/{\omega_{i,j,t}} \sim \mbox{N}(0,1)$ with $\omega_{i,j,t} \sim \mbox{PG} \left(1, \alpha_{i} + \beta_{j} - \gamma \,  d_{i,j,t}  \right)$, where 
 $ d_{i,j,t} = (\bm{a}_{i,t}, \bm{b}_j) $ when $T=1$ and  $ d_{i,j,t} = d_{i,t}  = (\bm{a}_{i,t}, \mT) $  when $T>1$, where $\mbox{PG} (b,c)$ is a P\'{o}lya-Gamma distribution with $b > 0$ and $c \in \mathbb{R}$ \citep{polson:13}.


    \item Sample $\alpha_{i}$ from its full conditional distribution: 
    \[ 
\alpha_{i} \sim \mbox{N}\left(m_\alpha, {v_\alpha^2} \right),
   \]
  \begin{align*}
         m_\alpha & =   \left(\frac{1}{\sigma^2_\alpha} + \sum_j \sum_t\, \omega_{i,j,t} \right)^{-1} \, \left( \sum_j \sum_t\, K_{i,j,t}  - \sum_j \sum_t\, \omega_{i,j,t} \, \beta_{j,t}  + \sum_j \sum_t \, \omega_{i,j,t} \, \gamma \, d_{i,j,t}  \right) \\
         {v_\alpha^2} &=  \left(\frac{1}{\sigma^2_\alpha} + \sum_j \sum_t \, \omega_{i,j,t} \right)^{-1}, 
  \end{align*} 

    \item Sample $\sigma_{\alpha}^2$ from the Inverse Gamma distribution 
    \[
    \sigma_{\alpha}^2 \sim \mbox{Inv-Gamma}\left(a_{\sigma_{\alpha}} + \frac{N}{2}, b_{\sigma_{\alpha}} + \frac{\sum_j \sum_t\alpha_i^2}{2}\right),
    \]

    \item Sample $\gamma$ from its full conditional distribution: 
    \[ 
\gamma \sim \mbox{Half-Normal}\left(m_\gamma, v^2_\gamma \right),
   \]
 {\small 
  \begin{align*}
         m_\gamma & = \left(\frac{1}{\sigma^2_\gamma} +  \sum_i \sum_j \sum_t \, \omega_{i,t,} \, d_{i,j,t}^2 \right)^{-1} \, \\
         \times & \left ( \sum_i \sum_j \sum_t \,K_{i,j,t} \, d_{i,j,t}   + \sum_i \sum_j \sum_t \, \omega_{i,j,t} \, \alpha_{i,t} \, d_{i,j,t}   + \sum_i \sum_j \sum_t \, \omega_{i,j,t} \, \beta_{j,t} \, d_{i,j,t} \right)  \\
         {v_\gamma^2} &=  \left(\frac{1}{\sigma^2_\gamma} +  \sum_i \sum_j \sum_t \, \omega_{i,t,} \, d_{i,j,t}^2 \right)^{-1}. 
  \end{align*} 
  }
     \item Propose $\lambda_{i,t}^\star$ from a symmetric proposal distribution and accept the proposal with probability
    \[
\min\left(1,\; \frac{f\Big(\lambda_{i,t}^\star \mid \bm{K},\, \bm{\omega},\,  \bm{A},\, \bm{B},\, \bm{\beta},\,  \gamma \Big)} {f\Big(\lambda_{i,t}^{(l)} \mid \bm{K},\, \bm{\omega},\, \bm{A},\, \bm{B},\, \bm{\beta},\, \gamma \Big)}\right).
\]

    \item Propose $\bm{a}_{i,1}^\star$ from a symmetric proposal distribution and accept the proposal with probability
\[
    \min \left(  1,\; {\frac{f\Big(\bm{a}_i^\star \mid \bm{K},\, \bm{\omega},\,  \bm{A}_{-i},\, \bm{B},\, \bm{\beta},\, \la, \, \gamma  \Big)} {f \Big( \bm{a}_i^{(l)} \mid \bm{K},\, \bm{\omega},\,  \bm{A}_{-j},\, \bm{B},\, \bm{\beta},\, \la, \,\gamma \Big) } }  \right) ,
\]

where $\bm{A}_{-i} = (\bm{a}_1, \dots, \bm{a}_{i-1}, \bm{a}_{i+1}, \dots, \bm{a}_n)$.

    \item Propose $\bm{b}_j^\star$ from a symmetric proposal distribution and accept the proposal with probability
    \[
\min\left(1,\; \frac{f\Big(\bm{b}_j^\star \mid \bm{K},\, \bm{\omega},\,  \bm{A},\, \bm{B}_{-j},\, \bm{\beta},\, \la,\, \gamma  \Big)} 
    {f\Big(\bm{b}_j^{(l)} \mid \bm{K},\, \bm{\omega},\,  \bm{A},\, \bm{B}_{-j},\, \bm{\beta},\, \la,\, \gamma  \Big)} \right)   ,
    \]
    where $\bm{B}_{-j} = (\bm{b}_1, \dots, \bm{b}_{j-1}, \bm{b}_{j+1}, \dots, \bm{b}_q)$.

\hide{     

\item Sample $\mT$ from its full conditional distribution: 
    \[
  \mT \sim \mbox{MVN}_q\left(\bm{m}_{\mT},\; {v_{\mT}^2} \, \bm{I}_q \right)
\]
  \begin{align*}
         \bm{m}_{\mT} & = \left(n  + \frac{1} {{\sigma_{{\mT}}}^2}\right)^{-1} \, \left( \sum_{j=1}^p \, \bm{b}_j \right)\s 
         \\
         {v_{\mT}^2} &= \left(\frac{1} {{\sigma_{{\mT}}}^2} + n \right)^{-1}. 
  \end{align*} 

}

\item Sample $\pi_{i,t}$ from its full conditional distribution: 
\[
\pi_{i,t} \sim \mbox{Beta} \left(a_\pi + r_{i,t}, \, b_\pi + (1- r_{i,t}) \right),
\]
where $\pi_{i,t} \coloneqq \mbP(r_{i,t} = 1 \mid \pi_{i,t})$ and $r_{i,t} \in \{0,1 \}$ is sampled from its full conditional distribution:
\[
r_{i,t} \sim \mbox{Bernoulli} 
\left( \left( 1 + \frac{ (1- \pi_{i,t})}{\pi_{i,t}}  
\frac{\left(2\pi \sigma_0^2 \right)^{-\frac{1}{2}} \exp \left( -\frac{1}{2\sigma_0^2}  \left(\logit(\lambda_{i,t}) - \mu_0 \right)^2 \right)   }  
{\left(2\pi \sigma_1^2 \right)^{-\frac{1}{2}} \exp \left( -\frac{1}{2\sigma_1^2}  \left(\logit(\lambda_{i,t}) - \mu_1 \right)^2 \right)   } \right)^{-1}  
\right).
\]
\end{enumerate}

As proposal distributions,
we use multivariate Gaussians centered at the current values of the quantities in question,
with diagonal variance-covariance matrices.
The variances are set to achieve acceptance rates between .3 and .4. 

\hide{

To address the identifiability issues reviewed in Section 2.4 of the manuscript,
we constrain the standard deviation of 
$Z \coloneqq 
(\bm{a}_{1,1}^\prime, \dots, \bm{a}_{1,T}^\prime, \dots, \bm{a}_{n,1}^\prime, \dots, \bm{a}_{n,T}^\prime,\, \bm{b}_{1}^\prime, \dots, \bm{b}_{p}^\prime)$ to be $1$.

}

\section{Details on prior and algorithm specification}
\label{appendix:estimation}

We provide additional details on the priors and Markov chain Monte Carlo algorithms used in the simulations and applications.
The hyperparameters were chosen so that the priors spread most over the mass over the most plausible subsets of the parameters space.

\subsection{Section \ref{sec:simulations}: Simulation results}
\label{appendix:sim}

\begin{enumerate} 

\item[] Section \ref{sec:sim_scenario1}

$q=1$:
    \begin{itemize}
        \item MCMC iterations: 45,000, and burn-in period: 30,000

\item Prior values:  
$\sigma_{\beta} = 5$,
$a_{\sigma_\alpha} = 1$, 
$b_{\sigma_\alpha} = 1$, 
$\sigma_{\gamma} = 2$, 
$\sigma_{a} = 1$, 
$\sigma_{b} = 1$,
$\mu_0 = -2$, 
$\sigma_0 = 1$, 
$\mu_1 = 0$, 
$\sigma_1 = 2$, 
$a_\pi = 1$, 
$b_\pi = 1$.
 
\item[] Standard deviations of the Gaussian proposal distributions,
centered at the current values of the parameters: 
1.7 ($\bm{a}_{1,t}$); 
.6 ($\bm{b}_{j}$);
and 5 ($\lambda_{i,t}$). 
\end{itemize}    

$q=2$:

    \begin{itemize}
        \item MCMC iterations: 65,000, and burn-in period: 50,000

\item Prior values:  
$\sigma_{\beta} = 5$,
$a_{\sigma_\alpha} = 1$, 
$b_{\sigma_\alpha} = 1$, 
$\sigma_{\gamma} = 2$, 
$\sigma_{a} = 1$, 
$\sigma_{b} = 1$,
$\mu_0 = -2$, 
$\sigma_0 = 1$, 
$\mu_1 = 0$, 
$\sigma_1 = 2$, 
$a_\pi = 1$, 
$b_\pi = 1$.
 
\item[] Standard deviations of the Gaussian proposal distributions,
centered at the current values of the parameters: 
1.4 ($\bm{a}_{1,t}$); 
.8 ($\bm{b}_{j}$);
and 5 ($\lambda_{i,t}$). 
\end{itemize}

$q=3$:

    \begin{itemize}
        \item MCMC iterations: 85,000, and burn-in period: 70,000

\item Prior values:  
$\sigma_{\beta} = 5$,
$a_{\sigma_\alpha} = 1$, 
$b_{\sigma_\alpha} = 1$, 
$\sigma_{\gamma} = 2$, 
$\sigma_{a} = 1$, 
$\sigma_{b} = 1$,
$\mu_0 = -2$, 
$\sigma_0 = 1$, 
$\mu_1 = 0$, 
$\sigma_1 = 2$, 
$a_\pi = 1$, 
$b_\pi = 1$.
 
\item[] Standard deviations of the Gaussian proposal distributions,
centered at the current values of the parameters: 
1 ($\bm{a}_{1,t}$); 
.6 ($\bm{b}_{j}$);
and 5 ($\lambda_{i,t}$). 
\end{itemize}

$q=4$:

    \begin{itemize}
        \item MCMC iterations: 85,000, and burn-in period: 70,000

\item Prior values:  
$\sigma_{\beta} = 5$,
$a_{\sigma_\alpha} = 1$, 
$b_{\sigma_\alpha} = 1$, 
$\sigma_{\gamma} = 2$, 
$\sigma_{a} = 1$, 
$\sigma_{b} = 1$,
$\mu_0 = -2$, 
$\sigma_0 = 1$, 
$\mu_1 = 0$, 
$\sigma_1 = 2$, 
$a_\pi = 1$, 
$b_\pi = 1$.
 
\item[] Standard deviations of the Gaussian proposal distributions,
centered at the current values of the parameters: 
.9 ($\bm{a}_{1,t}$); 
.3 ($\bm{b}_{j}$);
and 5 ($\lambda_{i,t}$). 
\end{itemize}

\item[] Section \ref{sec:sim_scenario2}:

$q=1$:

    \begin{itemize}
        \item MCMC iterations: 45,000, and burn-in period: 30,000

\item Prior values:  
$\sigma_{\beta} = 5$,
$a_{\sigma_\alpha} = 1$, 
$b_{\sigma_\alpha} = 1$, 
$\sigma_{\gamma} = 2$, 
$\sigma_{a} = 1$, 
$\sigma_{b} = 1$,
$\mu_0 = -2$, 
$\sigma_0 = 1$, 
$\mu_1 = 0$, 
$\sigma_1 = 2$, 
$a_\pi = 1$, 
$b_\pi = 1$.
 
\item[] Standard deviations of the Gaussian proposal distributions,
centered at the current values of the parameters: 
1.8 ($\bm{a}_{1,t}$); 
.1 ($\bm{b}_{j}$);
and 5 ($\lambda_{i,t}$). 
\end{itemize}

$q=2$:

    \begin{itemize}
        \item MCMC iterations: 65,000, and burn-in period: 50,000

\item Prior values:  
$\sigma_{\beta} = 5$,
$a_{\sigma_\alpha} = 1$, 
$b_{\sigma_\alpha} = 1$, 
$\sigma_{\gamma} = 2$, 
$\sigma_{a} = 1$, 
$\sigma_{b} = 1$,
$\mu_0 = -2$, 
$\sigma_0 = 1$, 
$\mu_1 = 0$, 
$\sigma_1 = 2$, 
$a_\pi = 1$, 
$b_\pi = 1$.
 
\item[] Standard deviations of the Gaussian proposal distributions,
centered at the current values of the parameters: 
1.4 ($\bm{a}_{1,t}$); 
.3 ($\bm{b}_{j}$);
and 5 ($\lambda_{i,t}$). 
\end{itemize}

$q=3$:

    \begin{itemize}
        \item MCMC iterations: 85,000, and burn-in period: 70,000

\item Prior values:  
$\sigma_{\beta} = 5$,
$a_{\sigma_\alpha} = 1$, 
$b_{\sigma_\alpha} = 1$, 
$\sigma_{\gamma} = 2$, 
$\sigma_{a} = 1$, 
$\sigma_{b} = 1$,
$\mu_0 = -2$, 
$\sigma_0 = 1$, 
$\mu_1 = 0$, 
$\sigma_1 = 2$, 
$a_\pi = 1$, 
$b_\pi = 1$.
 
\item[] Standard deviations of the Gaussian proposal distributions,
centered at the current values of the parameters: 
.8 ($\bm{a}_{1,t}$); 
.15 ($\bm{b}_{j}$);
and 5 ($\lambda_{i,t}$). 
\end{itemize}

$q=4$:

    \begin{itemize}
        \item MCMC iterations: 85,000, and burn-in period: 70,000

\item Prior values:  
$\sigma_{\beta} = 5$,
$a_{\sigma_\alpha} = 1$, 
$b_{\sigma_\alpha} = 1$, 
$\sigma_{\gamma} = 2$, 
$\sigma_{a} = 1$, 
$\sigma_{b} = 1$,
$\mu_0 = -2$, 
$\sigma_0 = 1$, 
$\mu_1 = 0$, 
$\sigma_1 = 2$, 
$a_\pi = 1$, 
$b_\pi = 1$.
 
\item[] Standard deviations of the Gaussian proposal distributions,
centered at the current values of the parameters: 
.7 ($\bm{a}_{1,t}$); 
.08 ($\bm{b}_{j}$);
and 5 ($\lambda_{i,t}$). 
\end{itemize}

\item[] Section \ref{sec:interaction}: 

    \begin{itemize}
        \item MCMC iterations: 25,000, and burn-in period: 15,000

\item Prior values:  
$\sigma_{\beta} = 5$,
$a_{\sigma_\alpha} = 1$, 
$b_{\sigma_\alpha} = 1$, 
$\sigma_{\gamma} = 2$, 
$\sigma_{a} = 1$, 
$\sigma_{b} = 1$,
$\mu_0 = -2$, 
$\sigma_0 = 1$, 
$\mu_1 = 0$, 
$\sigma_1 = 2$, 
$a_\pi = 1$, 
$b_\pi = 1$.
 
\item[] Standard deviations of the Gaussian proposal distributions,
centered at the current values of the parameters: 
.2 ($\bm{a}_{1,t}$); 
.05 ($\bm{b}_{j}$);
and 2 ($\lambda_{i,t}$). 
\end{itemize}

\end{enumerate}

\subsection{Section \ref{sec:math}: Application: online educational assessments}\label{appendix:math}

\begin{enumerate}

\item[] Section \ref{sec:math_spike}:

$q=1$:
    \begin{itemize}
        \item MCMC iterations: 45,000, and burn-in period: 30,000

\item Prior values:  
$\sigma_{\beta} = 5$,
$a_{\sigma_\alpha} = 1$, 
$b_{\sigma_\alpha} = 1$, 
$\sigma_{\gamma} = 2$, 
$\sigma_{a} = 1$, 
$\sigma_{b} = 1$,
$\mu_0 = -2$, 
$\sigma_0 = 1$, 
$\mu_1 = 0$, 
$\sigma_1 = 2$, 
$a_\pi = 1$, 
$b_\pi = 1$.
 
\item[] Standard deviations of the Gaussian proposal distributions,
centered at the current values of the parameters: 
1.4 ($\bm{a}_{1,t}$); 
.2 ($\bm{b}_{j}$);
and 6 ($\lambda_{i,t}$).


\end{itemize}    

$q=2$:
    \begin{itemize}
        \item MCMC iterations: 45,000, and burn-in period: 30,000

\item Prior values:  
$\sigma_{\beta} = 5$,
$a_{\sigma_\alpha} = 1$, 
$b_{\sigma_\alpha} = 1$, 
$\sigma_{\gamma} = 2$, 
$\sigma_{a} = 1$, 
$\sigma_{b} = 1$,
$\mu_0 = -2$, 
$\sigma_0 = 1$, 
$\mu_1 = 0$, 
$\sigma_1 = 2$, 
$a_\pi = 1$, 
$b_\pi = 1$.
 
\item[] Standard deviations of the Gaussian proposal distributions,
centered at the current values of the parameters: 
.8 ($\bm{a}_{1,t}$); 
.1 ($\bm{b}_{j}$);
and 5 ($\lambda_{i,t}$).


\end{itemize}    

$q=3$:
    \begin{itemize}
        \item MCMC iterations: 65,000, and burn-in period: 50,000

\item Prior values:  
$\sigma_{\beta} = 5$,
$a_{\sigma_\alpha} = 1$, 
$b_{\sigma_\alpha} = 1$, 
$\sigma_{\gamma} = 2$, 
$\sigma_{a} = 1$, 
$\sigma_{b} = 1$,
$\mu_0 = -2$, 
$\sigma_0 = 1$, 
$\mu_1 = 0$, 
$\sigma_1 = 2$, 
$a_\pi = 1$, 
$b_\pi = 1$.
 
\item[] Standard deviations of the Gaussian proposal distributions,
centered at the current values of the parameters: 
.6 ($\bm{a}_{1,t}$); 
.1 ($\bm{b}_{j}$);
and 5 ($\lambda_{i,t}$).


\end{itemize}    

$q=4$:
    \begin{itemize}
        \item MCMC iterations: 65,000, and burn-in period: 50,000

\item Prior values:  
$\sigma_{\beta} = 5$,
$a_{\sigma_\alpha} = 1$, 
$b_{\sigma_\alpha} = 1$, 
$\sigma_{\gamma} = 2$, 
$\sigma_{a} = 1$, 
$\sigma_{b} = 1$,
$\mu_0 = -2$, 
$\sigma_0 = 1$, 
$\mu_1 = 0$, 
$\sigma_1 = 2$, 
$a_\pi = 1$, 
$b_\pi = 1$.
 
\item[] Standard deviations of the Gaussian proposal distributions,
centered at the current values of the parameters: 
.6 ($\bm{a}_{1,t}$); 
.07 ($\bm{b}_{j}$);
and 5 ($\lambda_{i,t}$).


\end{itemize}

\end{enumerate}

\subsection{Section \ref{sec:depression}: Application: mental health}\label{appendix:depression}

\begin{enumerate} 

\item[] Section \ref{sec:dep_spike}:

$q=1$:
    \begin{itemize}
        \item MCMC iterations: 65,000, and burn-in period: 50,000

\item Prior values:  
$\sigma_{\beta} = 5$,
$a_{\sigma_\alpha} = 1$, 
$b_{\sigma_\alpha} = 1$, 
$\sigma_{\gamma} = 2$, 
$\sigma_{a} = 1$, 
$\sigma_{b} = 1$,
$\mu_0 = -2$, 
$\sigma_0 = 1$, 
$\mu_1 = 0$, 
$\sigma_1 = 2$, 
$a_\pi = 1$, 
$b_\pi = 1$.
 
\item[] Standard deviations of the Gaussian proposal distributions,
centered at the current values of the parameters: 
1.4 ($\bm{a}_{1,t}$); 
.2 ($\bm{b}_{j}$);
and 6 ($\lambda_{i,t}$).


\end{itemize}

$q=2$:
    \begin{itemize}
        \item MCMC iterations: 65,000, and burn-in period: 50,000

\item Prior values:  
$\sigma_{\beta} = 5$,
$a_{\sigma_\alpha} = 1$, 
$b_{\sigma_\alpha} = 1$, 
$\sigma_{\gamma} = 2$, 
$\sigma_{a} = 1$, 
$\sigma_{b} = 1$,
$\mu_0 = -2$, 
$\sigma_0 = 1$, 
$\mu_1 = 0$, 
$\sigma_1 = 2$, 
$a_\pi = 1$, 
$b_\pi = 1$.
 
\item[] Standard deviations of the Gaussian proposal distributions,
centered at the current values of the parameters: 
.7 ($\bm{a}_{1,t}$); 
.1 ($\bm{b}_{j}$);
and 5 ($\lambda_{i,t}$).


\end{itemize}

$q=3$:
    \begin{itemize}
        \item MCMC iterations: 65,000, and burn-in period: 50,000

\item Prior values:  
$\sigma_{\beta} = 5$,
$a_{\sigma_\alpha} = 1$, 
$b_{\sigma_\alpha} = 1$, 
$\sigma_{\gamma} = 2$, 
$\sigma_{a} = 1$, 
$\sigma_{b} = 1$,
$\mu_0 = -2$, 
$\sigma_0 = 1$, 
$\mu_1 = 0$, 
$\sigma_1 = 2$, 
$a_\pi = 1$, 
$b_\pi = 1$.
 
\item[] Standard deviations of the Gaussian proposal distributions,
centered at the current values of the parameters: 
.8 ($\bm{a}_{1,t}$); 
.1 ($\bm{b}_{j}$);
and 5 ($\lambda_{i,t}$).


\end{itemize}    

$q=4$:
    \begin{itemize}
        \item MCMC iterations: 200,000, and burn-in period: 185,000

\item Prior values:  
$\sigma_{\beta} = 5$,
$a_{\sigma_\alpha} = 1$, 
$b_{\sigma_\alpha} = 1$, 
$\sigma_{\gamma} = 2$, 
$\sigma_{a} = 1$, 
$\sigma_{b} = 1$,
$\mu_0 = -2$, 
$\sigma_0 = 1$, 
$\mu_1 = 0$, 
$\sigma_1 = 2$, 
$a_\pi = 1$, 
$b_\pi = 1$.
 
\item[] Standard deviations of the Gaussian proposal distributions,
centered at the current values of the parameters: 
.8 ($\bm{a}_{1,t}$); 
.1 ($\bm{b}_{j}$);
and 5 ($\lambda_{i,t}$).


\end{itemize}

\end{enumerate} 

\section{Convergence diagnostics}
\label{appendix:convergence}

To detect non-convergence of the Markov chains used for approximating the posterior,
we use
\bi
\item trace plots of parameters (Supplement \ref{sec:trace});
\item the multivariate Gelman-Rubin potential scale reduction factor of \citet{vats:18a},
which can be viewed as a stable version of the multivariate Gelman-Rubin convergence diagnostic and comes with a convergence criterion (Supplement \ref{appendix:gr}).
\ei

\subsection{Trace plots}
\label{sec:trace}

\subsubsection{Section \ref{sec:math}: Application: online educational assessments}

   \begin{figure}[hptb]
        \centering
        \begin{tabular}{ccc}
        \includegraphics[width=0.3 \textwidth]{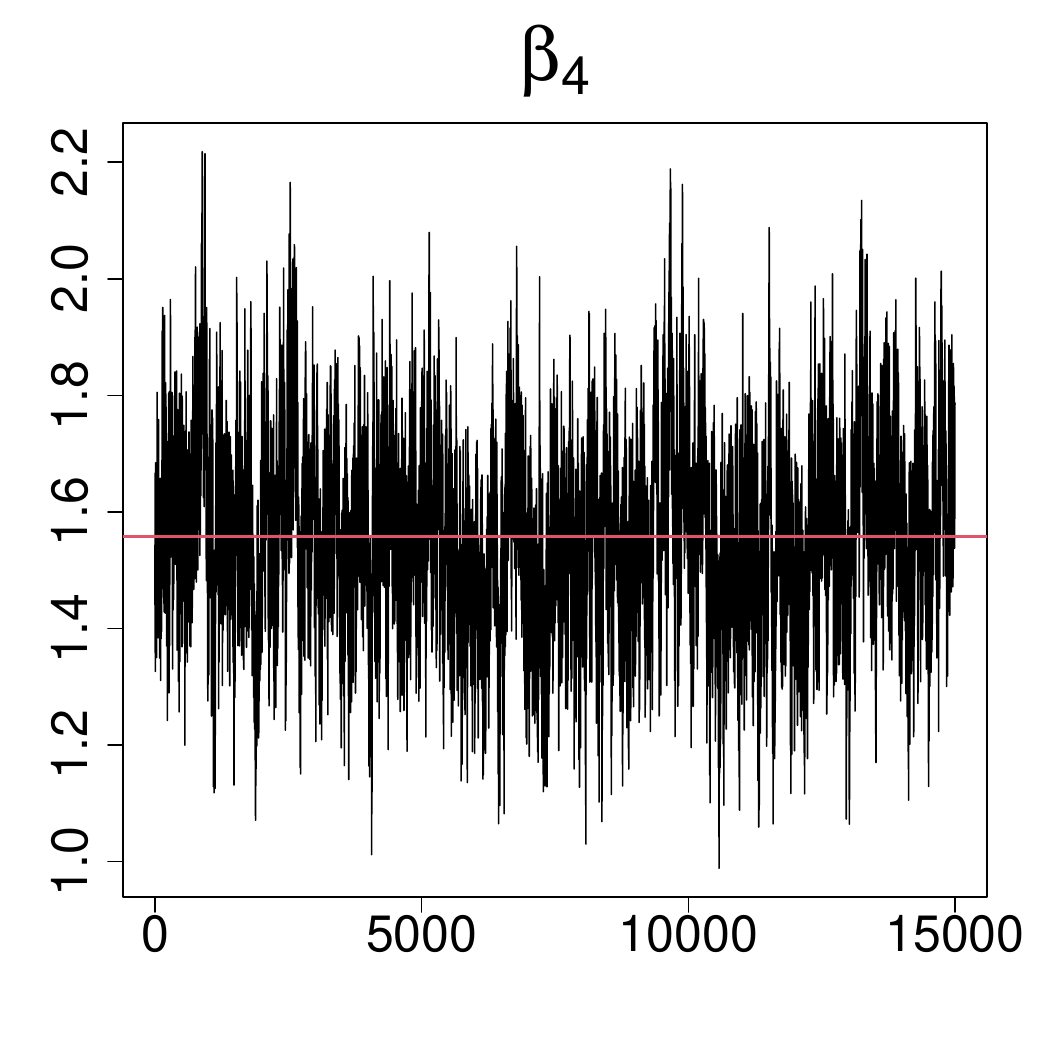} 
       &  \includegraphics[width=0.3 \textwidth]{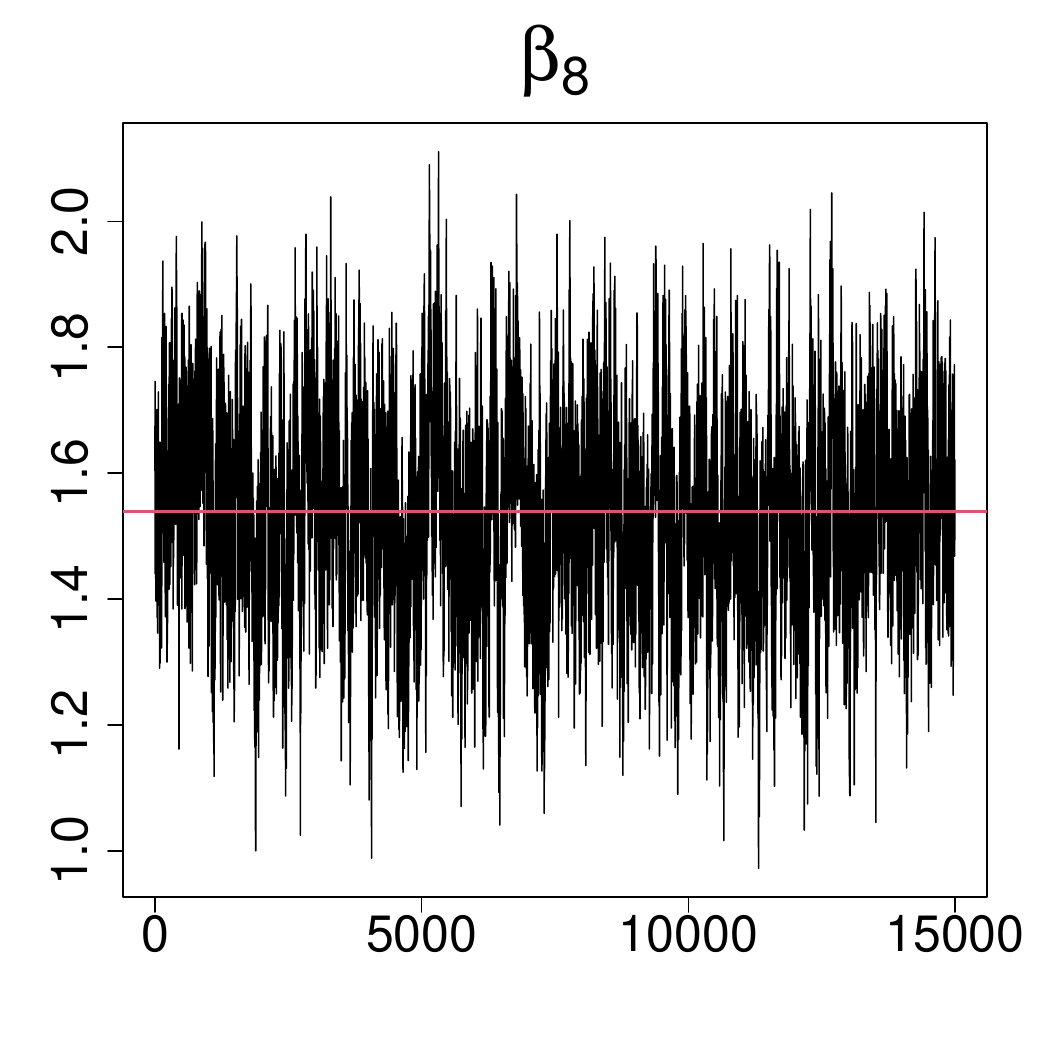} 
    &    \includegraphics[width=0.3 \textwidth]{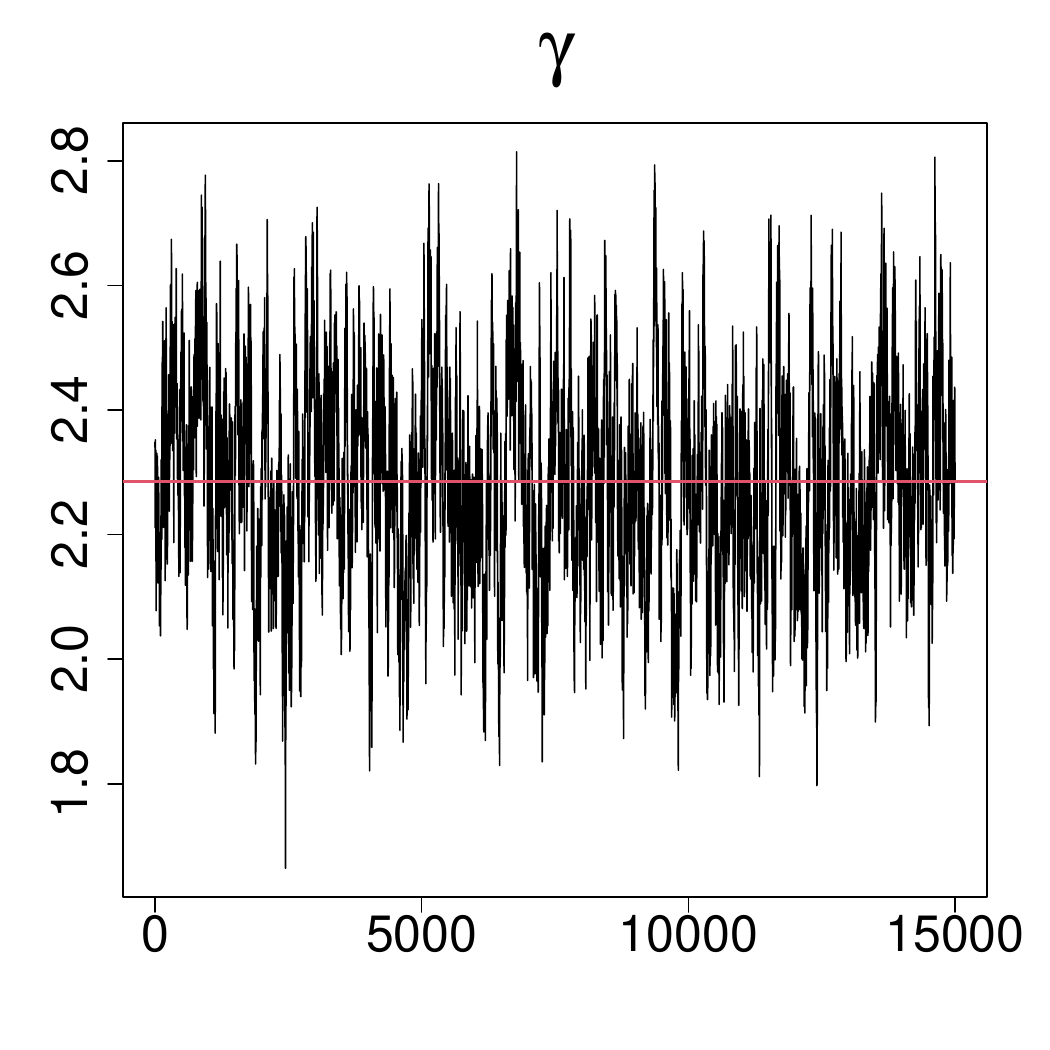} \\
    \includegraphics[width=0.3 \textwidth]{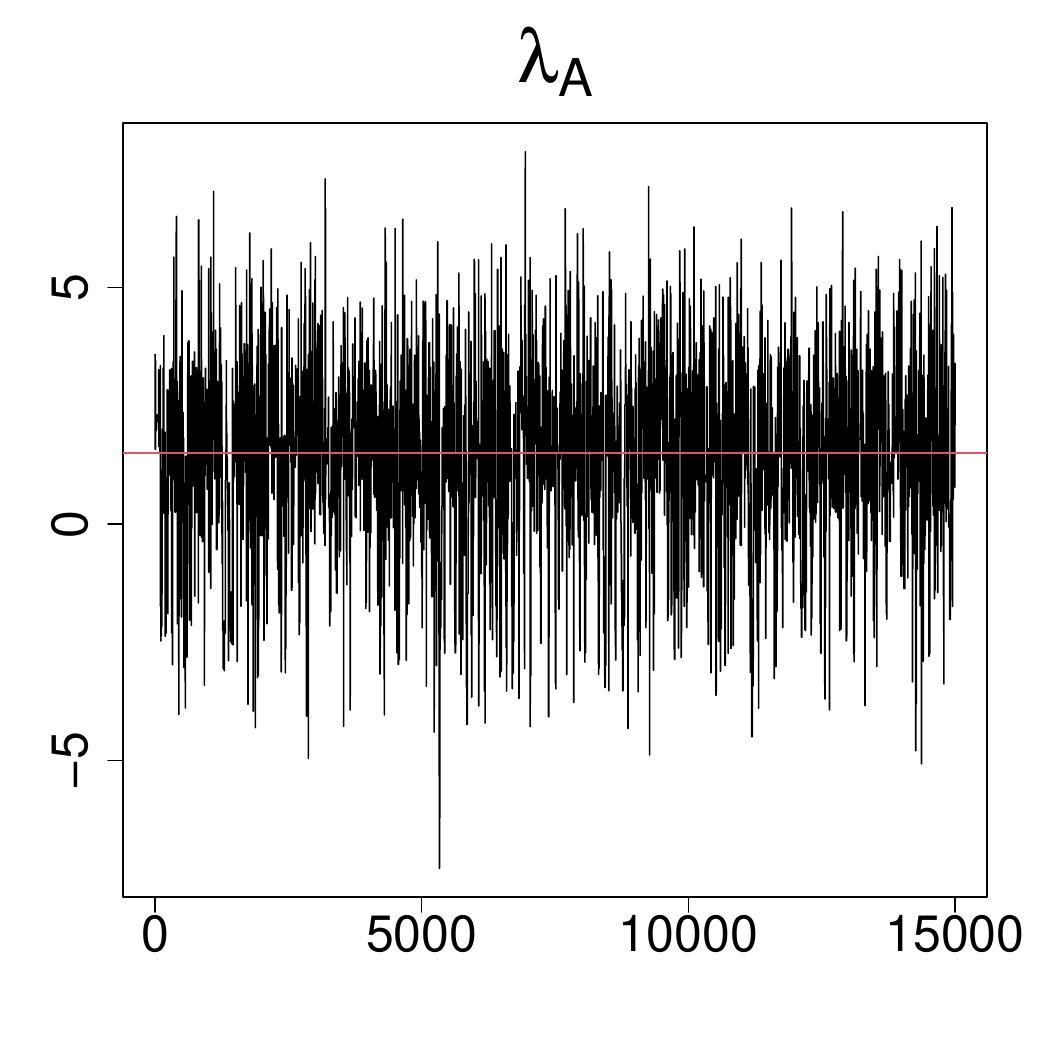}  
    &
     \includegraphics[width=0.3 \textwidth]{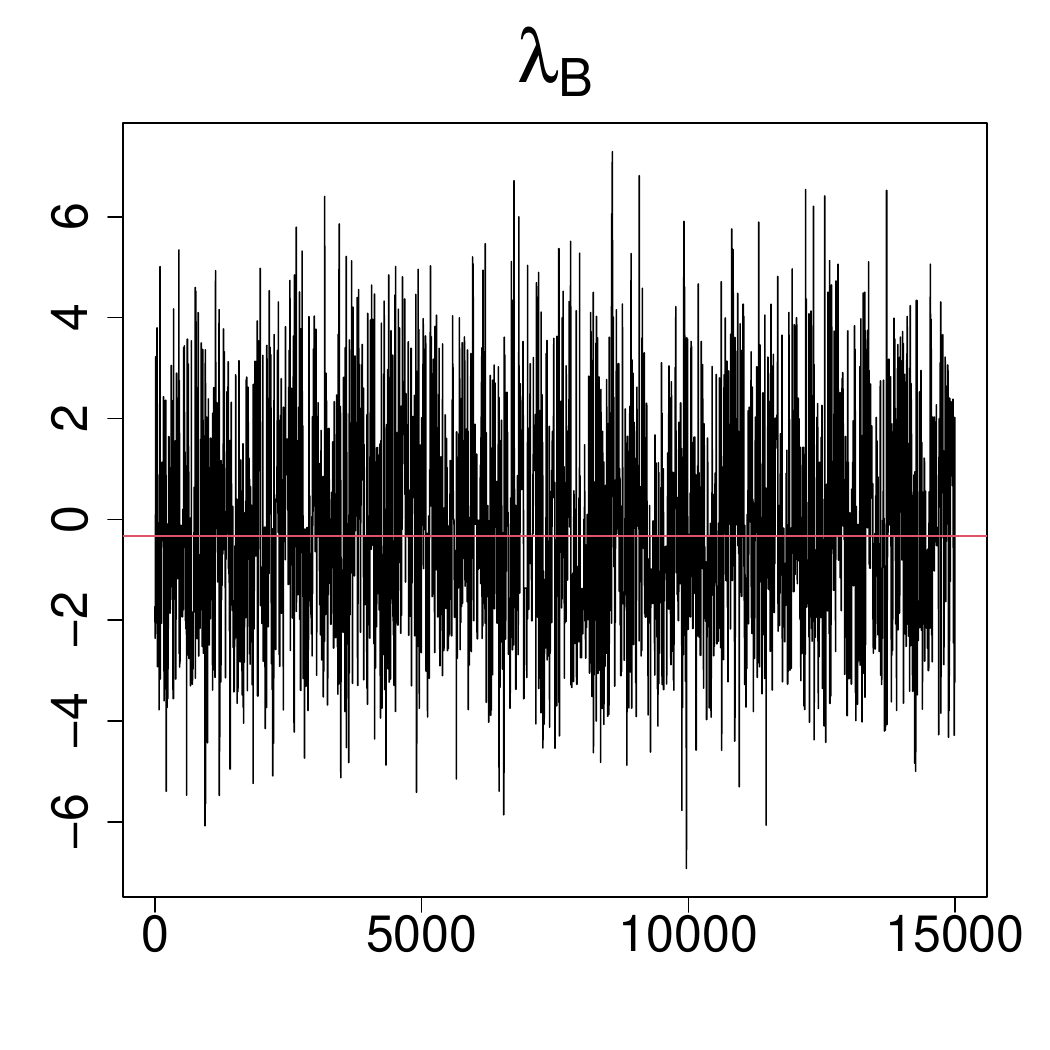} 
     & 
      \includegraphics[width=0.3 \textwidth]{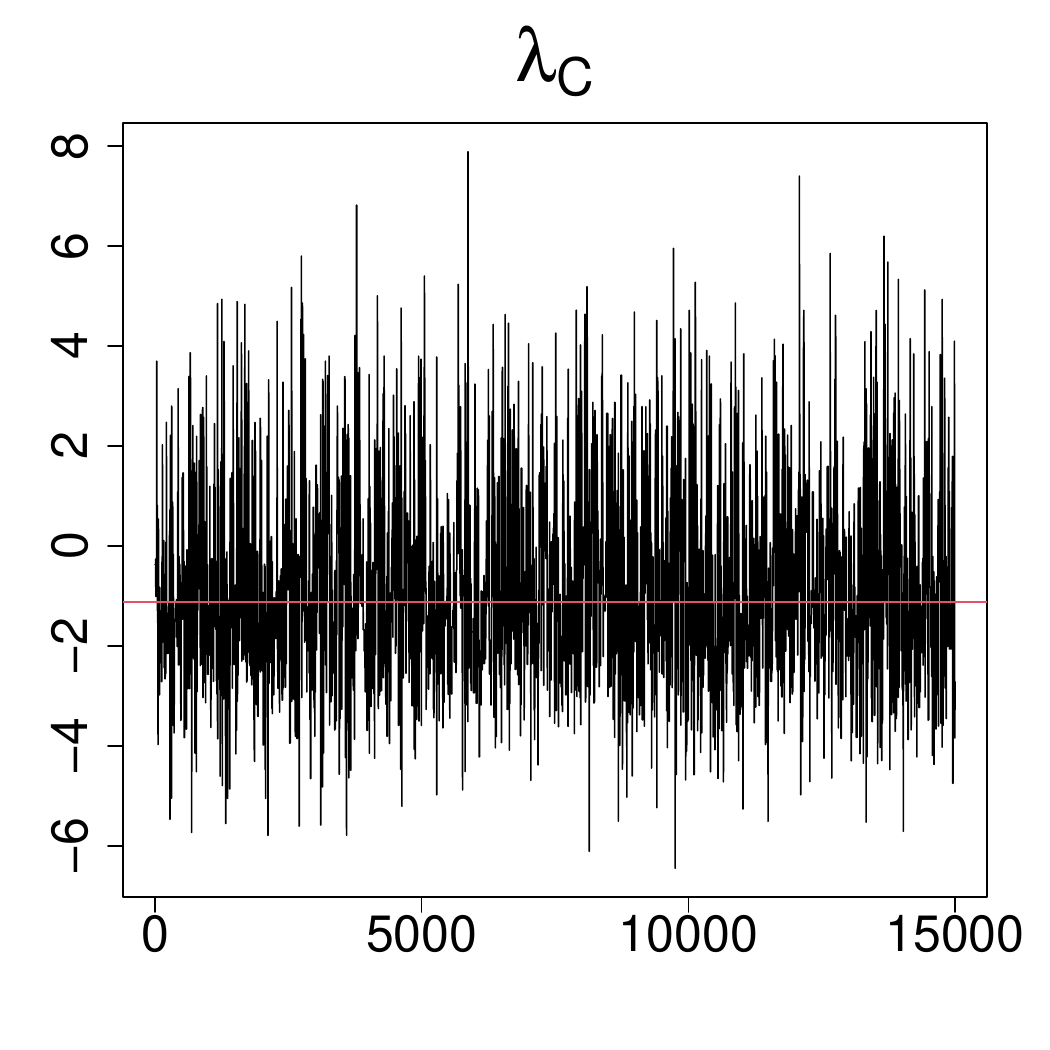} 
        \\ 
                \end{tabular}
\caption{Trace plots of select parameters from the latent process model with $q=1$ in Section \ref{sec:math_spike}. 
The red-colored horizontal lines indicate posterior medians.}
        \label{fig:trace_math_basic_spike1}
    \end{figure}

   \begin{figure}[hptb]
        \centering
        \begin{tabular}{ccc}
        \includegraphics[width=0.3 \textwidth]{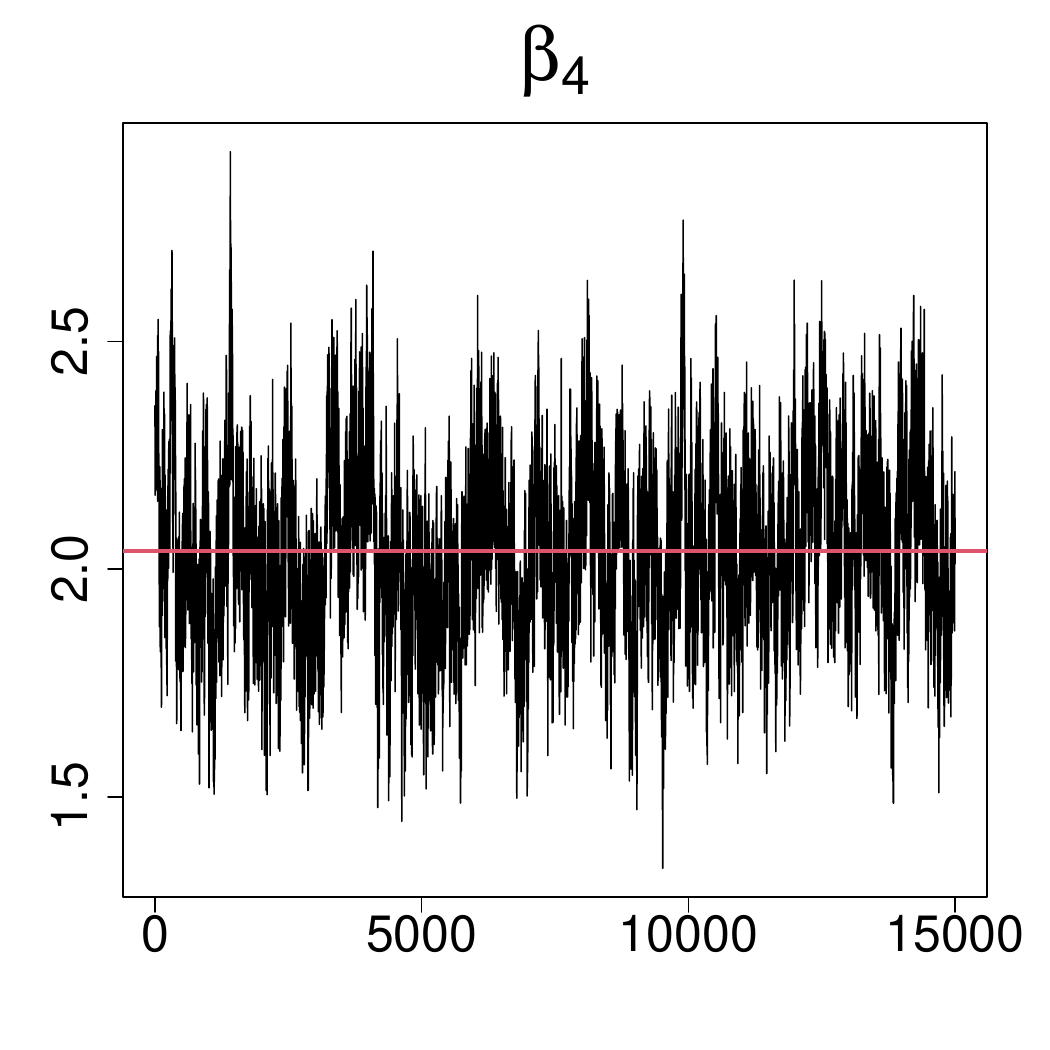} 
       &  \includegraphics[width=0.3 \textwidth]{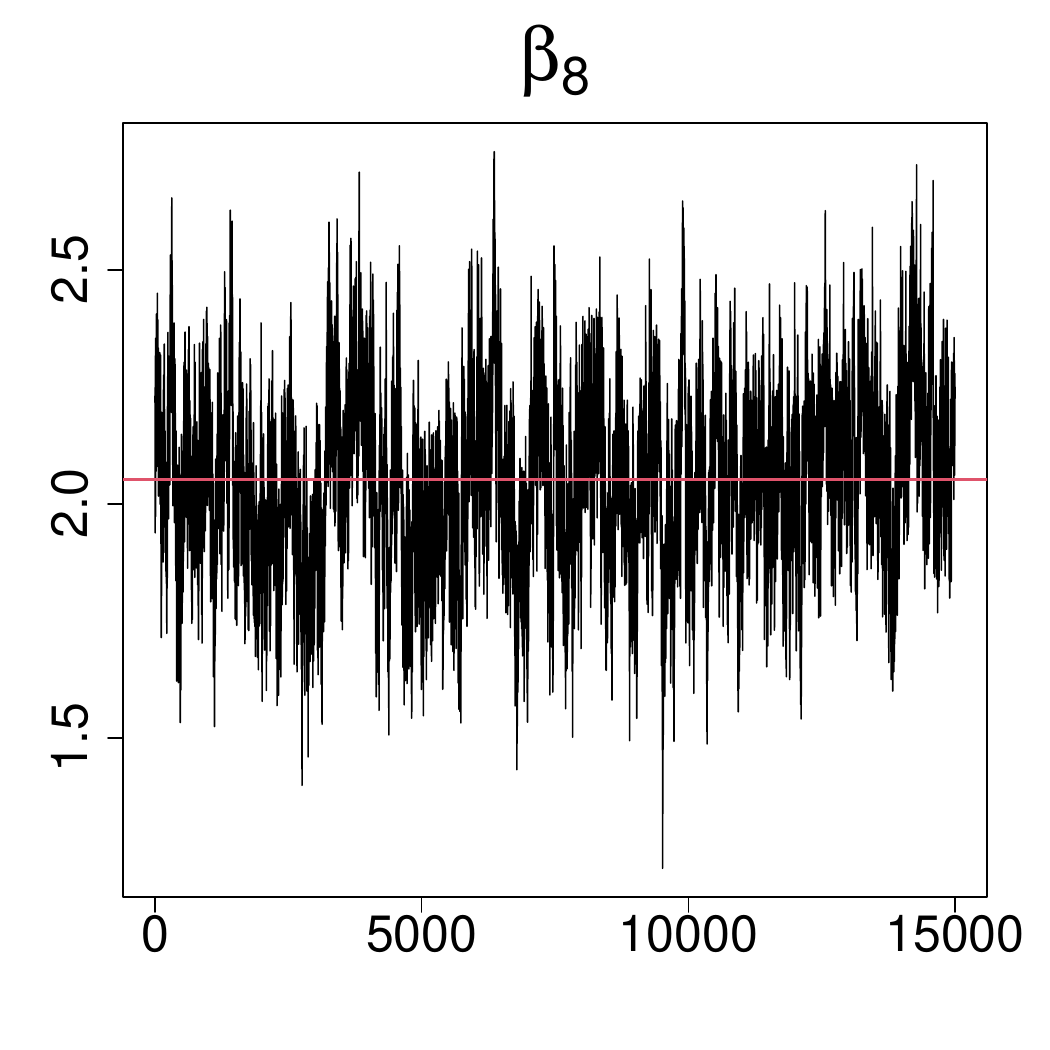} 
    &    \includegraphics[width=0.3 \textwidth]{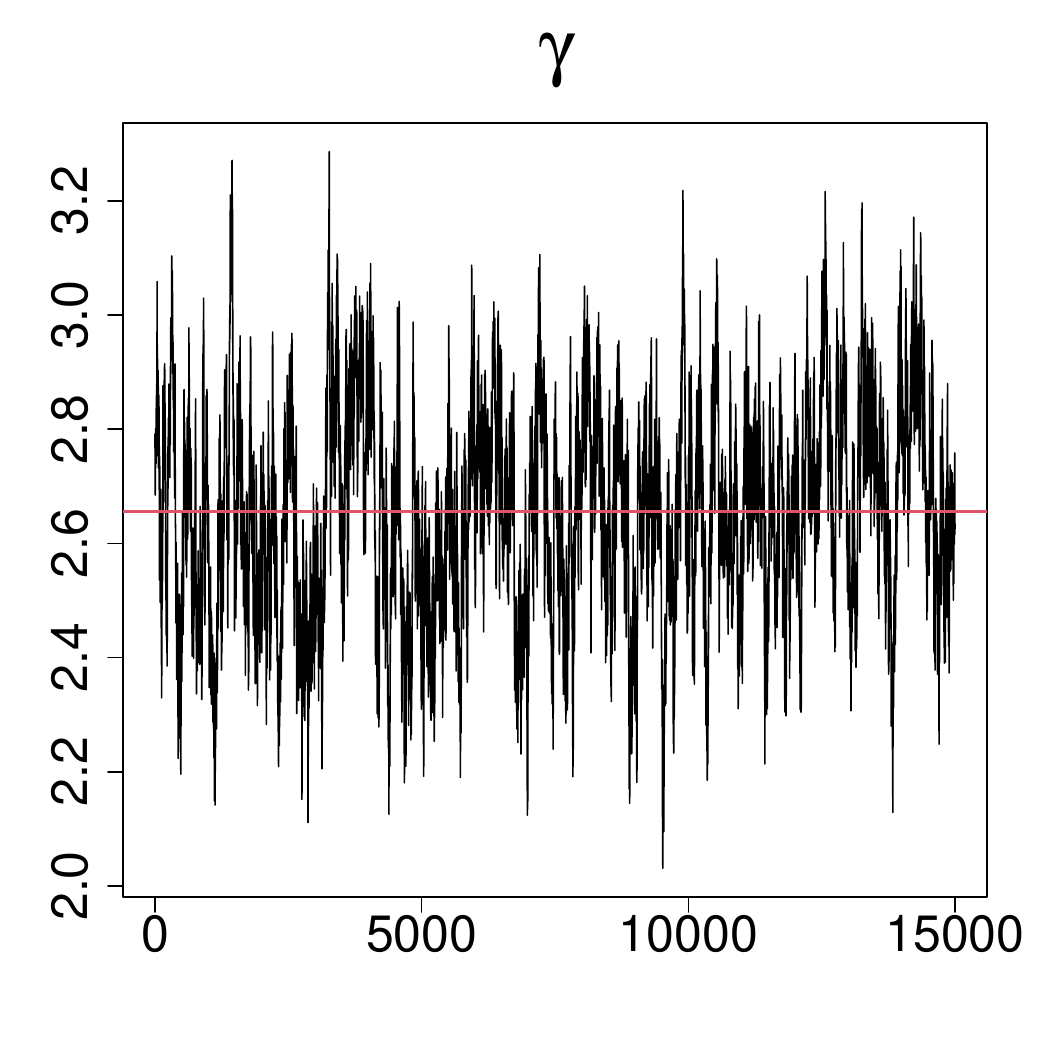} \\
    \includegraphics[width=0.3 \textwidth]{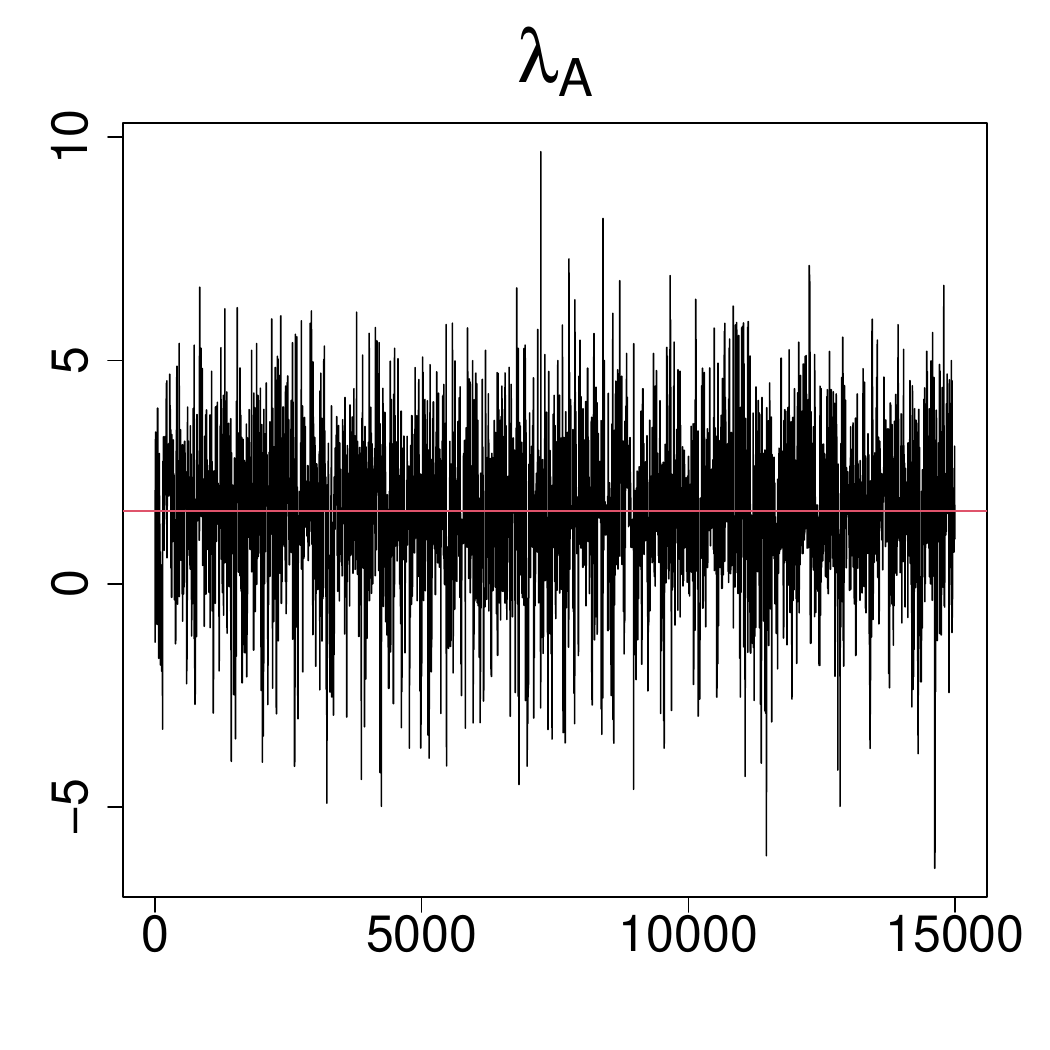}  
    &
     \includegraphics[width=0.3 \textwidth]{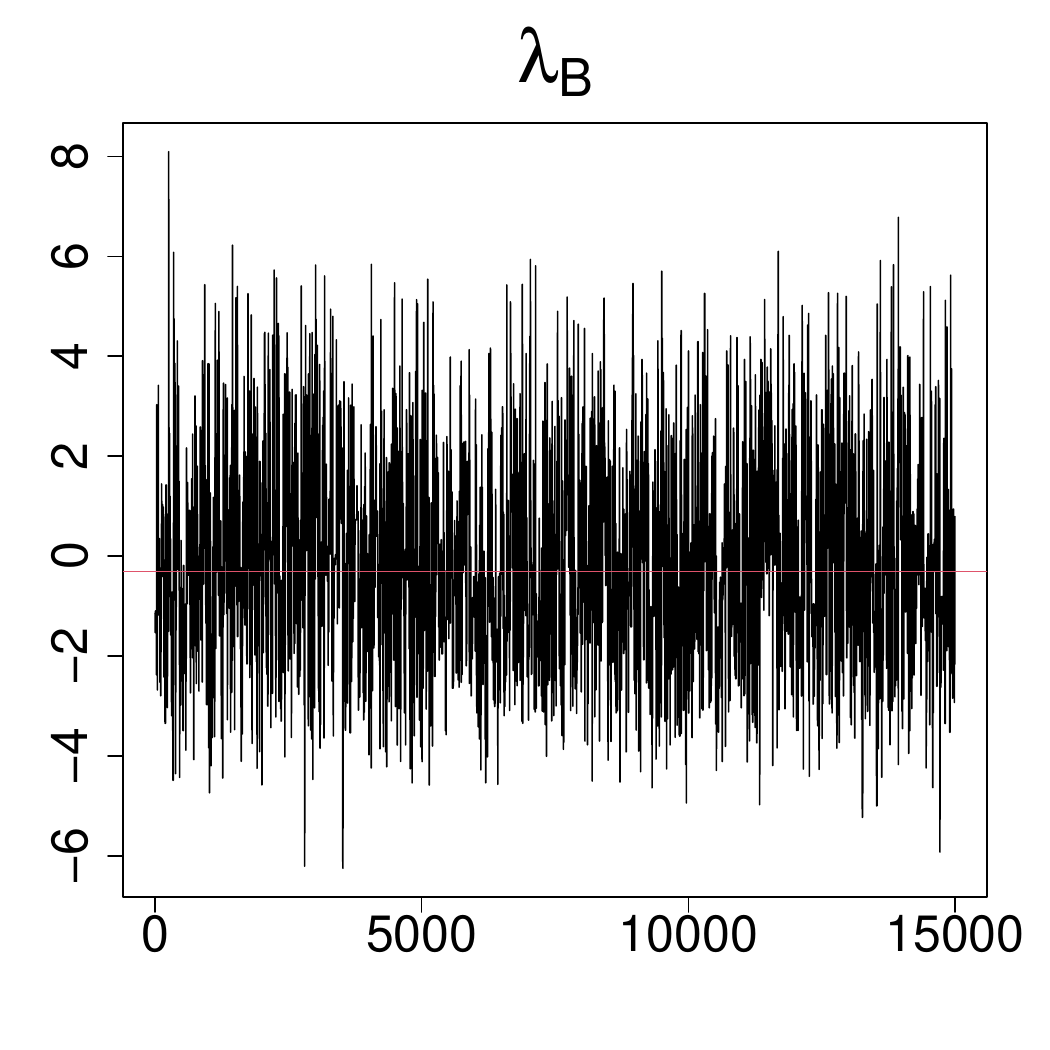} 
     & 
      \includegraphics[width=0.3 \textwidth]{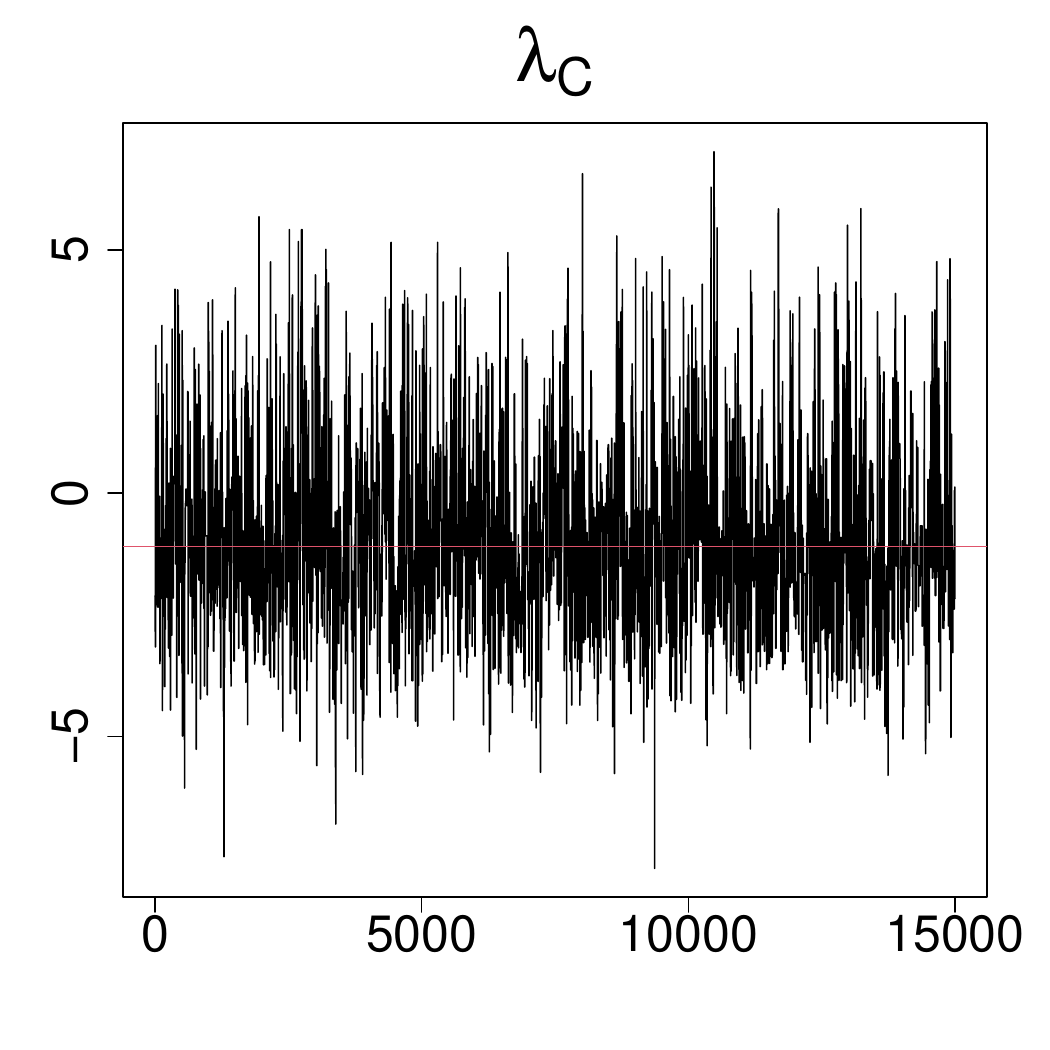} 
        \\ 
                \end{tabular}
\caption{Trace plots of select parameters from the latent process model with $q=2$ in Section \ref{sec:math_spike}. 
The red-colored horizontal lines indicate posterior medians.}
        \label{fig:trace_math_basic_spike2}
    \end{figure}

   \begin{figure}[hptb]
        \centering
        \begin{tabular}{ccc}
        \includegraphics[width=0.3 \textwidth]{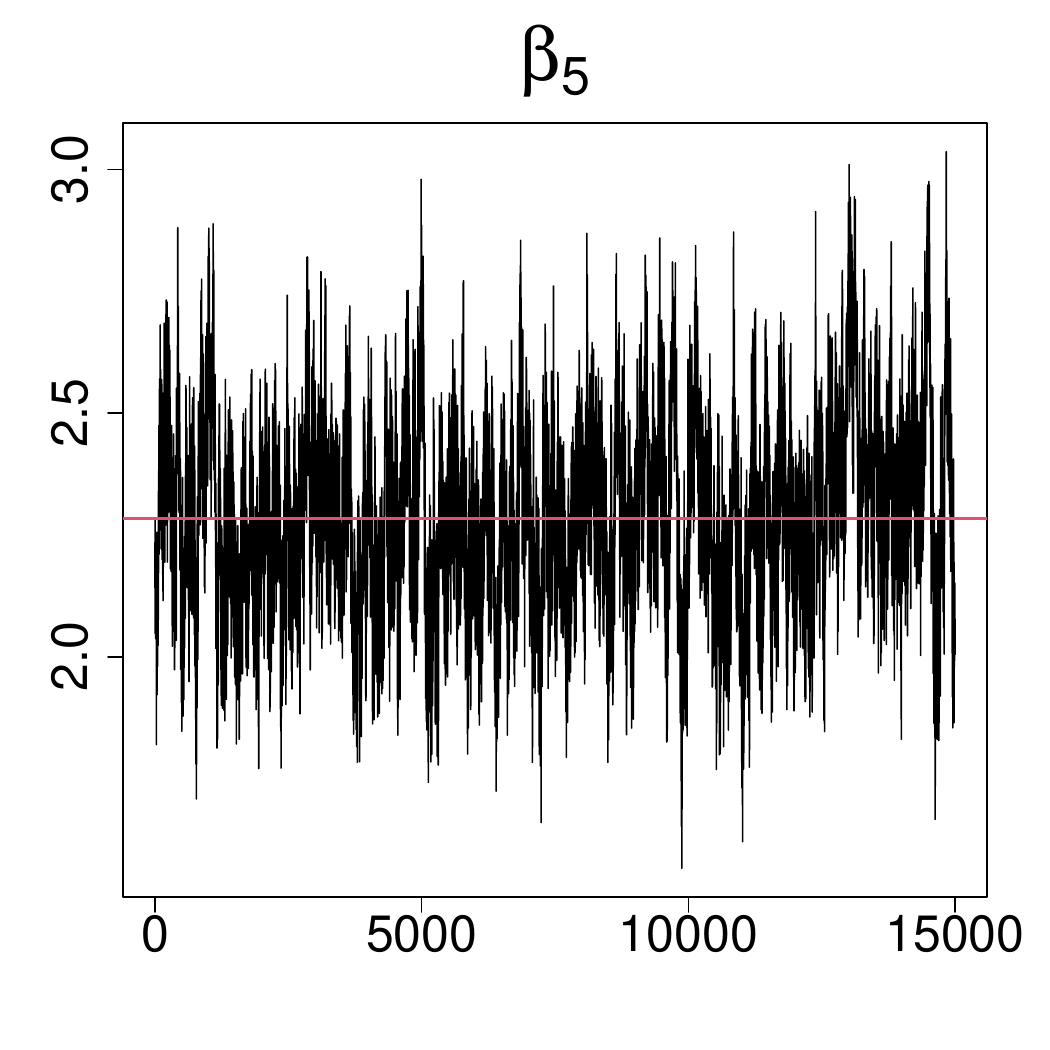} 
       &  \includegraphics[width=0.3 \textwidth]{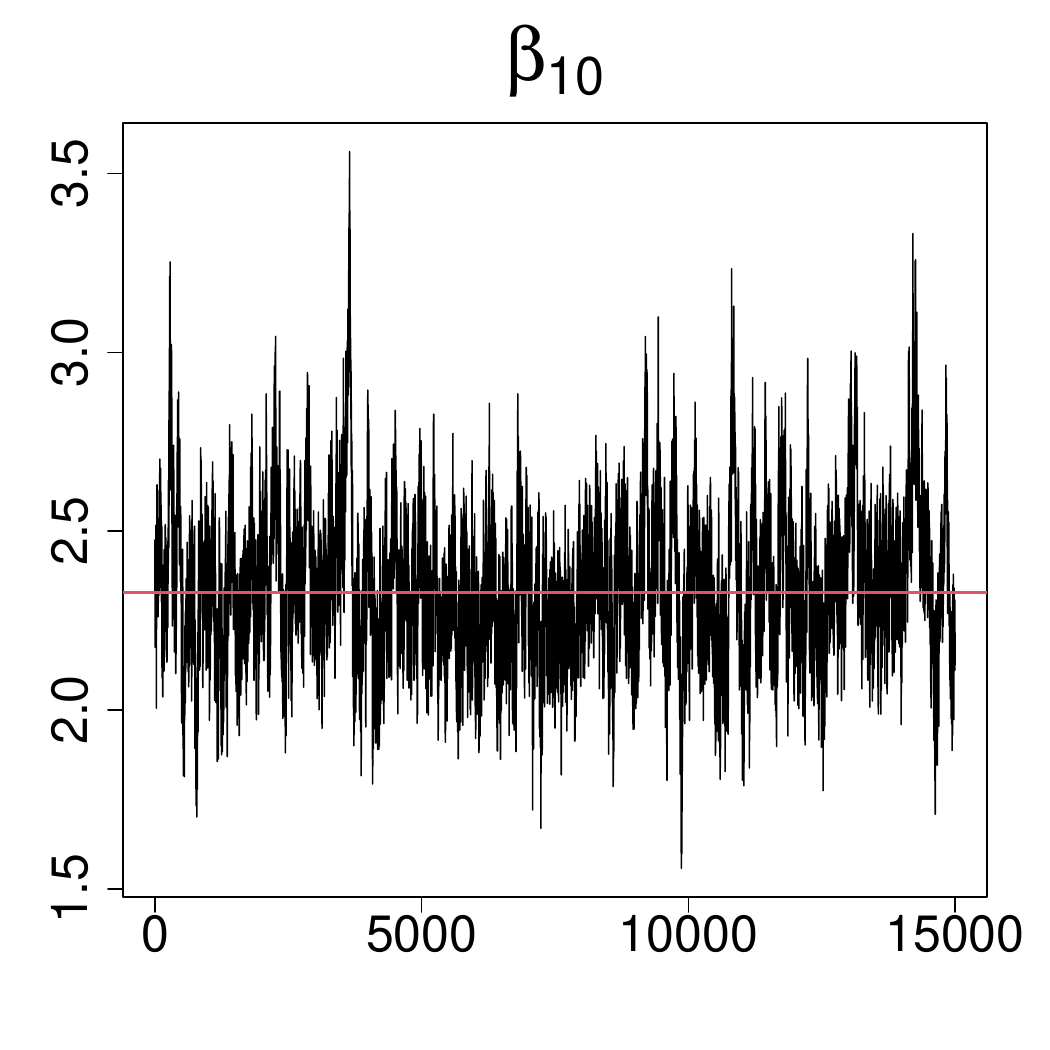} 
    &    \includegraphics[width=0.3 \textwidth]{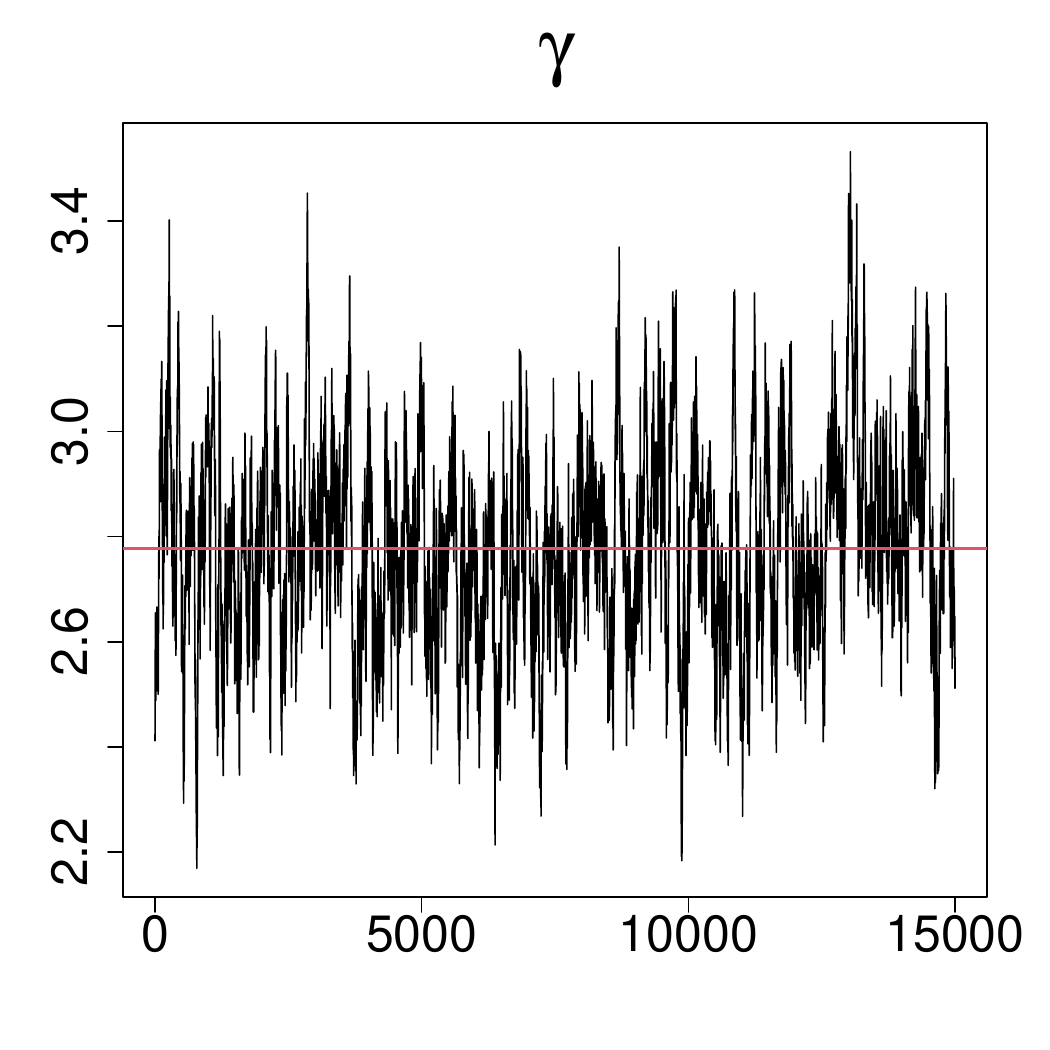} \\
    \includegraphics[width=0.3 \textwidth]{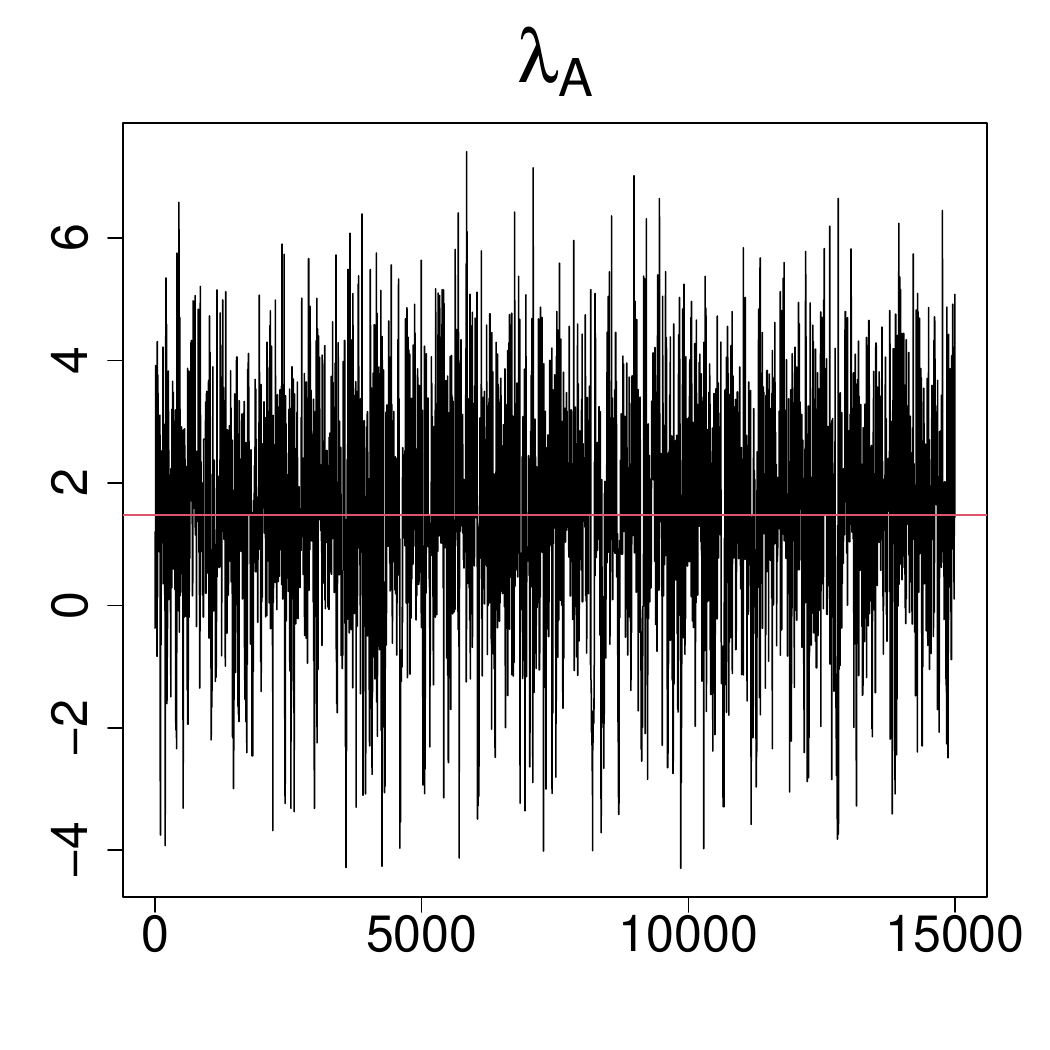}  
    &
     \includegraphics[width=0.3 \textwidth]{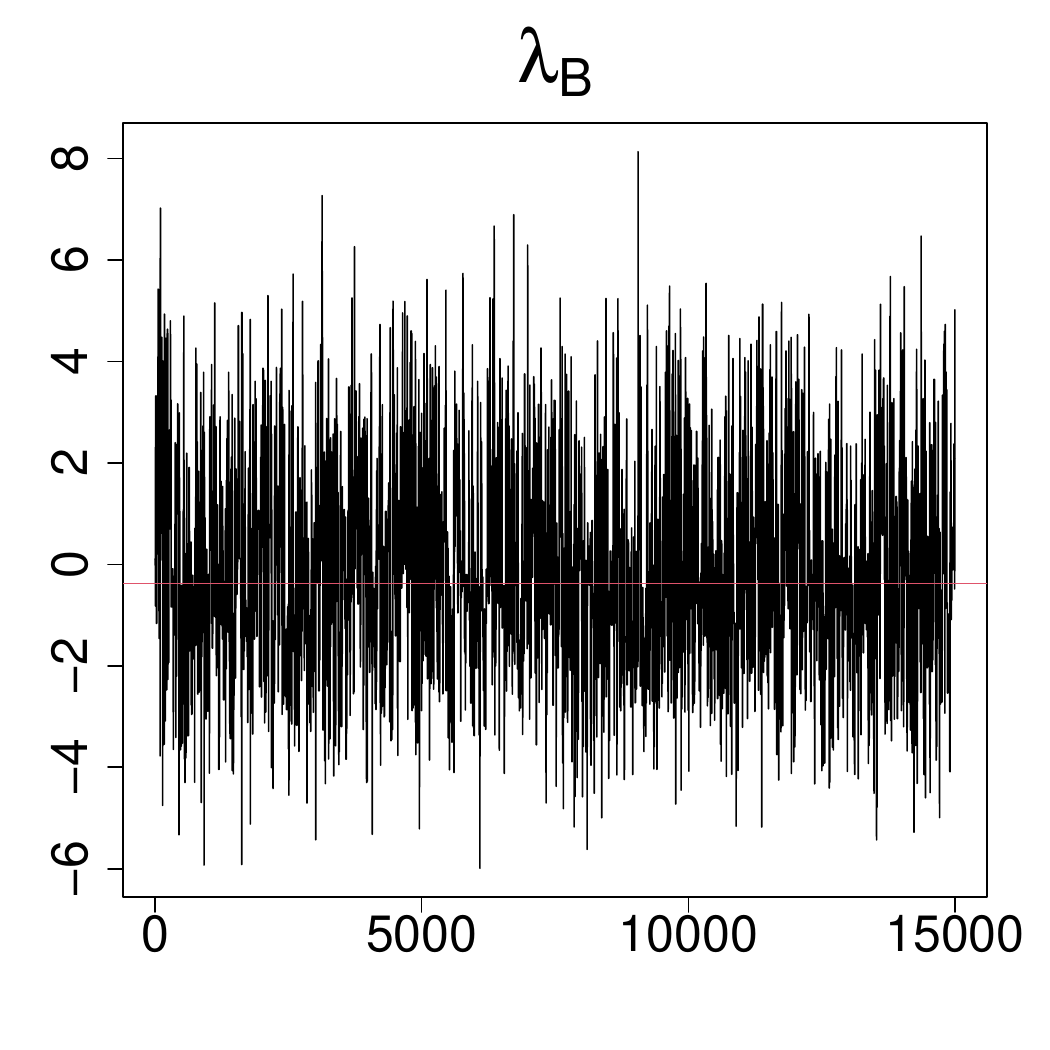} 
     & 
      \includegraphics[width=0.3 \textwidth]{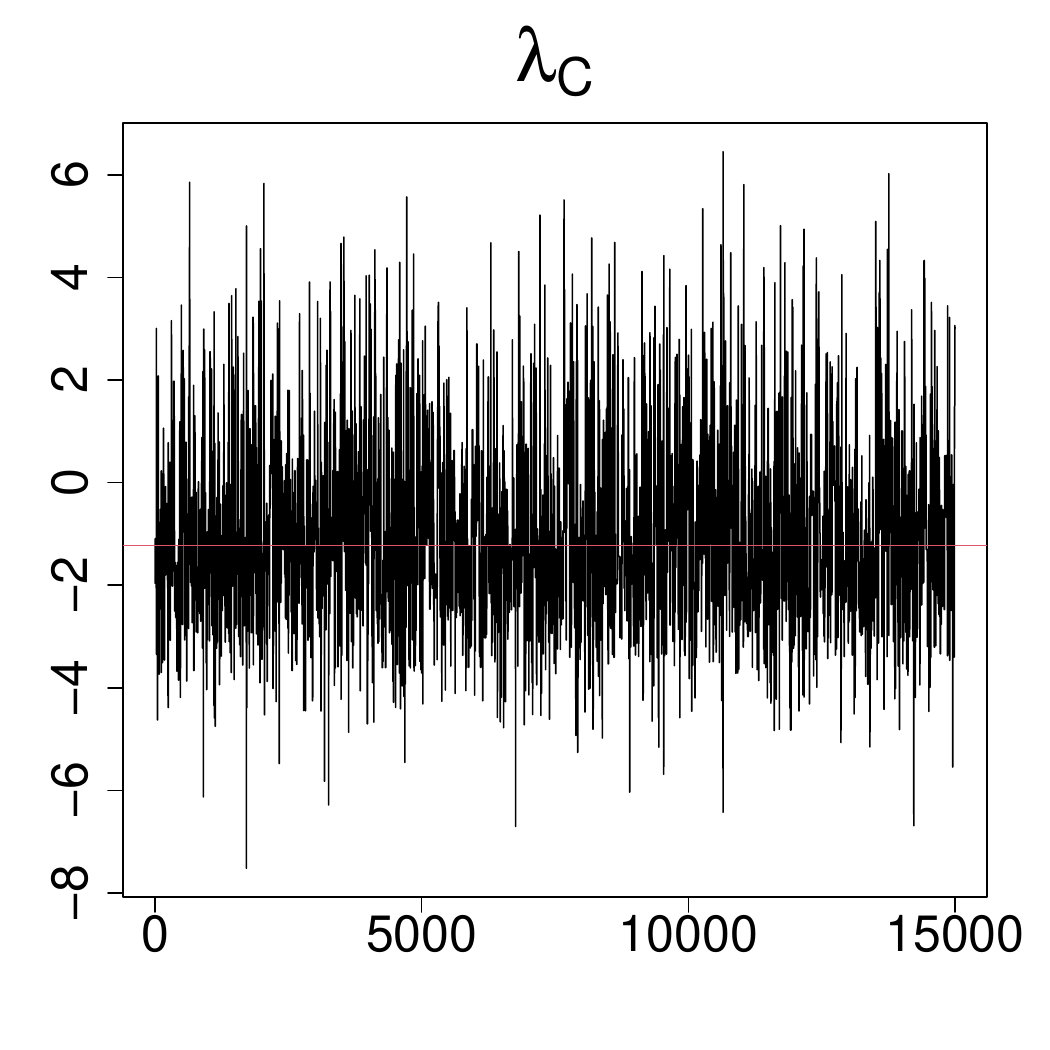} 
        \\ 
                \end{tabular}
\caption{Trace plots of select parameters from the latent process model  with $q=3$ in Section \ref{sec:math_spike}. 
The red-colored horizontal lines indicate posterior medians.}
        \label{fig:trace_math_basic_spike3}
    \end{figure}

   \begin{figure}[hptb]
        \centering
        \begin{tabular}{ccc}
        \includegraphics[width=0.3 \textwidth]{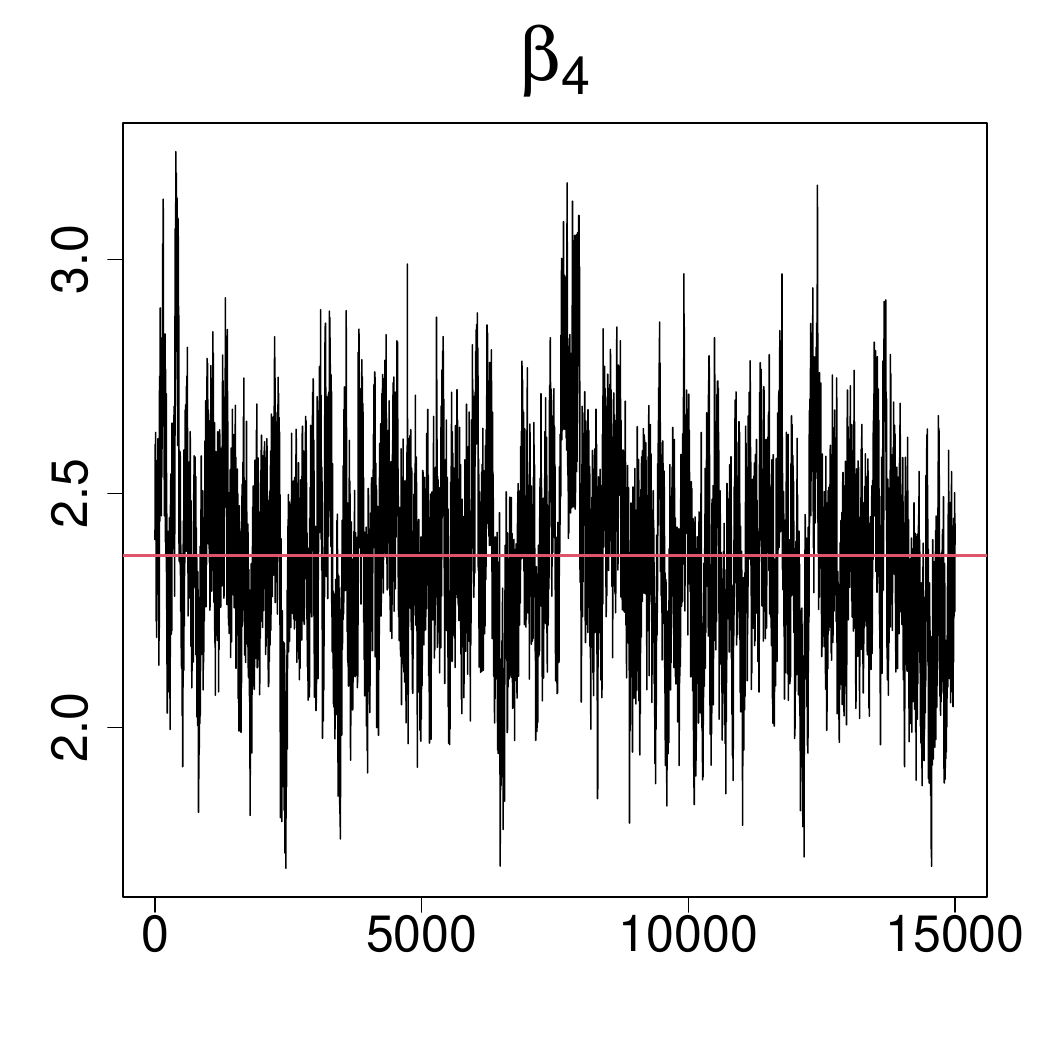} 
       &  \includegraphics[width=0.3 \textwidth]{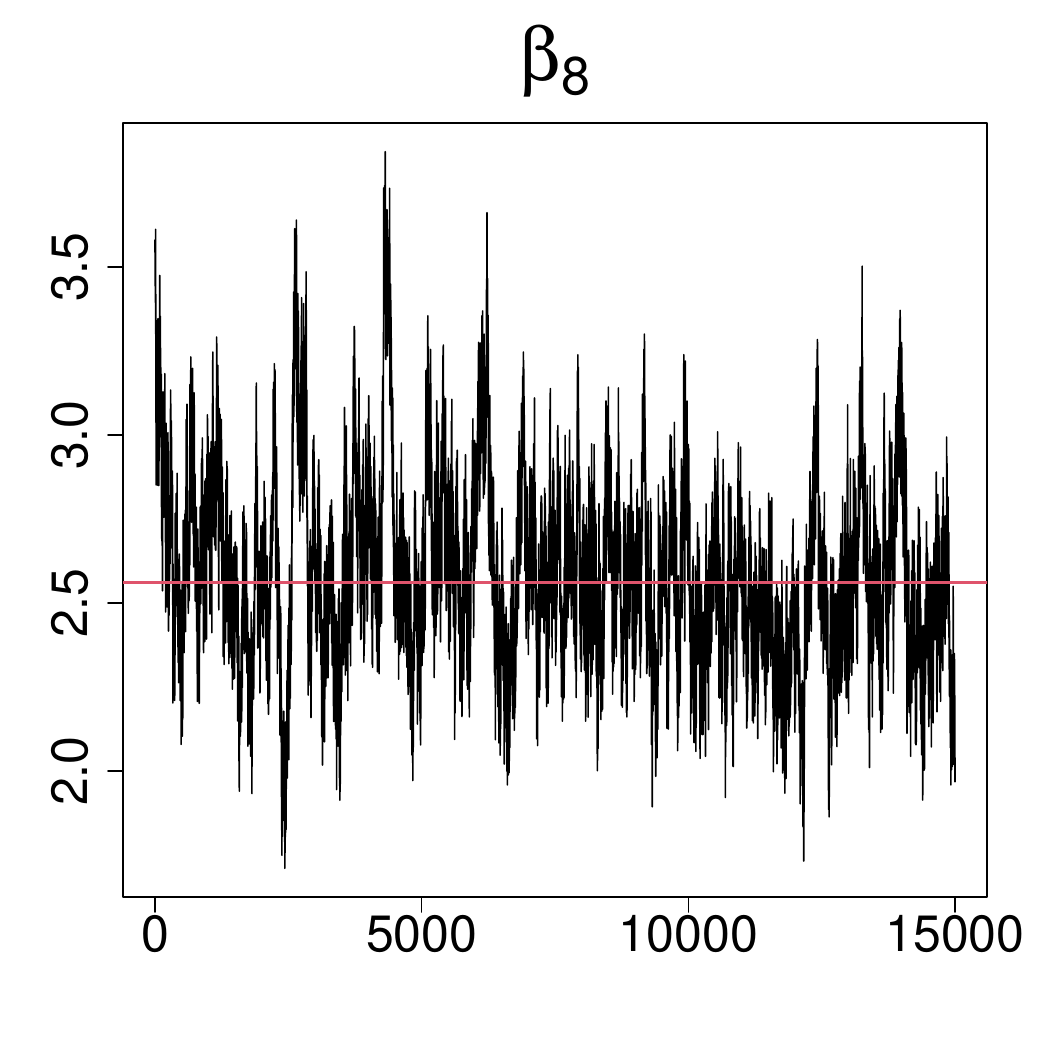} 
    &    \includegraphics[width=0.3 \textwidth]{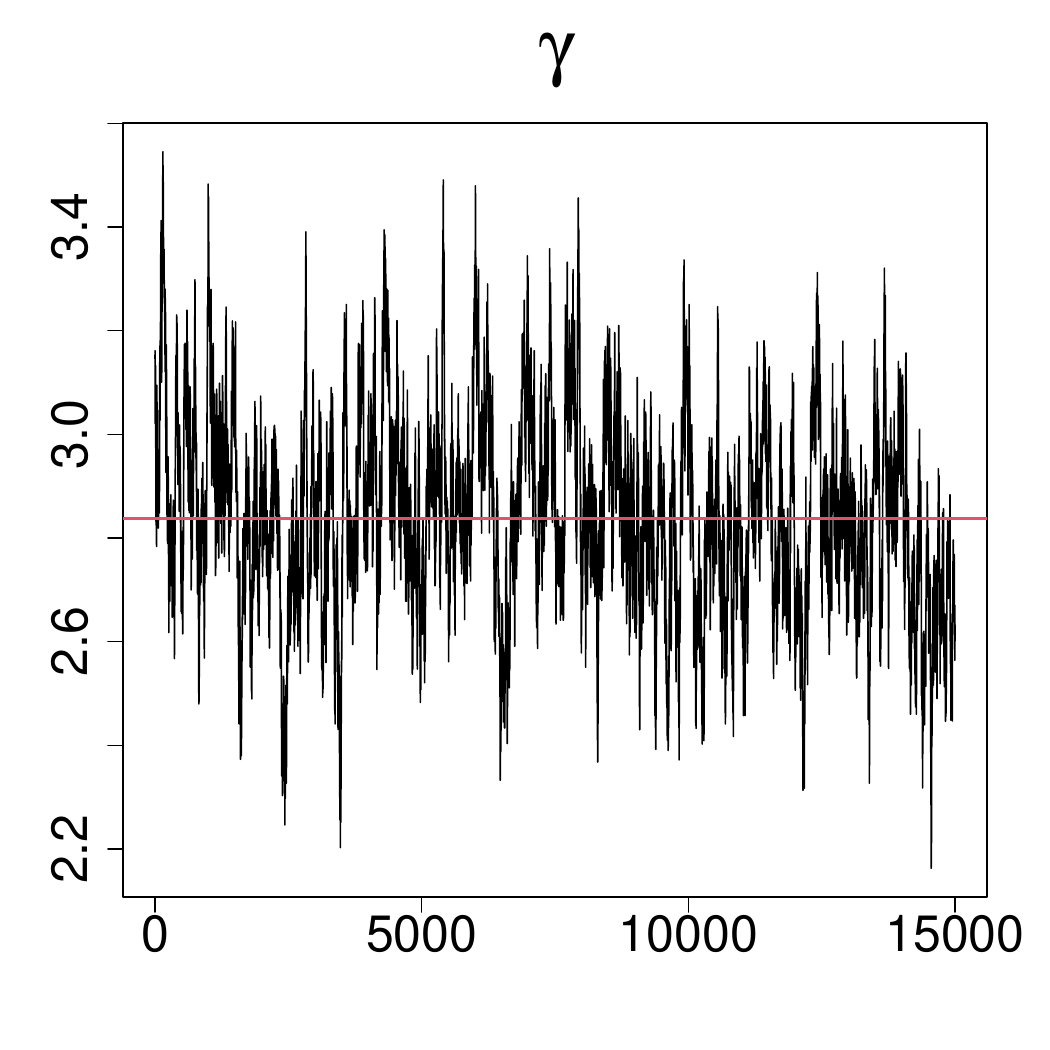} \\
    \includegraphics[width=0.3 \textwidth]{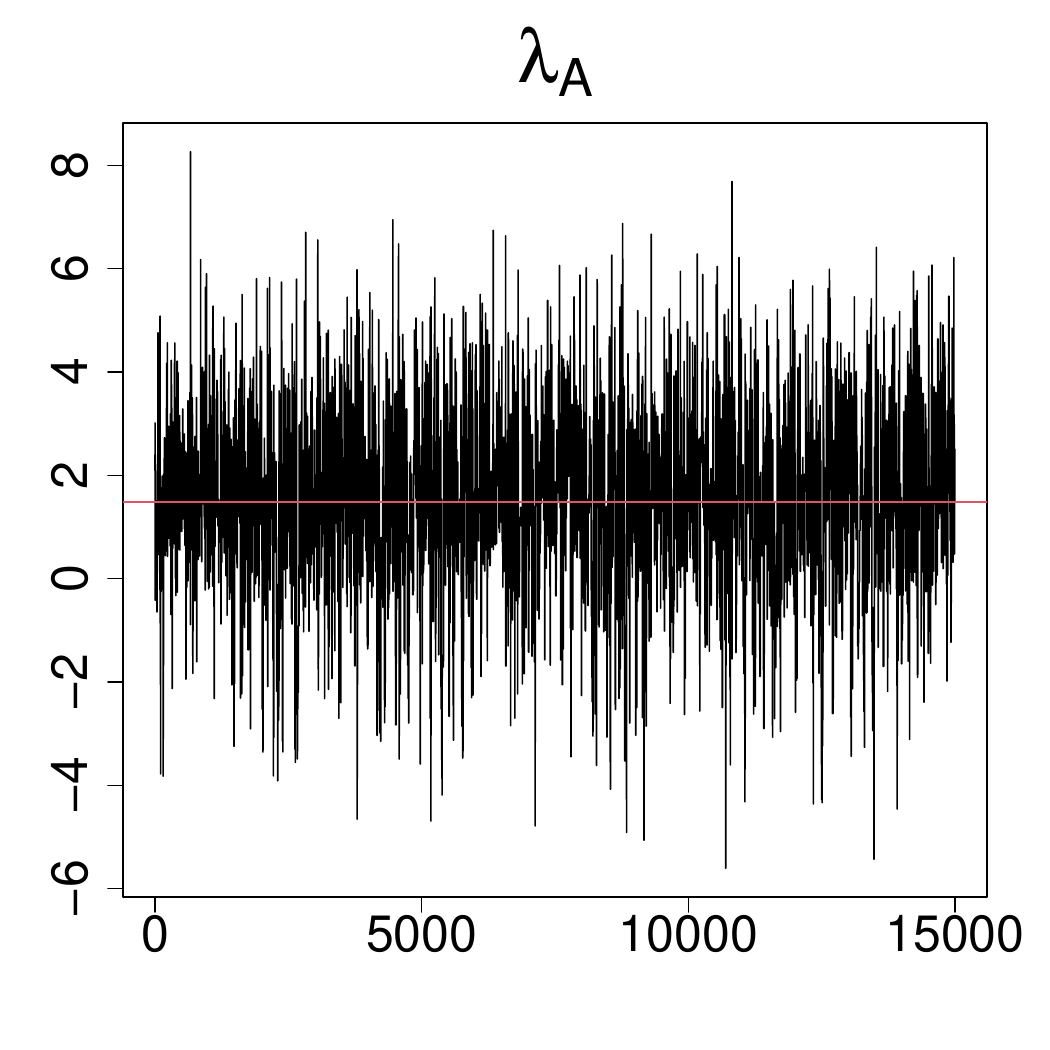}  
    &
     \includegraphics[width=0.3 \textwidth]{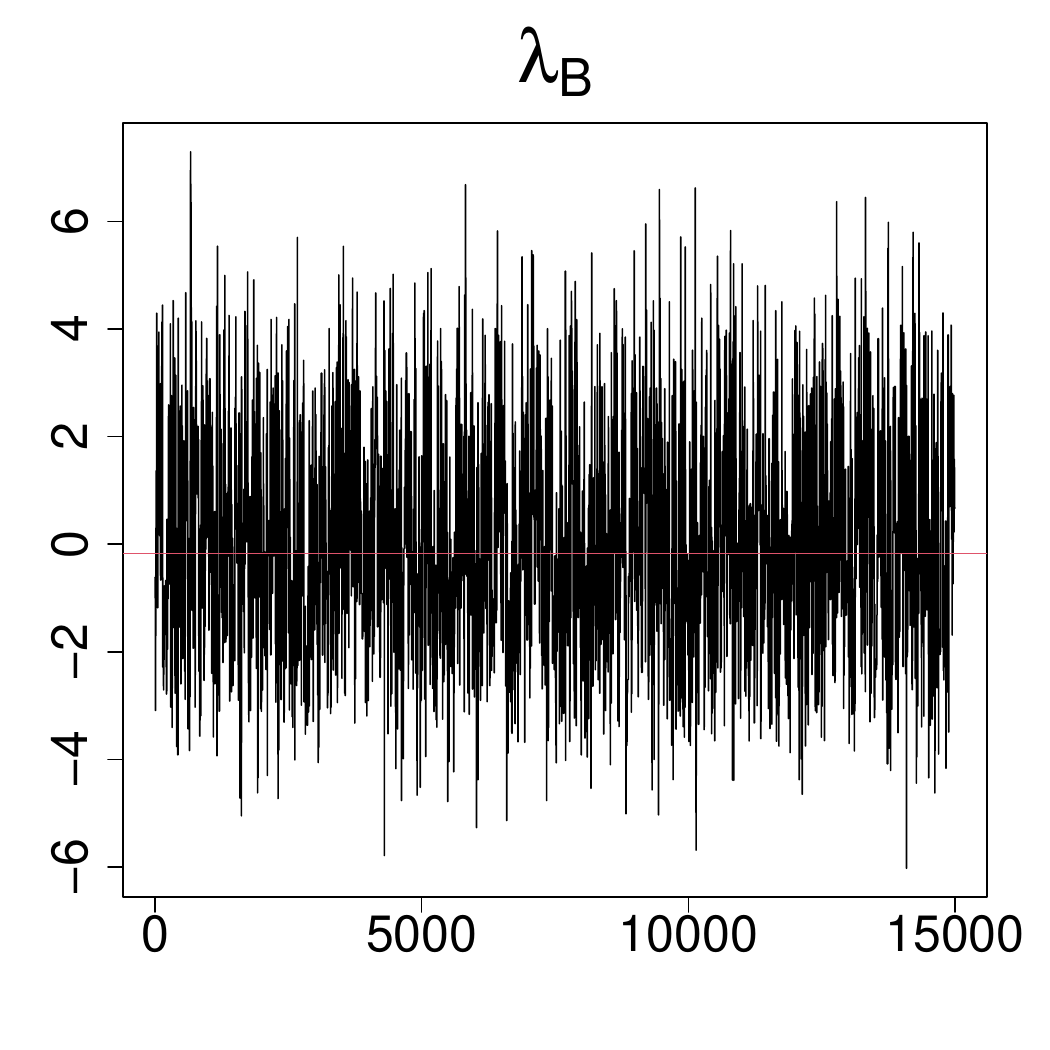} 
     & 
      \includegraphics[width=0.3 \textwidth]{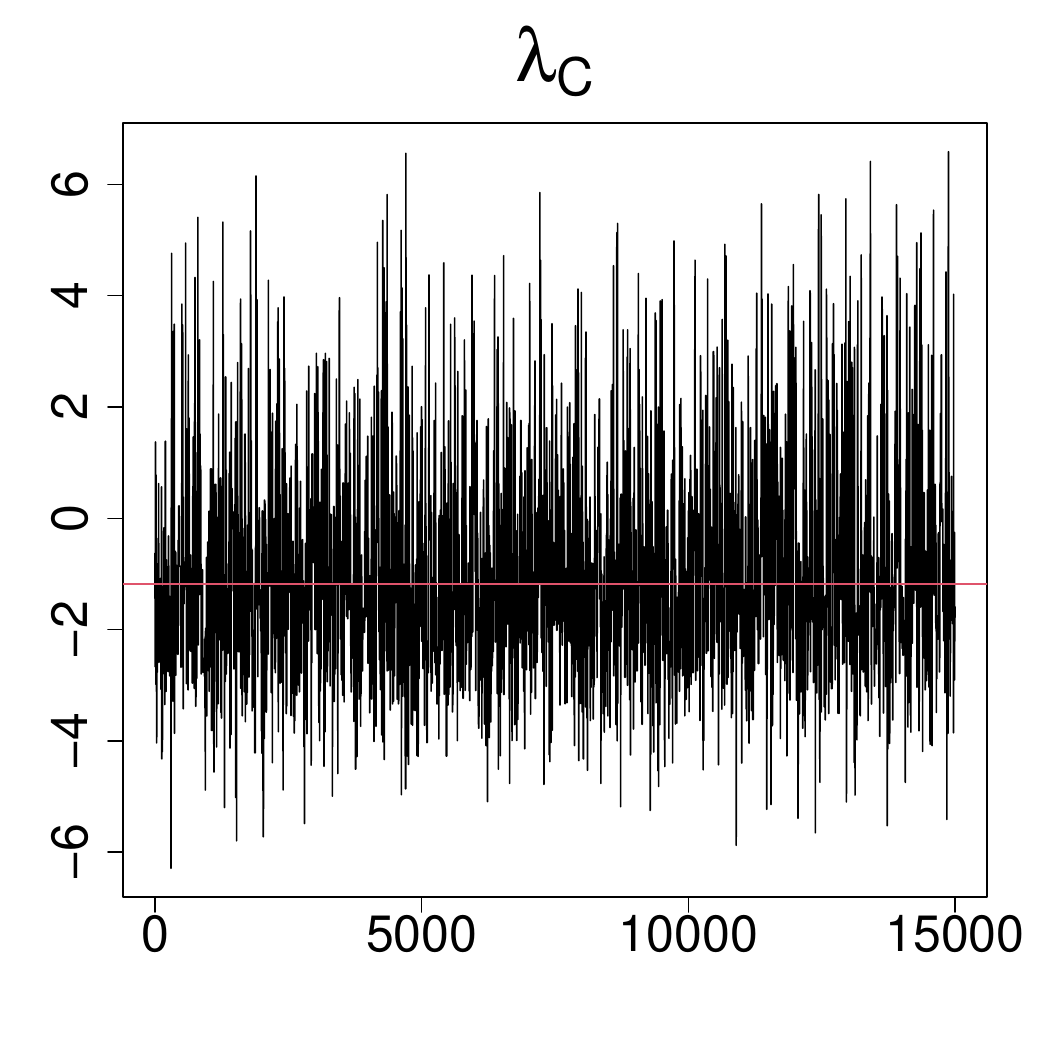} 
        \\ 
                \end{tabular}
\caption{Trace plots of select parameters from the latent process model  with $q=4$ in Section \ref{sec:math_spike}. 
The red-colored horizontal lines indicate posterior medians.}
        \label{fig:trace_math_basic_spike4}
    \end{figure}

\clearpage 

\subsubsection{Section \ref{sec:depression}: Application: mental health}

   \begin{figure}[hptb]
        \centering
        \begin{tabular}{ccc}
        \includegraphics[width=0.3 \textwidth]{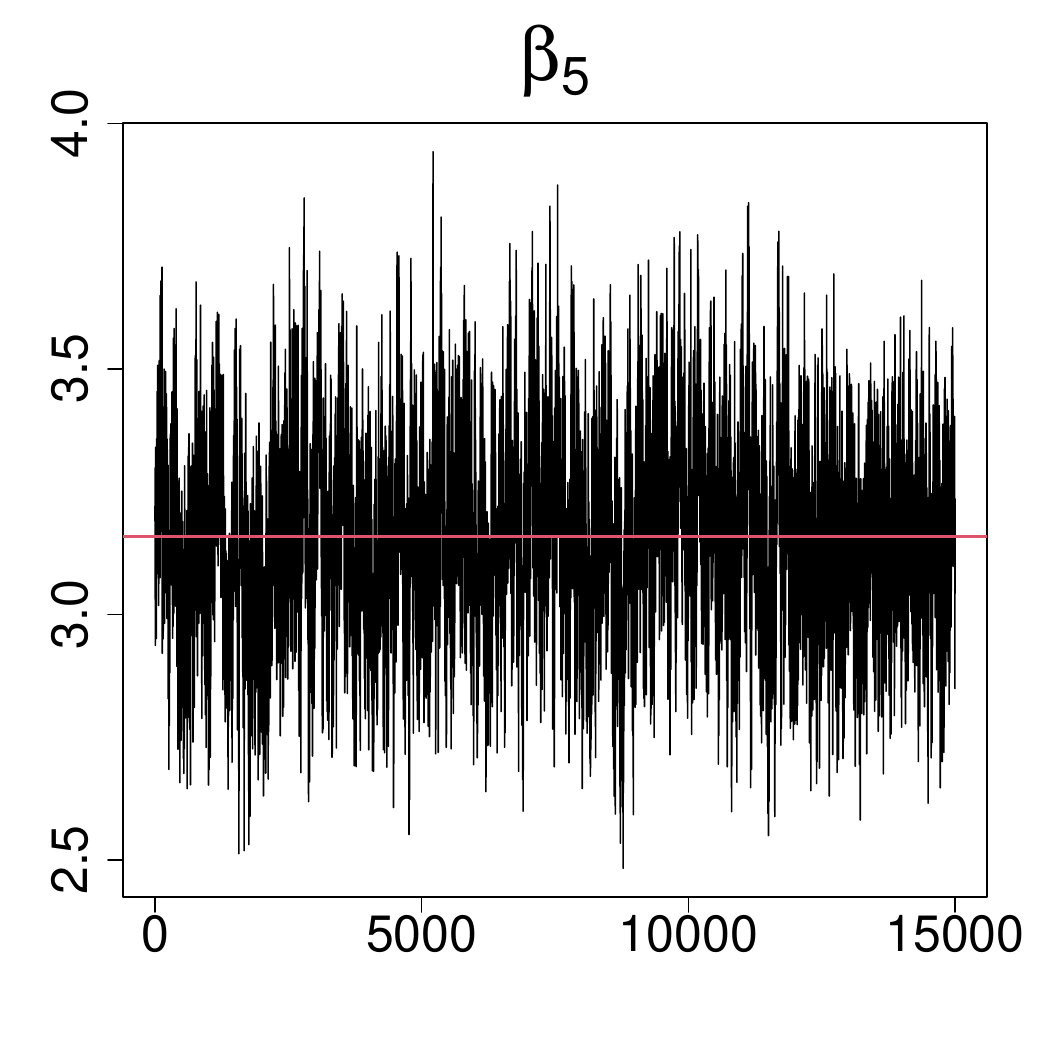} 
       &  \includegraphics[width=0.3 \textwidth]{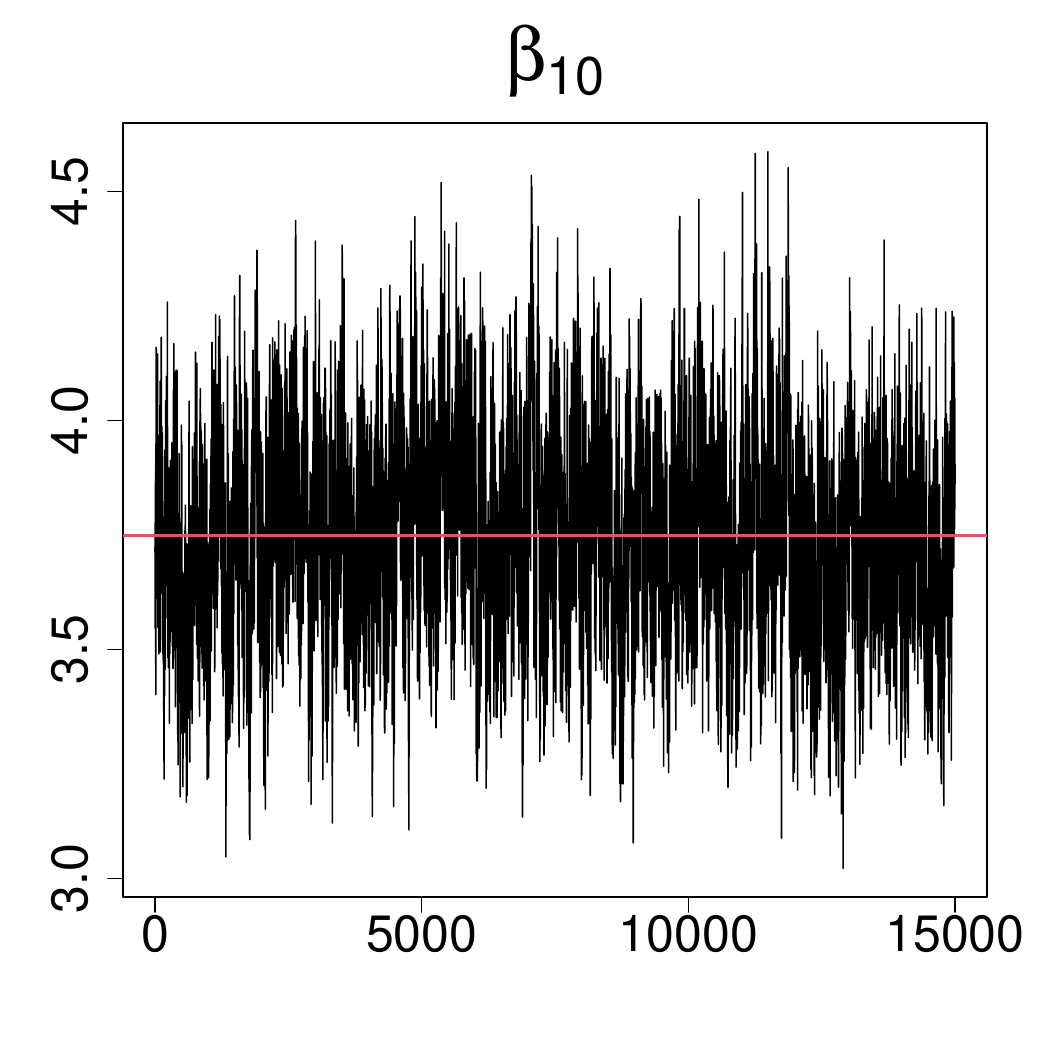} 
    &    \includegraphics[width=0.3 \textwidth]{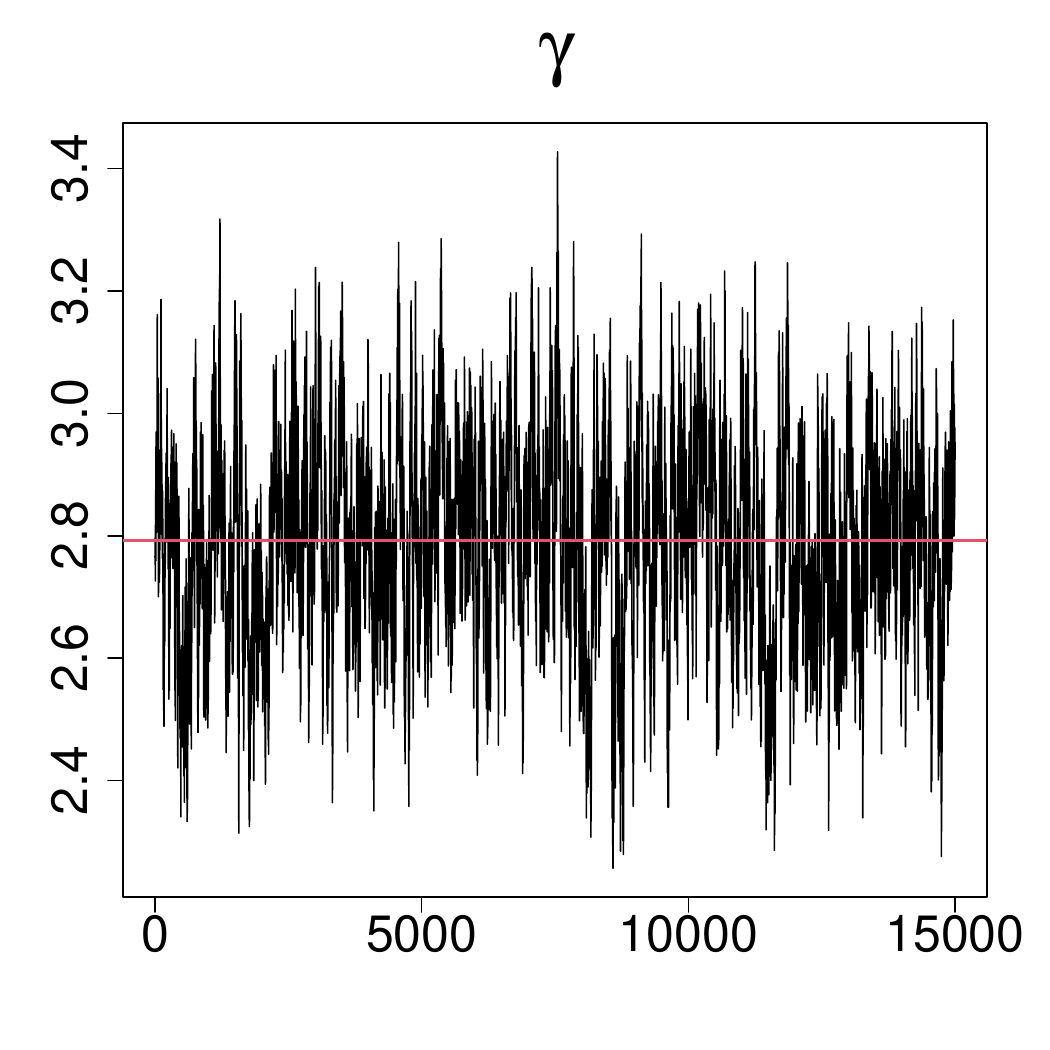} \\
    \includegraphics[width=0.3 \textwidth]{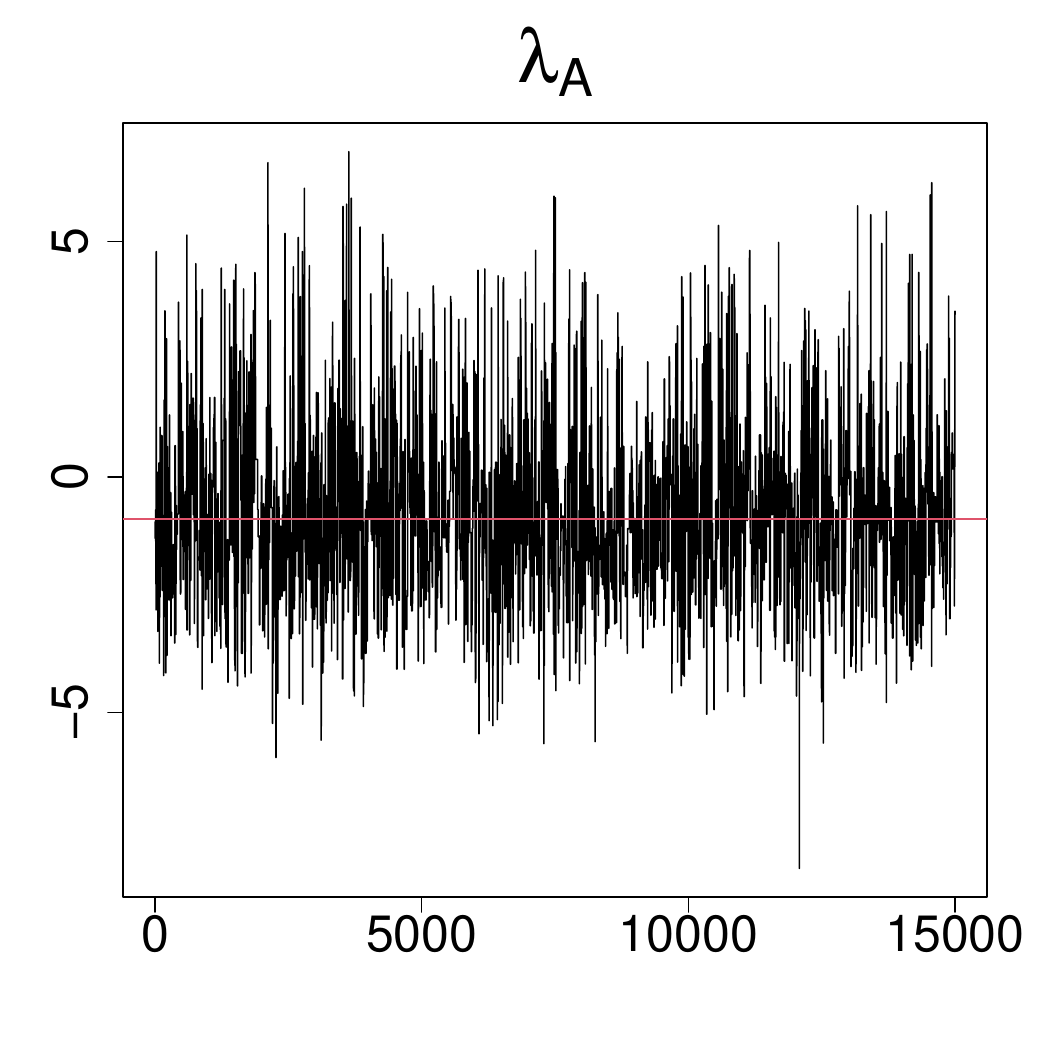}  
    &
     \includegraphics[width=0.3 \textwidth]{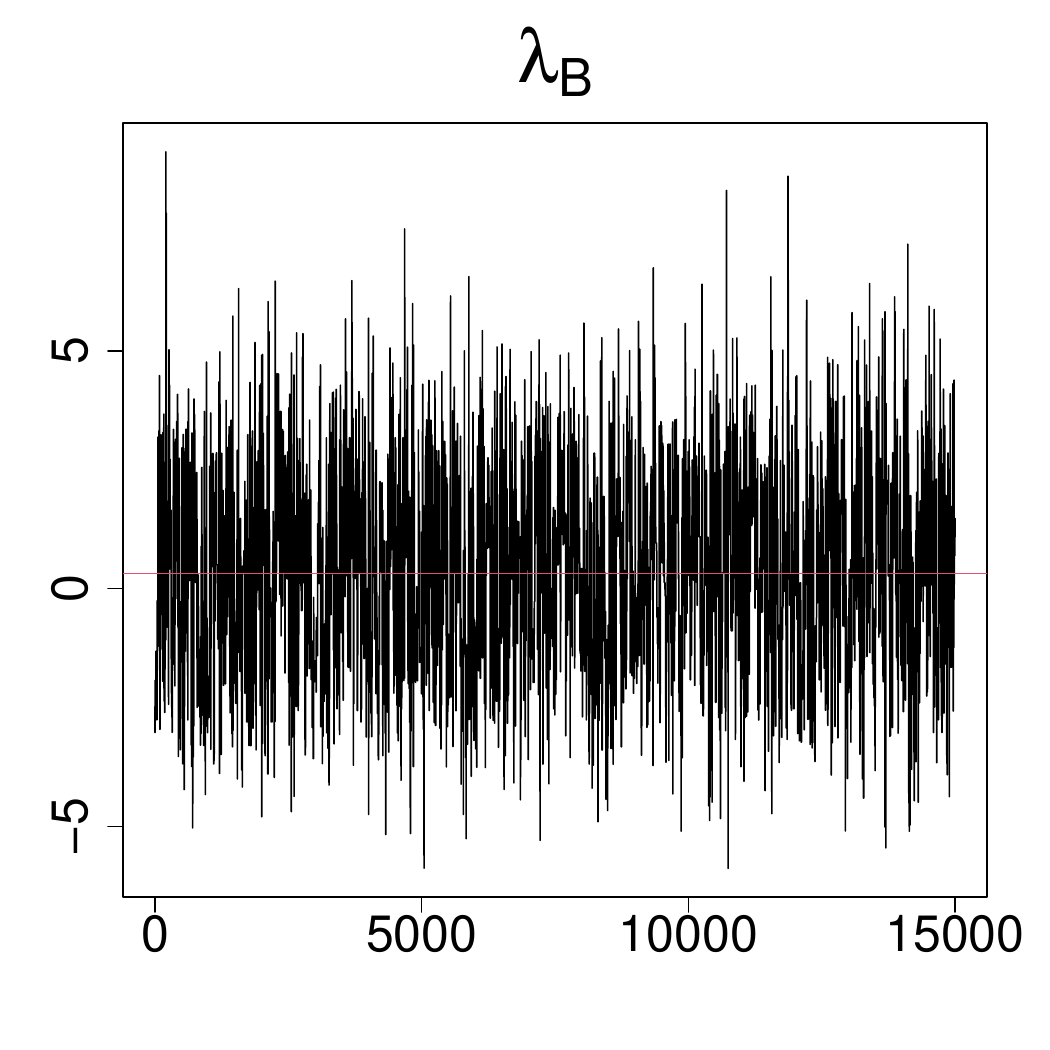} 
     & 
      \includegraphics[width=0.3 \textwidth]{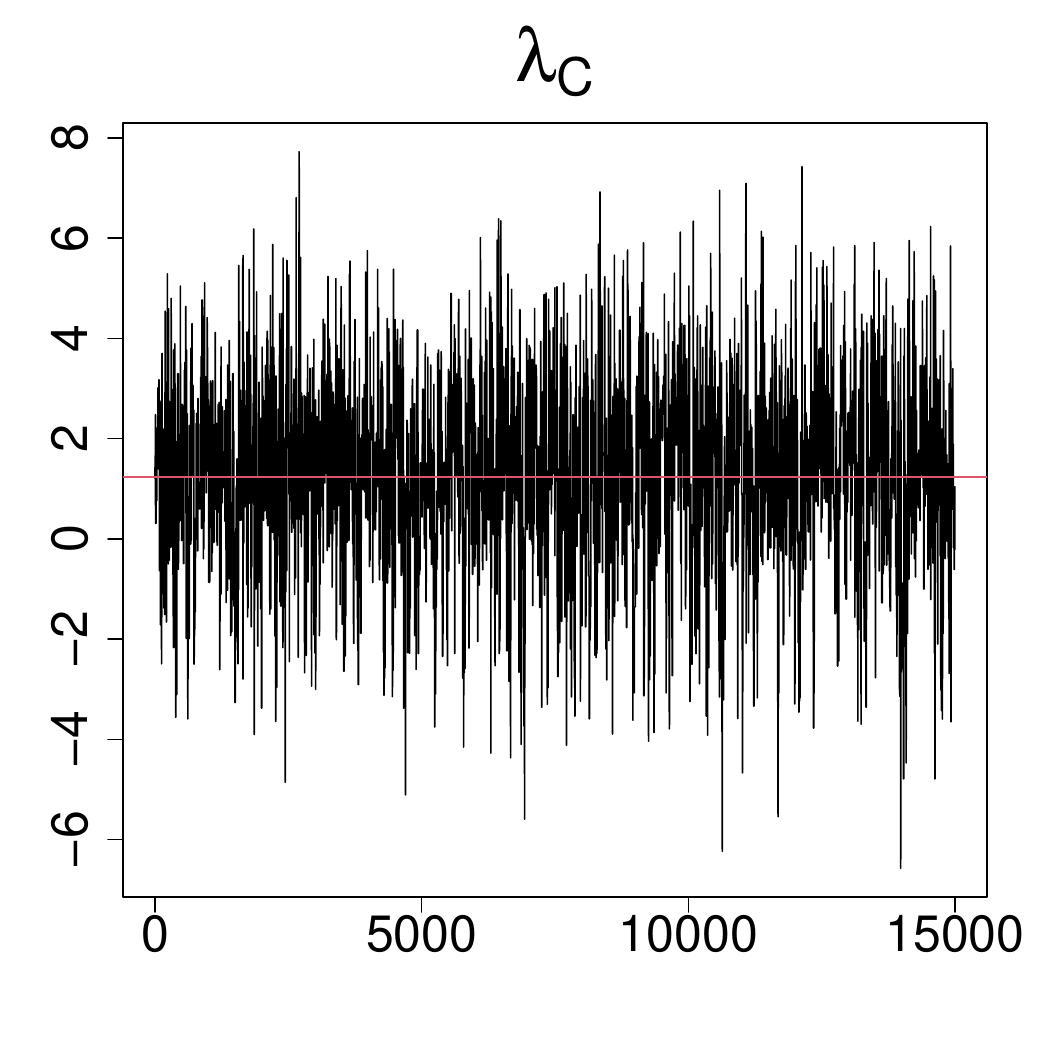} 
        \\ 
                \end{tabular}
\caption{Trace plots of select parameters from the latent process model  with $q=1$ in Section \ref{sec:dep_spike}. 
The red-colored horizontal lines indicate posterior medians.}
        \label{fig:trace_dep_basic_spike1}
    \end{figure}

   \begin{figure}[hptb]
        \centering
        \begin{tabular}{ccc}
        \includegraphics[width=0.3 \textwidth]{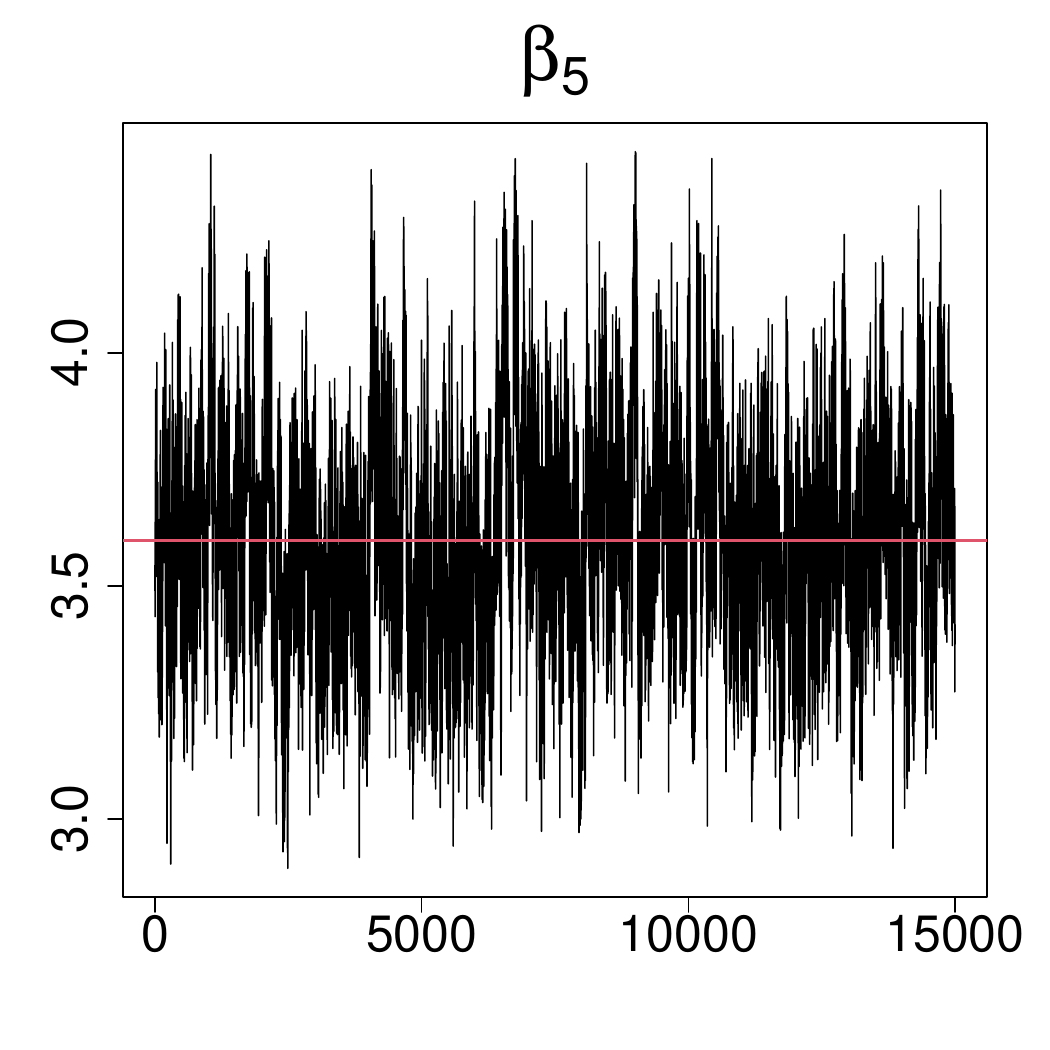} 
       &  \includegraphics[width=0.3 \textwidth]{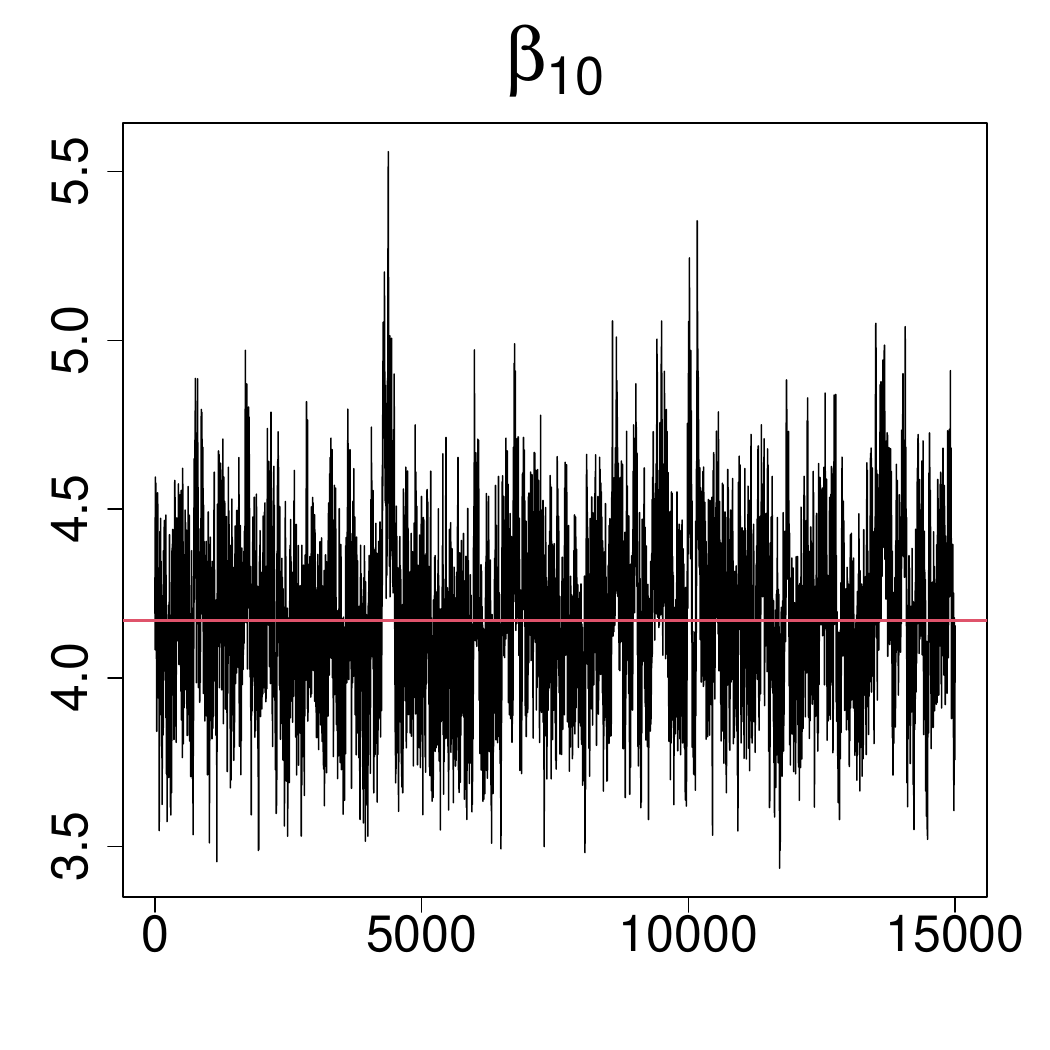} 
    &    \includegraphics[width=0.3 \textwidth]{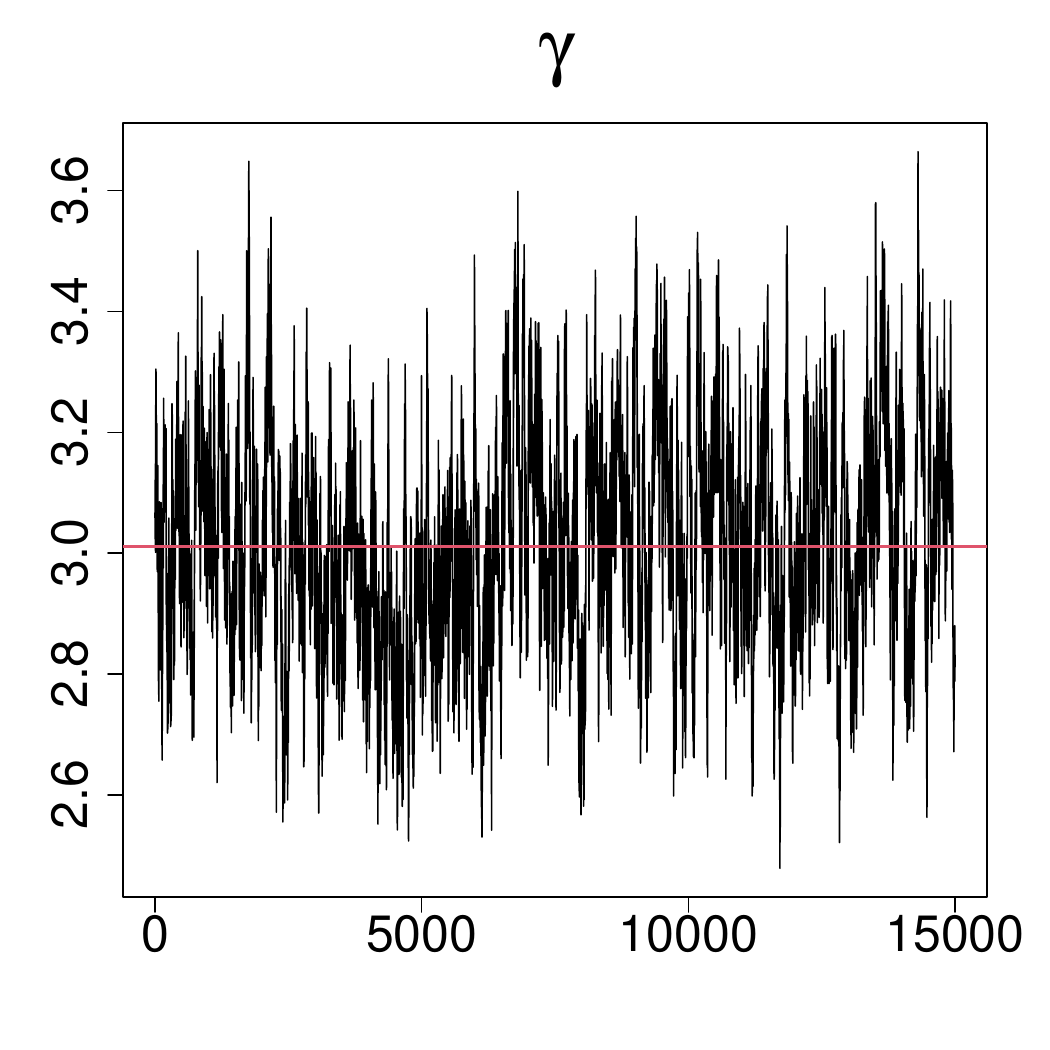} \\
    \includegraphics[width=0.3 \textwidth]{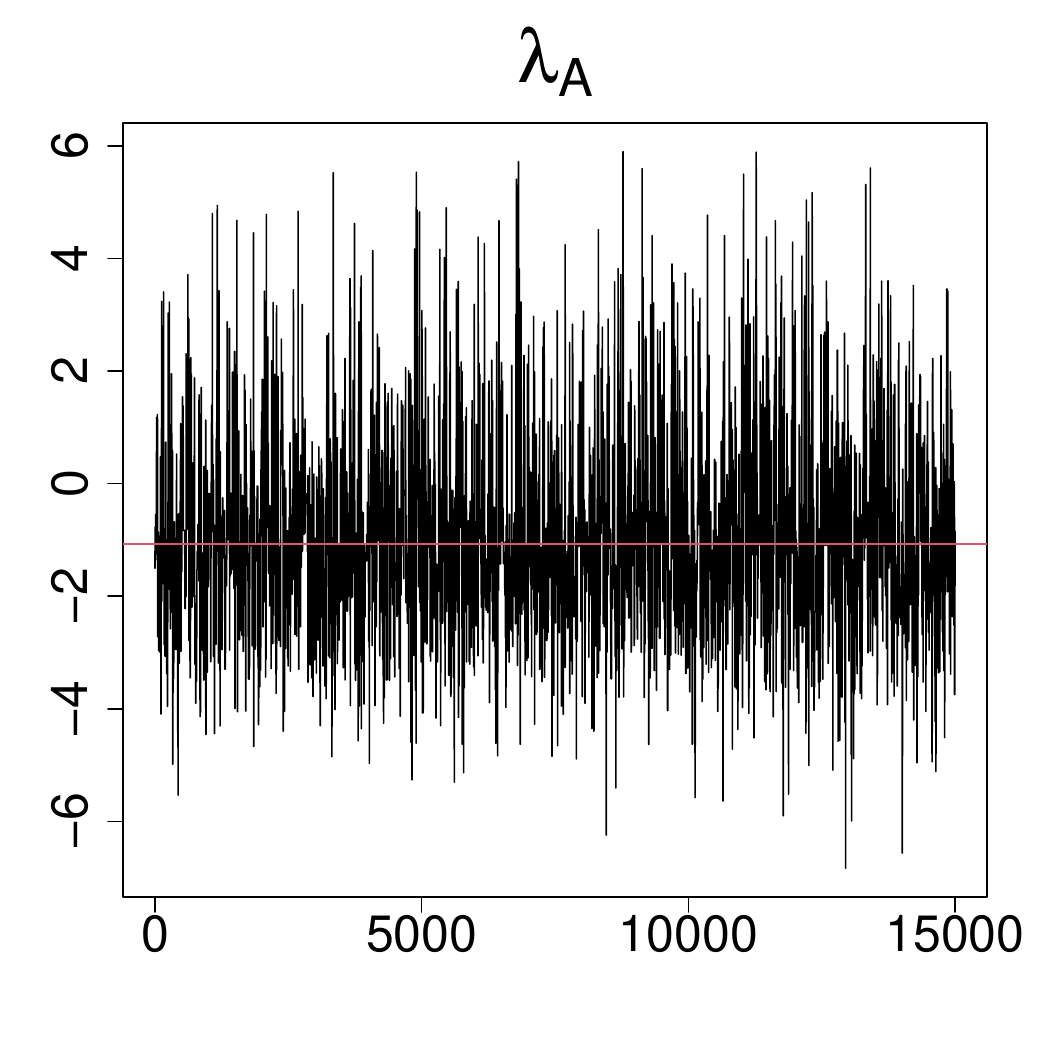}  
    &
     \includegraphics[width=0.3 \textwidth]{figure/dep_la2_poly_spike_1d.pdf} 
     & 
      \includegraphics[width=0.3 \textwidth]{figure/dep_la3_poly_spike_1d.pdf} 
        \\ 
                \end{tabular}
\caption{Trace plots of select parameters from the latent process model  with $q=2$ in Section \ref{sec:dep_spike}. 
The red-colored horizontal lines indicate posterior medians.}
        \label{fig:trace_dep_basic_spike2}
    \end{figure}

   \begin{figure}[hptb]
        \centering
        \begin{tabular}{ccc}
        \includegraphics[width=0.3 \textwidth]{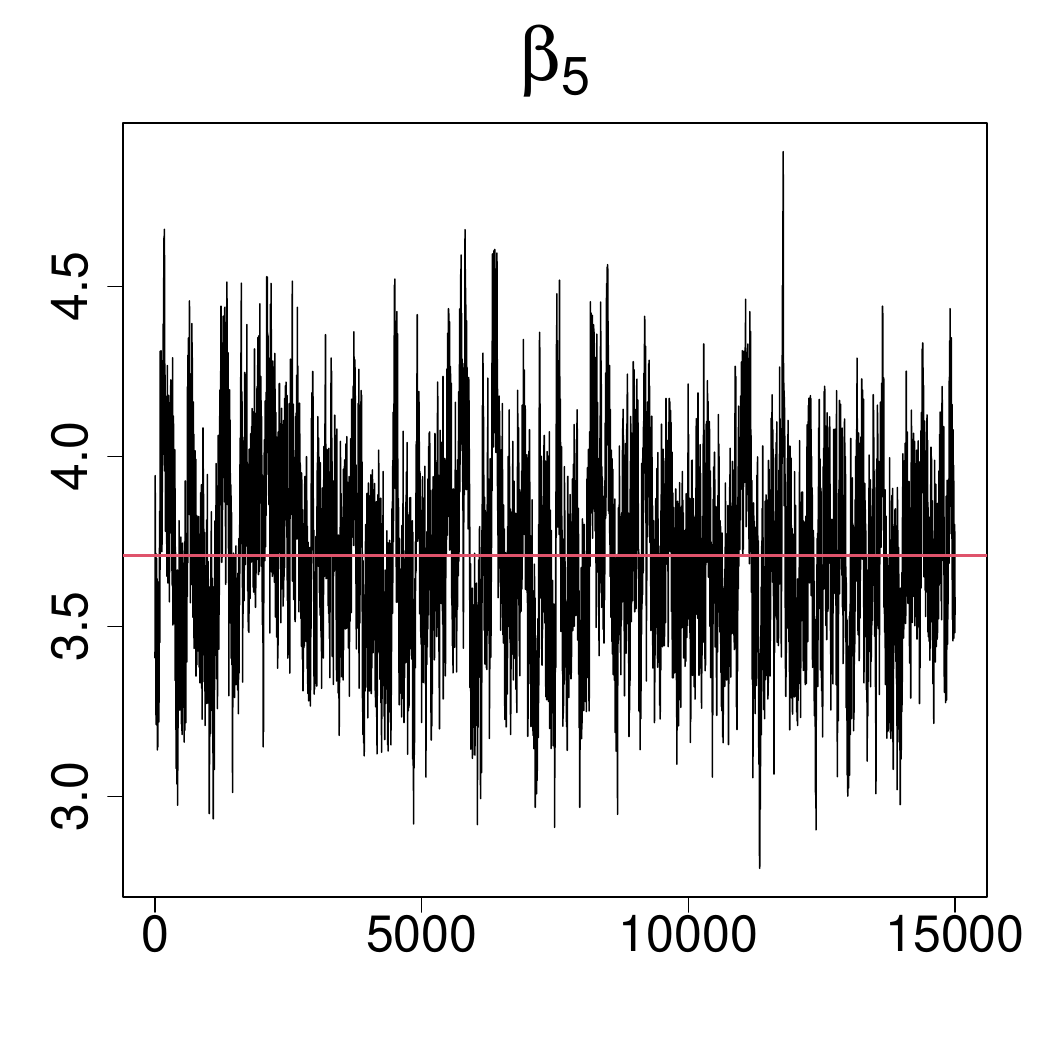} 
       &  \includegraphics[width=0.3 \textwidth]{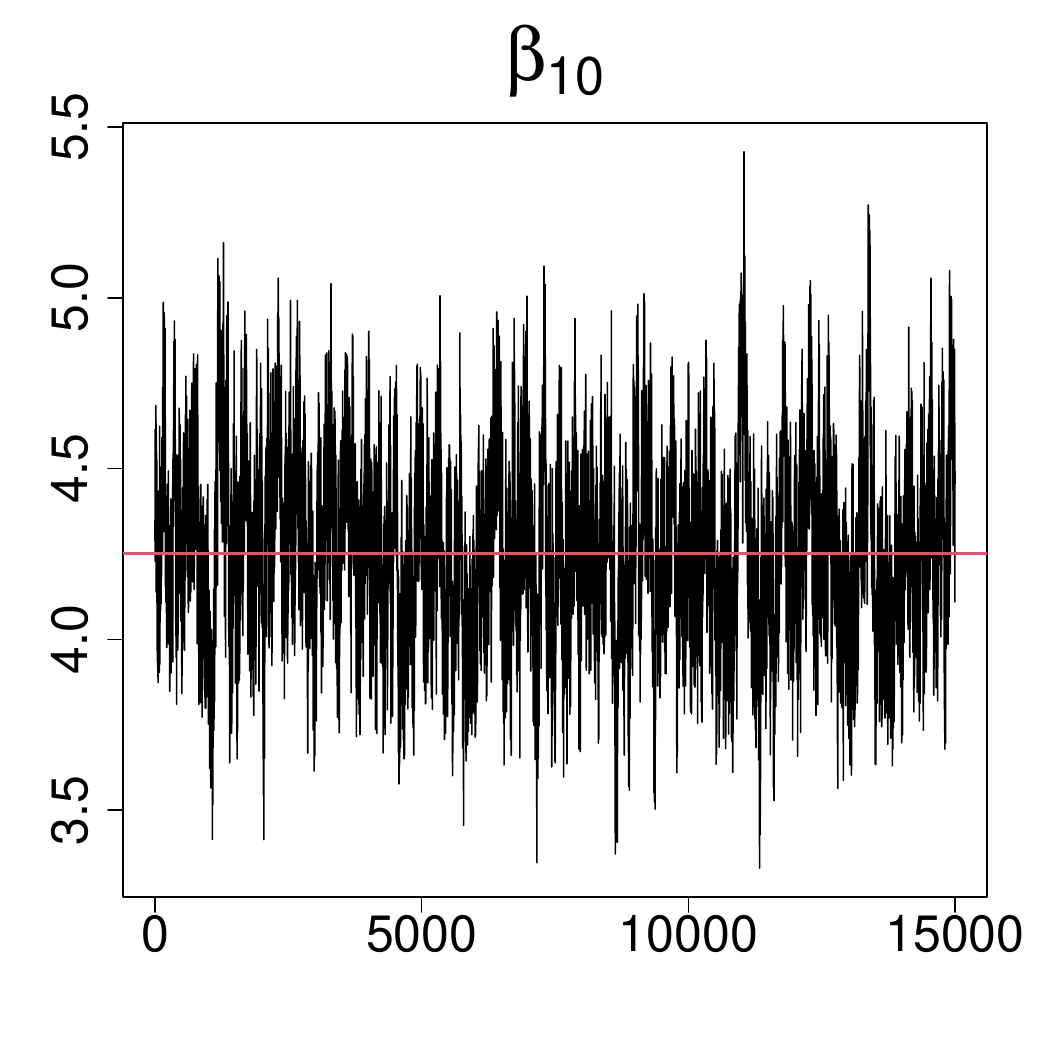} 
    &    \includegraphics[width=0.3 \textwidth]{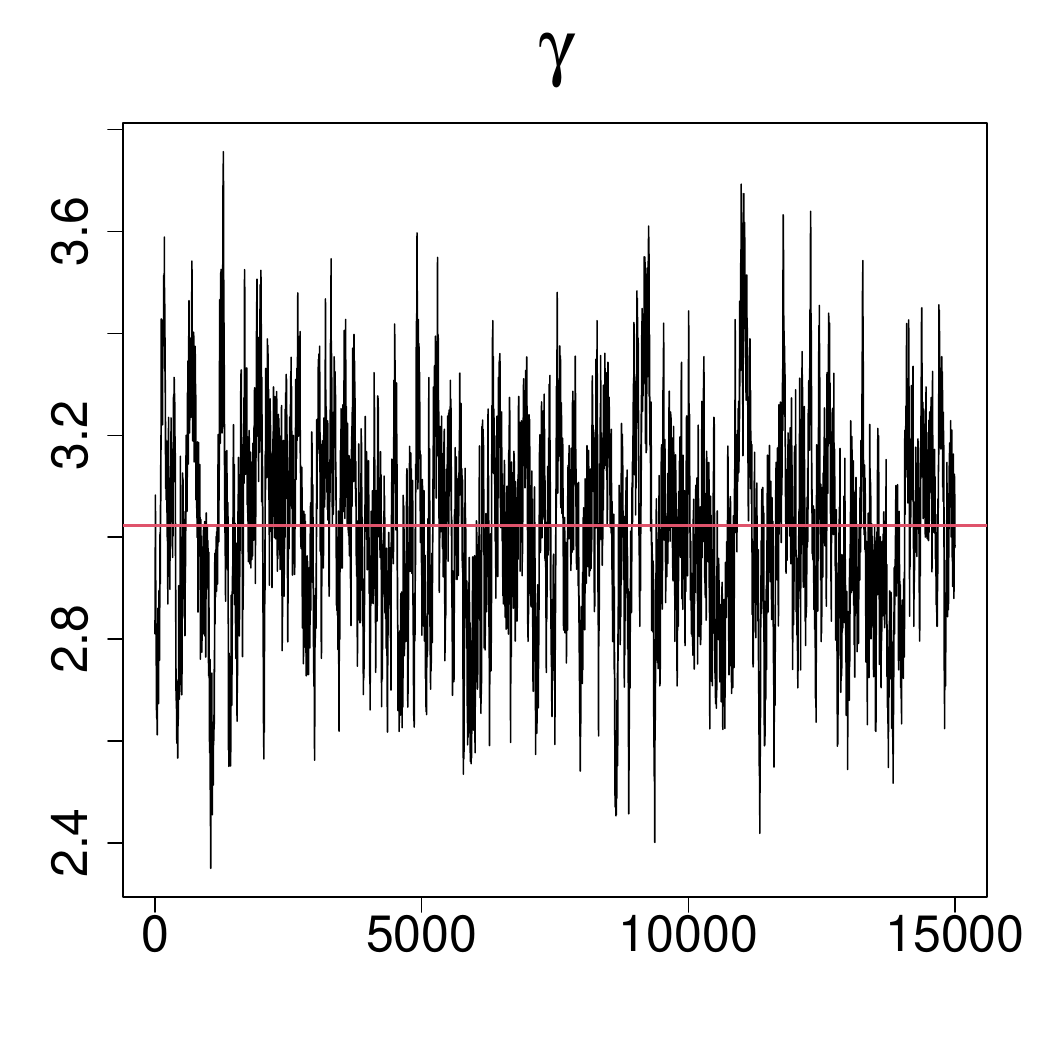} \\
    \includegraphics[width=0.3 \textwidth]{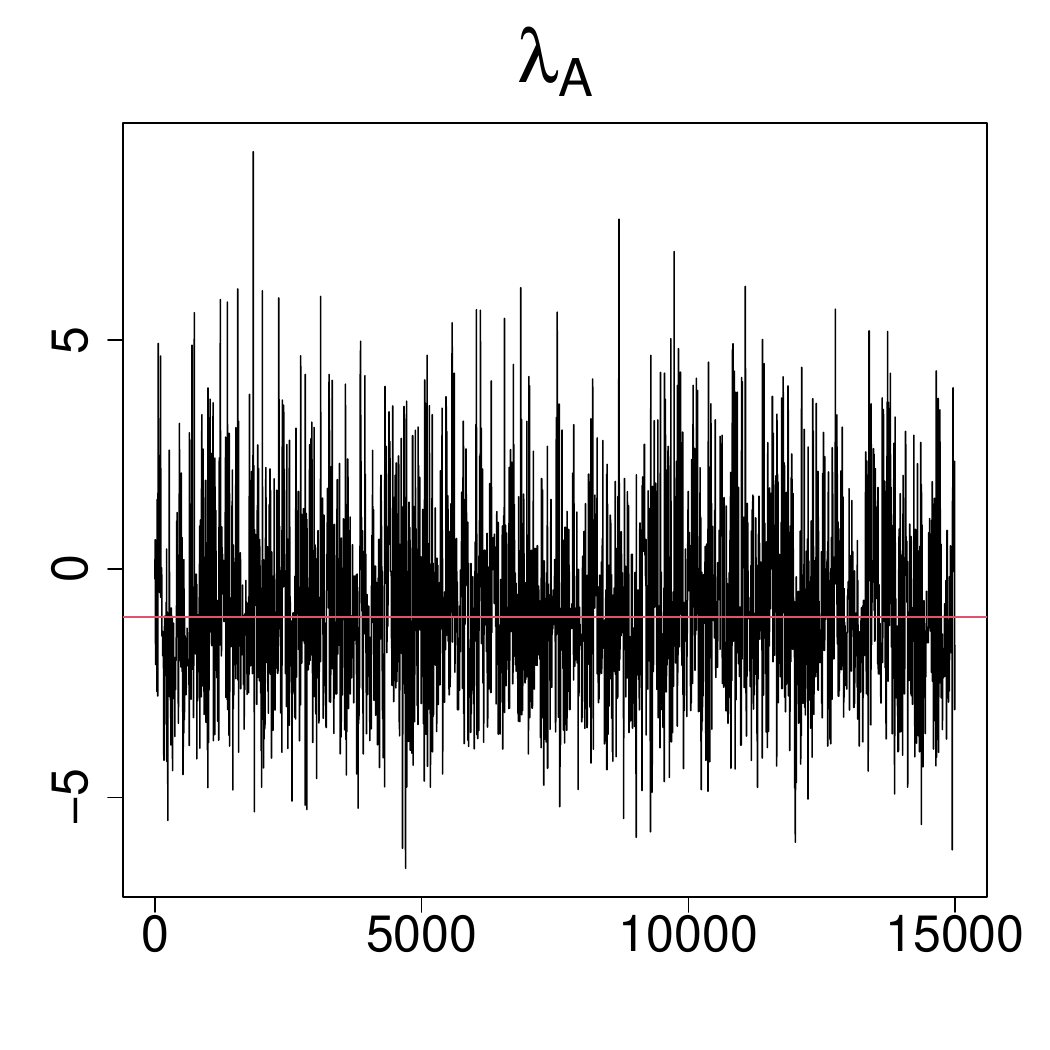}  
    &
     \includegraphics[width=0.3 \textwidth]{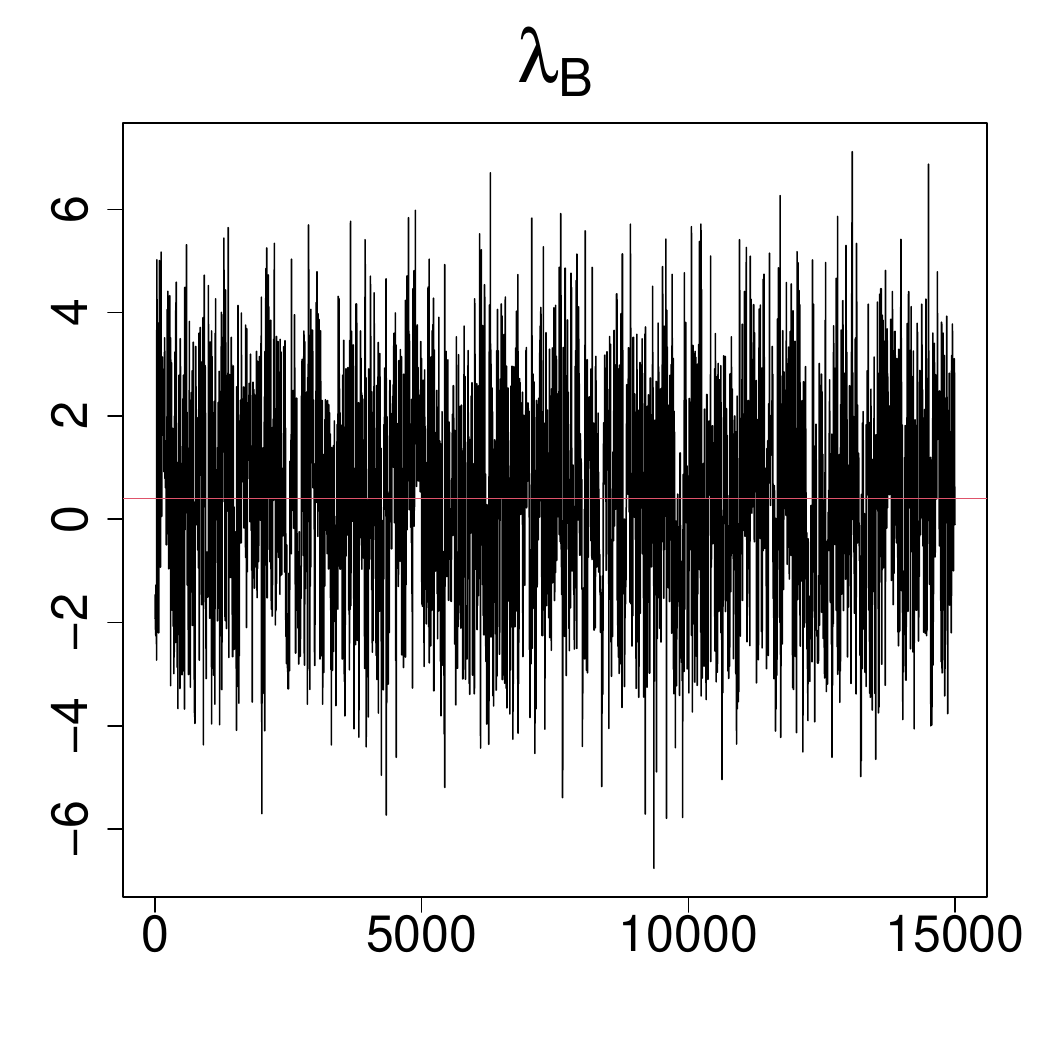} 
     & 
      \includegraphics[width=0.3 \textwidth]{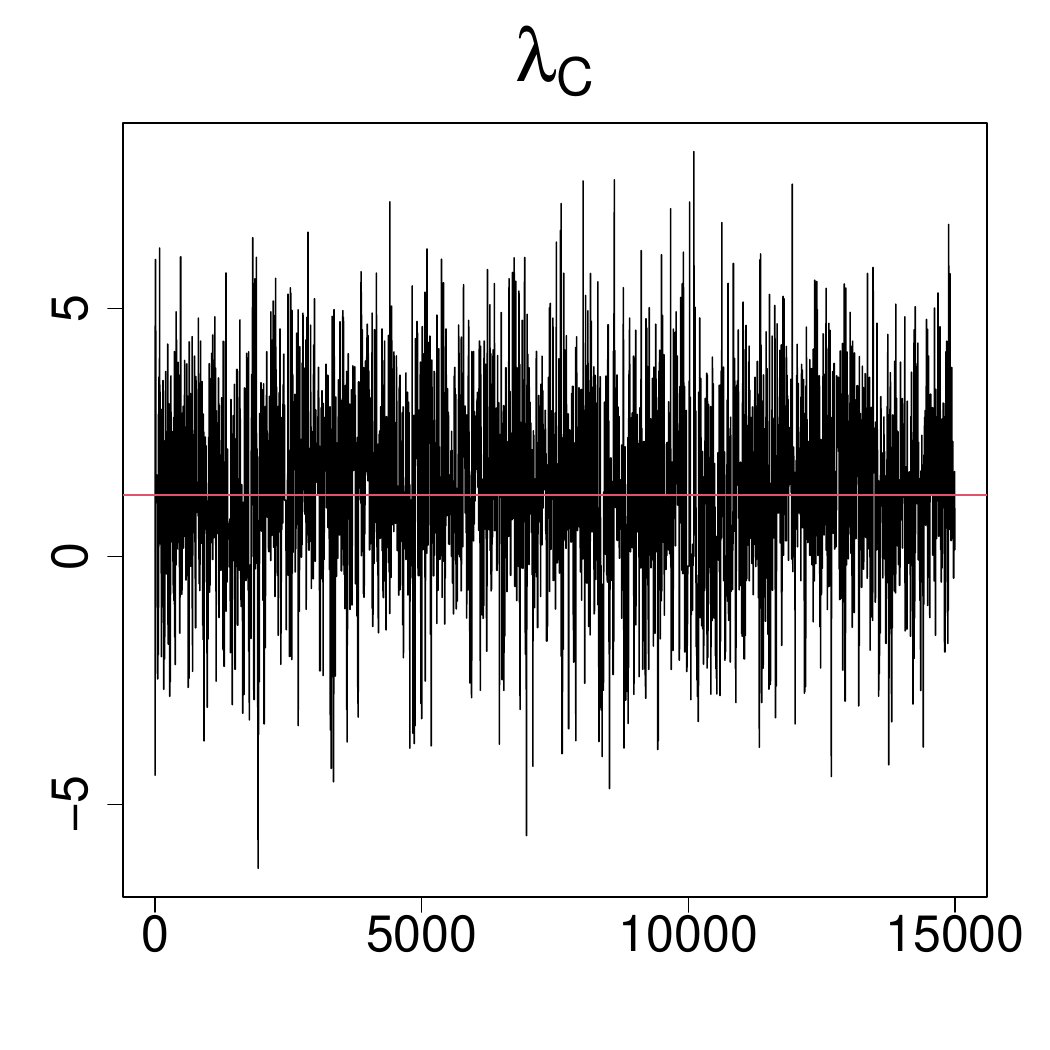} 
        \\ 
                \end{tabular}
\caption{Trace plots of select parameters from the latent process model  with $q=3$ in Section \ref{sec:dep_spike}. 
The red-colored horizontal lines indicate posterior medians.}
        \label{fig:trace_dep_basic_spike3}
    \end{figure}

   \begin{figure}[hptb]
        \centering
        \begin{tabular}{ccc}
        \includegraphics[width=0.3 \textwidth]{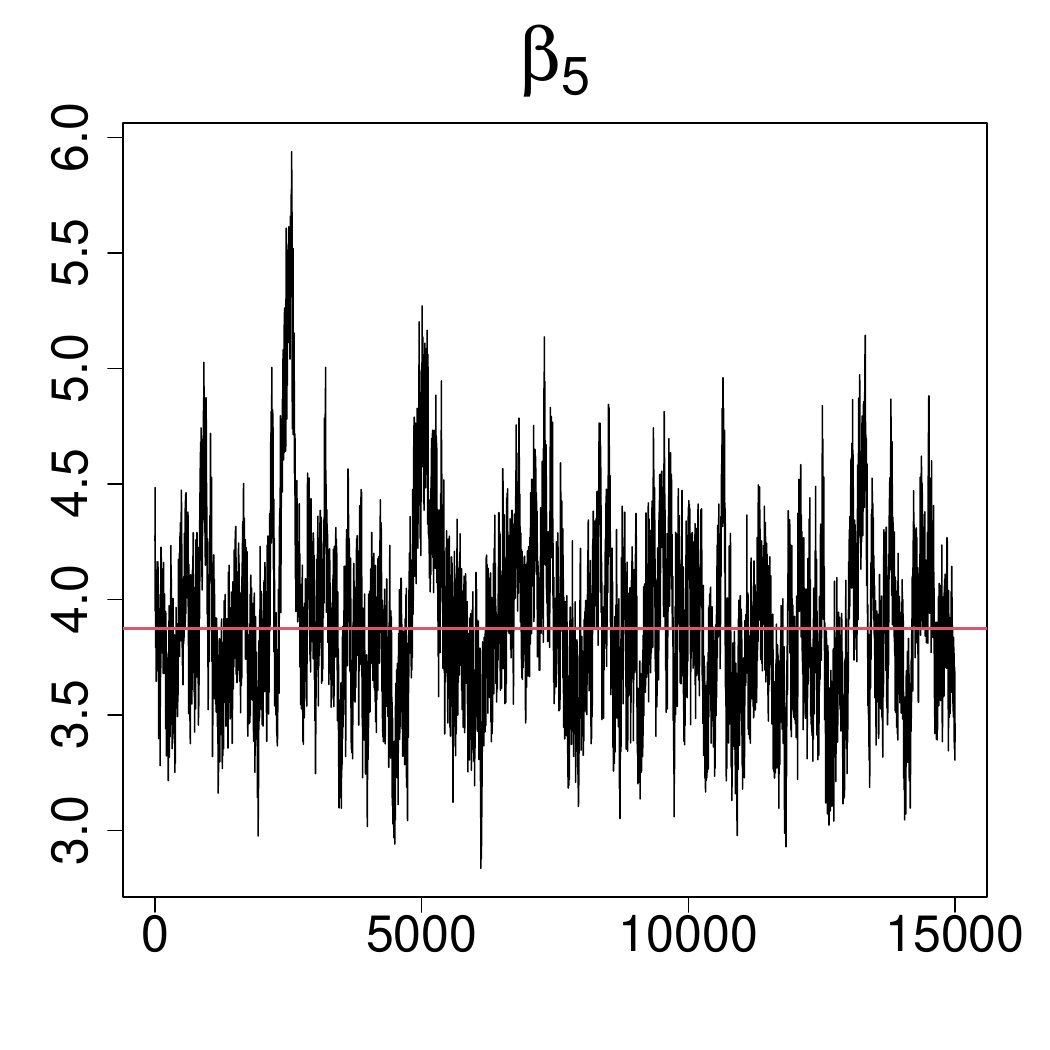} 
       &  \includegraphics[width=0.3 \textwidth]{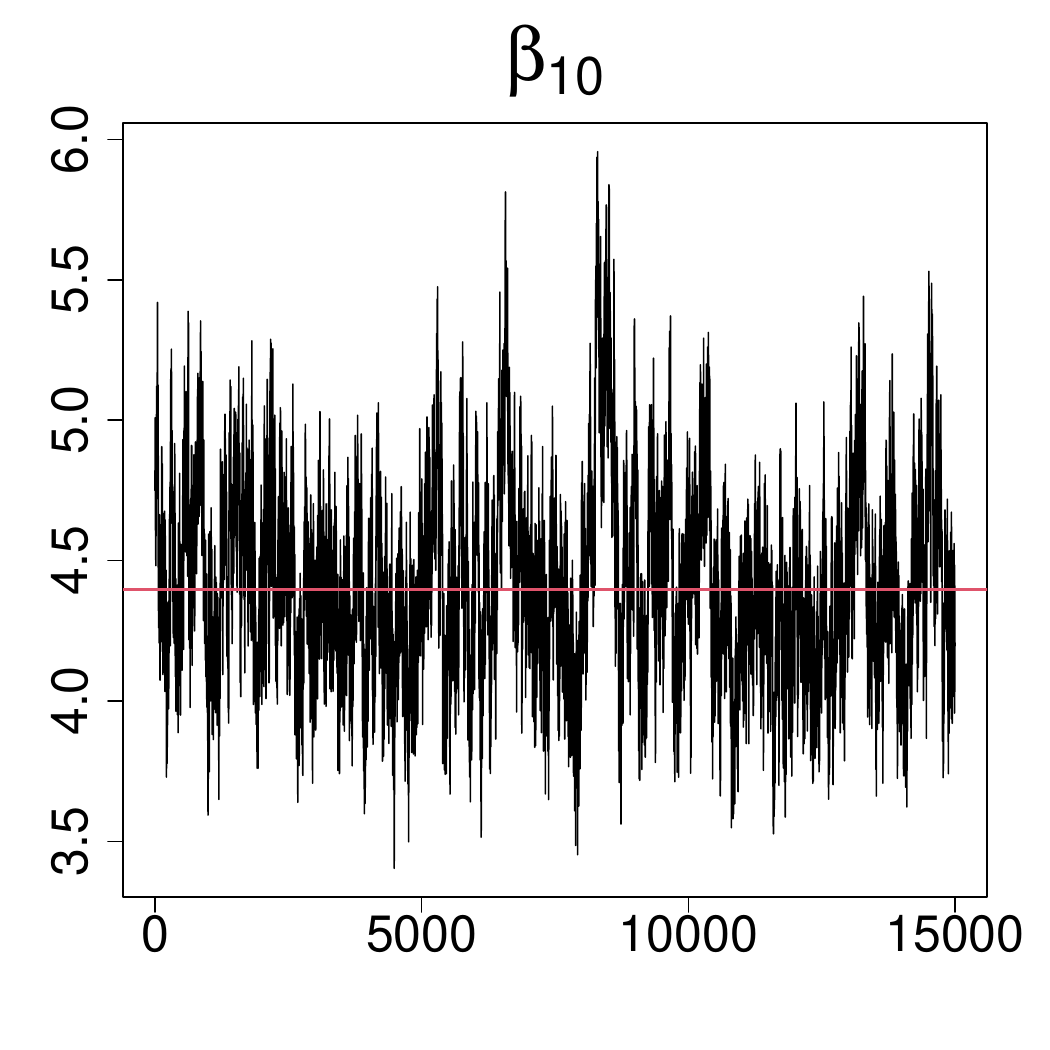} 
    &    \includegraphics[width=0.3 \textwidth]{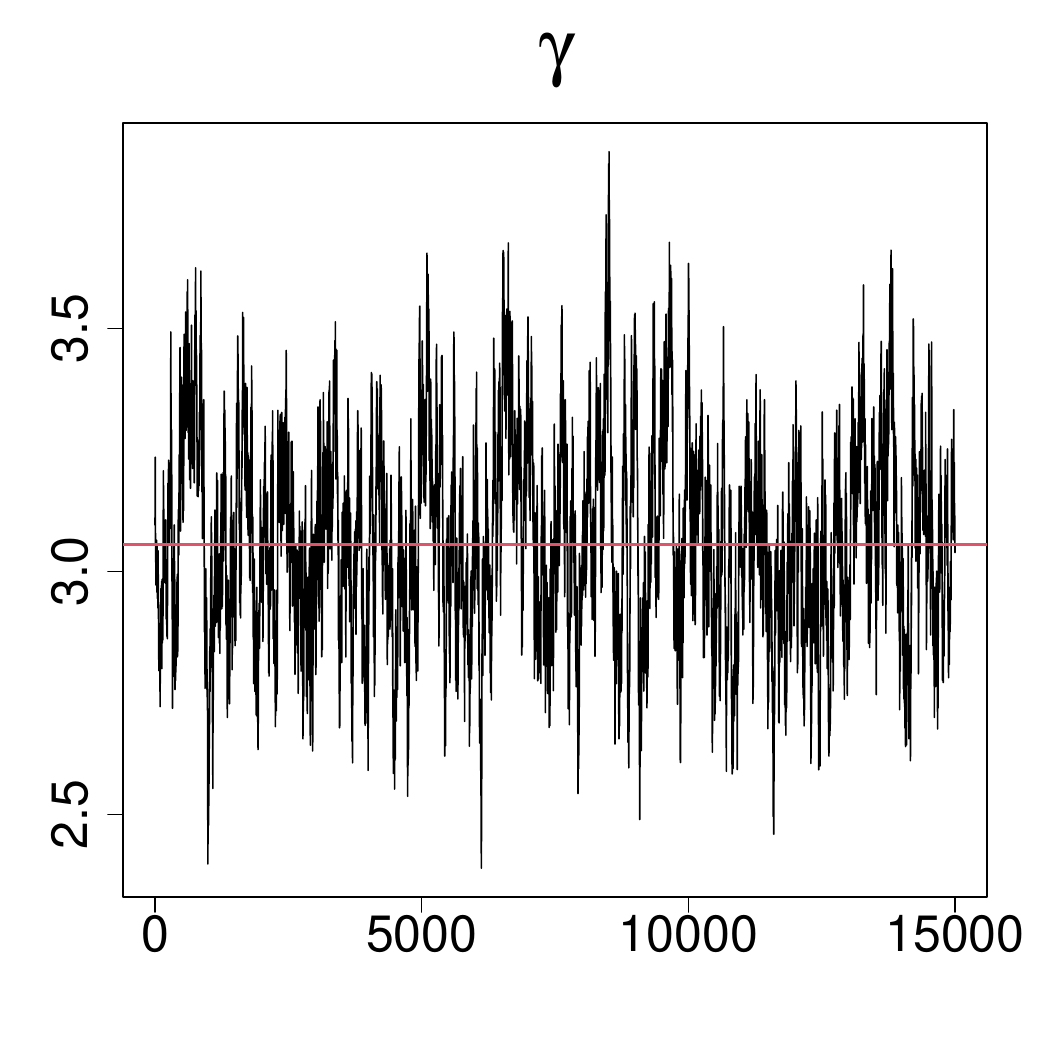} \\
    \includegraphics[width=0.3 \textwidth]{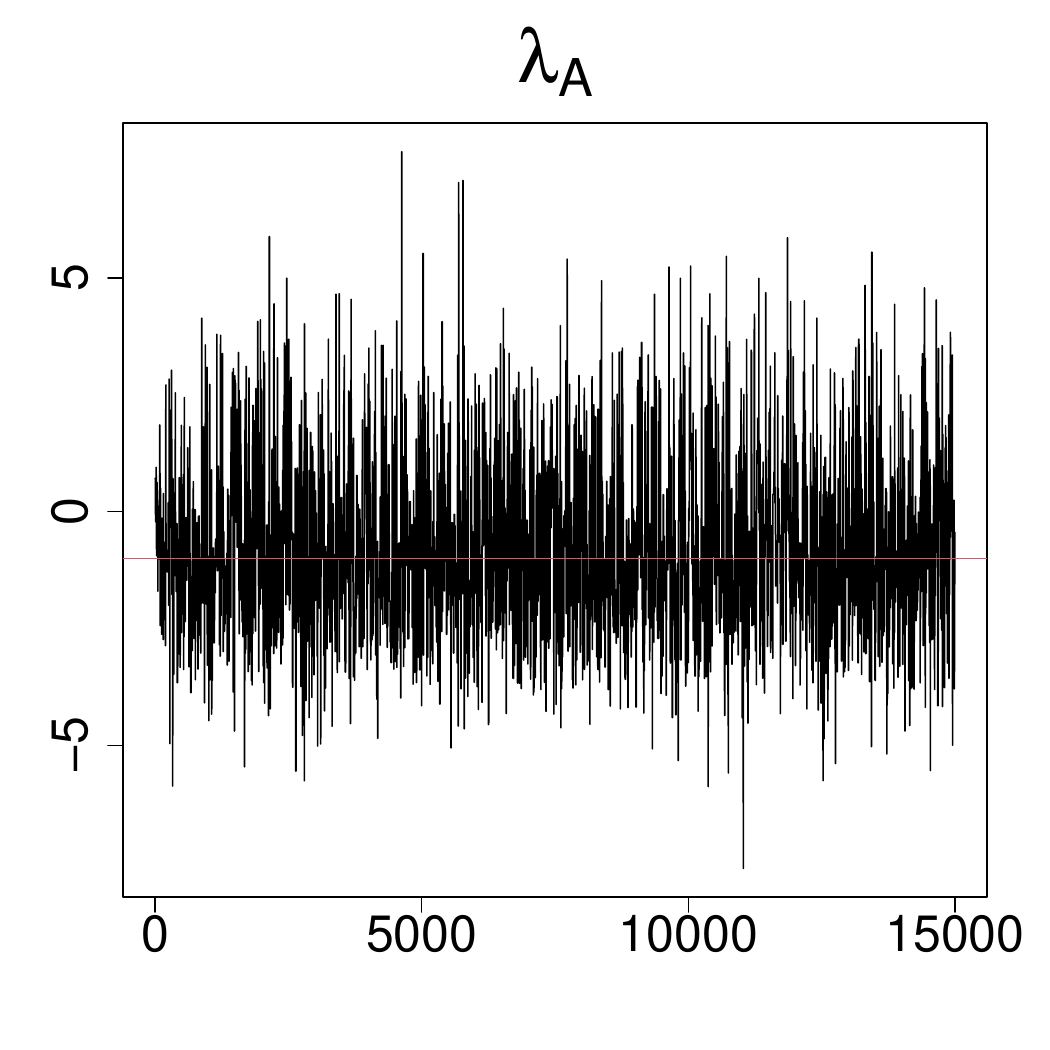}  
    &
     \includegraphics[width=0.3 \textwidth]{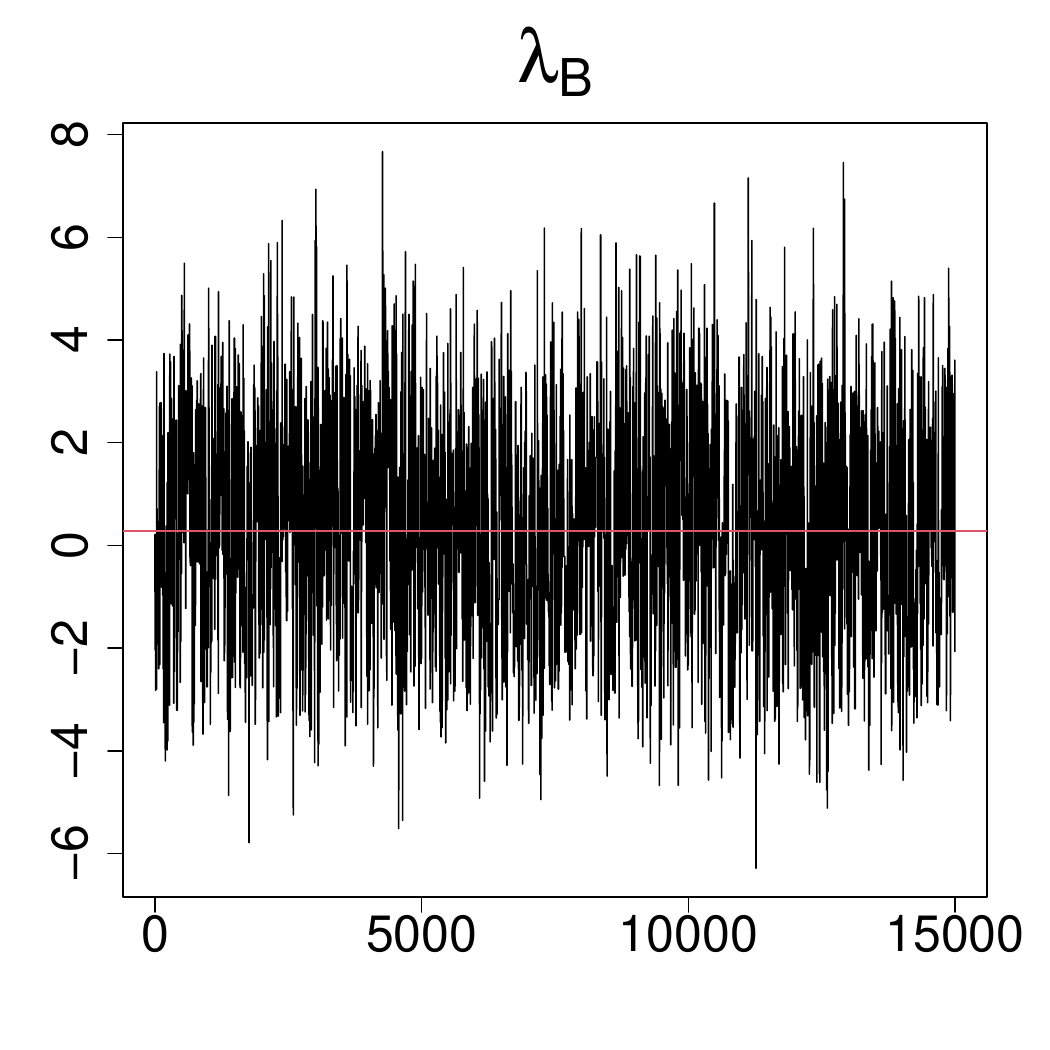} 
     & 
      \includegraphics[width=0.3 \textwidth]{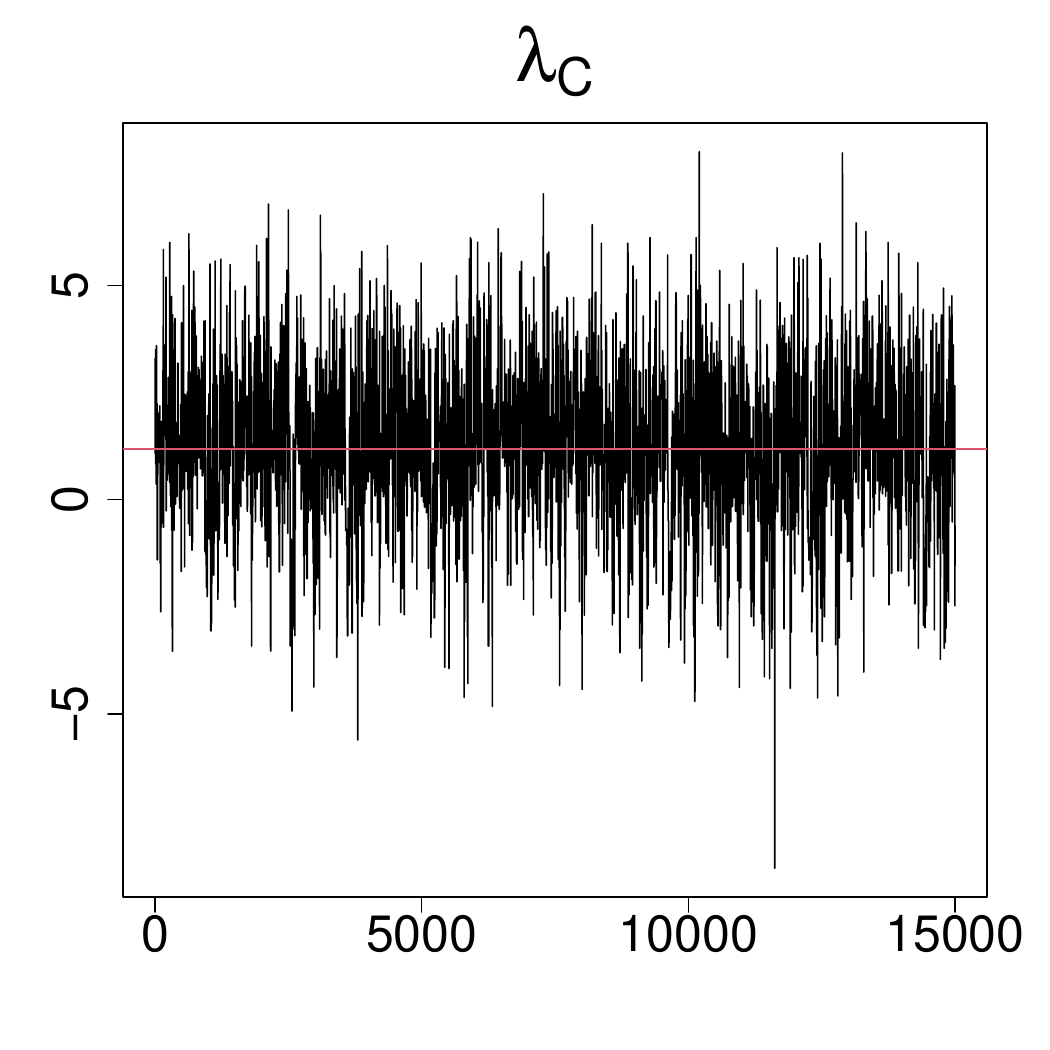} 
        \\ 
                \end{tabular}
\caption{Trace plots of select parameters from the latent process model  with $q=4$ in Section \ref{sec:dep_spike}. 
The red-colored horizontal lines indicate posterior medians.}
        \label{fig:trace_dep_basic_spike4}
    \end{figure}

\clearpage

\subsection{Multivariate Gelman-Rubin Potential Scale Reduction Factor}\label{appendix:gr}

\s

As a convergence diagnostic,
we use the multivariate Gelman-Rubin potential scale reduction factor (PSRF) stated in Equation (10) of \citet[][p.~522]{vats:18a} \citep[see also][]{vats:18b}.
The multivariate Gelman-Rubin PSRF can be viewed as a stable version of the Gelman-Rubin convergence diagnostic \citep{gelman:92} and has the additional advantage of providing a principled approach for determining whether a Markov chain did not converge.
We use the PSRF cutoff suggested by \citeauthor{vats:18a} in Example 1 on page 523 of \citet{vats:18a}.
If the PSRF exceeds the PSRF cutoff,
there is reason to believe that one or more Markov chains did not converge and more samples are required.
According to Table \ref{tab:gr_summary}, 
the multivariate Gelman-Rubin PSRF is less than the PSRF cutoff in all applications,
based on $5$ Markov chains with starting values chosen at random,
so there are no signs of non-convergence.

\begin{table}[hpbt]
\centering
\begin{tabular}{ccccccc}
\hline 
& \# parameters & \multicolumn{4} {c} {PSRF} & PSRF cutoff\\ 
&               & $q=1$ &  $q=2$ & $q=3$ & $q=4$ & \\
\hline 
Section \ref{sec:math_spike}  & 852 &  1.000227 & 1.000214 & 1.000208& 1.000208 &  1.000342 \\
Section \ref{sec:dep_spike}  & 526 &  1.000303 & 1.000268 & 1.000276 & 1.000280   & 1.000336 \\
\hline   
\end{tabular}
\caption{The multivariate Gelman-Rubin PSRF and PSRF cutoff in Sections \ref{sec:math_spike} and 
and \ref{sec:dep_spike}.
The results are based on $5$ Markov chains with starting values chosen at random.
}
\label{tab:gr_summary}
\end{table}

\clearpage 

\section{Additional Results}\label{appendix:add}


\begin{table}[tbh]
    \centering
    \begin{tabular}{c cccccc}
    \hline 
                  & Minimum & 25\% Percentile & Median & Mean & 75\% Percentile & Maximum \\ 
    \hline 
    $q=1$:  &&&&&& \\
       $\lambda$  &   0.133&  0.330&  0.477&  0.472&  0.606&  0.886 \\
       $\alpha$ &   -0.985& -0.593& -0.237& -0.039&  0.339&  1.911  \\
       $\beta$  &   -0.703&  0.637&  1.327&  1.123&  1.721&  2.529 \\
       \hline 
     $q=2$: &&&&&& \\
       $\lambda$  &  0.099& 0.290& 0.456& 0.448& 0.608& 0.881  \\
       $\alpha$  &  -1.042& -0.584& -0.229& -0.033&  0.362&  1.847 \\
       $\beta$  &  -0.263&  1.152&  1.810&  1.613&  2.215&  3.038  \\
       \hline   
     $q=3$: &&&&&& \\
       $\lambda$  &  0.095& 0.276& 0.444& 0.442& 0.602& 0.873  \\
       $\alpha$  &   -1.095& -0.577& -0.213& -0.032&  0.348&  1.857  \\
       $\beta$  &  -0.022&  1.474&  2.045&  1.876&  2.484&  3.294 \\
       \hline     
     $q=4$: &&&&&& \\
       $\lambda$  &  0.099&  0.298&  0.453&  0.449&  0.593&  0.878  \\
       $\alpha$ &   -1.024& -0.570& -0.231& -0.0362&  0.309&  1.834   \\
       $\beta$  &  0.071& 1.633& 2.171& 2.029& 2.694& 3.466  \\
       \hline 
    \end{tabular}
    \caption{Summary of the posterior medians of $\lambda$, $\theta$, and $\beta$}  from the latent process model for Application: online educational assessments in Section \ref{sec:math_spike}. Pearson correlations between the posterior medians of $\lambda$ across models with different latent space dimensions ranged from .990 to .994 and Spearman rank-order correlations ranged from .991 to .995. 
    \label{tab:math_parameter1}
\end{table}

\begin{table}[tbh]
    \centering
    \begin{tabular}{c ccc}
    \hline 
                  & Median & 2.5 Percentile & 97.5 Percentile  \\ 
    \hline 
    $q=1$: &&& \\
       $\gamma$  &  2.284& 1.978& 2.605   \\
       $\sigma_\alpha$ & 1.054& 0.962& 1.156   \\
       \hline 
     $q=2$: &&& \\
       $\gamma$  &   2.656& 2.303& 3.003  \\
       $\sigma_\alpha$ &  1.038& 0.948& 1.138  \\
       \hline   
      $q=3$: &&& \\
       $\gamma$  &     2.777& 2.429& 3.169 \\
       $\sigma_\alpha$ &  1.035& 0.943& 1.134 \\
       \hline     
     $q=4$: &&& \\
       $\gamma$  &   2.837& 2.470& 3.234  \\
       $\sigma_\alpha$ & 1.031& 0.941& 1.131   \\
       \hline 
    \end{tabular}
    \caption{Summary (posterior median and 95\% credible intervals) of $\gamma$ and $\sigma_\alpha$  from the latent process model for Application: online educational assessments in Section \ref{sec:math_spike}}
    \label{tab:math_parameter2}
\end{table}

     

\clearpage


\begin{table}[htb]
    \centering
    \begin{tabular}{c cccccc}
    \hline 
                  & Minimum & 25\% Percentile & Median & Mean & 75\% Percentile & Maximum \\ 
    \hline 
    $q=1$:  &&&&&& \\
       $\lambda$  &  0.069& 0.295& 0.554& 0.506& 0.701& 0.918 \\
       $\alpha$ &   -1.050& -0.504& -0.104&  0.002&  0.455&  1.075  \\
       $\beta$  &  1.894 &  2.360&   3.066&   2.983&   3.606&   3.939  \\
       \hline 
      $q=2$: &&&&&& \\
       $\lambda$  &  0.062&  0.245&  0.569&  0.495&  0.709&  0.914  \\
       $\alpha$  &  -1.213& -0.523&  0.015&  0.005&  0.553&  1.131 \\
       $\beta$  &  2.283&   2.766 &  3.493&   3.410&   4.023&   4.395   \\
       \hline   
      $q=3$: &&&&&& \\
       $\lambda$  &  0.062& 0.275& 0.565& 0.502& 0.704& 0.907  \\
       $\alpha$  &   -1.274& -0.589&  0.023&  0.001&  0.594&  1.197   \\
       $\beta$  &   2.266&   2.830&   3.568&   3.499&   4.142&   4.556  \\
       \hline     
     $q=4$: &&&&&& \\
       $\lambda$  &  0.069&  0.304&  0.552&  0.511&  0.706&  0.905  \\
       $\alpha$ &  -1.321& -0.641&  0.024& -0.005&  0.607&  1.204   \\
       $\beta$  & 2.256&   3.138&   3.801&   3.715&   4.390&   4.797  \\
       \hline 
    \end{tabular}
    \caption{Summary of the posterior medians of $\lambda$, $\theta$, and $\beta$}  from the latent process model for Application: mental health in Section \ref{sec:dep_spike}. Pearson correlations between the posterior medians of $\lambda$ across models with different latent space dimensions ranged from .988 to .996 and Spearman rank-order correlations ranged from .988 to .993. 
    \label{tab:dep_parameter1}
\end{table}

\begin{table}[tbh]
    \centering
    \begin{tabular}{c ccc}
    \hline 
                  & Median & 2.5 Percentile & 97.5 Percentile  \\ 
    \hline 
    $q=1$: &&& \\
       $\gamma$  & 2.791& 2.456& 3.109   \\
       $\sigma_\alpha$ & 1.018& 0.903& 1.147  \\
       \hline 
      $q=2$: &&& \\
       $\gamma$  &   3.010& 2.683& 3.385  \\
       $\sigma_\alpha$ & 1.031& 0.916& 1.162  \\
       \hline   
       $q=3$: &&& \\
       $\gamma$  &     3.023& 2.646& 3.407  \\
       $\sigma_\alpha$ &  1.051& 0.935& 1.181  \\
       \hline     
     $q=4$: &&& \\
       $\gamma$  &   3.055& 2.688& 3.477  \\
       $\sigma_\alpha$ & 1.065& 0.948& 1.195   \\
       \hline 
    \end{tabular}
    \caption{Summary (posterior median and 95\% credible intervals) of $\gamma$ and $\sigma_\alpha$  from the latent process model for  Application: mental health in Section \ref{sec:dep_spike}.  }
    \label{tab:dep_parameter2}
\end{table}

\begin{figure}[htbp]
    \centering
    \begin{tabular}{cccc}
  G1 &    G2 &   G3 \\
 \includegraphics[width=0.3 \textwidth]{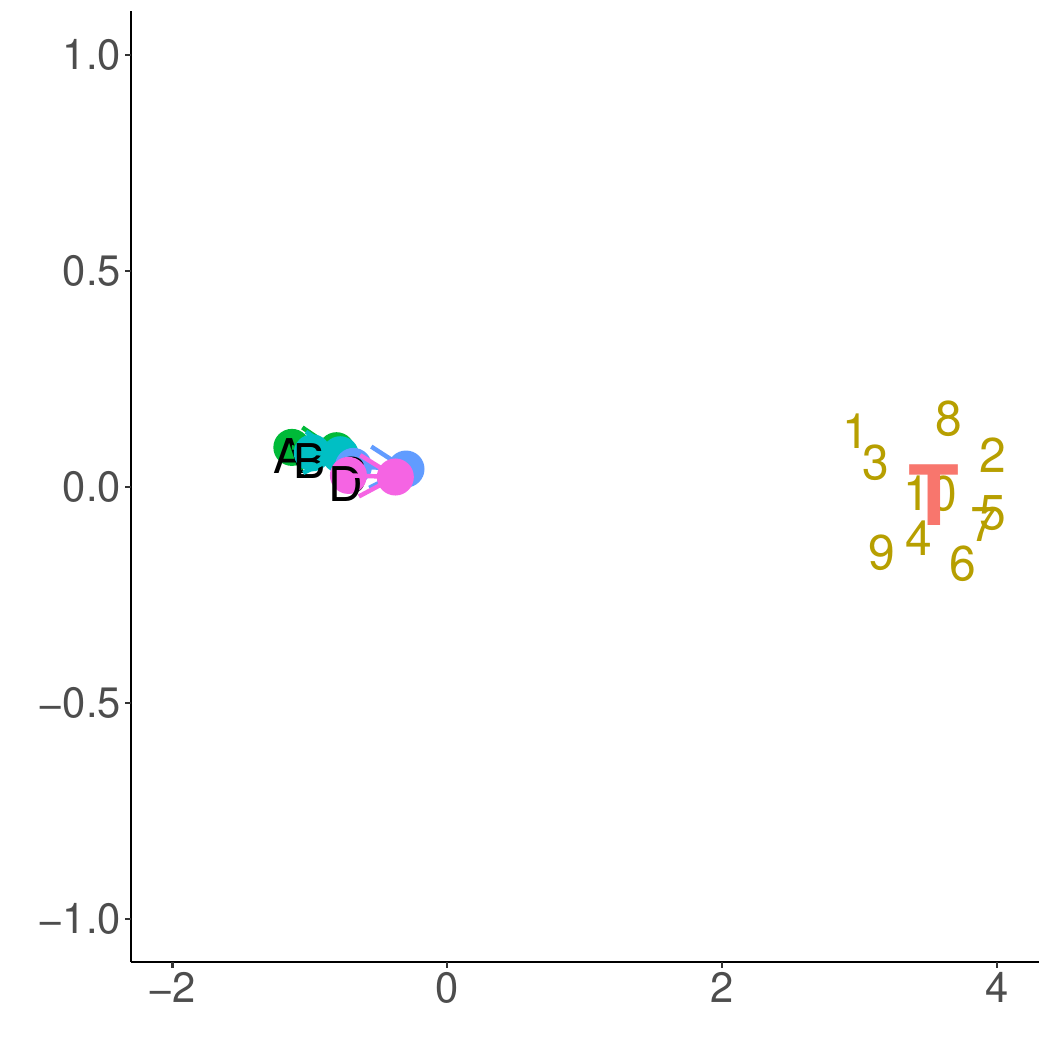}
  &   \includegraphics[width=0.3 \textwidth]{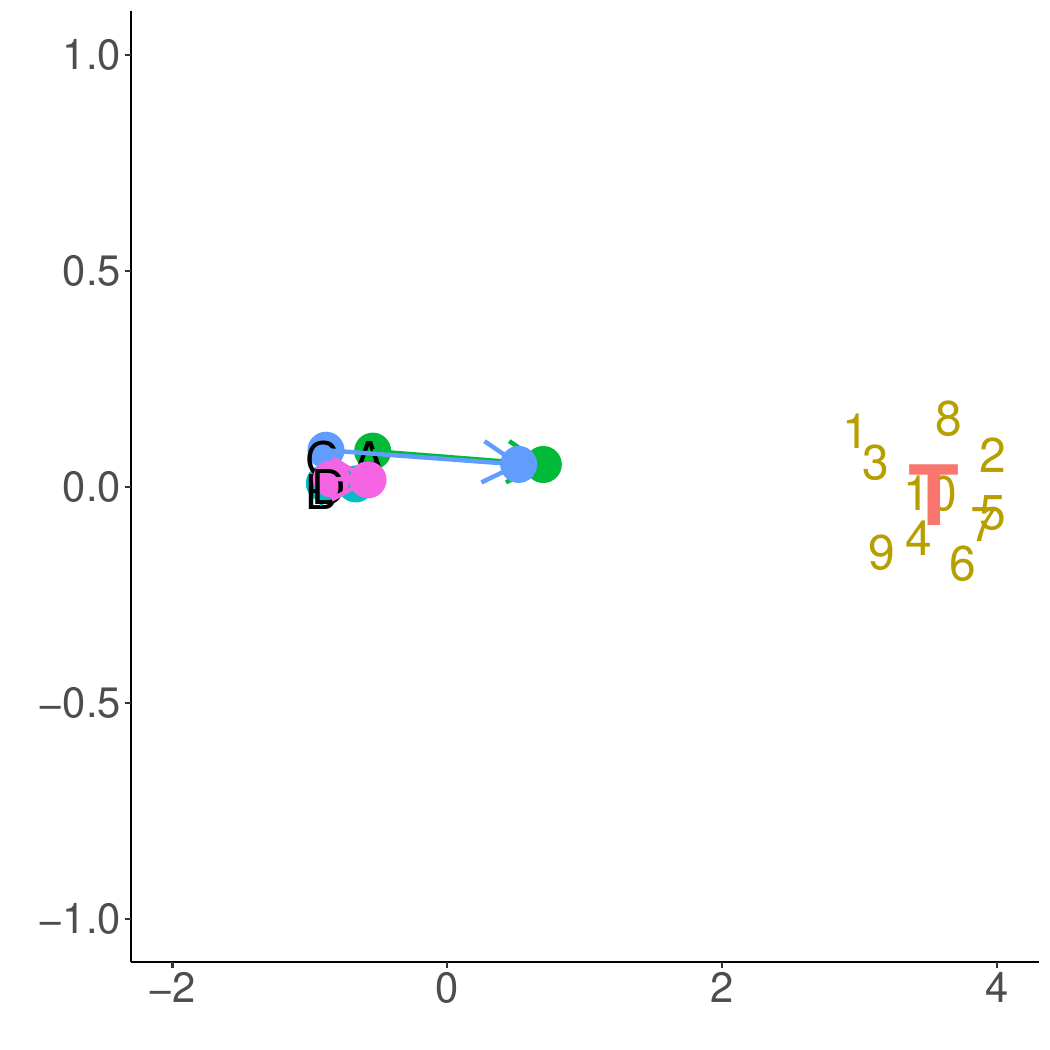} 
  &  \includegraphics[width=0.3 \textwidth]{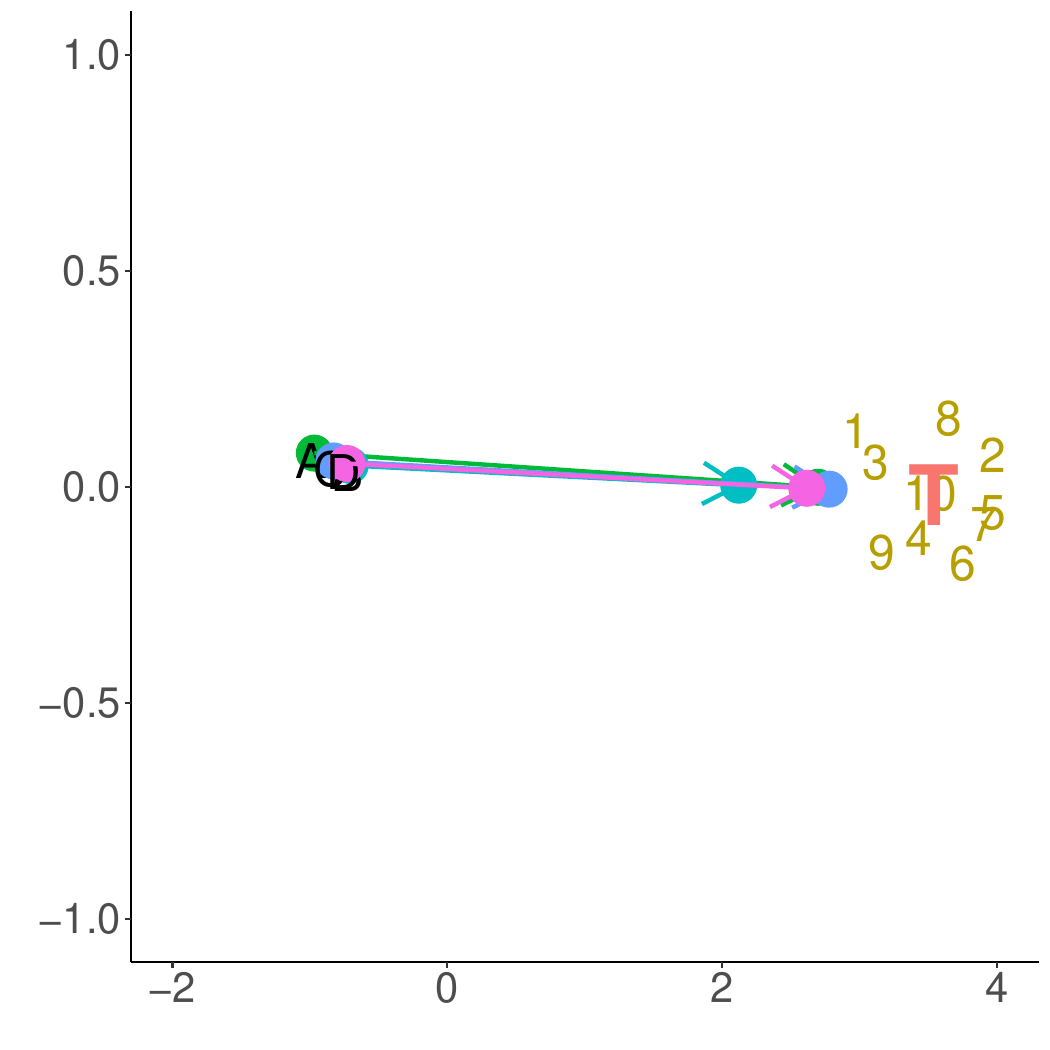}  \\
       \end{tabular}
    \caption{Illustrations of interaction maps for Simulation Condition 1 with $n=300$. One simulated data set was randomly selected. G1, G2, and G3 indicate the three respondent groups considered in the simulation condition with different progress rates (trivial, medium, and large progress, respectively). Four people are randomly selected from each group for illustration purposes, but the patterns are generally similar across all respondents in each group, as supported by the summary results reported in the simulation study section of the manuscript. }
    \label{fig:sim1}
\end{figure}

\begin{figure}[htbp]
    \centering
    \begin{tabular}{ccc}   
    G1 & G2 &\\
     \includegraphics[width=0.3 \textwidth]{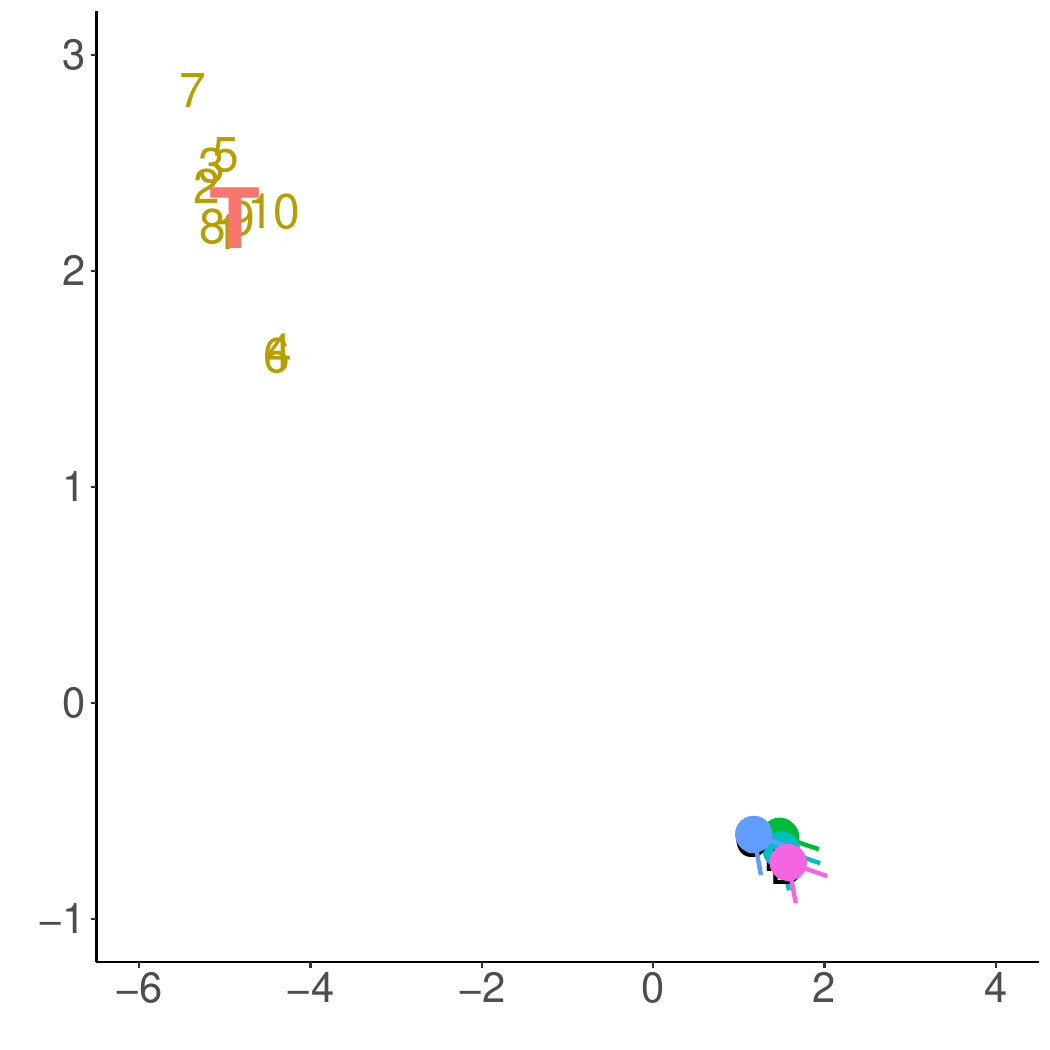} &
     \includegraphics[width=0.3 \textwidth]{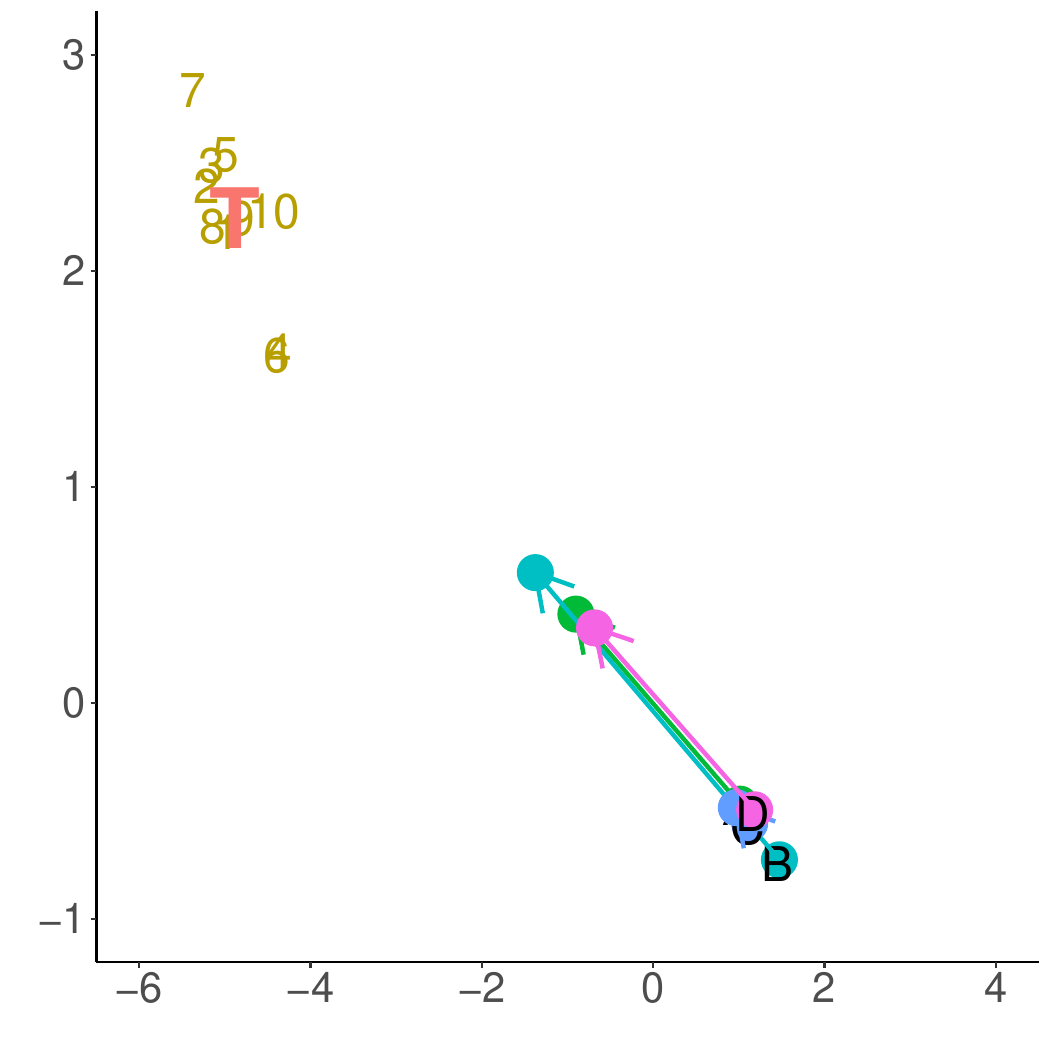} &\\
   G3 & G4 &\\
     \includegraphics[width=0.3 \textwidth]{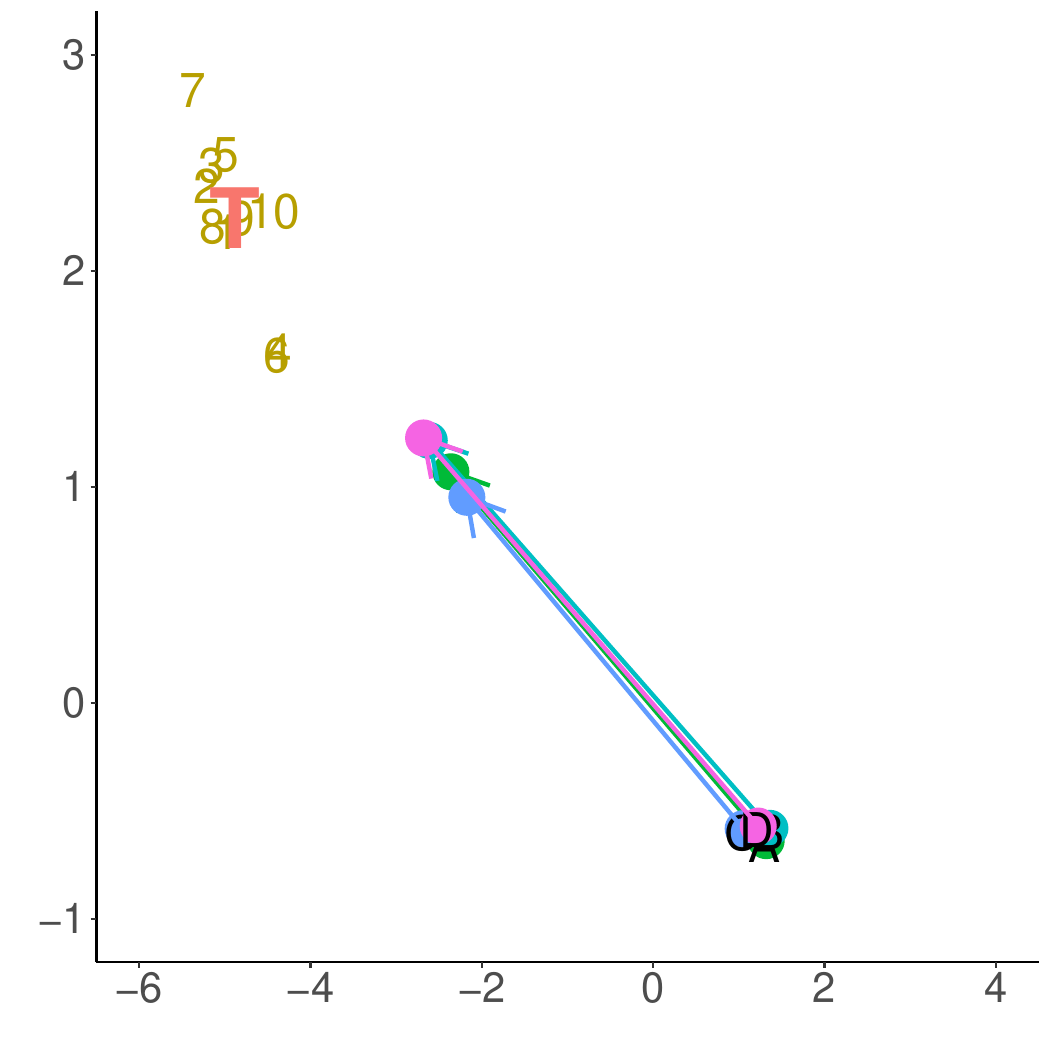} &
     \includegraphics[width=0.3 \textwidth]{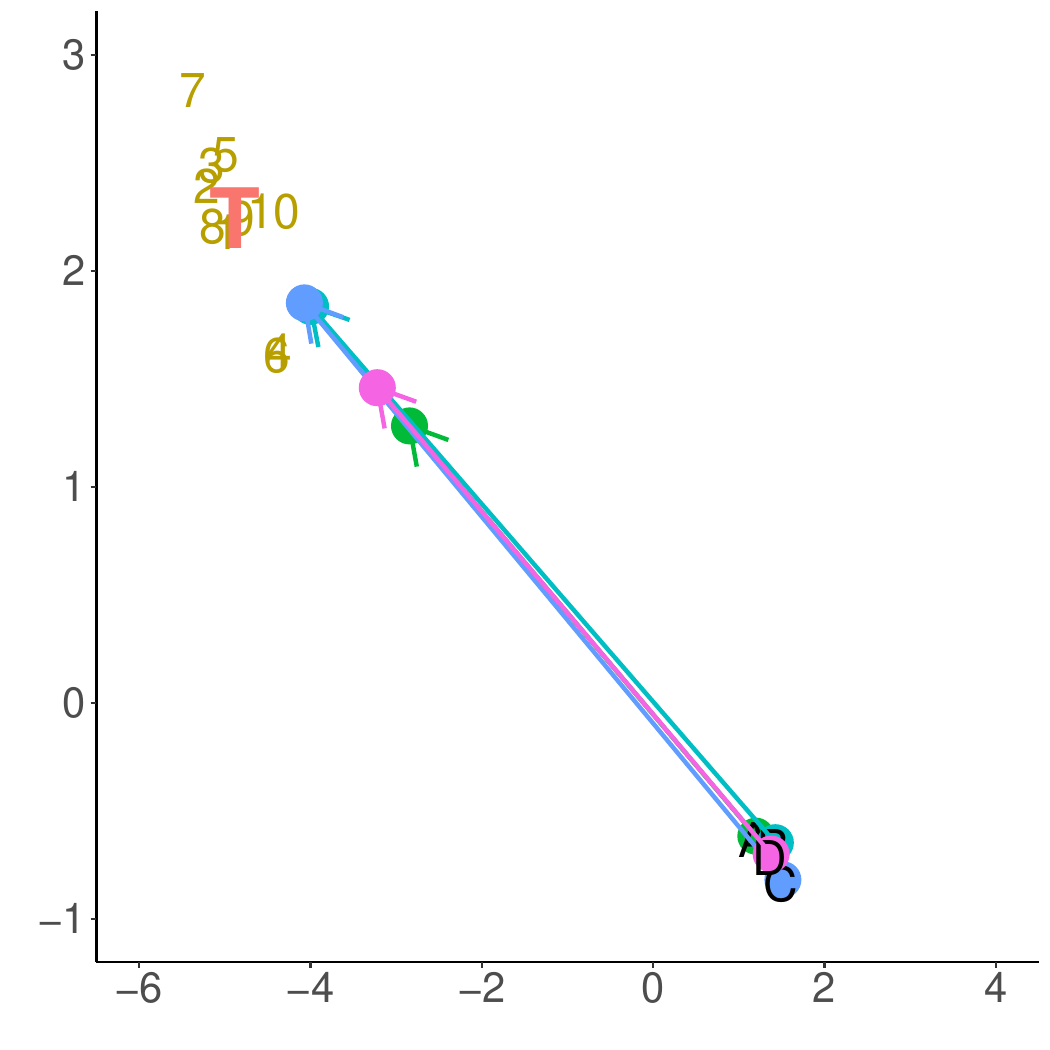} &\\
       \end{tabular}
    \caption{Illustrations of interaction maps for Simulation Condition 2 with $n=600$. One simulated data set was randomly selected in each condition. G1, G2, G3, and G4 indicate the four groups considered in the simulation condition with different progress rates from trivial to large. Four people are randomly selected from each group for illustration purposes, but the patterns are generally similar across all respondents in each group, as supported by the summary results reported in the simulation study section of the manuscript. }
    \label{fig:sim2}
\end{figure}

\end{appendix}

\clearpage 

\end{document}